\documentclass[a4paper,twoside,12pt]{book}
\usepackage[english]{babel}
\usepackage[latin1]{inputenc}
\usepackage{fancyhdr}   
\usepackage{latexsym}
\usepackage{amsmath}
\usepackage{graphicx}
\usepackage{subfigure}
\usepackage{setspace}
\usepackage{amssymb}    
\usepackage{feynmf}     
\usepackage{url}        
\usepackage{multirow}
\usepackage{textpos}
\usepackage{bm}
\newcommand{\beq}{\begin{equation}}
\newcommand{\eeq}{\end{equation}}
\newcommand{\be}{\begin{equation}}
\newcommand{\ee}{\end{equation}}
\newcommand{\bem}{\begin{displaymath}}
\newcommand{\eem}{\end{displaymath}}
\newcommand{\bey}{\begin{eqnarray}}
\newcommand{\eey}{\end{eqnarray}}

\newcommand{\p}{\partial}
\newcommand{\no}{\nonumber}
\newcommand{\bef}{\begin{fmfgraph*}}
\newcommand{\eef}{\end{fmfgraph*}}

\newcommand{\xb}{x_{Bj}}

\newcommand{\spettN}{P_N^A(k,E)}

\newcommand{\sez}{\frac{d^{2}\sigma(q,\nu)}{d{\Omega}_2\,d{\nu}}}

\makeatletter
\def\cleardoublepage{\clearpage\if@twoside \ifodd\c@page\else
\thispagestyle{empty}
\newpage
\if@twocolumn\hbox{}\newpage\fi\fi\fi}
\makeatother
\begin{document}
    \pagestyle{empty}




    \null \vspace{-1.1cm}

    \begin{figure}[h]

      \begin{center}

      \includegraphics[scale=0.657]{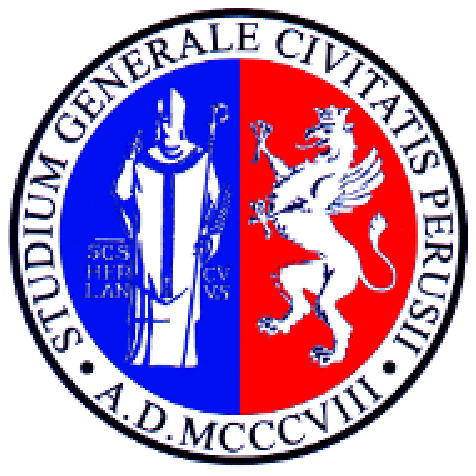}

      \end{center}

    \end{figure}

    \vspace{-0.85cm}



    \begin{center}


    {\large    Universit\`a degli Studi di Perugia                  \\ \medskip

               Facolt\`a di Scienze Matematiche, Fisiche e Naturali \\ \medskip

               Dottorato di Ricerca in Fisica XXII Ciclo}                           \\ \medskip

    \vskip 1 cm

     {\Large Tesi di Dottorato \\ \bigskip}

     {\huge \bf Theoretical investigation of two- and three-body short range correlations in inclusive electron scattering off nuclei at high momentum transfer\\
         \medskip
        }

    \vskip 1.3 cm

    {\large {Candidata} \\ \medskip

     \sl Dott.ssa \bf Chiara Benedetta Mezzetti \rm \\ \medskip }

    \vskip 1.5 cm

    \begin{tabular}{ccc}

    {\large  Relatore}  &\hspace{0.2cm} & {\large  Coordinatore}

     \\

    {\large \sl Prof. \bf Claudio Ciofi degli Atti  \rm}  &\hspace{0.2cm} &

    {\large \sl Prof. \bf Maurizio Busso \rm} \\
    \\
    \\



    \end{tabular}

    \end{center}

    \vskip 1.0 cm \centerline{\large Anno Accademico 2008-2009}

\newpage
%
%
\unitlength=1mm
\setcounter{totalnumber}{5}   
%
%
\pagestyle{fancy}     
\fancyhead[LE]{\nouppercase{\bf \leftmark}}    
\fancyhead[RE,LO]{}      
\fancyhead[RO]{\nouppercase{\fancyplain{}{\bf \rightmark}}}     
\fancyfoot[C]{\thepage}    
%
%
\renewcommand{\headrulewidth}{0.4pt}
\renewcommand{\footrulewidth}{0pt}
%
%
\tableofcontents
\pagestyle{plain}
\chapter*{Introduction}
\addcontentsline{toc}{chapter}{Introduction}
Obtaining information on short range correlations (SRC) in nuclei is a primary goal of modern nuclear physics \cite{quarkcosmos}. Interest in  SRC  stems not only from the necessity to firmly establish the limits of validity of the standard model of nuclei, i.e. a non relativistic description in terms of two- and three-nucleon interactions, but also from the impact that the knowledge of the detailed mechanism of SRC would
have  in understanding  the role played by  quark degrees of freedom in hadronic matter and the properties of the latter in dense configurations \cite{FSS}. Recently, evidence of SRC has been provided by new experimental data on inclusive [$A(e,e')X$] \cite{FSS2,Egiyan2} and exclusive [$A(e,e'pN)X$ and $A(p,pN)X$] lepton and hadron scattering  off nuclei at high momentum transfer ($Q^2 \gtrsim 1$ $GeV^2$) (see Ref. \cite{Subedi} and references therein quoted). In inclusive scattering  the observation  of a scaling behavior of the ratio of the cross section on heavy nuclei to that on the Deuteron \cite{FSS2}, for values of the Bjorken scaling variable $1.4 \lesssim x_{Bj} \lesssim 2$, and to that on  $^3He$ \cite{Egiyan2},  for $2 \lesssim x_B \lesssim 3$, has been interpreted as evidence that the electron probes two- and three-nucleon correlations in complex nuclei similar to the ones occurring in the two- and three-nucleon  systems \cite{mark1,FS}.
By combining the results of inclusive and exclusive experiments, a convincing experimental evidence of SRC in nuclei have been eventually found. It should be pointed out, however, that in inclusive experiments statistics in the region of three-nucleon short range correlations is very poor and, at the same time, a general framework to describe these correlations is still lacking; moreover, in exclusive experiments, the found evidence of SRC is limited to the $^{12}C$ nucleus. For these reasons it is not only necessary to extend experimental measurements to other nuclei, as planned e.g. at the Thomas Jefferson National Accelerator Facility (JLab), but it is also urgent to improve our theoretical knowledge on the nature of two- and three-nucleon correlations.  The aim of this Thesis is to critically review the present theoretical and experimental knowledges on SRC in nuclei, and, at the same time, to provide a theoretical framework within which to coherently treat two- and three-nucleon correlations. As already pointed out, most of our experimental knowledge on two- and three-nucleon correlations comes from exclusive and inclusive experiments on lepton and hadron scattering off nuclei at high momentum transfer. It should be pointed out, however, that whereas exclusive processes can directly access the relative and center of mass motions of a correlated pair in a nucleus \cite{theoryexclusive,Schiavilla,ACM2}, obtaining information on these quantities from inclusive scattering is, in principle,  more difficult.  Various approaches based on scaling  concepts have  therefore been proposed, going from the scaling behavior of the cross section ratio plotted versus  $x_{Bj}$, to the scaling behavior of the ratio of the nuclear to the nucleon cross sections plotted versus proper scaling variables; among the latter, a process that has been most investigated in the past is the so called Y-scaling, for it is believed that this may  represent a powerful  tool to extract the high momentum part of the nucleon momentum distribution which is governed by  SRC \cite{CPS3,CPS,Day2}.  In this Thesis the concepts of $Y$-scaling will be critically reanalyzed,  mainly because of: i) the lack of a general consensus about the usefulness of such a concept, and ii) a strong renewal of interest in Y-scaling owing to recent  experimental data on $A(e,e')X$ reactions from the JLab \cite{arrington,newdata}.
We will show that the analysis of inclusive scattering in terms of proper Y-scaling variables could indeed provide  useful information on SRC; to this end, following the suggestion of Refs. \cite{CW,CFW1,CFW2}, a new approach to Y-scaling and its usefulness will be illustrated in detail.
\\In the first part of the Thesis, mainly in Chapter $1$, we present the necessary formalism to introduce the original part, which is illustrated in Chapter $2$-$6$. In more detail, the structure is as follows:
\begin{enumerate}
\item In Chapter 1, the realistic many-body problem of nuclei and cold hadronic matter is recalled, in order to introduce the concept of SRC;  an overview of the experimental evidence of SRC is given, and the relevance of SRC in various fields of Physics is illustrated.
\item In Chapter 2, the basic features of the spectral function and the nucleon momentum distributions, which are two basic quantities of our new approach to inclusive cross sections, are recalled.
\item In Chapter 3, the formalism of inclusive electron scattering off nuclei and the general expression of the cross section is presented, within the plane wave impulse approximation (PWIA), and by taking into account the final state interaction (FSI) of the knocked out nucleon with the residual system ($A-1$).
\item In Chapter 4, the inclusive process is analyzed in terms of $Y$-scaling: three different scaling variables are introduced, each one describing a particular process occurring in electron scattering off nuclei.
\item In Chapter 5, the results of our calculations of the $\xb$ dependence of the inclusive cross section ratios in terms of PWIA and FSI are presented.
\end{enumerate} 

\cleardoublepage
%
%
\pagestyle{fancy}
\pagenumbering{arabic}    
\chapter{Short range correlations in nuclei}
\section*{Introduction}
We will now introduce what short range correlations are and what is their relevance in modern physics.
\section{The realistic many-body problem of nuclei and cold hadronic matter} \label{sec:realistic_manybody}
\begin{figure}
\begin{center}
\subfigure
{\includegraphics[scale=0.7]{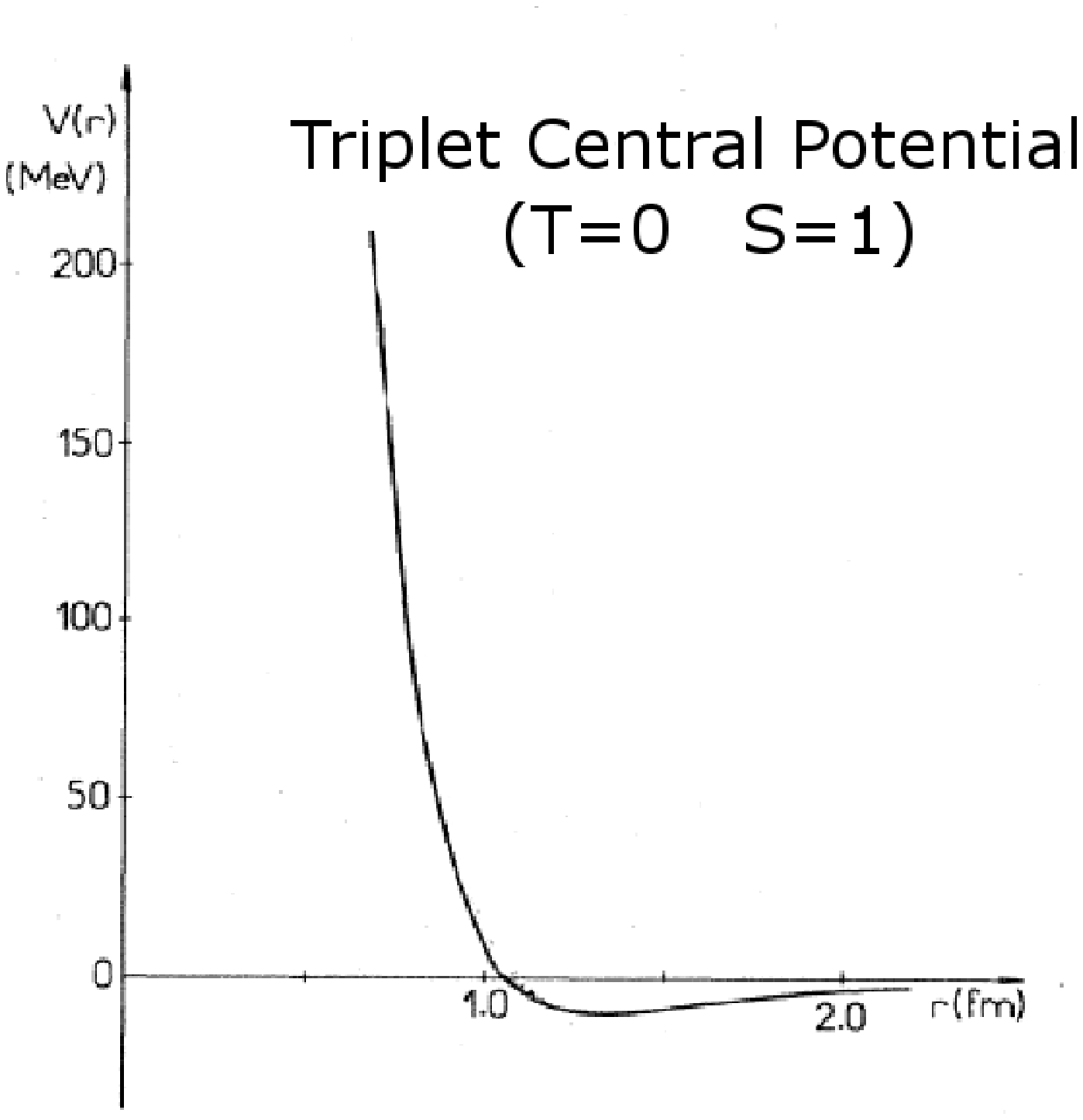}}\\
\subfigure{
\includegraphics[scale=0.7]{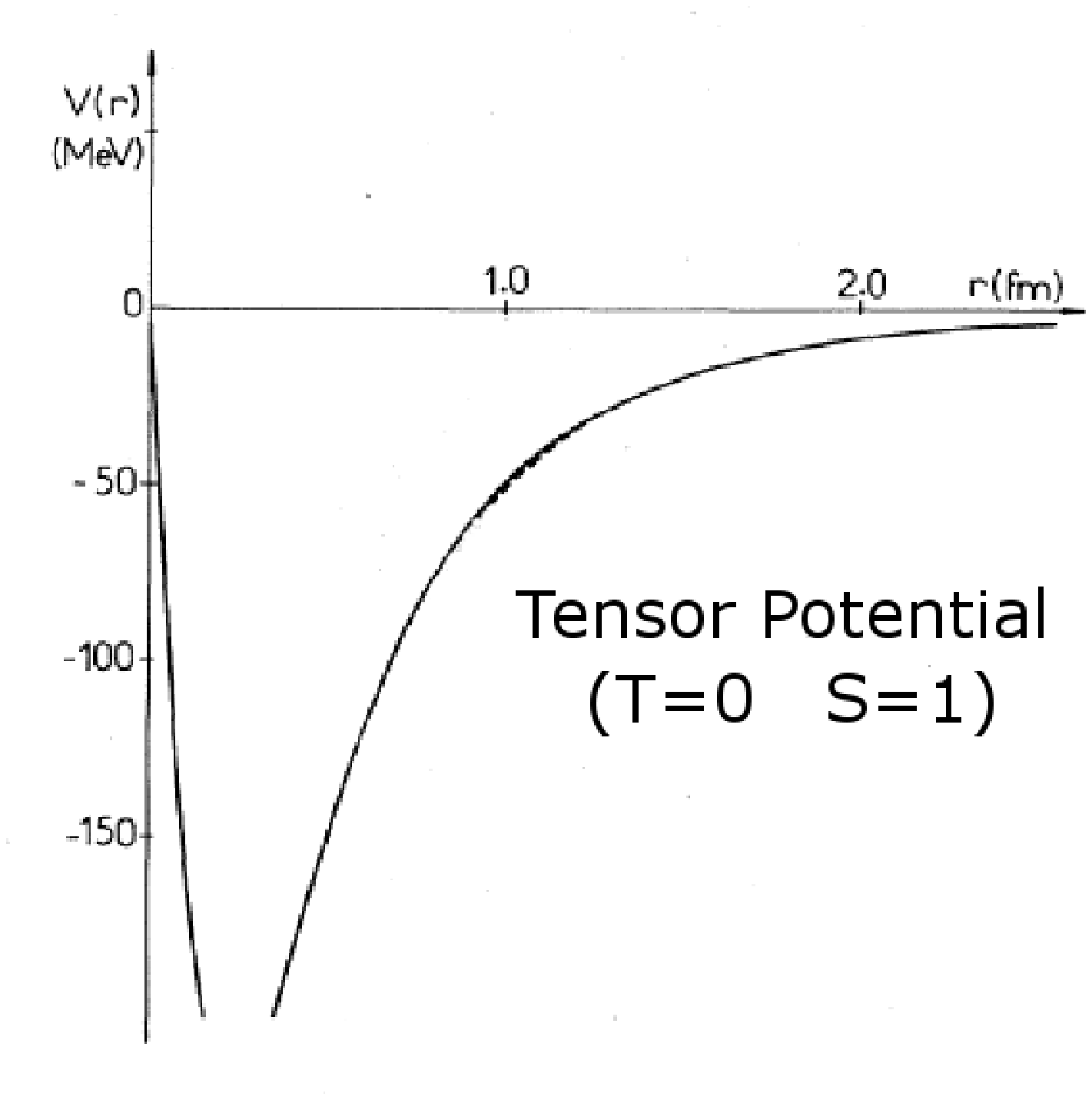}}
\caption{The Paris nucleon-nucleon potential in triplet ($S=1$) central and tensor states. After Ref. \cite{Lacombe}.}
\label{potential2}
\end{center}
\end{figure}

\begin{figure}[h]
\begin{center}
\includegraphics[scale=0.55,angle=270]{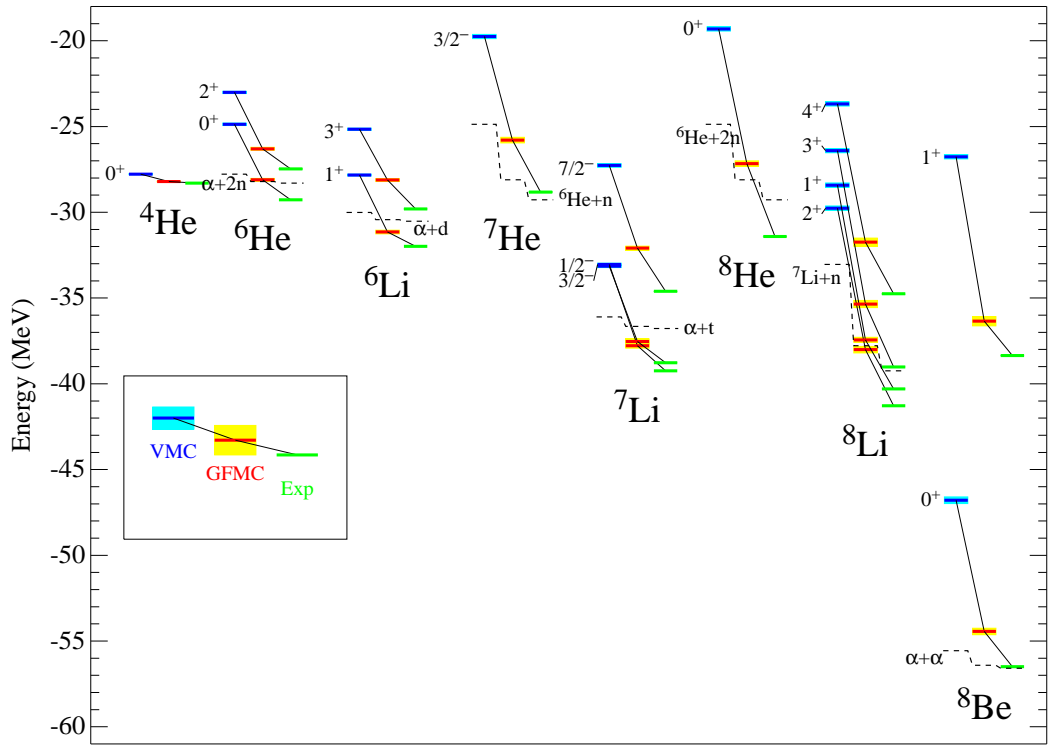}
\caption{Variational Monte Carlo (VMC) and Green Function Monte Carlo (GFMC) energies using the two-body AV18 \cite{Pie} and three-body UIX \cite{Carlson} interactions compared with experiment (Exp). Black dashed lines show the indicated breakup thresholds for each method. The Monte Carlo statistical errors are shown by the light blue and yellow bands. After Ref. \cite{WiringaMC}.}
\label{Wiri}
\end{center}
\end{figure}

In what follows, we will consider a bound system of $Z$ protons and $N$ neutrons, with $A=Z+N$. We will simply call such a system a nucleus $A$.
What exposed in the following equally well applies to both nuclei and cold hadronic matter, the latter being a system composed of an infinite number of bound nucleons (e.g. neutron stars).
As stressed in Ref. \cite{PH}, when particles interact with each other through the intervening mechanism of a field, the description of their dynamical behavior in terms of instantaneous potentials is only an approximate nature, and the two-body $\hat{v}_2(x_i,x_j)$, three-body $\hat{v}_3(x_i,x_j,x_k)$,\ldots, $A$-body potentials may be regarded as successive stages of this approximation. Thus, considering the nucleus $A$ as a non relativistic quantum-mechanical system, its quantum states are the solution of the nuclear many-body problem, represented by the following Schr$\ddot{o}$dinger equation
\beq \label{Schrodinger}
    \hat{H}(x_1\ldots x_A)\Psi_A^n(x_1\ldots x_A)=E_n\Psi_A^n(x_1\ldots x_A)
\eeq
with Hamiltonian
\beq \label{hamiltonian}
    \hat{H}(x_1\ldots x_A)=-\frac{\hbar^2}{2m_N}\: \sum_{i=1}^A\: \hat{\nabla}^2_i + \sum_{i< j=1}^A\:\hat{v}_2(x_i,x_j) + \sum_{i< j < k=1}^A\:\hat{v}_3(x_i,x_j,x_k)+\ldots
\eeq
Here $x_i\equiv\{\textbf{r}_i,\textbf{s}_i, \textbf{t}_i\}$ denotes the generalized coordinate of the $i-th$ nucleon, which includes its radial coordinate $\textbf{r}_i$, spin $\textbf{s}_i$ and isospin $\textbf{t}_i$; $n$ stands for the set of quantum numbers of the state under consideration; $m_N$ is the nucleon mass.
\\Solving the nuclear many-body problem is not an easy task; the reasons are manyfold:
\begin{itemize}
\item many-body forces are unknown;
\item the two-body potential obtained from the analysis of nucleon-nucleon (NN) scattering data is very complicated, owing to its spin, isospin and tensor dependences \cite{WSS};
\item it is not yet clear what is the role (if any) played by the quark-gluon structure of the nucleon in the description of nuclear properties;
\item it is not yet clear to which extent the nucleons bound in a nucleus retain the same properties as the free ones;
\item last but not least: can nucleon motion in a nucleus be considered within the non relativistic approximation?
\end{itemize}
It has however been demonstrated in Ref. \cite{PH} that, independently of the detail of the field, the $m$-body potentials, in systems governed by the strong force, can be written as follows
\beq
    \left(m\mbox{-body potentials}\right) \simeq\left(\frac{v_N}{c}\right)^{m-2}\times\: \left(two\mbox{-body potentials}\right)
\eeq
where $m=\{3,\ldots,A\}$, and $v_N$ is the \emph{average} nucleon velocity which, using the uncertainty principle and the  known nuclear dimension,  can be estimated to be $v_N\sim 0.02c$, which means that, to a large extent, nuclear systems can be considered as non relativistic systems bound by two- and, at most, three-nucleon interactions. Such a conclusion is confirmed by
a wealth of experimental information on basic properties of nuclei (e.g. the dependence of their radii upon $A^{1/3}$, which, in turns, leads to a constant value of the binding energy per nucleon and to the constance of the volume, the similarity between the magnetic moments of odd nuclei and the nucleon magnetic moments, etc.) which lead to the conclusion that the atomic nucleus can, on the average, be described as an incompressible low density system (with density $\rho_0 \sim 0.17\:N/fm^3$) composed of non relativistic nucleons interacting mainly via the same two-nucleon strong force acting between free nucleons plus, at most, we reiterate, three-nucleon forces. The nuclear many-body problem thus reduces to what has been called the \emph{standard model of nuclei} \cite{Ciofi}, described by the following Schr$\ddot{o}$dinger equation
\bey \label{Schrodinger_23}
    \left[-\frac{\hbar^2}{2m_N}\: \sum_{i=1}^A \hat{\nabla}^2_i + \sum_{i< j=1}^A \hat{v}_2(x_i,x_j)
    + \sum_{i< j < k=1}^A \hat{v}_3(x_i,x_j,x_k)\right]\Psi_A^n(x_1\ldots x_A)\no \\
    =E_n\Psi_A^n(x_1\ldots x_A)
\eey
Even in this simplified form, Eq. (\ref{Schrodinger_23}) is difficult to solve, due to the complicated structure of the two-nucleon interaction. For such a reason,
%
in the past half century, various phenomenological models have been proposed to explain the structure of nuclei, and the nuclear shell model (SM), for which Maria G\"{o}ppert-Mayer and Hans D. Jensen were awarded by the Nobel Prize in 1963, turned out to be the most successful one \cite{Bohr}. In the simplest version of this model, the independent particle shell model, the nucleus is described as an ensemble of independent nucleons which move in an average potential filling, according to the Pauli Exclusion Principle, proper shell model states. Moreover, neutrons and protons occupy all states below the Fermi level, leaving the above states empty.
More technically, one says that the \emph{occupation probability} of states below the Fermi level is one, and above the Fermi level is zero.
\\The Hamiltonian of the system reduces to
\beq\label{hamiltonianSM}
    \hat{H}_0(x_1\ldots x_A)=-\frac{\hbar^2}{2m_N}\: \sum_{i=1}^A\: \hat{\nabla}^2_i + \sum_{i=1}^A\:\hat{V}(r_i)
\eeq
 and the Schr$\ddot{o}$dinger equation is
\beq \label{SchrodingerSM}
    \sum_{i=1}^A\:\left[-\frac{\hbar^2}{2m_N}\: \hat{\nabla}^2_i+\:\hat{V}(r_i)\right]\phi_0=\epsilon_0\:\phi_0
\eeq
$\Phi_0$ being a Slater determinant.
In the most refined SM description, various types of \emph{residual interactions} are added to Eq. (\ref{hamiltonianSM}); these include, for example, the spin orbit interaction, non spherical single particle potentials to account for the deviation from the spherical shape of classes of nuclei in the periodic table, and others.
The success of the advanced SM in reproducing many properties of nuclei was awarded in $1975$ by a Nobel Prize to A. Bohr, B. Mottelson and L. Rainwater. In the advanced shell model, a prominent role is played by the so-called \emph{long range correlations}, whose main effect is to partly deplete the occupation probability of the states below the Fermi level, making the states above the Fermi level partially occupied. It should however  be pointed out that the main feature of the independent particle and advanced shell models, is the independent particle motion. This fact is difficult to reconcile with one of the main features of the realistic NN interaction, namely the strong repulsive core at relative distances of the order of $0.5-0.6\:fm$, which is one of the facts which makes the solution of the nuclear many-body problem (\ref{Schrodinger_23}) a very difficult one.
\\The realistic two-body interaction, which explains two-body bound and scattering data, has the following form \cite{WSS}
\beq \label{v2}
    \hat{v}_2(x_i,x_j)=\sum_{n=1}^{N} v^{(n)}(r_{ij})\:\hat{O}_{ij}^{(n)}
\eeq
where $r_{ij}\equiv |\textbf{r}_i-\textbf{r}_j|$ is the relative distance of nucleons $i$ and $j$, and $n$, ranging up to $N=18$, labels the  state-dependent operator $\hat{O}_{ij}^{(n)}$, whose first six components are defined as follows
\bey
    \hat{O}_{ij}^{(1)}&\equiv& \hat{O}_{ij}^{c}=1  \\
    \hat{O}_{ij}^{(2)}&\equiv& \hat{O}_{ij}^{\sigma}=\bm{\sigma}_i \cdot \bm{\sigma}_j \\
    \hat{O}_{ij}^{(3)}&\equiv& \hat{O}_{ij}^{\tau}=\bm{\tau}_i \cdot \bm{\tau}_j  \\
    \hat{O}_{ij}^{(4)}&\equiv& \hat{O}_{ij}^{\tau}=\left(\bm{\sigma}_i \cdot \bm{\sigma}_j\right)\:
    \left(\bm{\tau}_i \cdot \bm{\tau}_j\right) \\
    \hat{O}_{ij}^{(5)}&\equiv& \hat{O}_{ij}^{(t)}=\hat{S}_{ij} \\
    \hat{O}_{ij}^{(6)}&\equiv& \hat{O}_{ij}^{(t)}=\hat{S}_{ij}\:\left(\bm{\tau}_i \cdot \bm{\tau}_j\right)\: .
\eey
Here
\beq
    \hat{S}_{ij}=3\left(\bm{\sigma}_i\cdot {\hat{r}}_{ij}\right)\:\left(\bm{\sigma}_j\cdot {\hat{r}}_{ij}\right)-
    \left(\bm{\sigma}_i\cdot \bm{\sigma}_{j}\right)
\eeq
is the tensor operator, and $\bm{\sigma}_i$ and $\bm{\tau}_i$ are the spin and isospin of the $i$-th nucleon of the pair, respectively.
It can be seen that the two-body interaction exhibits a strong state dependence. Particular worth being mentioned are the strong short range repulsion and  the tensor attractive interaction in $T=0$ and $S=1$ states, as shown in Fig. \ref{potential2}.
%
These features of the NN interaction generate, as we shall see in more detail in the following, strong correlations between nucleons, which are not present in a shell model picture.
\\Recently, important progress has been made in solving the many-body Schr\"o\-din\-ger equation (\ref{Schrodinger_23}), which is rewritten by considering only the ground state, denoting  $\Psi_A^0 \equiv \Psi_A$ and $E_0\equiv E_A$, namely
\beq \label{hamiltoniancorr}
    \left[ -\frac{\hbar^2}{2m_N}\: \sum_{i=1}^A\: \hat{\nabla}^2_i + \sum_{i< j=1}^A\:\hat{v}_2(x_i,x_j) + \sum_{i< j < k=1}^A\:\hat{v}_3(x_i,x_j,x_k) \right]\Psi_A=E_A\:\Psi_A\: .
\eeq
The most advanced approaches to the solution of the nuclear many-body problem rely on the numerical integration of Eq. (\ref{hamiltoniancorr}) by Monte Carlo techniques \cite{WiringaMC}, in particular the Variational Monte Carlo (VMC) method, which is used to optimize the expectation value of observables by adjusting a trial wave function, and the Green Function Monte Carlo method. By Monte Carlo techniques it was possible to solve Eq. (\ref{hamiltoniancorr}) for both the ground and excited states in the range $3 \leq A \leq 8$, as shown in Fig.  \ref{Wiri}.
For heavier nuclei the Monte Carlo methods become very difficult to apply, and alternative realistic approaches have been developed, such as the Hypernetted chain method with correlated wave function \cite{Fantorosa}, large shell model basis approach \cite{Vary}, the coupled-cluster theory \cite{Roth}, and the variational calculations with correlated wave functions \cite{CO}. In this Thesis we will consider the \emph{number conserving cluster expansion} approach with correlated wave functions \cite{ACM}. In this method, the variational principle is applied in minimizing the expectation value of the Hamiltonian
\beq \label{Hmedio}
    <\hat{H}>=\frac{<\Psi_A^v|\hat{H}|\Psi_A^v>}{<\Psi_A^v|\Psi_A^v>} \geq E_0
\eeq
with the trial nuclear wave function cast in the following form
\beq \label{wavefunc}
    \Psi_A^v=\hat{F}\:\phi_0
\eeq
where $\phi_0$ is the shell model mean field wave function, and
\beq \label{corr_op}
    \hat{F}=\hat{S}\:\prod_{i<j}\:\hat{f}(x_i,x_j)
\eeq
is the correlation operator; the latter is defined in terms of the symmetrization operator $\hat{S}$, and the two-body correlation function
\beq \label{correfunc}
    \hat{f}(x_i,x_j)=\sum_{n=1}^N \:f^{(n)}(r_{ij})\:\hat{O}^{(n)}_{ij}
\eeq
with the same operatorial dependence appearing in the two-body potential $\hat{v}_2(x_i,x_j)$ given by Eq. (\ref{v2}). In Ref. \cite{ACM}, a new effective method for the calculation of the expectation value of any quantum-mechanical operator $\hat{A}$ in the many-body ground state described by the wave function $\Psi_A^v$, i.e.
\beq \label{exp}
    <\hat{A}>=\frac{<\Psi_A^v|\hat{A}|\Psi_A^v>}{<\Psi_A^v|\Psi_A^v>}
\eeq
with $\Psi_A^v$ given by Eq. (\ref{wavefunc}), based upon the cluster expansion, is presented.
\\Introducing the quantity
\beq
    \hat{\eta}_{ij}\equiv \hat{f}_{ij}^\dag\:\hat{f}_{ij} -1
\eeq
the expectation value (\ref{exp}) can be written as follows
\bey
    <\hat{A}>&=&\frac{<\phi_0|\hat{F}^\dag\: \hat{A}\:\hat{F}|\phi_0>}{<\phi_0|\phi_0>}=
    \frac{<\phi_0\left|\prod_{i<j}\left(1+\hat{\eta}_{ij}\right)\:\hat{A}\right|\phi_0>}
    {<\phi_0\left|\prod_{i<j}\left(1+\hat{\eta}_{ij}\right)\right|\phi_0>}\no \\
    &=&\frac{<\phi_0\left|\left(1+\sum_{i<j}\hat{\eta}_{ij}+\sum_{(i<j)<(k<l)}\hat{\eta}_{ij}\hat{\eta}_{kl}+\ldots
    \right)\:\hat{A}\right|\phi_0>}
        {1+<\phi_0\left|\sum_{i<j}\hat{\eta}_{ij}\right|\phi_0>+\ldots}\: . \no \\
\eey
Expanding in series the denominator
\beq
    \frac{1}{1+x}=1-x+x^2-\ldots
\eeq
one gets
\bey
    <\hat{A}>&=&\left[ <\phi_0|\hat{A}|\phi_0>+ <\phi_0|\sum_{i<j}\hat{\eta}_{ij}\hat{A}|\phi_0>+ \ldots  \right]\no \\ &\times&
    \left[1- <\phi_0|\sum_{i<j}\hat{\eta}_{ij}|\phi_0>+<\phi_0|\sum_{i<j}\hat{\eta}_{ij}|\phi_0>^2+\ldots \right]
\eey
and collecting all terms containing the same number of function $\hat{\eta}_{ij}$, one obtains the infinite series
\beq
    <\hat{A}>=<\hat{A}>_0+<\hat{A}>_1+<\hat{A}>_2+\ldots+<\hat{A}>_n+\ldots
\eeq
At second order in $\eta$, one has, explicitly,
\bey
    <\hat{A}>_0&=&<\phi_0|\hat{A}|\phi_0> \\
    <\hat{A}>_1&=&<\phi_0|\sum_{i<j}\hat{\eta}_{ij}\:\hat{A}|\phi_0>-<\hat{A}>_0\: <\phi_0|\sum_{i<j}\hat{\eta}_{ij}|\phi_0>\\
    <\hat{A}>_2&=&<\phi_0|\sum_{(i<j)<(k<l)}\hat{\eta}_{ij}\hat{\eta}_{kl}\hat{A}|\phi_0>-
    <\phi_0|\sum_{i<j}\hat{\eta}_{ij}\hat{A}|\phi_0> \no \\
    &\times& <\phi_0|\sum_{i<j}\hat{\eta}_{ij}|\phi_0> \no \\
    &\times&
\left( <\phi_0|\sum_{(i<j)<(k<l)}\hat{\eta}_{ij}\hat{\eta}_{kl}|\phi_0>-<\phi_0|\sum_{i<j}\hat{\eta}_{ij}|\phi_0>^2
    \right)
\eey
\begin{figure}[h]
\begin{center}
\includegraphics[scale=1.2]{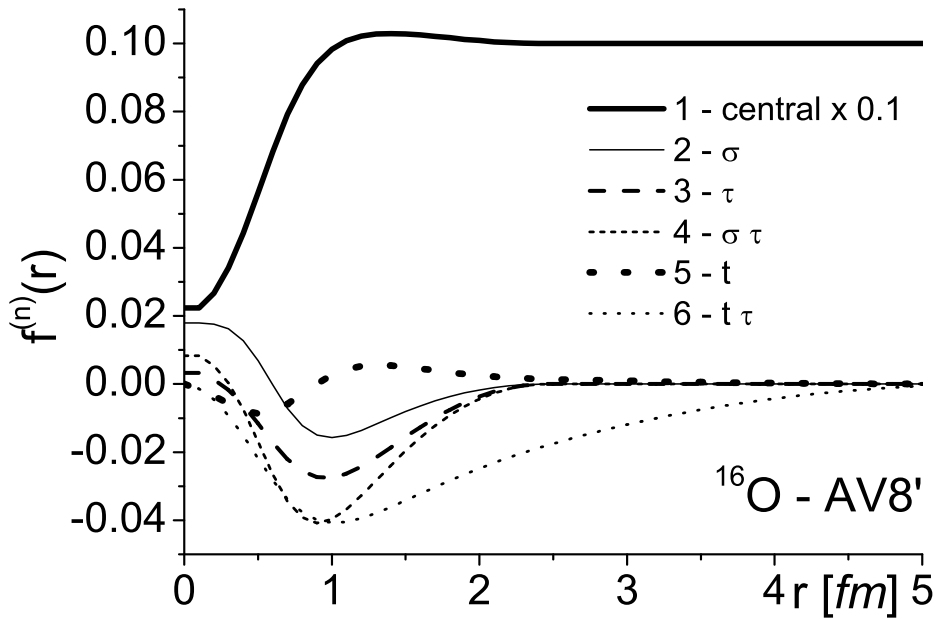}
\caption{The correlation functions (\ref{correfunc}) for $^{16}O$ \emph{vs.} the relative distance $r\equiv r_{ij}$, corresponding to the full Argonne $V8'$ interaction used in Ref. \cite{ACM}.}
\label{Fig_corfunc}
\end{center}
\end{figure}
\noindent
where the term of the order of $n$ contains $\hat{\eta}_{ij}$ $(\hat{f}_{ij})$ up to the $n$th ($2$dn) power.
In Ref. \cite{ACM}, the ground state properties of closed shell nuclei, $^4He$, $^{16}O$ and $^{40}Ca$, have been calculated by minimizing the ground state energy at $2$nd order in $\hat{\eta}_{ij}$ with respect to the single particle wave functions and the correlation functions $f^{(n)}(r_{ij})$. The results for the latter, corresponding to the $V8'$ interaction \cite{Fabrocini}, are shown in Fig.  \ref{Fig_corfunc} versus the distance $r$, in case of $^{16}O$.
It should be pointed out that:
\begin{itemize}
 \item at relative distances  significantly smaller than the average internucleon distance ($\sim 1.7\:fm$), nucleons feel the strong central repulsion and the tensor attraction;
%
%
\item  at large distances, the effects due to noncentral correlations vanish, the central component is equal to $1$,  and the mean field wave function in Eq. (\ref{wavefunc}) is recovered.
\end{itemize}
\begin{figure}[!h]
\centerline{\includegraphics[scale=0.83]{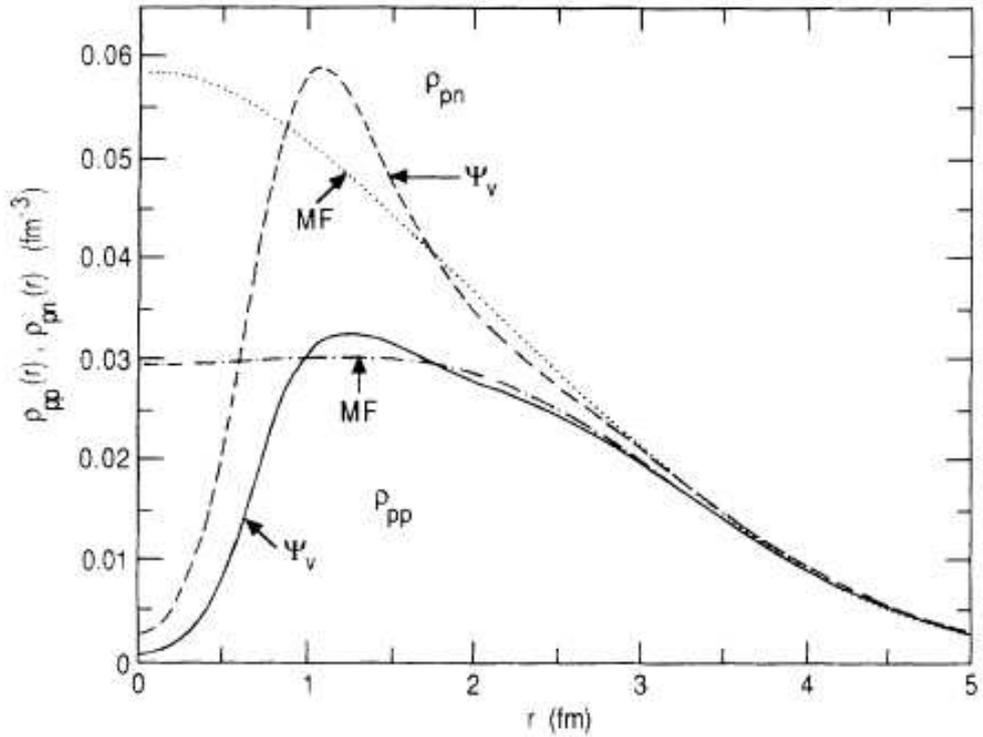}}
\caption{The two-body density distribution versus the relative distance $r\equiv r_{ij}$. Dot-dashed line: $p$-$p$ mean field (MF) contribution, calculated from $\phi_0$; solid line: $p$-$p$ correlated contribution, calculated from $\Psi_A^v\equiv \Psi_v$; dotted line: $p$-$n$ MF part; dashed line: $p$-$n$ correlated component.  After Ref. \cite{Wiringa}}
\label{Fig_2rho}
\end{figure}
\noindent
The net effect generated by the correlation function is demonstrated in Fig.  \ref{Fig_2rho}, where the calculated two-body density distribution for $p$-$n$ and $p$-$p$ pairs in $^{16}O$ resulting from the solution of Eq. (\ref{hamiltoniancorr}) with interaction (\ref{v2}), containing the Argonne $v_{14}$ two-nucleon \cite{v14} and Urbana $VII$ three-nucleon potential \cite{urbana}, is compared with the shell model density solution of Eq. (\ref{SchrodingerSM}). It can be seen that, as a result of the contrasting short range central repulsion and the intermediate range tensor attraction, the realistic two-body density strongly differs from the shell model density in the region $0 \leq r \leq 1.1 \: fm$; since the average internucleon distance is $\sim 1.7-2\:fm$, we will call the deviation from the shell model in the range $0 \leq r \leq 1.1 \: fm$ \emph{short range correlations} (SRC), always remembering that they are due to both short range central repulsion and intermediate range tensor attraction, with the latter acting only in $T=0$ and $S=1$ states, and thus enhancing the $n$-$p$ distributions. In what follows, it will be shown that SRC strongly depopulate the  states below the Fermi sea, creating highly excited virtual two-particle ($2p$)- two-hole ($2h$) states.
\section{Experimental investigation of long and short range correlations in nuclei}
\begin{figure}[!h]
\centerline{\includegraphics[scale=0.6]{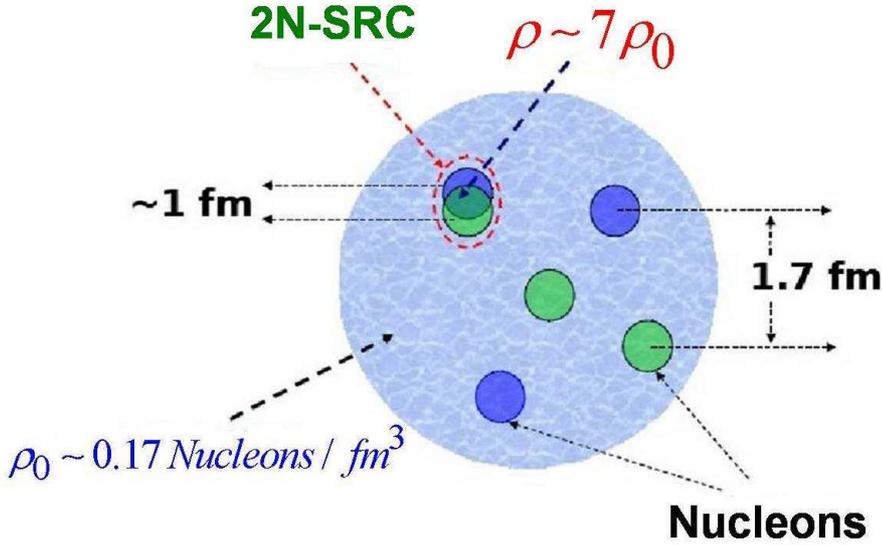}}
\caption{A simple cartoon of NN short range correlations. When the distance between two nucleons becomes smaller than the average internucleon distance, of order $\sim 1.7 \: fm$, nucleons result to be in a correlated pair, and their local density becomes comparable with those in the core of neutron stars, up to $\sim 7$ times the average nuclear density $\rho_0$.}
\label{Fig_SRC}
\end{figure}
A simple cartoon of NN SRC, as predicted by theoretical calculations, is depicted in Fig.  \ref{Fig_SRC}, and the question arises as to whether such a picture can be experimentally observed. The answer is a positive one, and will be given qualitatively in what follows.
%
%
\subsection{Exclusive lepton scattering: the $A(e,e'p)X$ reaction}
\begin{figure}[!h]
\begin{center}
\vskip-0.6cm
\subfigure[\label{lnf}]
{\includegraphics[scale=0.4]{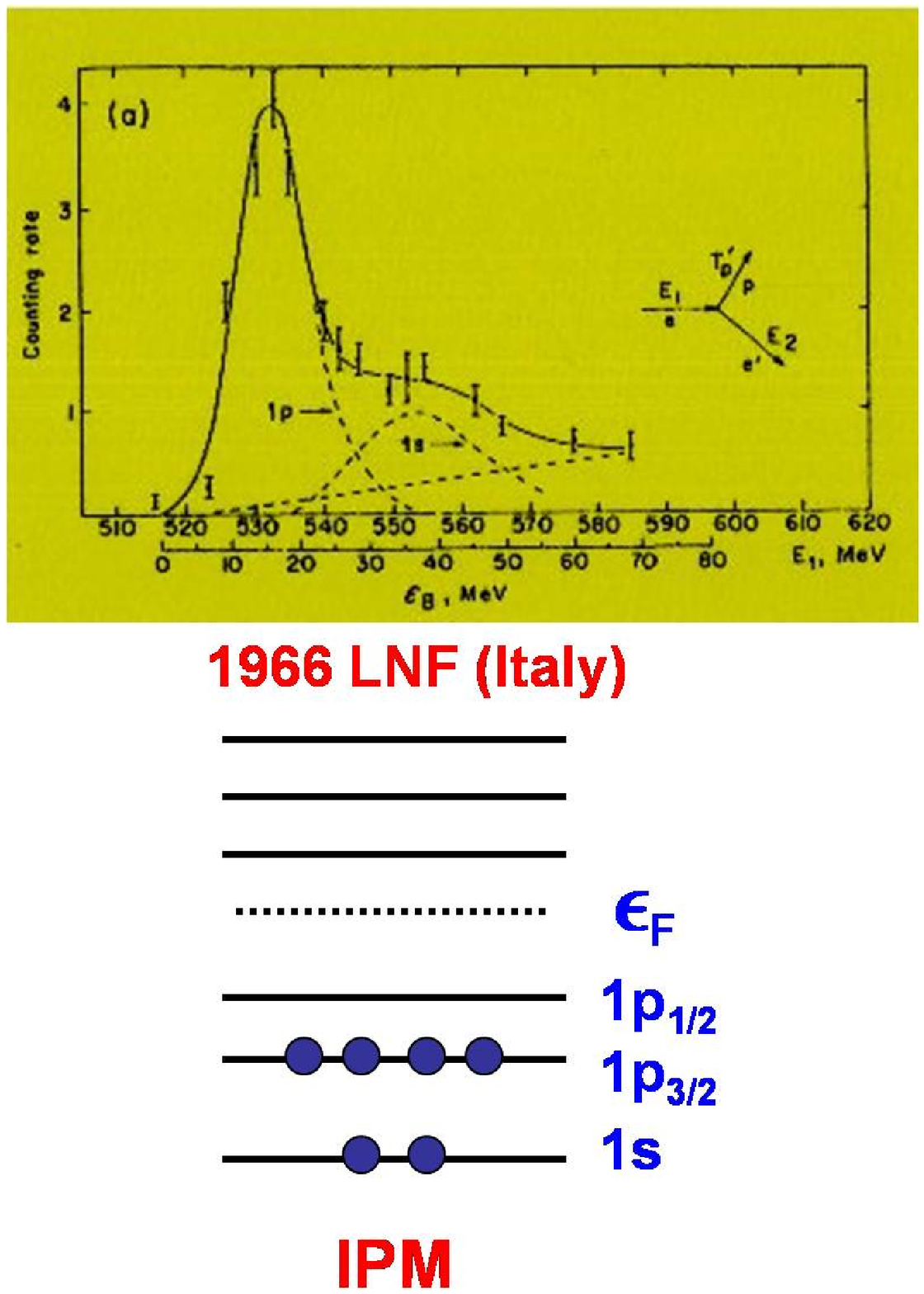}}\:
\subfigure[\label{nikhef}]{
\includegraphics[scale=0.4]{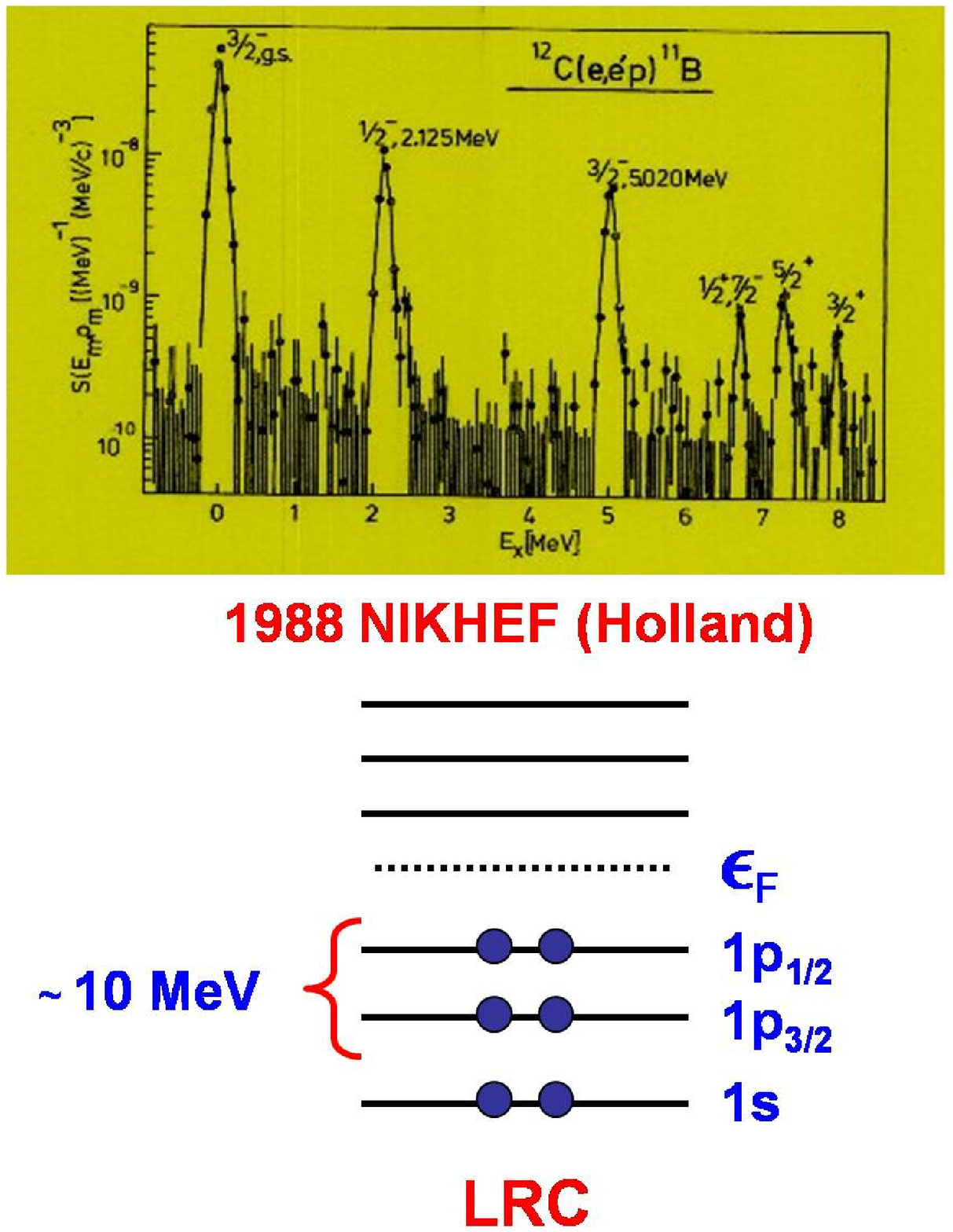}}
\vskip-0.2cm
\subfigure[\label{jlab}]{
\includegraphics[scale=0.4]{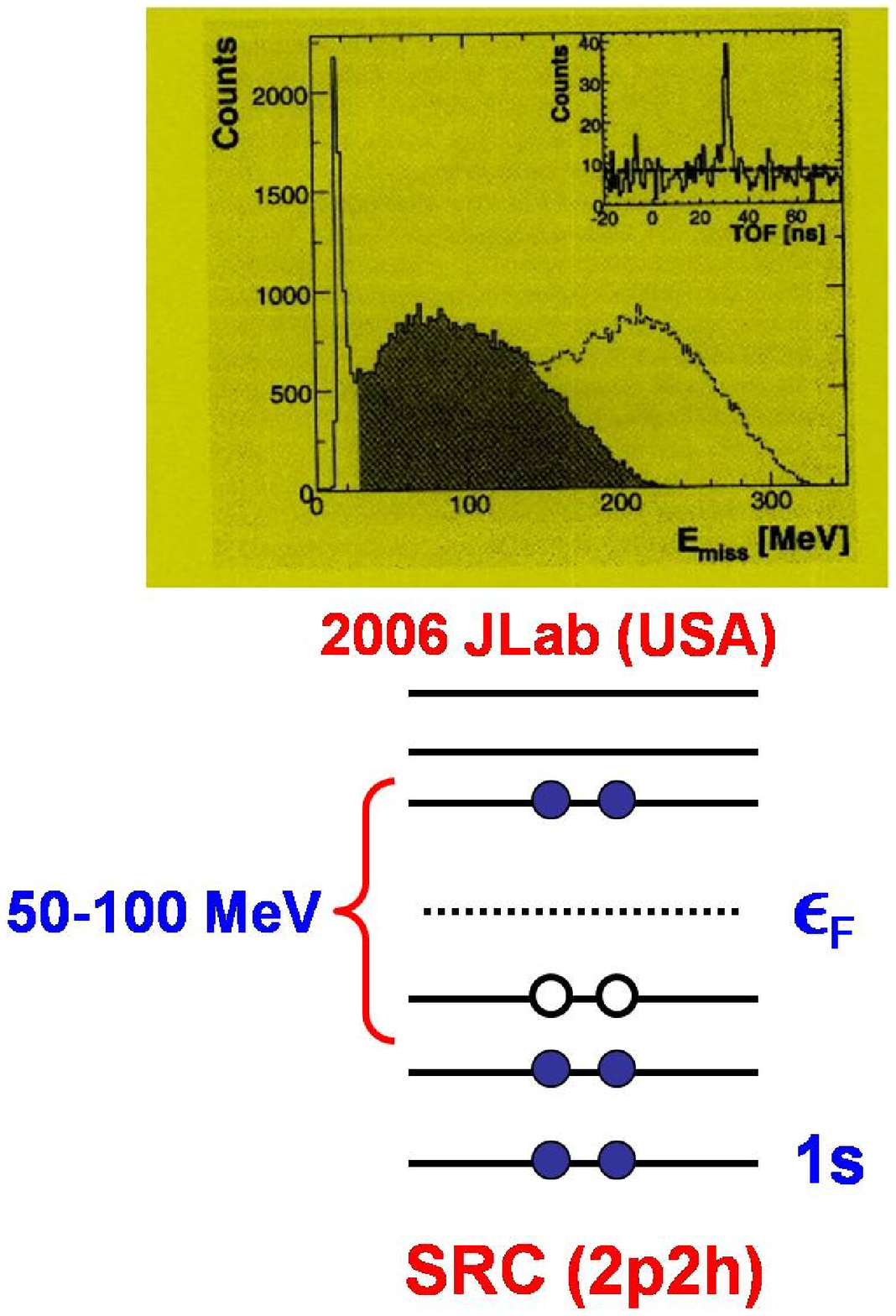}}
\vskip-0.4cm
\caption{\emph{Nuclear ionization} experiments $^{12}C(e,e'p)X$ performed at: (a) Laboratori Nazionali di Frascati (LNF) at low energy resolution and low removal energy $E_{min}<E\lesssim40\: MeV$ \cite{Amaldi}, (b) the National Institute for Nuclear Physics and High Energy Physics (NIKHEF) at high energy resolution and low  removal energy $E_{min}<E\lesssim20\: MeV$\cite{Nikhef} and (c) the Jefferson Laboratory (JLab) facilities at high values of the removal energy $E_{min}<E\lesssim200\: MeV$; the black area is attributed to SRC \cite{Shneor}.}
\label{Fig_longrange}
\end{center}
\end{figure}

\begin{figure}[!h]
\centerline{\includegraphics[scale=0.7]{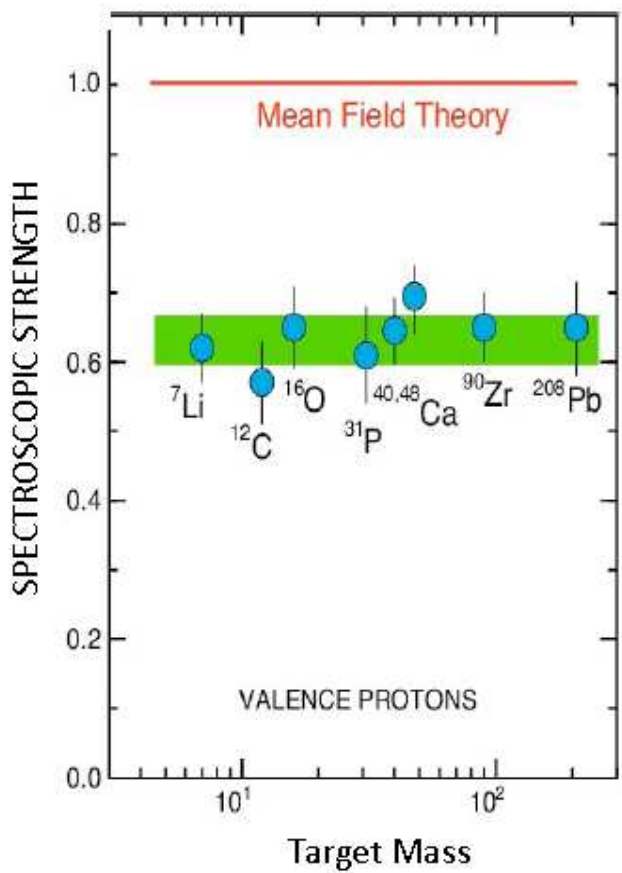}}
\caption{Spectroscopic strength for knocked out valence protons measured with the reaction $(e,e'p)$ relative to independent particle shell model prediction \cite{Lapikas}.}
\label{Fig_Lapikas}
\end{figure}

A way to learn about correlations in nuclei is represented by the nuclear reaction in which an energetic electron $e$, with energy $\epsilon_1$, knocks out from a nucleus $A$ a proton $p$, which is detected in coincidence with the scattered electron $e'$, with energy $\epsilon_2$. In this process, the $(A-1)$ nucleus is left in some excited states with energy $E^*=\nu-T_p-T_{A-1}$, where $\nu=\epsilon_1-\epsilon_2$ is the energy transfer (the energy lost by the electron in the scattering process) and $T_p$ and $T_{A-1}$ are the kinetic energies of the proton and the residual nucleus, respectively. This kind of \emph{nuclear ionization} experiments, initiated more than $40$ years ago \cite{Amaldi}, depends upon the number of protons in various shell model states (see $\S$\ref{sec:P0}) or, in other words, on the occupation probability in various shell model states, which, as we know, in the single particle shell model is one, for states below the Fermi sea (e.g. $1s_{1/2}$ and $p_{3/2}$ in $^{12}C$), and zero for states above the Fermi sea.
\\The situation for the $^{12}C$ nucleus is illustrated in Fig.  \ref{Fig_longrange}.
%
It can be seen that the first experiments, performed at low energy resolution, identified only two shell model states in $^{12}C$, namely $1p$ and $1s$, whereas the high energy resolution experiments \cite{Nikhef} demonstrated the occupation \cite{Amaldi} of both $p_{3/2}$ and $p_{1/2}$ shells, as a result of long range correlations (\emph{configuration mixing} \cite{Kurath}). The occupation numbers of the valence protons in various nuclei obtained from $(e,e'p)$ reactions are summarized in Fig.  \ref{Fig_Lapikas}.
It can be seen that the occupation probability of the least bound ('valence') nucleons is about $0.6-0.7$; it strongly deviates from one, i.e. from the prediction of the independent particle shell model. Such a large deviation cannot however be explained only by long range correlations which, as depicted in Fig.  \ref{Fig_longrange}, involve excitation energies of tenth of MeV. It has been argued that the deviation is evidence of SRC which, by populating  highly excited states, strongly deplete shell model states below the Fermi sea.
\\It should be pointed out that, for decades, understanding the role played by SRC in nuclei has been a very elusive problem of nuclear physics, due to the difficulties in isolating the signal of SRC. As a matter of fact, at low and medium energies, effects due to the final state interaction of the knocked out nucleon with the residual system, as well as effects from the excitation energy of hit nucleon, could produce the same final state as the one which is expected to be produced by an initially correlated pair. Investigating SRC thus requires high energy probes, in order to cover kinematical regions with
\beq \label{kin}
Q^2>1\:\left(GeV/c\right)^2 \quad , \qquad \xb=\frac{Q^2}{2m_N\nu}>1
\eeq
which are available in modern lepton and hadronic accelerators like, e.g., JLab (USA), GSI (Germany) and JPARC (Japan).  In Eq. (\ref{kin}), $Q^2$ is the four-momentum squared of the virtual photon, $\nu$ is the energy transfer, $m_N$ is the nucleon mass, and $\xb$ is the Bjorken scaling variable.
\\It is only recently that such experiments could be performed, and the excitation strength due to SRC could be observed (cf. Fig.  \ref{Fig_longrange}).
%
%
\subsection{Exclusive lepton scattering: the $A(e,e'pN)X$ reaction}\label{sec:Hig}
\begin{figure}[!h]
\centerline{\includegraphics[scale=1]{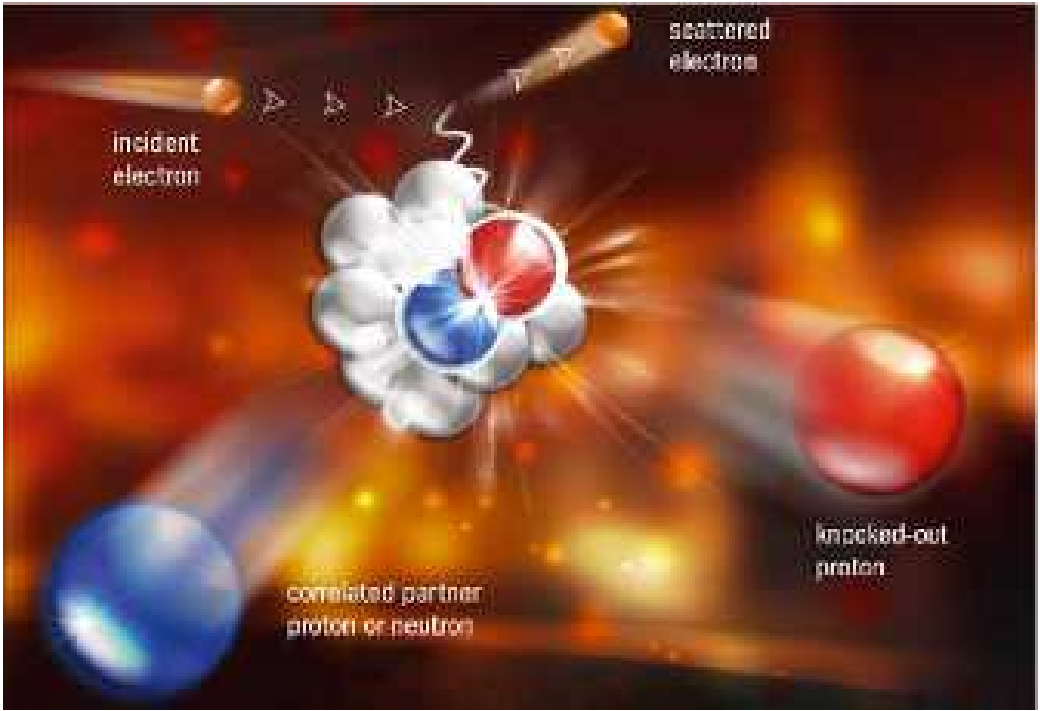}}
\caption{A simple cartoon to illustrate the $A(e,e'pN)X$ reaction. The incident electron couples to a nucleon-nucleon pair via a virtual photon. In the final state, the scattered electron and struck nucleon are detected along with the correlated nucleon that is ejected from the nucleus \cite{Subedi}.}
\label{Fig1.4}
\end{figure}

%
%
Experimentally, high momentum probes can knock out a proton off a nucleus, leaving the rest of the system nearly unaffected. If, on the other hand, the proton being struck is part of a correlated pair, the high relative momentum in the pair leads the correlated nucleon to recoil and be ejected and detected, as pictorially shown in Fig.   \ref{Fig1.4}. This triple coincidence experiments have been performed on $^{12}C$ targets, both at Brookhaven National Laboratory (BNL) using incident protons \cite{Aclander,Tang,Eli}, and at Jefferson Laboratory (JLab) by using electrons \cite{Subedi}.
\\The BNL experiment EVA (E850) studied the process
\begin{figure}
\begin{center}
\vskip-0.3cm
\subfigure[\label{process}]
{\includegraphics[scale=0.6]{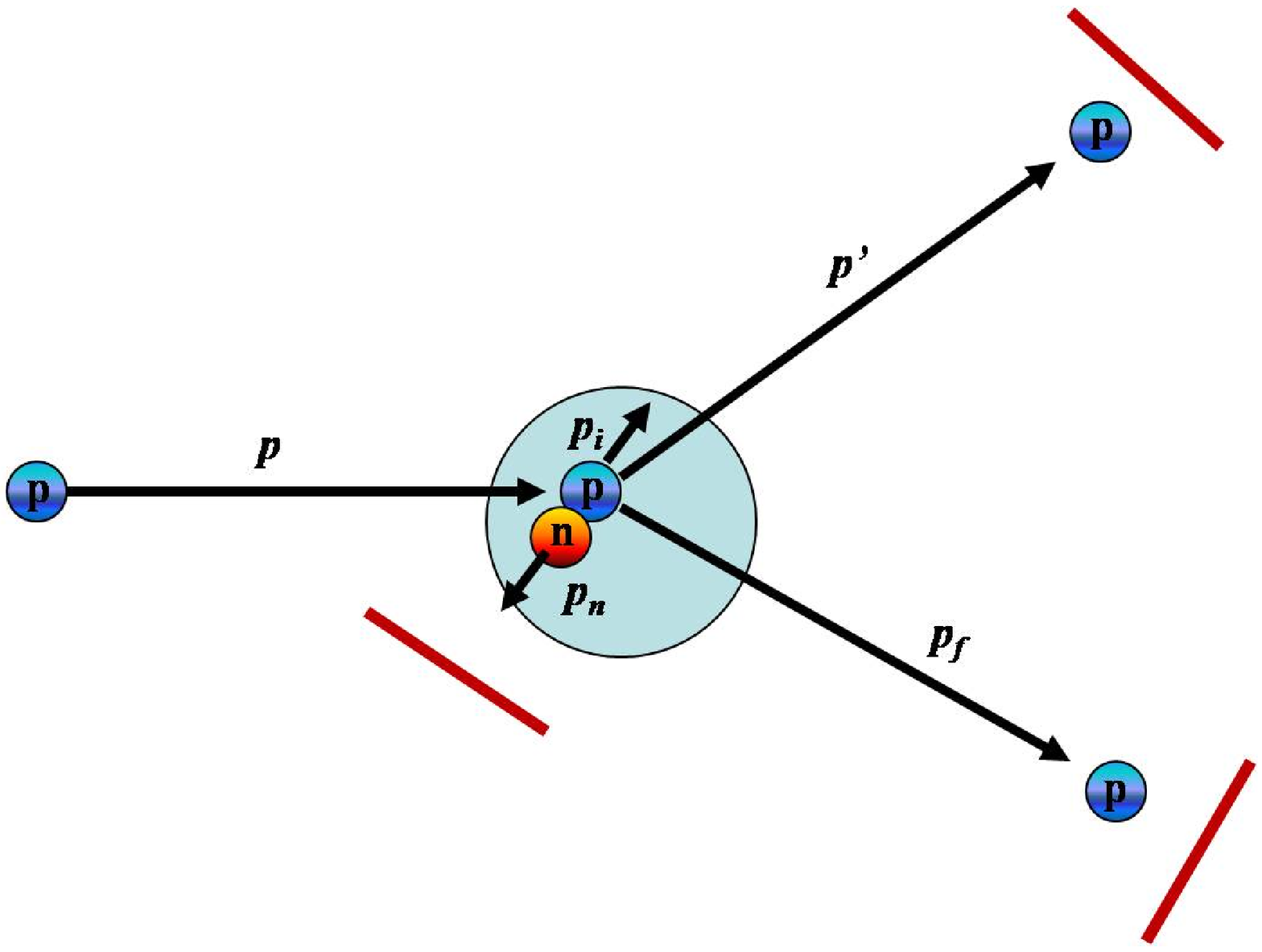}}\\
\vskip-0.1cm
\subfigure[\label{momenta}]{
\includegraphics[scale=0.6]{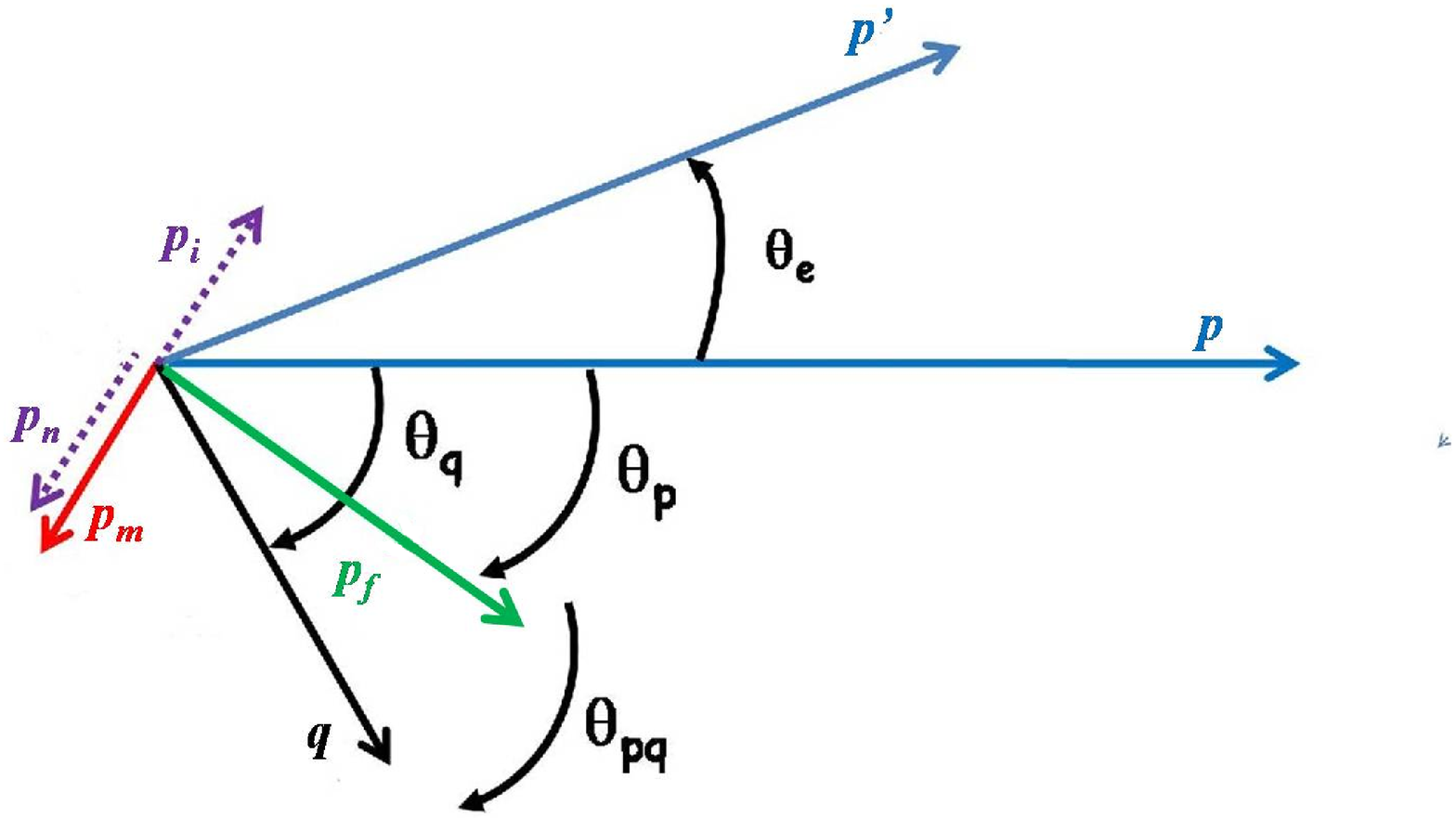}}
\caption{Two simple cartoons to illustrate the $^{12}C(p,p'pn)X$ reaction studied at BNL. (a) The incident and the struck protons, with initial momenta $\textbf{p}$ and $\textbf{p}_i$, and final momenta $\textbf{p}'$ and $\textbf{p}_f$, respectively, are detected in coincidence with a neutron with momentum $\textbf{p}_n$; (b) The same process, viewed also in terms of the missing momentum $\textbf{p}_m$ and the three-momentum transfer $\textbf{q}$. The notation for angles is self explaining.}
\label{Fig_pmiss}
\end{center}
\end{figure}
\beq
p+^{12}C \rightarrow 2p+n+X
\eeq
with proton beam momentum $\textbf{p}$ ranging from $6\:GeV/c$ up to $9\:GeV/c$.
\\In the final state, the scattered proton, with momentum $\textbf{p}'$, and the knocked out proton, with momentum
\beq
\textbf{p}_f\simeq \textbf{q}=\textbf{p}-\textbf{p}'
\eeq
where $\textbf{q}$ is the momentum transfer, are detected. By assuming that the knocked out proton leaves the nucleus without interacting with the residual system $(A-1)$, one has
\beq
\textbf{p}_f= \textbf{p}_i+\textbf{q} \: .
\eeq
By introducing the so-called \emph{missing momentum}
\beq
    \textbf{p}_m \equiv \textbf{q}-\textbf{p}_f
\eeq
one has
\beq
    \textbf{p}_m=\textbf{P}_{A-1}=-\textbf{p}_i
\eeq
as shown in Fig. \ref{Fig_pmiss}.
\\If the knocked out nucleon was initially correlated with a neutron, with the nucleus $(A-2)$ almost at rest, one should observe and detect the recoiling neutron with momentum
\beq
\textbf{p}_n =- \textbf{p}_i = \textbf{p}_m \: .
\eeq
Within these assumptions, by plotting the momentum $\textbf{p}_n$ of the recoiling neutron as a function of the cosine of the angle between $\textbf{p}_i$ and $\textbf{p}_n$, i.e.
\beq
    \cos\gamma=\frac{\textbf{p}_i \cdot \textbf{p}_n}{|\textbf{p}_i||\textbf{p}_n|}
\eeq
which is known as the number of directional correlations, one should observe, at high values of the momentum $\textbf{p}_n$, a strong back-to-back directional correlation between $\textbf{p}_i$ and $\textbf{p}_n$, due to SRC, which has indeed been observed, as shown in Fig. \ref{Fig_angle}.
\begin{figure}[!h]
\centerline{\includegraphics[scale=0.75]{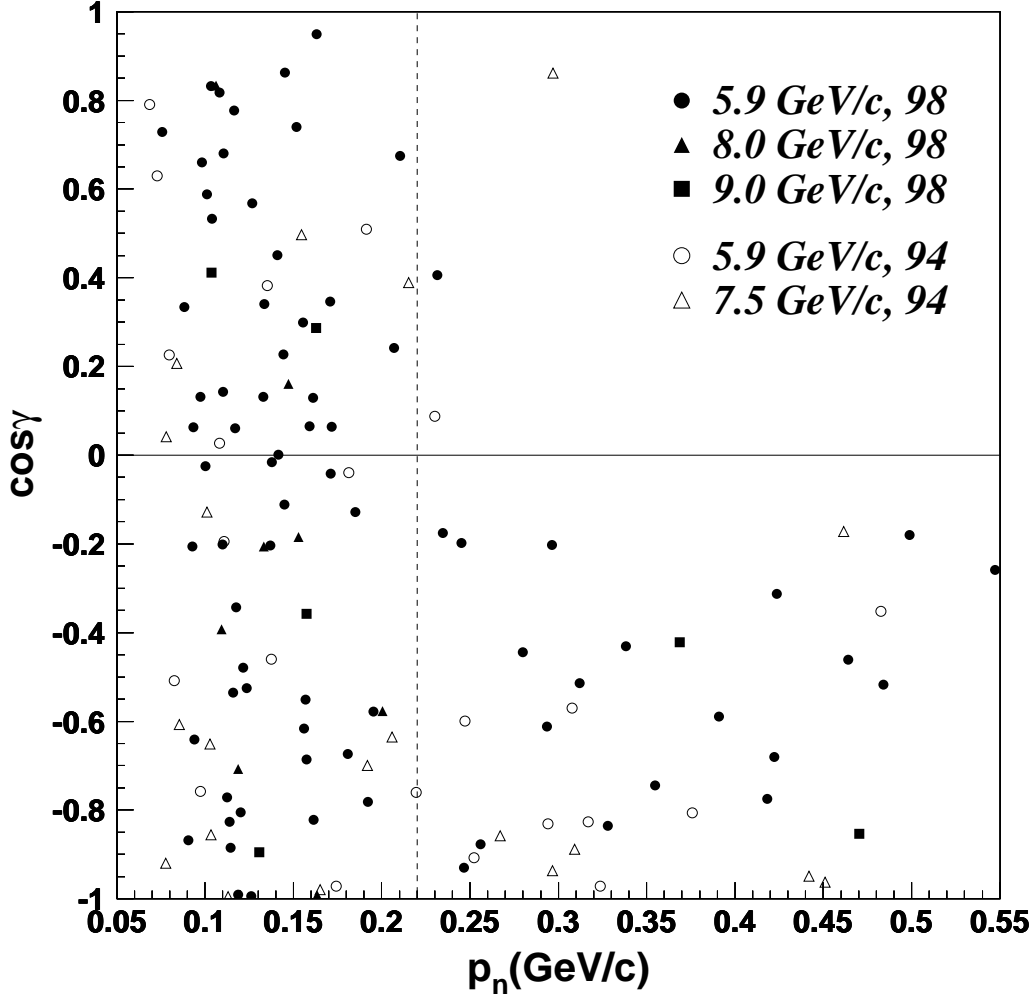}}
\caption{The correlation between $\textbf{p}_n$ and its direction $\gamma$ relative to $\textbf{p}_i$. Data labeled by $94$ and $98$ are from Refs. \cite{Aclander,Tang}. The dotted vertical line indicates the Fermi momentum $k_F=0.221\:GeV/c$. After Ref. \cite{Eli}.}
\label{Fig_angle}
\end{figure}
\\In Ref. \cite{Tang,Eli,Eli2}, the ratio
\beq \label{F}
    F=\frac{\mbox{Number of (p,2pn) events} (p_i,p_n > k_F)}{\mbox{Number of (p,2p) events} (p_i > k_F)}\: .
\eeq
has bee extracted from these data. Eq. (\ref{F}) represents the measure of correlation of backward neutrons with initial momentum $\textbf{p}_n \sim- \textbf{p}_i$, when $|\textbf{p}_n|,|\textbf{p}_i| \geq k_F$; by determining the ratio (\ref{F}) it has been demonstrated that $92_{-18}^{+8}\%$ of protons in $^{12}C$ with momenta $p_i\geq 275\:MeV/c$ are partners in $n$-$p$ SRC pairs.
\begin{figure}[!h]
\centerline{\includegraphics[scale=0.65]{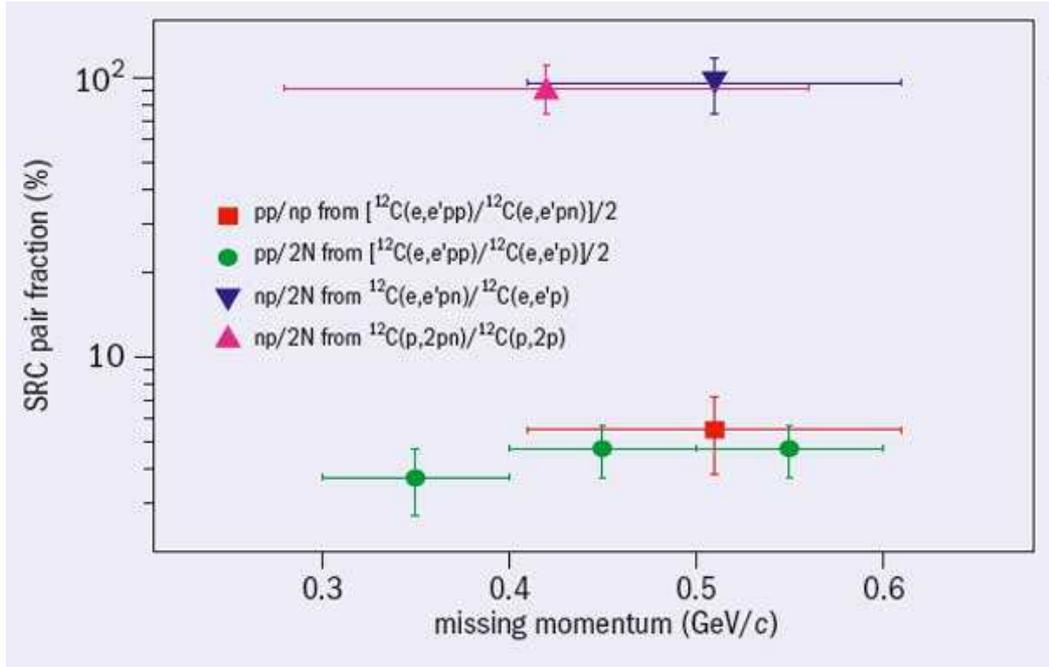}}
\caption{The fractions of SRC pair combination in $^{12}C$, resulting from $(e,e'pp)$ and $(e,e'pn)$ reactions performed at JLab \cite{Subedi}, as well as from BNL $(p,2pn)$ data \cite{Aclander,Tang,Eli}. The results show the dominance of $p$-$n$  pairs \cite{Subedi}.}
\label{Fig_pair}
\end{figure}
\\The Jefferson Lab experiment has demonstrated that nearly all nucleons in $^{12}C$ with momentum in the range $300$-$600\:MeV/c$ have a correlated nucleon partner with roughly equal and opposite momentum \cite{Subedi,Shneor}. By comparing $n$-$p$ to $p$-$p$ pairs yields, it has also been found that SRC are mainly due to $n$-$p$ pairs, whose probability results of the order of $18\pm5$, as shown in Fig.  \ref{Fig_pair} \cite{Subedi}. Calculations explain the magnitude of this $n$-$p$ to $p$-$p$ ratio as arising from the short range tensor part. Both experiments have shown that recoiling nucleons, with a momentum above the Fermi sea level in the nucleus, are part of a correlated pair, and both observed the same strength of $p$-$n$ correlations. This confirms that the process is accessing a universal property of nuclei, unrelated to the probe \cite{Cern}.
%
%
\subsection{Inclusive lepton scattering: the $A(e,e')X$ reaction} \label{sec:inclusive}
\begin{figure}[!h]
\centerline{\includegraphics[scale=0.7]{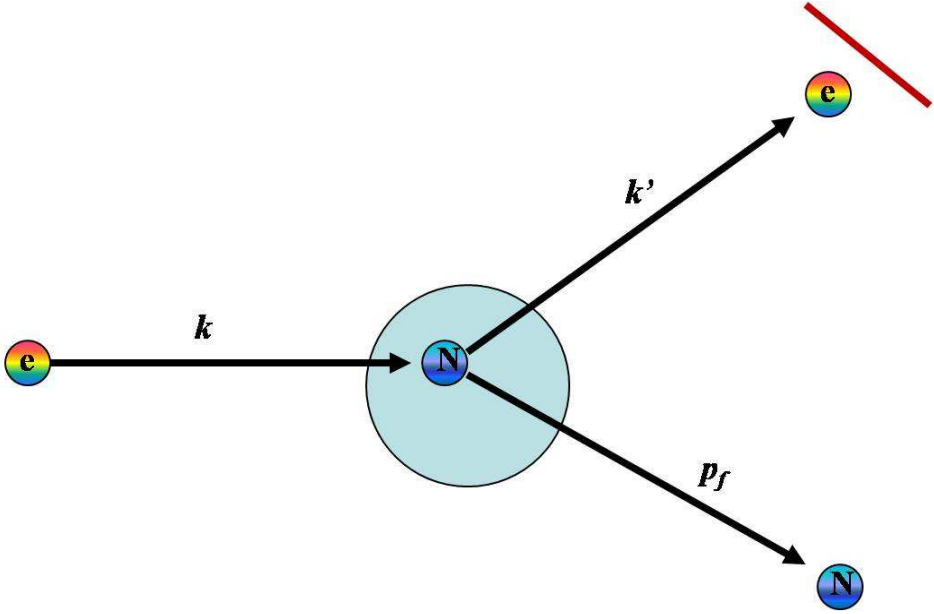}}
\caption{A simple picture of the $A(e,e')X$ process. An electron with momentum $\textbf{k}$ knocks out a nucleon $N$ with final momentum $\textbf{p}_f$ and, after interaction, is detected with momentum $\textbf{k}'$. The knocked out nucleon is not detected.}\label{Fig_inclusivo3}
\end{figure}

The high energy $A(e,e')X$ reaction depicted in Fig. \ref{Fig_inclusivo3}, i.e. the process in which only the scattered electron is detected, represents the simplest reaction to investigate SRC and, in particular, to measure the probabilities of SRC in nuclei.
\\In Ref. \cite{mark1,FS} it has been supposed that, at high momentum transfer, the inclusive cross section off a nucleus $A$ can be written as follows
\beq \label{FSratio}
    \sigma_A(\xb,Q^2)=\sum_{j=2}^A \: \frac{a_j(A)}{j} \: \sigma_j(\xb,Q^2)
\eeq
where $ \sigma_A(\xb,Q^2)$ is the electron-nucleus cross section, written in terms of the $4$-momentum transfer and the Bjorken scaling variable $\xb$, $ \sigma_j(\xb,Q^2)$ is the inclusive cross section off a correlated cluster of $j$-particles, and, eventually, the quantity  $a_j(A)$ is the probability of finding a nucleon in the cluster $j$. By this way, one expects that the inclusive cross section is dominated, at $1 \lesssim x_{Bj} \lesssim 2$, by the absorption of the virtual photon on a pair of correlated nucleons and by the elastic rescattering  in the continuum, whereas, at $2 \lesssim x_{Bj} \lesssim 3$, it is governed by three-nucleon correlations (3NC), and so on.
\\In the process under analysis, as shown in Fig. \ref{Fig_inclusivo3}, an electron beam with momentum $\textbf{k}$, knocks out from the nucleus $A$, a nucleon with final four-momentum given by the following energy and momentum conservation
\beq
p_f^2=\left(q+P_A-P_{A-1}\right)^2=m_N^2
\eeq
where $q$ is the four-momentum transfer, $P_A$ and $P_{A-1}$ are the momenta of the target nucleus and the residual system $(A-1)$, respectively, and $m_N$ is the nucleon mass;
\begin{figure}[!h]
\centerline{\includegraphics[scale=0.75]{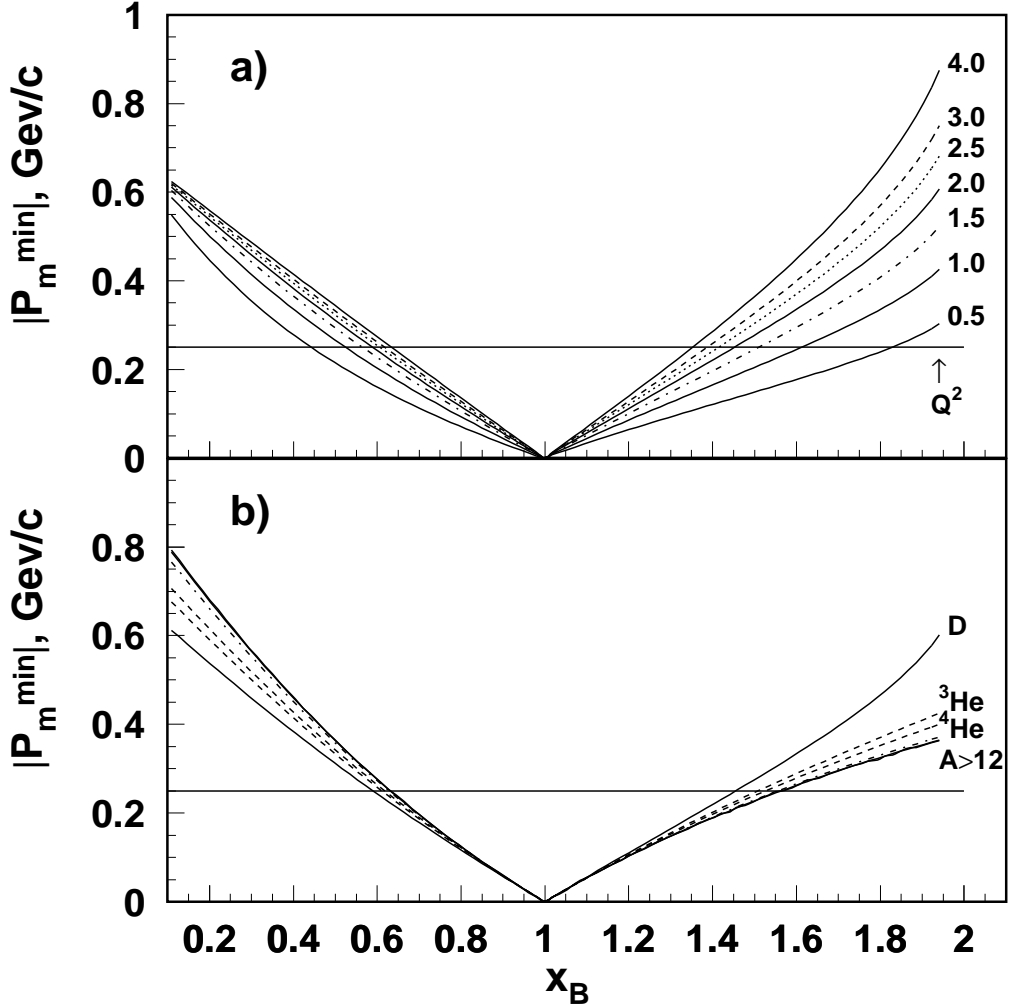}}
\caption{The minimum recoil momentum versus the Bjorken scaling variable $\xb$ for a) Deuteron, calculated at several $Q^2\:[GeV^2]$,and for b) different nuclei at $Q^2=2.0\:GeV^2$. Horizontal lines at $250\:MeV/c$ indicate the Fermi momentum typical of uncorrelated motion of nucleons in nuclei \cite{Egiyan}.}
\label{Fig_pmin}
\end{figure}
as already pointed out, after interaction only the scattered electron, with momentum $\textbf{k}'$, is detected. In Fig. \ref{Fig_pmin}, the minimum value of the \emph{missing momentum}
\beq
        \textbf{p}_{m}=\textbf{q}-\textbf{p}_f=\textbf{P}_{A-1}
\eeq
defined in terms of the the three-momentum transfer $\textbf{q}=\textbf{k}-\textbf{k}'$ and the momentum of the residual system $\textbf{P}_{A-1}$, is plotted versus the Bjorken scaling variable $\xb$.
It can be clearly seen that, for any nucleus $A$ and fixed $Q^2$, the minimum recoil momentum $|\textbf{p}_m^{min}|$ increases with $\xb$, and exceeds the average Fermi momentum in nucleus $A$ at $\xb > \xb^0$, the latter depending upon the nucleus \cite{Egiyan}. As already pointed out, SRC correspond to high momentum components of the nuclear wave function, therefore, with the gradual increase of $\xb$, the virtual photon should first probe high momentum configurations due to 2NC and then, following Eq. (\ref{FSratio}), at $\xb > 2$, it should probe 3NC \cite{mark1,FS}.
\\To avoid the difficulties due to the theoretical calculations of electron off-shell nucleon cross section, inclusive data from JLab-HallB have been analyzed \cite{Egiyan2,Egiyan} not directly in terms of cross sections, but  by taking the cross section ratios of $^4He$, $^{12}C$ and $^{56}Fe$ to $^3He$.
\begin{figure}[!h]
\centerline{\includegraphics[scale=0.85]{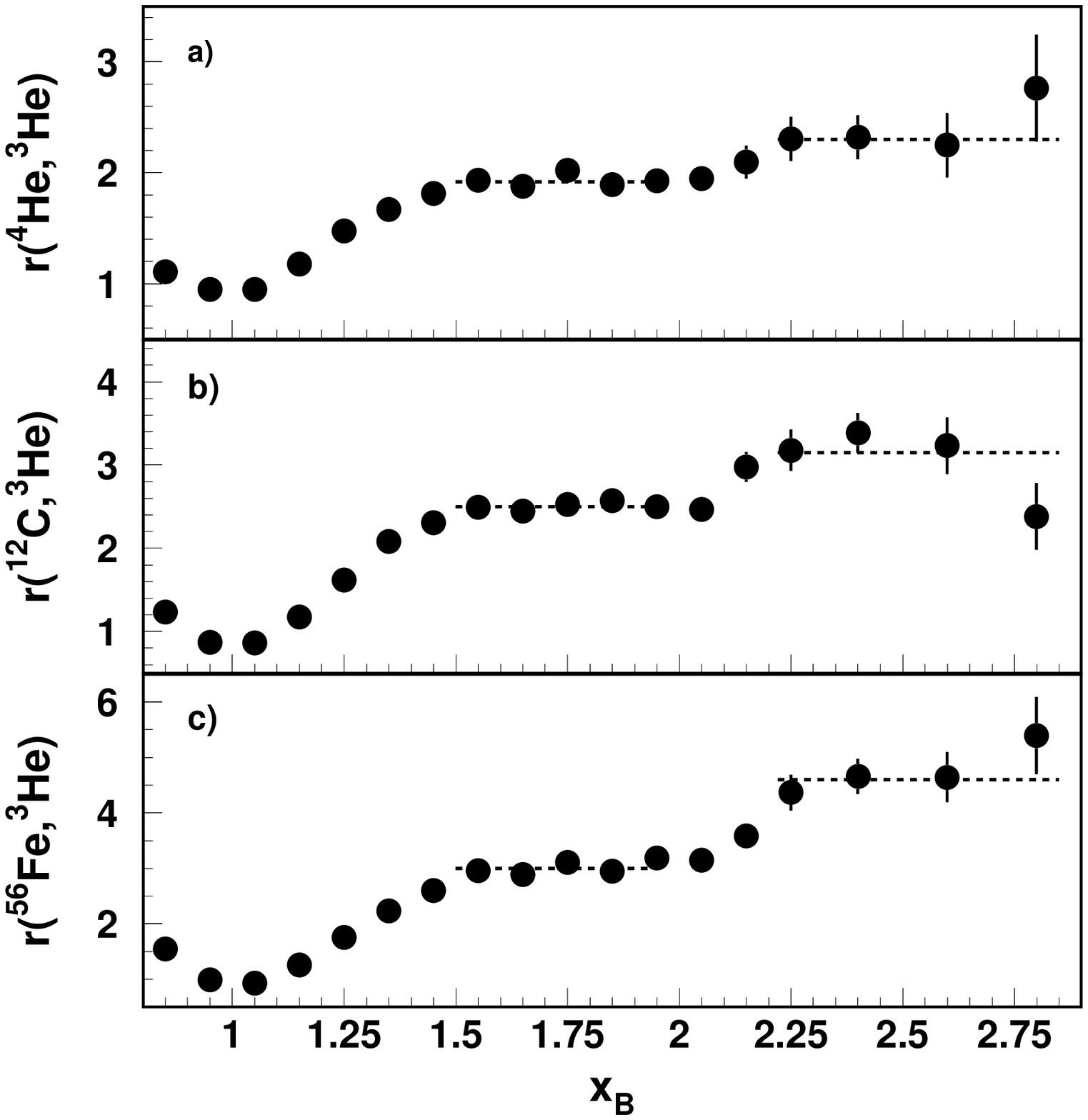}}
\caption{Weighted cross section ratios of a) $^4He$, b) $^{12}C$ and c) $^{56}Fe$ to $^3He$ as a function of $\xb$ for $Q^2> 1.4\:GeV^2$. The horizontal dashed lines indicate the 2N and 3N scaling regions used to calculate the per-nucleon probabilities of 2N and 3N SRC in nucleus $A$ relative to $^3He$. After Ref. \cite{Egiyan2}.}
\label{Fig_inclusive}
\end{figure}
\\The experimental cross section ratio of nucleus $A$ to the nucleus $^3He$
\beq \label{ratio_inclusive}
    r(A,^3He)=\frac{2\sigma_{ep}+\sigma_{en}}{Z\sigma_{ep}+N\sigma_{en}}\:\frac{\sigma(A)}{\sigma(^3He)}
\eeq
plotted versus the Bjorken scaling variable $\xb$, is shown in Fig.   \ref{Fig_inclusive}. In Eq. (\ref{ratio_inclusive}), $\sigma(A)$ and $\sigma(^3He)$ are the $A(e,e')X$ and $^3He(e,e')X$ inclusive cross sections, respectively, and $\sigma_{ep}$ and $\sigma_{en}$ are the elementary elastic electron-proton and electron-neutron scattering cross sections. Three different kinematical regions are clearly seen:
\begin{itemize}
\item $x_{Bj}\lesssim 1.5$: here the shape of the ratio is governed by the different behavior of the magnitude of the quasi elastic peak in different nuclei (higher peaks for light nuclei, and lower peaks for heavy nuclei);
\item $1.5 \lesssim x_{Bj}\lesssim 2$: this plateaux is interpreted as evidence of two-nucleon correlations (2NC), which in complex nuclei and in $^3He$ should differ only by a scale factor;
\item $2 \lesssim x_{Bj}\lesssim 3$: the presence of a second plateaux is interpreted as evidence of 3NC.
\end{itemize}
The relative per-nucleon probabilities $a_2(A,^3He)$ and $a_3(A,^3He)$ of finding, respectively, 2NC and 3NC in nuclei relative to $^3He$, extracted from the experimental results shown in Fig.  \ref{Fig_inclusive}, are listed in Table \ref{a2_a3}. Here also shown are the absolute (per-nucleon) values $a_2(A)$ and $a_3(A)$ of the same probabilities in nucleus $A$, calculated in Ref. \cite{Egiyan2} from
\bey
    a_2(A,^3He)=\frac{a_{2N}(A)}{a_{2N}(^3He)}\\
      a_3(A,^3He)=\frac{a_{3N}(A)}{a_{3N}(^3He)}
\eey
by using realistic wave functions of $^3He$ and Deuteron nuclei.
\begin{table} [!h]
\begin{center}
\begin{tabular}{||*{5}{c|}|}
\hline
\hline
 $A$ & $a_2(A,^3He)$ & $a_{2N}(A)$ $(\%)$ & $a_3(A,^3He)$ & $a_{3N}(A)$ $(\%)$\\
\hline
$3$ &        $1$ &        $8.0 \pm 1.6$      &      $1$        &      $0.18\pm0.06$ \\
\hline
$4$ & $1.93\pm0.01\pm0.03$ & $15.4 \pm 3.2$  & $2.33\pm0.12\pm0.04$ & $0.42\pm0.14$ \\
\hline
$12$ & $2.49\pm0.01\pm0.15$ & $19.8 \pm 4.4$  & $3.18\pm0.14\pm0.19$ & $0.56\pm0.21$ \\
\hline
$56$ & $2.98\pm0.01\pm0.18$ & $23.9 \pm 5.3$  & $4.63\pm0.19\pm0.27$ & $0.83\pm0.27$ \\
\hline
\hline
\end{tabular}
\caption{The relative per-nucleon probabilities $a_2(A,^3He)$ and $a_3(A,^3He)$ of 2NC and 3NC in nucleus $A$ relative to $^3He$, and the absolute value $a_{2N}(A)$ and $a_{3N}(A)$ of the same probabilities in nucleus $A$ (in $\%$), from Ref. \cite{Egiyan2}. Errors shown are statistical and systematic for $a_2$ and $a_3$, and are combined (but systematic dominated) for $a_{2N}$ and $a_{3N}$.}
\label{a2_a3}
\end{center}
\end{table}
\\It should be pointed out that no direct calculations of the inclusive cross section ratio shown in Fig.   \ref{Fig_inclusive} have been performed so far. These calculations would represent a relevant contribution towards the solution of the longstanding problem concerning the role played by SRC in nuclei.
\\In this Thesis, preliminary results of the calculation of the inclusive ratio $r(^{56}Fe/^3He)$ will be given in Chapter \ref{chap:results}, and a new approach \cite{CC} to inclusive electron scattering at high momentum transfer will be illustrated.
%
%
\section{Relevance of short range correlations in various fields}
As shown in Fig.  \ref{Fig_SRC}, SRC can lead, in small portion of time, to local densities in nuclei comparable with those in the core of neutron stars, i.e. up to $\sim 7$ times the average nuclear density $\rho_0 \sim 0.17\:N/fm^3$.
\begin{figure}[!h]
\centerline{\includegraphics[scale=0.6]{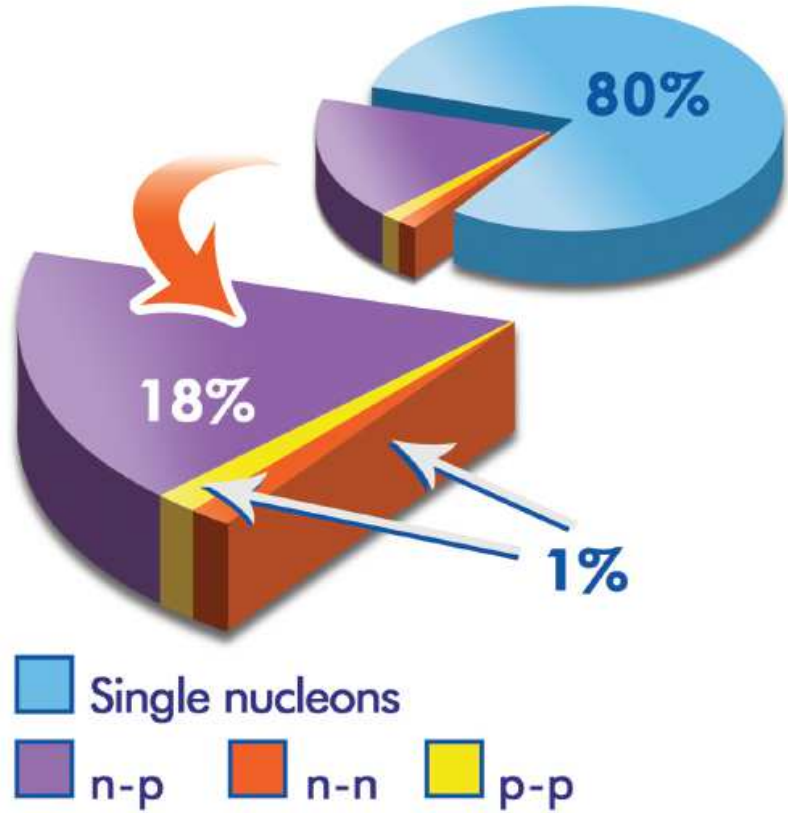}}
\caption{Probability of finding shell model and correlated nucleons in $^{12}C$. After Ref. \cite{Subedi}.}
\label{Fig_pizza}
\end{figure}
Moreover, combining the results of the various experiments we have illustrated, the probability of finding nucleons in $^{12}C$ which move in an average potential has been found to be of the order of $\sim 80\%$, whose $\sim 60$-$70\%$ due to shell model potential, whereas $\sim 10$-$20\%$ due to long range correlations; the remaining $20\%$ represents the SRC contributions, which are dominated by $p$-$n$ correlations, due to the strong tensor force which acts between a $n$-$p$ pair and does not act in $n$-$n$ and $p$-$p$ pairs, as shown in Fig.  \ref{Fig_pizza} \cite{Subedi}.
\\Obtaining information on SRC phenomena in nuclei would have a strong impact on various fields of physics, e.g. in particle-, nuclear- and astro-physics. Let us briefly illustrate some of them:
\begin{itemize}
\item[-]\underline{NN interaction}:  NN interaction processes at large and intermediate distances are well described in terms of meson exchange, as shown in Fig.  \ref{Fig_Wpotetial}, but such a picture does not allow a proper description of the strongly repulsive NN interaction at short distances. Therefore understanding SRC in nuclei should allow a deeper knowledge of the NN force \cite{Weise}.
    \begin{figure}[!h]
\centerline{\includegraphics[scale=0.6,angle=270]{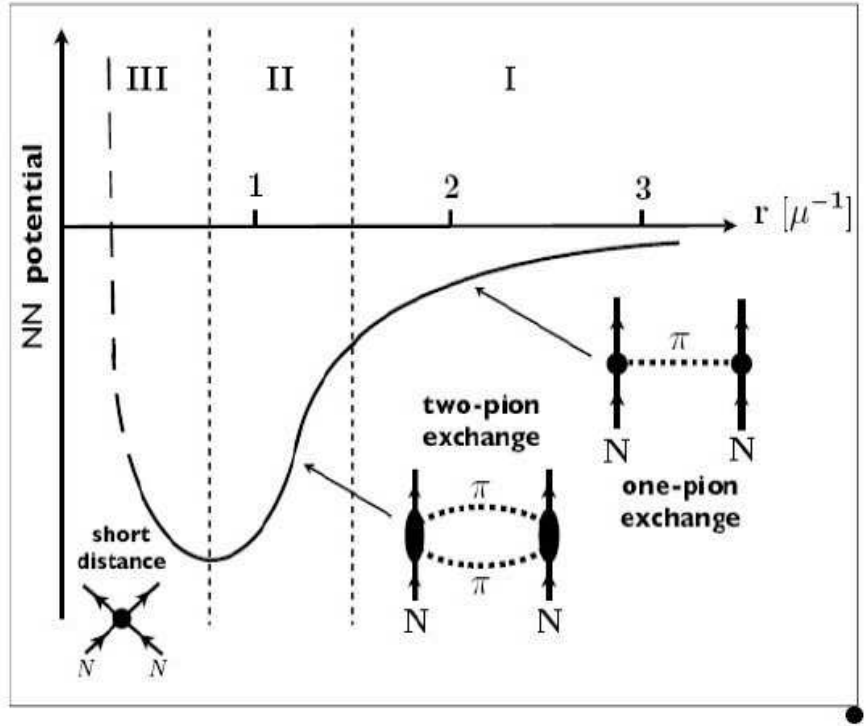}}
\caption{Hierarchy of scales governing the NN interaction. The distance $r$ is given in units of the pion Compton wavelength $\mu^{-1}\simeq 1.4 \: fm$. After Ref. \cite{Weise}.}
\label{Fig_Wpotetial}
\end{figure}
\item[-] \underline{Cold dense nuclear matter}: the density between correlated nucleons is  similar to the one in cold dense nuclear objects, as neutron stars, thus a deep understanding of SRC effects should lead to the formulation of a realistic equation of state for such systems. As shown in Fig.  \ref{Fig1.3}, the core of neutron stars, neglecting SRC, could be well approximated by two independent Fermi gases: the prevalent one constituted by neutrons, and the smallest one by protons. Even if the number of protons is small, the strong $n$-$p$ correlation  acts towards the coupling of the two Fermi liquids, thus affecting the equation of state. Moreover, it becomes worth analyzing the role played by SRC phenomena in the  URCA processes
    \bey
        n \rightarrow p+e+\bar{\nu}_e \\
        e+p \rightarrow n+\nu_e
    \eey
   which involve neutrino $\nu_e$ and antineutrino $\bar{\nu}_e$ emission, leading to changes in the physical properties of the system \cite{FSS}.
    \begin{figure}[!h]
\centerline{\includegraphics[scale=0.6]{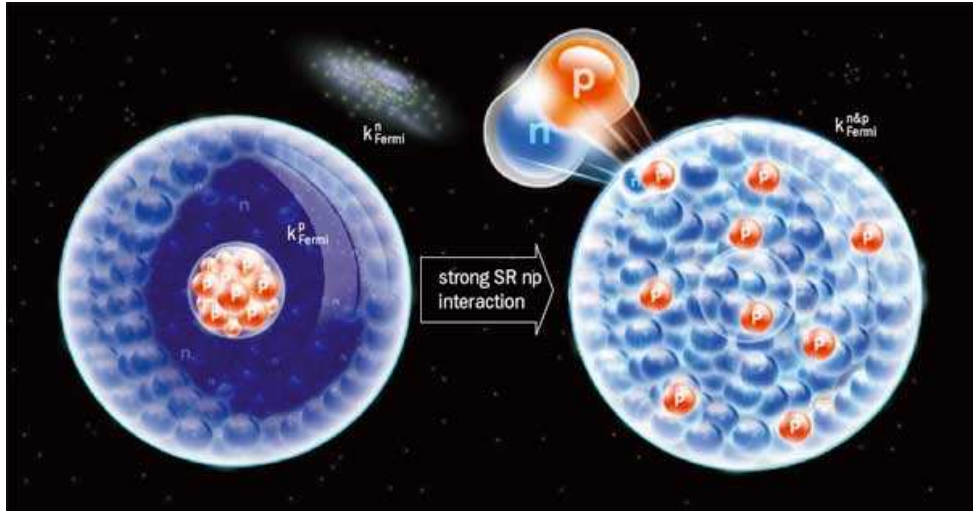}}
\caption{Illustration in momentum space of the momentum distribution, $k_F$, of protons and neutrons in neutron stars. The left side shows the case where the protons and neutrons weakly interact and can be approximated as separate Fermi spheres, with the neutron momentum much greater than the proton momentum. The figure on the right shows how strong neutron proton SRC cause protons to escape their Fermi sphere and have higher momentum then would otherwise be allowed. After Ref. \cite{Cern}.}
\label{Fig1.3}
\end{figure}
\item[-] \underline{Hadron properties}: another important problem in modern nuclear physics is given by the modifications of hadron properties such as masses and radii in the nuclear medium. The nuclear EMC effect, i.e. the change in the inclusive deep inelastic structure function of a nucleus relative to that of a free nucleon, induces such modifications, which are of fundamental importance in understanding implications of quantum chromodynamics (QCD) for nuclear physics. As recently shown, possible modifications of nucleon properties induced by the medium can be better studied by analyzing the short range properties of nuclei \cite{CKFS}.
\item[-] \underline{QCD}: another open problem in modern nuclear and particle physics which could be answered by investigating SRC, concerns the role played by quark and gluon degrees of freedom in nuclear matter at
        \begin{figure}[!h]
\centerline{\includegraphics[scale=0.7]{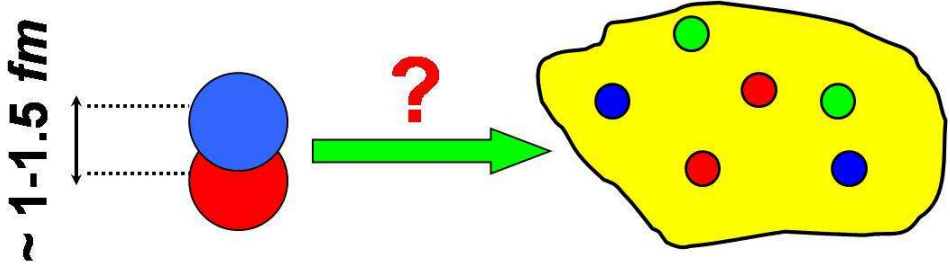}}
\caption{Distances between correlated nucleons are so close that non nucleonic degrees of freedom should reveal themselves.}
\label{Fig_bag}
\end{figure}
distances, as depicted in Fig. \ref{Fig_bag}, which are expected to be most relevant in this interaction range.
\item[-] \underline{High energy scattering processes}: recently it has been shown \cite{Marchino,ACK} that high energy diffraction effects in nuclei are strongly affected by SRC, which appear to be of the same order as Gribov inelastic shadowing \cite{gribov}. An example is given in Fig.  \ref{Fig_Marchino} by the total neutron-Nucleus cross section.
\begin{figure}[!h]
\centerline{\includegraphics[scale=1.3]{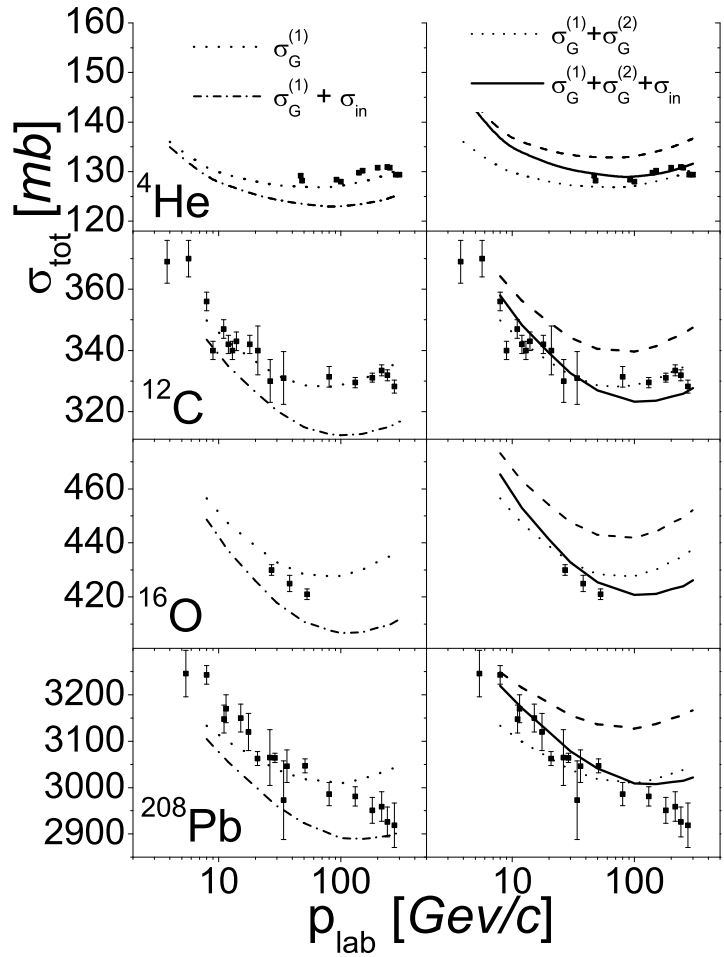}}
\caption{The neutron-nucleus total cross section, for $^{4}He$, $^{12}C$, $^{16}O$ and
  $^{208}Pb$. \textit{Left panel}: the result without the inclusion of correlations;
  dotted curves: shell model contribution;
  dot-dashed curves:
  shell model contribution plus inelastic shadowing effects. \textit{Right panel}:
  results with the inclusion of correlation.
  Dotted curves: shell model contribution;
  dashed curves: shell model contribution plus two-nucleon correlations;
 solid curves:  shell model contribution plus two-nucleon correlations
  plus inelastic shadowing effects. Experimental data from
  \cite{moniz,data02}. After Ref. \cite{Marchino}.}
\label{Fig_Marchino}
\end{figure}
\end{itemize}
A deep comprehension of SRC phenomena is mainly addressed to answer the following questions:
\begin{itemize}
\item What is the percentage of finding correlated nucleons in nuclei?
\item What is the relevance of three nucleon SRC in nuclei?
\item What is the ratio of pp to nn pairs?
\item Are tensor forces relevant for SRC?
\item Are these nucleons different from free nucleons?
\item Which type of analysis should be used in order to obtain information  on SRC?
\end{itemize}
The increasing interest of the scientific community towards the comprehension of SRC in nuclei can be stressed by the large number of conferences and workshops organized in the last year. Some examples are shown in Figs. \ref{Fig_APS}-\ref{Fig_GSI}.
\begin{figure}[!h]
\centerline{\includegraphics[scale=0.65,angle=90]{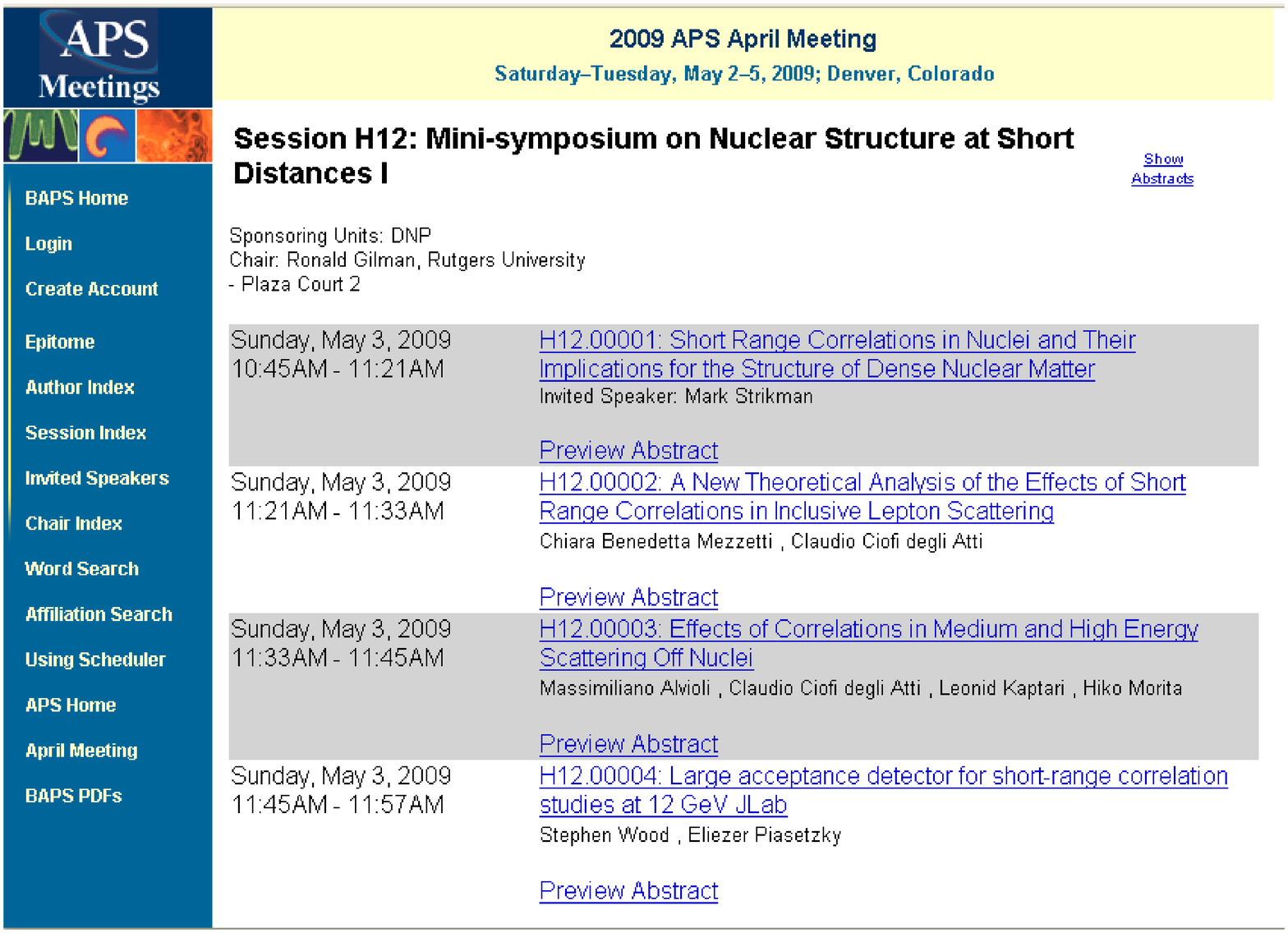}}
\caption{$2009$ \textbf{Mini-symposium on Nuclear Structure at Short Distances I}, held within the \emph{American Physical Society April Meeting}, May $2-5$, $2009$, Denver (USA) \cite{CC,APS}.}
\label{Fig_APS}
\end{figure}
\begin{figure}[!h]
\centerline{\includegraphics[scale=0.7]{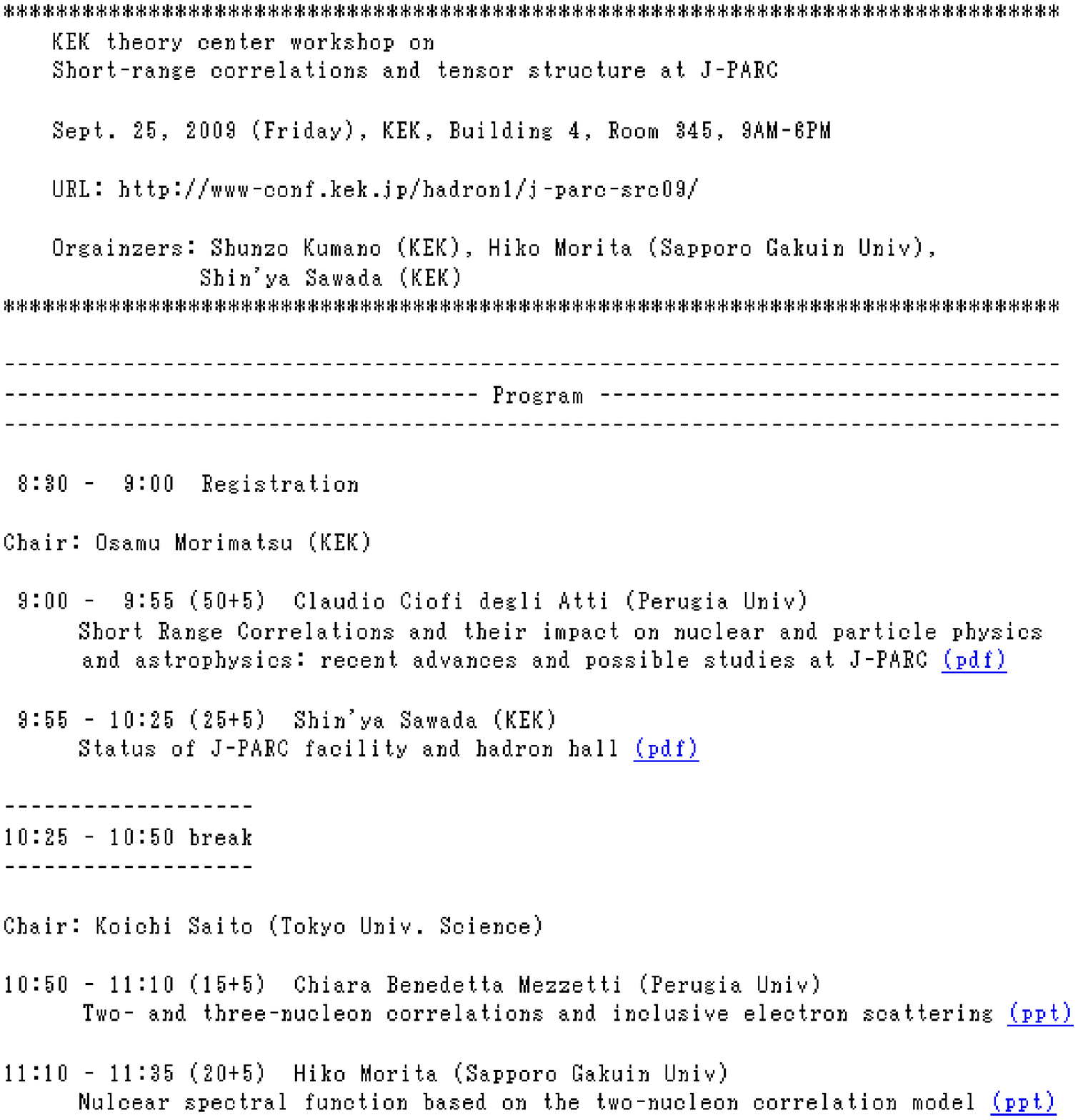}}
\caption{\textbf{Short-range correlation and tensor structure at J-PARC}, September $25$, $2009$, KEK Tsukuba (Japan) \cite{KEK}.}
\label{Fig_KEK}
\end{figure}
\begin{figure}[!h]
\centerline{\includegraphics[scale=0.6, angle=90]{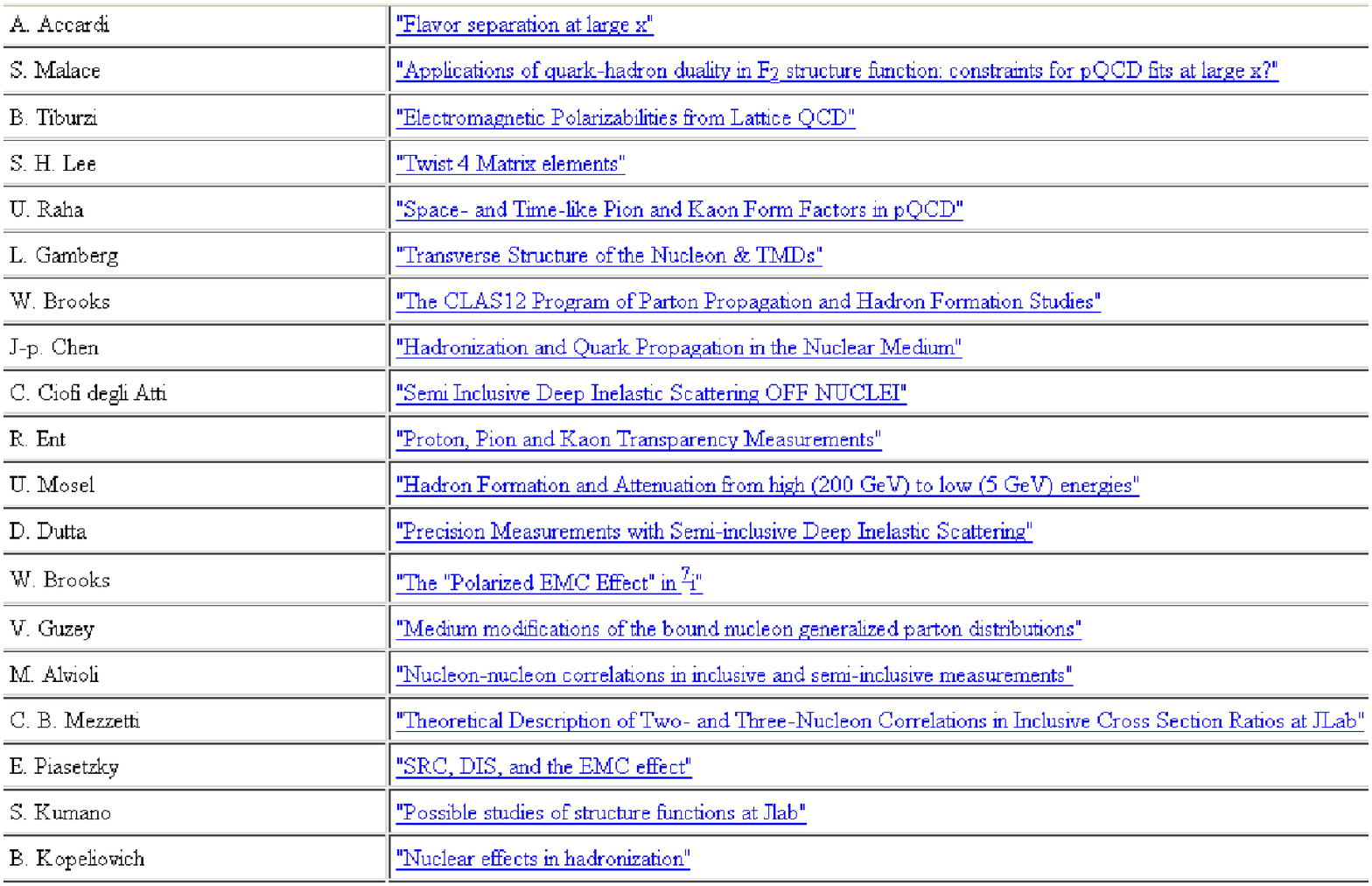}}
\caption{\textbf{The Jefferson Laboratory Upgrade to $12$ GeV}, September $14$ - October $16$ $\&$ October $26$ - November $20$, $2009$, INT Seattle (USA) \cite{INT}.}
\label{Fig_INT}
\end{figure}
\begin{figure}[!h]
\vskip-0.3cm
\centerline{\includegraphics[scale=0.67]{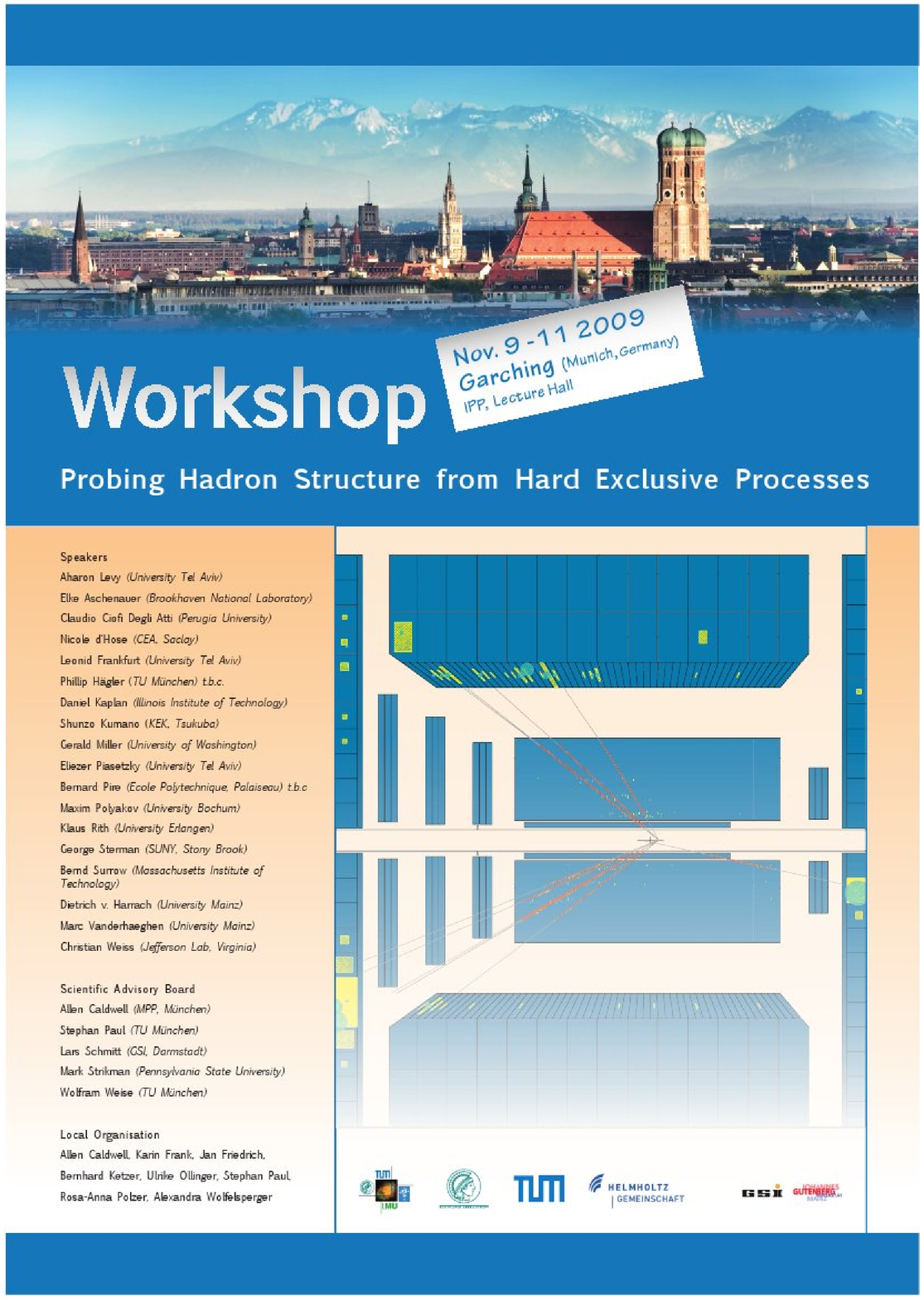}}
\caption{\textbf{Probing Hadron Structure from Hard Exclusive Processes}, November $9-11$, $2009$, Garching (Germany) \cite{GSI}.}
\label{Fig_GSI}
\end{figure} 
\cleardoublepage
\chapter{The spectral function and the nucleon momentum distributions: two- and three-nucleon correlations} \label{ch:PkE_nmd}
\section*{Introduction}
In order to introduce our new approach to the inclusive cross section, it is necessary to recall some basic concepts and general properties about the spectral function and the nucleon momentum distributions.
\section{The spectral function}
The one-body spectral function, which is defined as follows (see e.g. \cite{MB})
\bey \label{PkE}
    P^A(k,E)&=&\frac{1}{2J+1}\sum_{M,\sigma}<\Psi_A\left|a_{\mathbf{k}\sigma}^\dagger\:
    \delta\left(E-\left( H-E_A\right) \right)a_{\mathbf{k}\sigma} \right|\Psi_A>
\eey
represents the joint probability distribution of finding in the target nucleus a nucleon with momentum $k\equiv |\textbf{k}|$ and removal energy
\beq
    E=E_{min}+E_{A-1}^*=m_N+M_{A-1}-M_A+E_{A-1}^* \: .
\eeq
Here, $a_{\mathbf{k}\sigma}^\dagger$ and $a_{\mathbf{k}\sigma}$ are creation and annihilation operators of a nucleon with momentum $\mathbf{k}$ and spin $\sigma$, respectively; $H$ is the intrinsic Hamiltonian of $A$ interacting nucleons; $\Psi_A^0$ is the ground state eigenfunction with eigenvalue $E_A$, total angular momentum $J$ and projection $M$; eventually, $m_N$ is the mass of the nucleon, $M_A$ the mass of the target nucleus, and $M_{A-1}$ the mass of the system $(A-1)$, with excitation energy $E_{A-1}^*$. In what follows, for ease of presentation, the absolute value of a vector $\textbf{a}$ will be indicated as $a \equiv |\textbf{a}|$.
\\By placing in Eq. (\ref{PkE}) the completeness relation
\beq \label{completeness}
    \sum_f |\Psi_{A-1}^f><\Psi_{A-1}^f|=1
\eeq
the spectral function becomes
\bey \label{Pf}
    P^A(k,E)&=&\frac{1}{2J+1}\sum_{M,\sigma}\sum_f\left|<\Psi_{A-1}^f|a_{\mathbf{k}\sigma}|\Psi_A^0>
    \right|^2\:\delta\left(E-\left( E_{A-1}^f-E_A\right) \right)\nonumber\\
    &=&\frac{1}{(2\pi)^3}\: \frac{1}{(2J+1)} \: \sum_{M,\sigma} \sum_f \: \left|\int e^{\imath \textbf{k}\cdot \textbf{z}} G_f^{M\sigma}(\mathbf{z})d\textbf{z}\right|^2 \no \\
    &\times& \:\delta\left(E-(E_{A-1}^f-E_A)\right)
\eey
where $\Psi_{A-1}^f$ is the intrinsic eigenfunction of the final state $f$ of the Hamiltonian $H_{A-1}$ with eigenvalue
\beq
E_{A-1}^f\equiv |E_{A-1}|+E_{A-1}^*
\eeq
and $G_f^{M\sigma}(\mathbf{z})$ is the overlap integral
\beq
    G_f^{M\sigma}(\mathbf{z})=<\chi_\sigma^{1/2},\Psi_{A-1}^f(\textbf{x},\ldots,\textbf{y})|
   \Psi_A(\textbf{x},\ldots,\textbf{y},\textbf{z})>
\eeq
where $\chi_\sigma^{1/2}$ is the two-component Pauli spinor of the nucleon.
Since the set of states $f$ also includes the continuum states of the residual $(A-1)$-nucleon system, the sum over $f$ in Eq. (\ref{Pf}) stands for summation over the discrete states of the $(A-1)$ system and integration over the continuum states. Thus the spectral function exactly includes all final states interactions in the states of the $(A-1)$ system, the only plane wave being that describing the relative motion of the knocked out nucleon and the $(A-1)$ system.
\\The spectral function obeys the normalization condition
\beq \label{norm}
    \int P^A(k,E)d\textbf{k}\:dE=1
\eeq
and, owing to Eq. (\ref{PkE}), it can be written as follows
\beq \label{decomposition}
    P^A(k,E)=P_0^A(k,E)+P_1^A(k,E)
\eeq
where the contributions from different final nuclear states have been explicitly separated out, namely
\bey \label{PkE0}
    P_0^A(k,E)&=& \frac{1}{(2\pi)^3}\: \frac{1}{(2J+1)} \: \sum_{M,\sigma} \sum_{f<c} \: \left|\int e^{\imath \textbf{k}\cdot \textbf{r}} G_\alpha^{M\sigma}(\textbf{r})d\textbf{r}\right|^2 \no \\
    &\times& \:\delta\left(E-(E_{A-1}^f-E_A)\right)
\eey
includes the ground and one-hole states of the ($A-1$)-nucleon system, and
\bey \label{PkE1}
    P_1^A(k,E) &=& \frac{1}{(2\pi)^3}\: \frac{1}{(2J+1)} \: \sum_{M,\sigma} \sum_{f > c} \: \left|\int e^{\imath \textbf{k}\cdot \textbf{r}} G_\alpha^{M\sigma}(\textbf{r})d\textbf{r}\right|^2     \no \\
    &\times& \:\delta\left(E-(E_{A-1}^f-E_A)\right)
\eey
more complex highly excited configurations generated in the target ground state by NN correlations. Here $f<c$ ($f>c$) means that all final states of the residual system below (above) the continuum threshold are considered.
\section{The nucleon momentum distribution}\label{sec:nmd}
By definition, the calculation of the spectral function requires the knowledge of the whole set of final states $\Psi_{A-1}^f$. The calculation of the nucleon momentum distribution requires, on the contrary, only the knowledge of the ground state wave function. As a  matter of fact, the momentum distribution is defined as follows
\beq
    n^A(k)=\frac{1}{2\pi^2}\: \int d\textbf{z} \: d\textbf{z}' \: e^{\imath \: \textbf{k}\cdot (\textbf{z}-\textbf{z}')} \: \rho(\textbf{z},\textbf{z}')
\eeq
where
\beq \label{rho}
    \rho(\textbf{z},\textbf{z}')=\int d\textbf{x}\ldots d\textbf{y}\:
    \left[ \Psi_A^0(\textbf{x}\ldots \textbf{y},\textbf{z}) \right]^* \Psi_A^0(\textbf{x}\ldots \textbf{y},\textbf{z}')
\eeq
is the non diagonal one-body density matrix.
\\The spectral function and the momentum distributions are related by the momentum sum rule
\beq \label{sumrule}
    n^A(k)=4\pi\:\int_{E_{min}}^{+\infty} P^A(k,E)\,dE
\eeq
which is readily obtained by inserting the completeness relation (\ref{completeness}) in Eq. (\ref{rho}).
\\Within the decomposition rule given by Eq. (\ref{decomposition}), one has
\beq
      n^A(k)=n_0^A(k)+n_1^A(k)
\eeq
with
\beq
    n_0^A(k)=4\pi\:\int_{E_{min}}^{+\infty} P_0^A(k,E)\,dE =
    \frac{1}{2\pi^2}\:\frac{1}{(2J+1)}\:\sum_{M,\sigma}\left| \int e^{\imath \textbf{k}\cdot \textbf{z}}\: G_0^{M\sigma}(\textbf{z})d\textbf{z}\right|^2
\eeq
and
\beq
    n_1^A(k)=4\pi\:\int_{E_{min}}^{+\infty} P_1^A(k,E)\,dE
    =
    \frac{1}{2\pi^2}\:\frac{1}{(2J+1)}\:\sum_{M,\sigma}\sum_{f\neq0}\left| \int e^{\imath \textbf{k}\cdot \textbf{z}}\: G_f^{M\sigma}(\textbf{z})d\textbf{z}\right|^2\:.
\eeq
The integral of the nucleon momentum distributions yields the spectroscopic factors (or occupation probabilities)
\beq \label{S0}
    S_0 \equiv \int_0^{+\infty} dk\: k^2 \: n_0^A(k)  
\eeq
and
\beq \label{S1}
    S_1 \equiv \int_0^{+\infty} dk\: k^2 \: n_1^A(k)
\eeq
which, owing to the normalization condition
\beq \label{nmdnorm}
    \int_0^{+\infty} dk\: k^2 \: n^A(k)=1
\eeq
satisfy the relation
\beq
    S_0+S_1=1 \: .
\eeq
Note that, in the independent particle shell model description, as will be discussed in more detail in $\S$\ref{sec:P0}, the spectral function $P_0^A(k,E)$  can be written as follows
\beq \label{P0}
    P^A_0(k,E)=\frac{1}{4\pi A}\sum\displaylimits_{\alpha} A_\alpha n_\alpha(k)\,\delta(E-|\epsilon_\alpha|)
\eeq
where
\beq
    \int_0^{+\infty} k^2\:dk\: n_\alpha(k)=1
\eeq
and $A_\alpha$ is the number of nucleons in the state $\alpha$ ($A=\sum_\alpha \:A_\alpha$) with removal energy $\epsilon_\alpha$ and nucleon momentum distribution $n_\alpha(k)$.
\begin{figure}[!h]
\centerline{\includegraphics[scale=0.9]{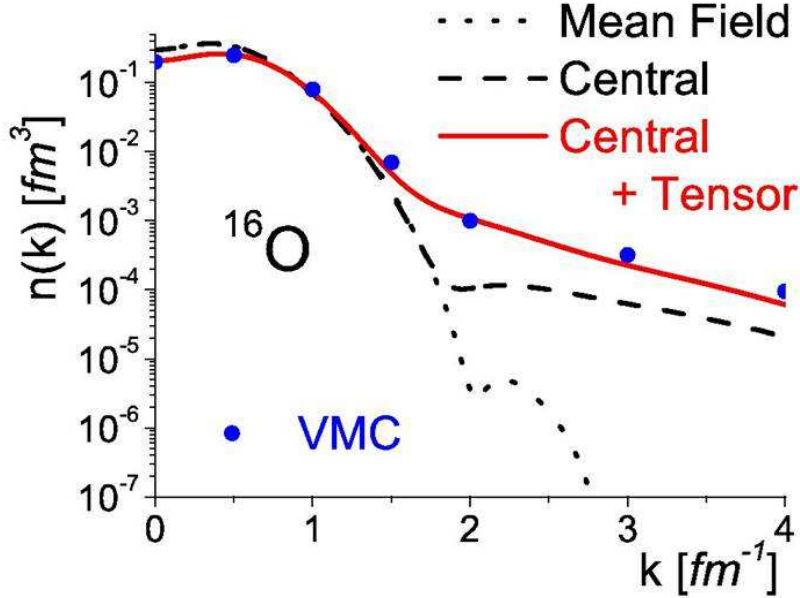}}
\caption{The nucleon momentum distribution $n(k)\equiv n^{16}(k)$ \emph{vs.} the momentum $k$. The dotted line is the mean field contribution, the dashed line the one arising from central forces, and the long line includes both central and tensor forces. The full dotts are the results of the Variational Monte Carlo approach \cite{Wiringa}. After Ref. \cite{ACM}.}
\label{Fig_nmdcorre}
\end{figure}
\noindent
\\In Fig. \ref{Fig_nmdcorre}, the correlated nucleon momentum distributions resulting from the cluster expansion techniques discussed in \S \ref{sec:realistic_manybody} \cite{ACM} are compared with the mean field component and the variational Monte Carlo calculations \cite{Wiringa}, in case of $^{16}O$. It can be clearly seen that the mean field distributions almost totally exhaust the low momentum part of $n^A(k)$, and drop to zero at $k\geq 1.5-2\:fm^{-1}$; on the contrary, because of NN correlations, the high momentum tails are entirely governed by tensor forces, acting in $T=0$ and $S=1$ states which, at high momenta, are several orders of magnitude larger than the predictions from shell model calculations.
\begin{figure}[!h]
\centerline{\includegraphics[scale=1.2]{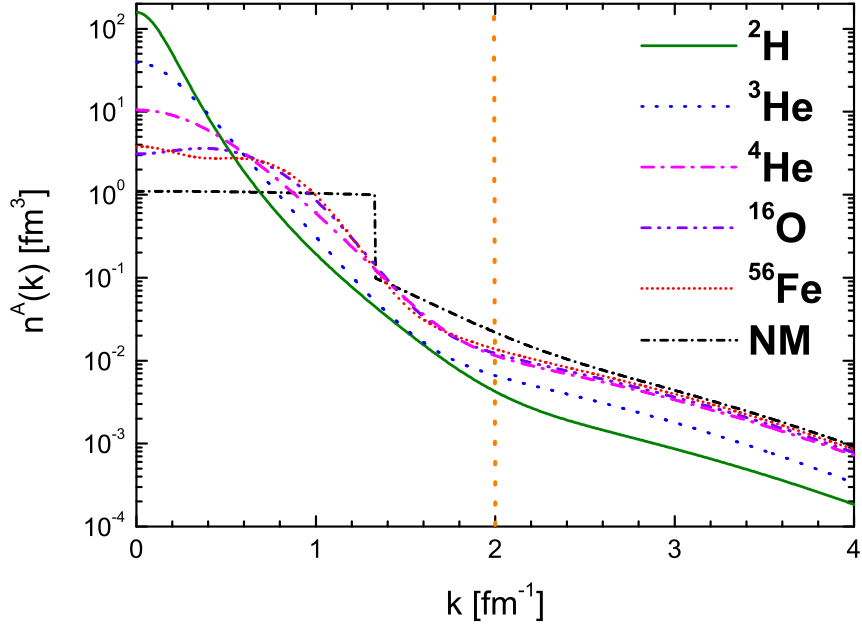}}
\vskip -0.3cm \caption{The nucleon momentum distributions $n^A(k)$ for
nuclei ranging from $^2H$ to Nuclear Matter ($NM$).
It can be seen that, at high values of the momentum $k$, $n^A(k)$ can be considered as a
rescaled version of the momentum distributions of $^2H$. After Ref. \cite{ciosim,urbana}.}
\label{Fig_nmd}
\end{figure}
\begin{figure}[!h]
\centerline{\includegraphics[scale=1.2]{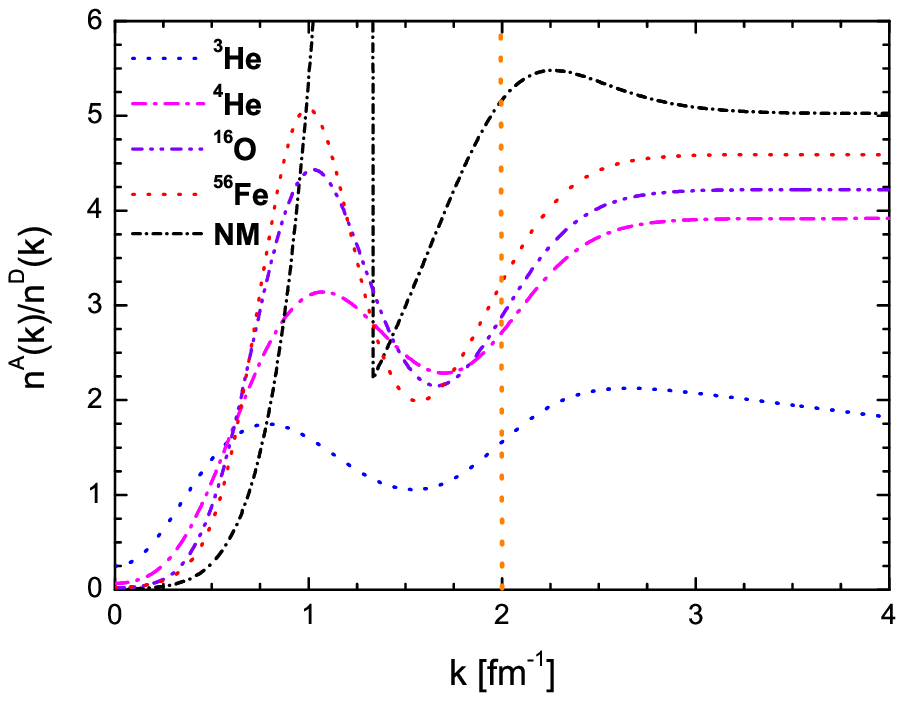}}
\vskip -0.3cm \caption{The ratio $n^A(k)$ to the Deuteron momentum distribution $n^D(k)$. After Ref. \cite{ciosim}.}\label{Fig_nmdratio}
\end{figure}
Moreover, as shown in Fig.  \ref{Fig_nmd} and, more clearly, in Fig. \ref{Fig_nmdratio}, apart from a scaling factor, the nucleon momentum distributions $n^A(k)$ seem to be almost independent of the atomic weight $A$; thus, a simple model for the nucleon momentum distributions $n^A(k)$ could be written as follows \cite{ciosim}
\bey
  \label{n0SM}
   n_0^A(k) &=& \frac{1}{4\pi A}\: \sum_{\alpha<\alpha_F} A_\alpha\:\widetilde{n}_{\alpha}(k)  \\
   \label{nkscale}
    n_1^A(k) &=& C^A\:n^D(k)
\eey
where, because of the effects of correlations, which depopulate states below the Fermi sea, one has
\beq
    \int_0^{+\infty} k^2\:dk\: n_0^A(k)<1
\eeq
and
\beq
    \int_0^{+\infty} k^2\:dk\: n_1^A(k)>0 \:.
\eeq
Here the low momentum component, owing to (\ref{sumrule}), is nothing but Eq. (\ref{P0}) integrated over the removal energy $E$, with the shell model momentum distribution $n^A_{\alpha}$ replaced by the modified ones which takes into account the effects of correlations; on the contrary, the high momentum part is nothing but the Deuteron nucleon momentum distribution $n^D(k)$ rescaled by a constant $C^A$, which depends on the nucleus $A$ under consideration, and whose values are listed in Table \ref{CA}. We call such a behavior \emph{Deuteron scaling}.
\begin{table} [!h]
\large
\begin{center}
\begin{tabular}{||*{9}{c|}|}
\hline
\hline
 $Nucleus$ & $^3He$ & $^4He$ & $^{12}C$ & $^{16}O$ & $^{40}Ca$ & $^{56}Fe$ & $^{208}Pb$ & $NM$ \\
\hline
  $C^A$    & $1.9$  & $3.8$ &   $4.0$ &    $4.2$ &    $4.4$ &     $4.5$ &    $4.8$ &     $4.9$ \\
\hline
\hline
\end{tabular}
\caption{Values of the constant $C^A$ appearing in Eq. (\ref{nkscale}).}
\label{CA}
\end{center}
\end{table}
\\\emph{Deuteron scaling} is even more pronounced in the the two-nucleon momentum distributions, defined as follows
\beq
    n^{N_1\:N_2}(\textbf{k}_1,\textbf{k}_2)=\frac{1}{(2\pi)^6}\:\int\:d\textbf{z}_1\:d\textbf{z}_2\:d\textbf{z}_1'
    d\textbf{z}_2'\:e^{\imath\;\textbf{k}_1(\textbf{z}_1-\textbf{z}_1')}\:
    e^{\imath\;\textbf{k}_2(\textbf{z}_2-\textbf{z}_2')}\:\rho^{N_1\:N_2}\left(
    \textbf{z}_1,\textbf{z}_2,\textbf{z}_1',\textbf{z}_2'\right)
\eeq
which have recently been calculated in Refs. \cite{Schiavilla,ACM}.
\\The knowledge of $n^{N_1\:N_2}(\textbf{k}_1,\textbf{k}_2)$ allows one to calculate the relative and center of mass (CM) momentum distributions of a two-nucleon pair $N_1\:N_2$, i.e.
\bey\label{nrel}
    n_{rel}^{N_1\:N_2}(\textbf{k}_{rel})=\int d\textbf{k}_{CM}\: n^{N_1\:N_2}
    \left(\textbf{k}_{rel}+\frac{\textbf{k}_{CM}}{2}, -\textbf{k}_{rel}+\frac{\textbf{k}_{CM}}{2}  \right) \\ \label{nCM}
    n_{CM}^{N_1\:N_2}(\textbf{k}_{CM})=\int d\textbf{k}_{rel}\: n^{N_1\:N_2}
    \left(\textbf{k}_{rel}+\frac{\textbf{k}_{CM}}{2}, -\textbf{k}_{rel}+\frac{\textbf{k}_{CM}}{2}  \right)
\eey
where
\beq
    \textbf{k}_{rel}\equiv \frac{\textbf{k}_1-\textbf{k}_2}{2} \qquad \qquad
    \textbf{k}_{CM}\equiv \textbf{k}_1+\textbf{k}_2
\eeq
are the relative and CM momenta of the pair $N_1$-$N_2$, respectively, with $\textbf{k}_1$ and $\textbf{k}_2$ being measured from the CM of the system, and
\beq
    \rho^{N_1\:N_2}\left(
    \textbf{z}_1,\textbf{z}_2,\textbf{z}_1',\textbf{z}_2'\right)\equiv
    \int\:d\textbf{x}\ldots\textbf{y}\:
    \left[\Psi_A^0\left(\textbf{x}\ldots \textbf{y},\textbf{z}_1,\textbf{z}_2\right)\right]^*
    \: \Psi_A^0\left(\textbf{x}\ldots \textbf{y},\textbf{z}_1',\textbf{z}_2'\right)
\eeq
is the off-diagonal two-body density matrix.
\\\emph{Deuteron scaling} can be clearly seen in Figs. \ref{Fig_Rocco}, \ref{Fig_Rocco2} and \ref{Fig_2nmd}, where
\begin{figure}
\begin{center}
\includegraphics[scale=0.6,angle=270]{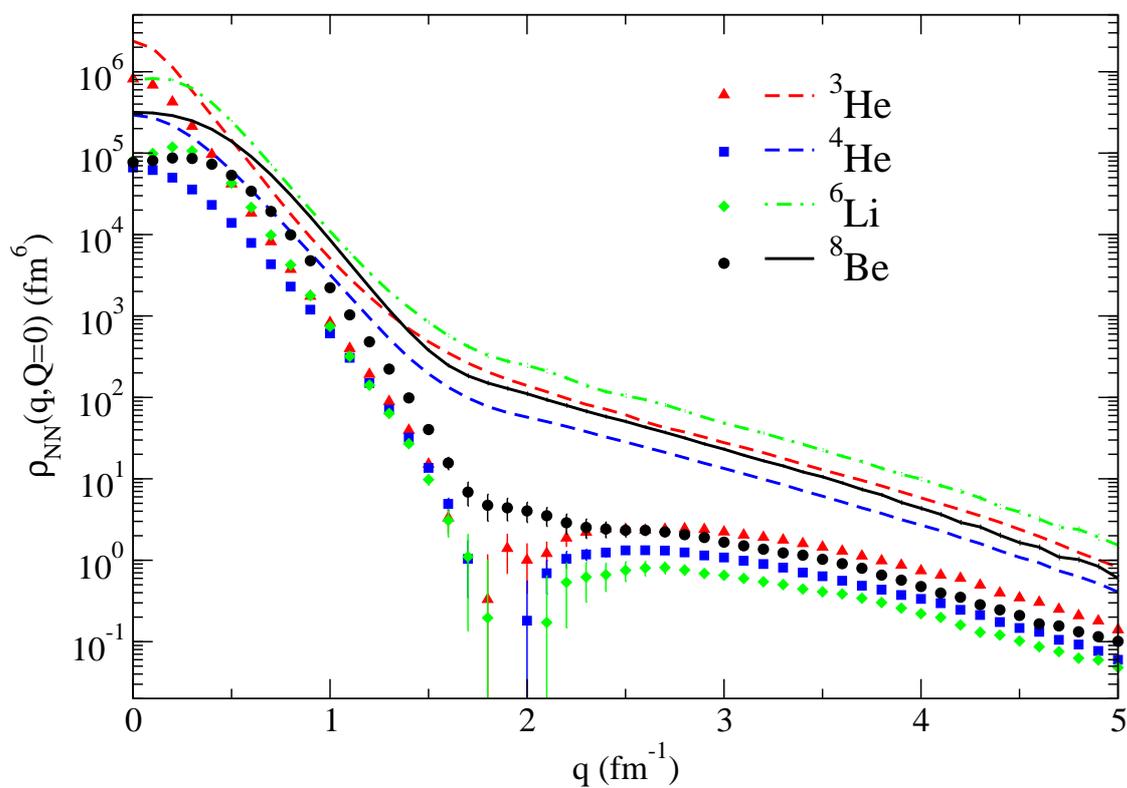}
\caption{The $n$-$p$ (lines) and $p$-$p$ (symbols) two-nucleon momentum distributions $\rho_{NN}(q,Q)\equiv n(k_{rel},k_{CM})$ in various nuclei as functions of the relative momentum $q\equiv k_{rel}$ at vanishing total pair momentum $Q\equiv k_{CM}$. After Ref. \cite{Schiavilla}.}
\label{Fig_Rocco}
\end{center}
\end{figure}
\begin{figure}
\begin{center}
\includegraphics[scale=0.6,angle=270]{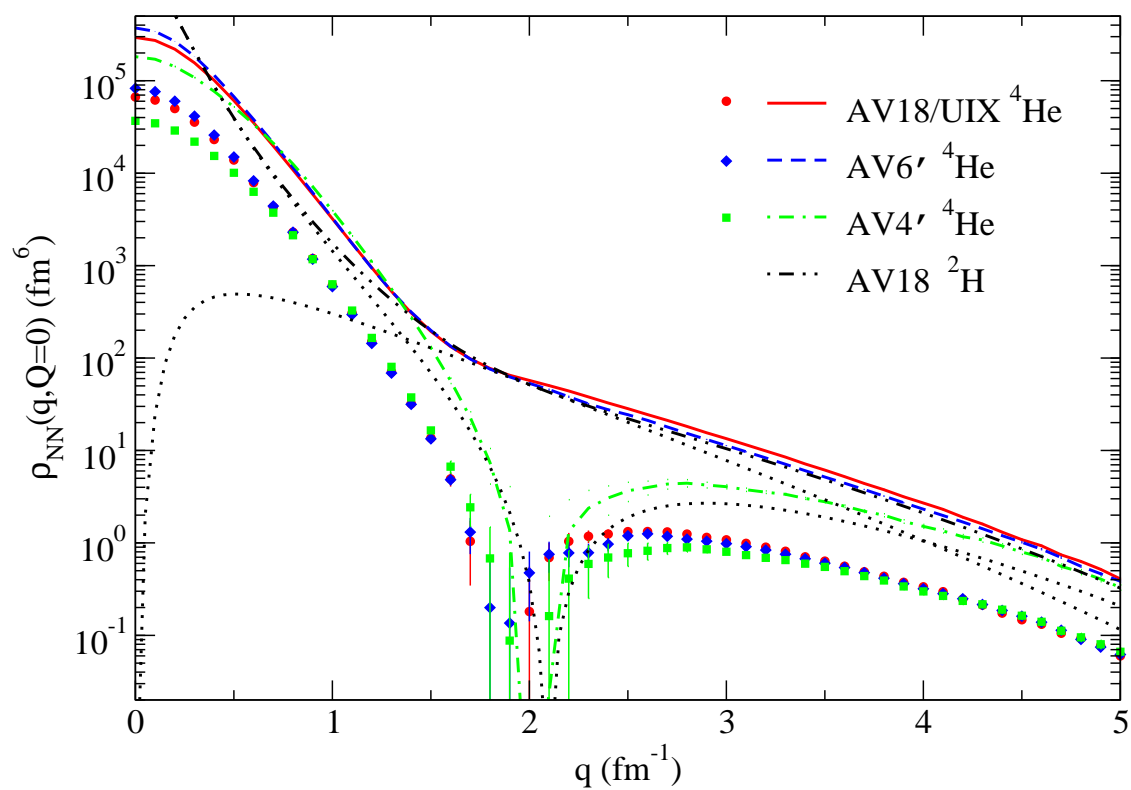}
\caption{The $n$-$p$ (lines) and $p$-$p$ (symbols) momentum distributions in $^4He$ obtained with different Hamiltonians. Also shown is the scaled momentum distribution for the AV18 Deuteron; its separate S- and D-wave components are shown by dotted lines. After Ref. \cite{Schiavilla}.}
\label{Fig_Rocco2}
\end{center}
\end{figure}
\begin{figure}
\begin{center}
\includegraphics[scale=0.9]{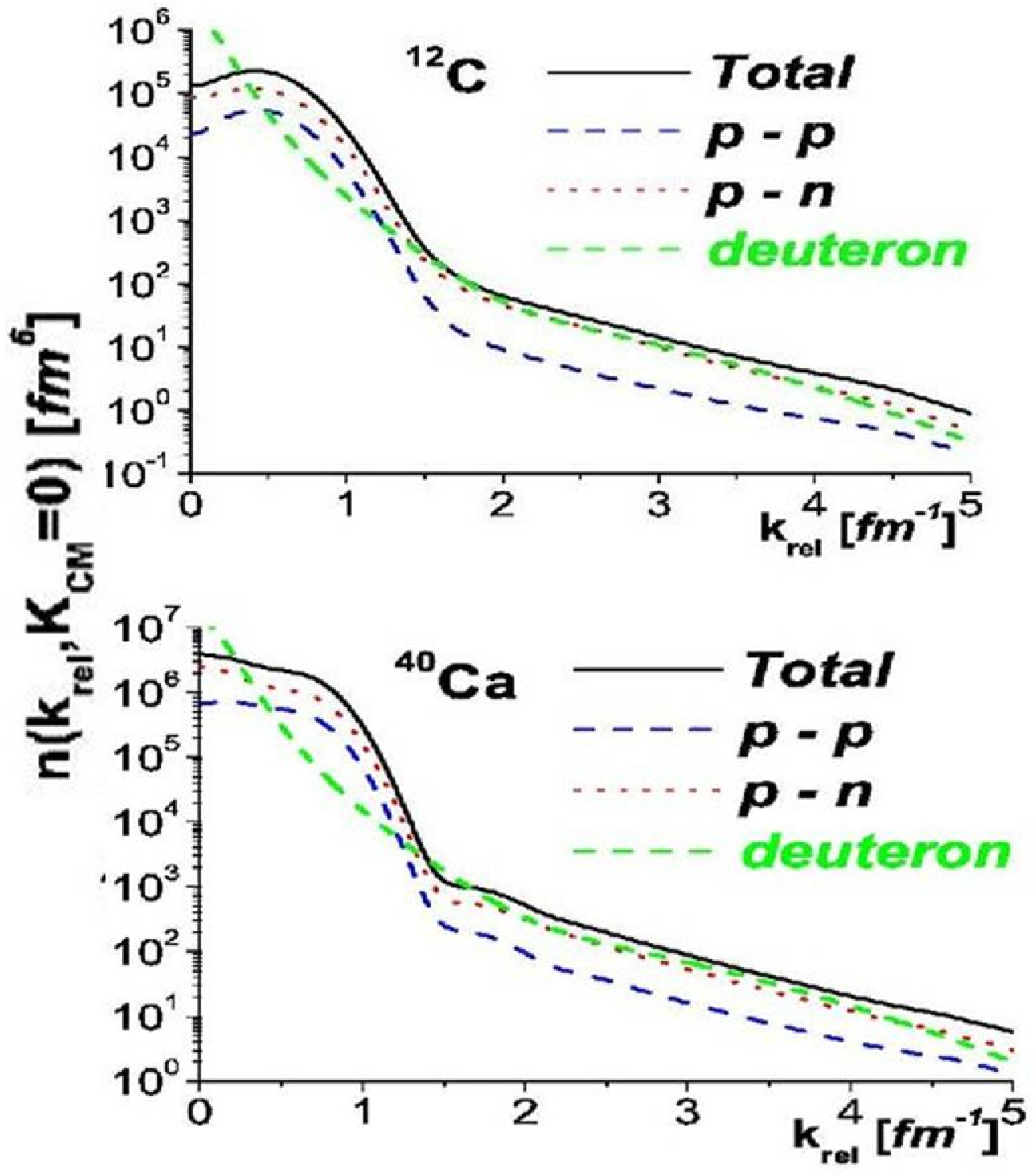}
\end{center}
\caption{The two-nucleon momentum distributions $n(k_{rel},k_{CM}=0)$ \emph{vs.} the momentum $k_{rel}$, for $^{12}C$ (upper panel) and $^{40}Ca$ (lower panel). After Ref. \cite{ACM2}.}
\label{Fig_2nmd}
\end{figure}
the quantity $n(\textbf{k}_{rel},\textbf{k}_{CM}=0)$, describing back-to-back nucleons, is plotted versus the relative momentum $\textbf{k}_{rel}$. These figures clearly illustrate the dominant role of the tensor forces in producing a substantial difference between $p$-$n$ and $p$-$p$ two-nucleon momentum distributions, both in few-nucleon systems and complex nuclei.
\\In Ref. \cite{ACM2}, the following ratio
\beq \label{proba}
    P_{pN}=\frac{\int_a^b\:dk_{rel}k_{rel}^2\:n_{pN}(k_{rel},0)}
    {\int_a^b\:dk_{rel}k_{rel}^2\:\left[n_{pp}(k_{rel},0)+n_{pn}(k_{rel},0)\right]}
\eeq
which represents the percentage probability of finding a $p$-$n$ pair in the nucleus $A$, has been calculated. The results, listed in Table \ref{pp_pn}, show that $P_{pN}$ is proportional to the percentage of $p$-$N$ pairs, when integration runs over the whole range of $k_{rel}$; on the contrary,
\begin{table} [!h]
\begin{center}
\begin{tabular}{||*{5}{c|}|}
\hline
\hline
 $A$ & $P_{pp} \: (\%)$ $[0,\infty]$ & $P_{pn} \: (\%)$  $[0,\infty]$ & $P_{pp} \: (\%)$ $[1.5,3.0]$ & $P_{pn} \: (\%)$ $[1.5,3]$ \\
\hline
$4$ & $19.7$ & $81.3$ & $2.9$ & $97.1$   \\
\hline
$12$ & $30.6$ & $69.4$ & $13.3$ & $86.7$   \\
\hline
$16$ & $29.5$ & $70.5$ & $10.8$ & $89.2$   \\
\hline
$40$ & $31.0$ & $69.0$ & $24.0$ & $76.0$   \\
\hline
\hline
\end{tabular}
\caption{The $p$-$p$ and $p$-$n$ percentage probability given by Eq. (\ref{proba}) evaluated in the momentum range shown in square brackets in $fm^{-1}$. After Ref. \cite{Trieste}.}
\label{pp_pn}
\end{center}
\end{table}
when integration is limited to the correlation region, the percentage of $p$-$n$ pairs results much larger than that of $p$-$p$ pairs, which is a clear consequence of the effects of the tensor forces acting between a proton and a neutron.
\subsection{The saturation of the momentum sum rule}
\begin{figure}[!h]
\centerline{\includegraphics[scale=0.5]{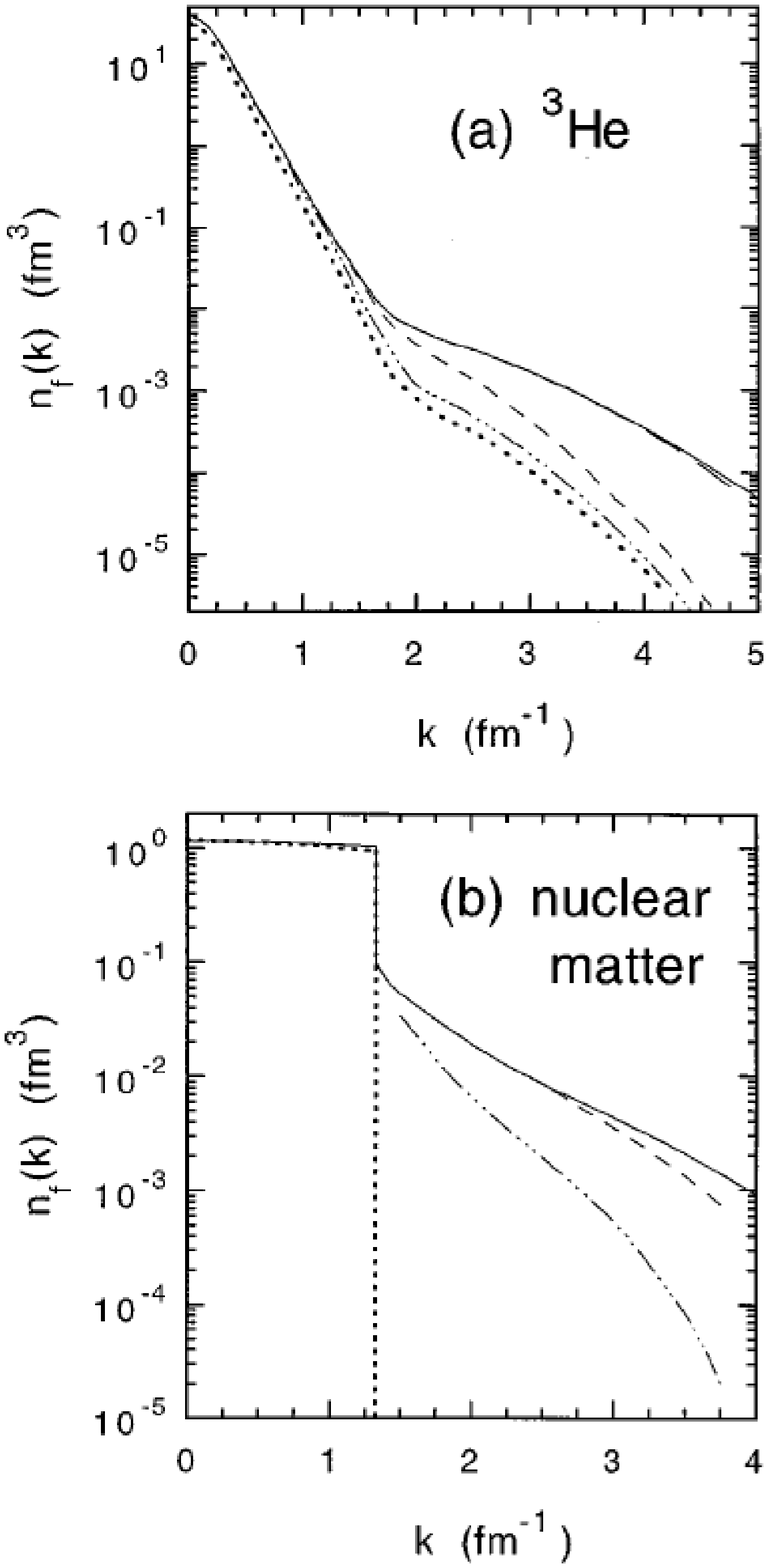}}
\caption{The saturation of the momentum sum rule in $^3He$ (a) and infinite nuclear matter (b). The dotted and solid lines correspond to the momentum distribution $n_0^A(k)$ and to the total momentum distribution $n^A(k)$, respectively. In case of $^3He$ the dot-dashed, dashed, and long dashed lines correspond to Eq. (\ref{saturation}) calculated at $E_f=17.75,55.5,305.5\:MeV$, whereas for nuclear matter the dot-dashed and dashed lines correspond to $E_f=100$ and $300\:MeV$, respectively. The spectral function for $^3He$ is from Ref. \cite{CPS2}, and for nuclear matter from  Ref. \cite{BFF,BFF2}. After Ref. \cite{ciosim}.}\label{Fig_satura}
\end{figure}
A relevant relationship between high momentum and high removal energy components can be obtained by considering the partial momentum distribution \cite{CPS2}
\beq \label{saturation}
    n_f^A(k)\equiv 4\pi\:\int_{E_{min}}^{E_f}\:dE\:P^A(k,E)
\eeq
where the upper limit of integration $E_f$ can be varied from $E_{min}$ to $\infty$. By definition, Eq. (\ref{saturation}) represents that part of $n^A(k)$ which is due to final $(A-1)$-nucleon states with $E \leq E_f$. In the limit $E_f \rightarrow \infty$ one gets
\beq
    n_f^A(k) \rightarrow n^A(k)
\eeq
and the momentum sum rule given by Eq. (\ref{sumrule}) is recovered. Thus the behavior of $n_f^A(k)$ as a function of $E_f$ provides information on the saturation of the momentum sum rule and the relevance of binding effects.
\\The saturation of the momentum sum rule for $^3He$  \cite{CPS2} and Nuclear Matter \cite{CLS} is shown in Fig. \ref{Fig_satura}, using realistic spectral functions. It can be clearly seen that, at $k<1.5\:fm^{-1}$, the momentum sum rule is saturated already at values of $E_f$ very close to $E_{min}$, whereas at $k>1.5\:fm^{-1}$, on the contrary, the momentum sum rule is saturated only when high values of $E_f$ are considered. This can be explained in terms of the spectral function $P^A(k,E)$ appearing in Eq. (\ref{saturation}) which, at low momenta, is dominated by its component $P_0^A(k,E)$, whose strength is almost totally concentrated at low values of the removal energy; at high momenta, on the contrary, it depends upon $P_1^A(k,E)$,  spread, for a given $k$, over a wide range of values of $E$.
\subsection{Probabilities of independent particle and correlated momentum components}
Let us define
\beq \label{pro0}
    \widetilde{S}_0\equiv \int_{k^*}^{+\infty}\:dk\:k^2\:n_0^A(k)
\eeq
and
\beq\label{pro1}
    \widetilde{S}_1\equiv \int_{k^*}^{+\infty}\:dk\: k^2\:n_1^A(k)
\eeq
as the probability of finding independent particle and correlated momentum components in nucleon momentum distributions, with $k^*$ ranging from $0$ to $\infty$. When $k^*=0$, the spectroscopic factors given by Eqs. (\ref{S0}) and (\ref{S1}) are recovered.
\\The values of these two quantities, calculated for different values of $k^*$ and for different nuclei, are listed in Table \ref{prob}.
\begin{table} [!h]
\normalsize
\begin{center}
\begin{tabular}{|c||c|c||c|c||c|c|}
\hline
&\multicolumn{2}{c|}{$^4He$}&\multicolumn{2}{c|}{$^{12}C$}&\multicolumn{2}{c|}{$^{56}Fe$}\\
\hline
 $k^*$ [fm$^{-1}$]& $\widetilde{S}_0$ & $\widetilde{S}_1$ & $\widetilde{S}_0$ & $\widetilde{S}_1$ & $\widetilde{S}_0$ & $\widetilde{S}_1$\\
\hline
$0.00$              & $0.80$ & $0.20$   & $0.80$ & $0.20$           & $0.80$ & $0.20$\\
\hline
$0.25$              & $0.75$ & $0.19$   & $0.78$ & $0.19$           & $0.78$ & $0.19$\\
\hline
$0.50$              & $0.55$ & $0.17$    & $0.69$ & $0.18$          & $0.69$ & $0.19$\\
\hline
$0.75$               & $0.31$ & $0.14$   & $0.48$ & $0.15$           & $0.45$ & $0.17$\\
\hline
$1.00$               & $0.14$ & $0.12$   & $0.24$ & $0.13$           & $0.17$ & $0.15$\\
\hline
$1.50$ & $1.5\:10^{-2}$ & $8.3 \:10^{-2}$ & $2.1\:10^{-2}$ & $8.5 \:10^{-2}$ & $2.7\:10^{-3}$ & $0.11$ \\
\hline
$2.00$ & $7.4\:10^{-4}$ & $6.1\:10^{-2}$  & $6.2\:10^{-4}$ & $6.1 \:10^{-2}$ & $2.5\:10^{-6}$ & $7.5 \:10^{-2}$ \\
\hline
\hline
\end{tabular}
\caption{Eqs. (\ref{pro0}) and (\ref{pro1}) calculated at different values of the momentum $k^*$, for $^{4}He$, $^{12}C$ and $^{56}Fe$.}
\label{prob}
\end{center}
\end{table}
It can be seen that:
\begin{itemize}
\item when the whole range of the momentum $k$ is considered, the probability of finding independent particle low momentum shell model components is much larger than the probability to find correlated high momentum components;
\item for $0.75 \lesssim k^* \lesssim 1.50 \: fm^{-1}$ the contribution from shell model and correlated nucleons is comparable;
\item for $k^* \gtrsim 1.50 \: fm^{-1}$ the probability of finding shell model nucleons is still different from zero, but orders of magnitude less than the probability due to correlated nucleons.
\end{itemize}
Let us now discuss in more detail the low and high momentum components of the nuclear wave function.
\section{Low momentum components and the mean field structure of the nuclear wave function} \label{sec:P0}
Eq. (\ref{PkE0}) yields the probability distribution that the final $(A-1)$-nucleon system is left into its ground state, i.e. with $E_{A-1}^f=0$ and $E=E_{min}$.  As already pointed out, the shell model spectral function can be written as follows
\beq \label{PkEsm}
    P_{SM}^{A}(k,E)=\frac{1}{4\pi A}\sum\displaylimits_{\alpha} A_\alpha n_\alpha^{SM}(k)\,\delta(E-|\epsilon_\alpha|)
\eeq
with $A_\alpha$ denoting the number of nucleons in the state $\alpha$ with removal energy $\epsilon_\alpha$ and nucleon momentum distribution $n_\alpha(k)$.
At the same time, one has
\beq
    P_1^{A}(k,E)=0
\eeq
In Eq. (\ref{PkEsm}), the sum over $\alpha$ runs only over the hole states of the target, which means that the occupation probability $S_\alpha$ is
\bey \label{Salpha}
    S_\alpha \equiv \int_0^\infty k^2\:dk \:n^{SM}_\alpha(k)
\left\{ \begin{array}{lr}
=1 &    \qquad\qquad {\mbox for }\: \alpha < \alpha_F  \\
=0 &   \qquad\qquad {\mbox for } \: \alpha > \alpha_F
    \end{array} \right.
\eey
The main effect of NN SRC is to deplete states below the Fermi level and to make the states above the Fermi level partially occupied. By such a mechanism
\beq
    P_0^A(k,E) \neq P_{SM}^A(k,E)
\eeq
and
\beq
    P_1^A(k,E) \neq 0
\eeq
Disregarding the finite width of the states below the Fermi level, the modified shell model contribution can be written as
\beq \label{PkEsm2}
    P_{SM}^A(k,E)=\frac{1}{4\pi A}\sum\displaylimits_{\alpha < \alpha_F} A_\alpha \widetilde{n}_\alpha(k)\,\delta(E-|\epsilon_\alpha|)
\eeq
where the occupation probability for hole states is
\beq
    S_\alpha \equiv \int_0^\infty k^2dk \widetilde{n}_\alpha(k)<1
\eeq
Eq. (\ref{PkEsm2}) drops down very quickly for $k>k_F$ and $E>E_F$, where  the spectral function behavior is governed by NN correlations, and thus by its correlated component $P_1^A(k,E)$.
%
%
\section{Two-nucleon correlations} \label{sec:2Nuc}
\begin{figure}[!h]
\begin{center}
\subfigure[\label{Fig_PkE}]
{\includegraphics[scale=0.59]{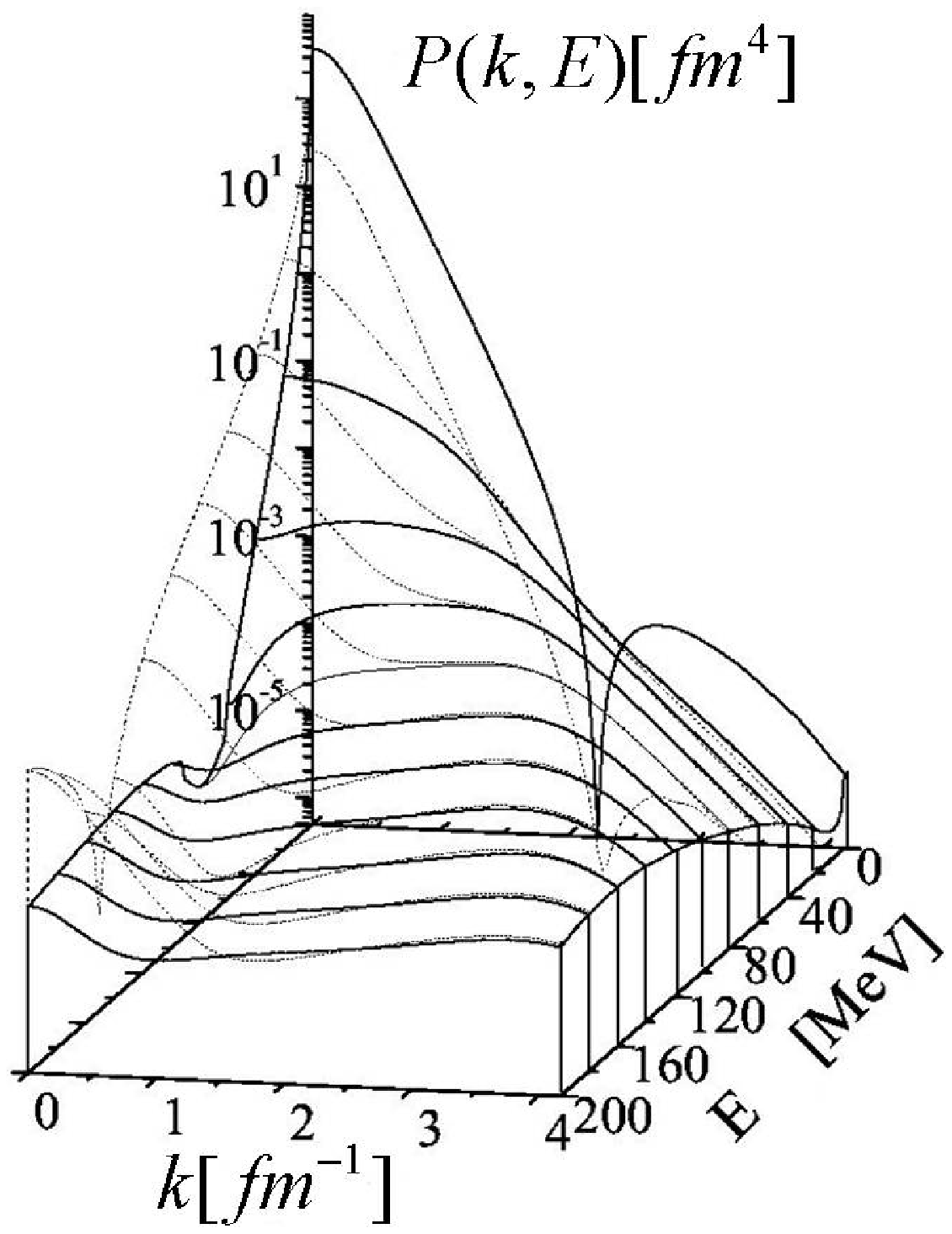}}\hskip-0.1cm
\subfigure[\label{Fig_Peakssss}]
{\includegraphics[scale=0.59]{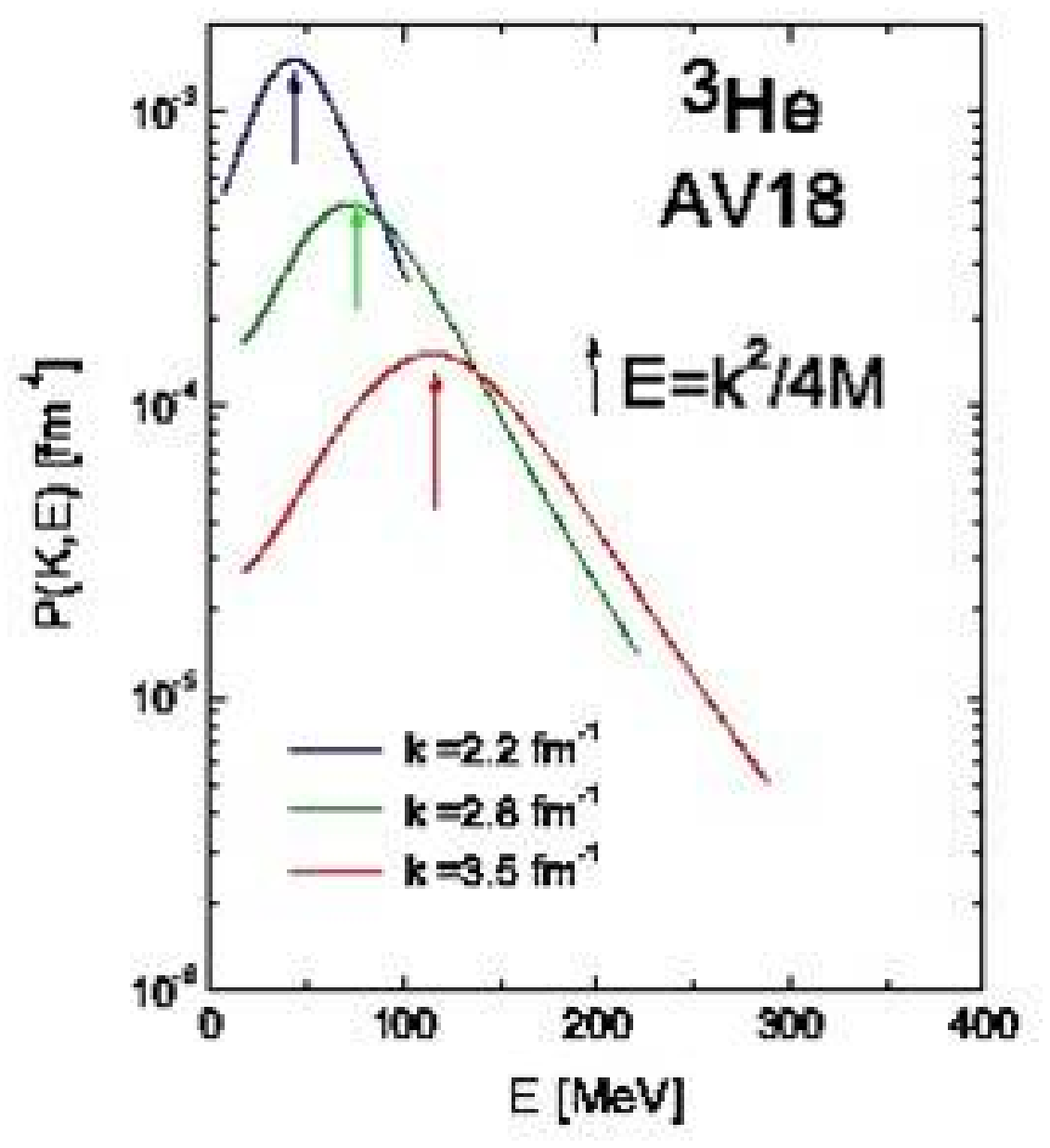}}
\caption{(a) The $^3He$ spectral function corresponding to the AV18 potential \cite{WSS}; (b) Some cuts of the spectral function $P(k,E)$ at high values of $k$ and $E$. After Ref. \cite{CK,CK2}.}
\end{center}
\end{figure}
In Fig.  \ref{Fig_PkE} the spectral function of $^3He$ corresponding to the AV18 interaction \cite{WSS} is shown versus the momentum $k$ and the removal energy $E$. A typical feature of the spectral function, common also to the spectral function of Nuclear Matter \cite{BFF,BFF2}, is that it exhibits, at high values of $k$ and $E$,  broad peaks located at
\beq \label{Epeak}
    E_{A-1}^* \simeq \frac{A-2}{A-1}\:\frac{k^2}{2m_N}
\eeq
and whose width increases with $k$, as clearly exhibited in Fig. \ref{Fig_Peakssss}, where some cuts of $P^A(k,E)$ are shown. Eq. (\ref{Epeak}) is generated by 2NC in nuclei, as will be demonstrated in what follows, by illustrating different model developed in the last few years, differing in the description of the nucleon configurations inside the nucleus.
\begin{figure}[!h]
\begin{center}
\includegraphics[scale=0.65]{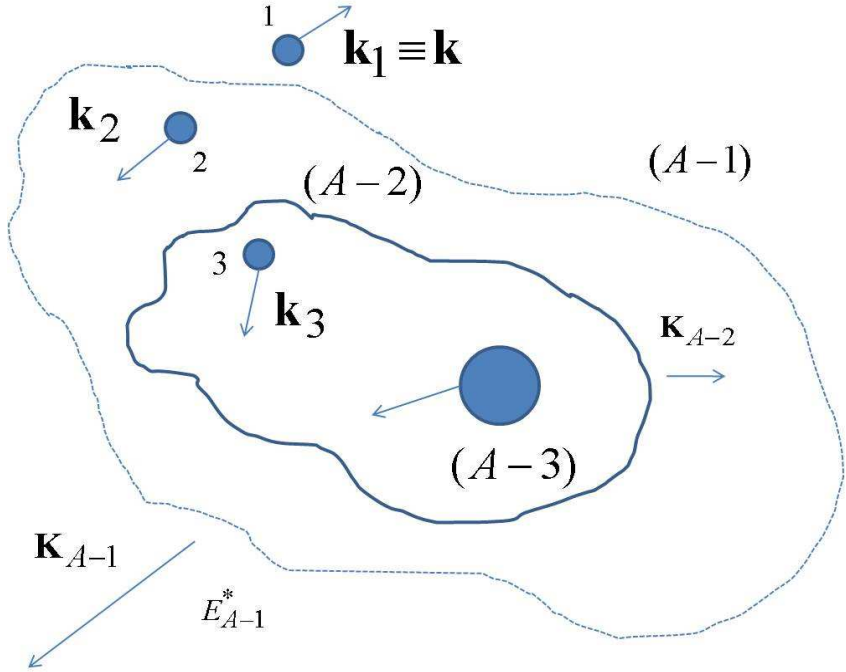}
\caption{A momentum description of a nucleus $A$. Nucleon '$1$', also called the ''active'' nucleon, has momentum $\textbf{k}_1 \equiv \textbf{k}$, and the nucleus $(A-1)$ consists of nucleon '$2$', with momentum $\textbf{k}_2$ and  nucleus $(A-2)$, with momentum $\textbf{K}_{A-2}$; the latter is defined in terms of nucleon '$3$' and nucleus $(A-3)$, with momenta $\textbf{k}_3$ and $\textbf{K}_{A-3}$, respectively.} \label{Fig_nucleus}
\end{center}
\end{figure}
To this end, in Fig. \ref{Fig_nucleus}, a simple cartoon of a target nucleus $A$ and its nucleon constituents is given.
%
%
\subsection{The naive two-nucleon correlation model}
\begin{figure}[!h]
\begin{center}
\includegraphics[scale=0.6]{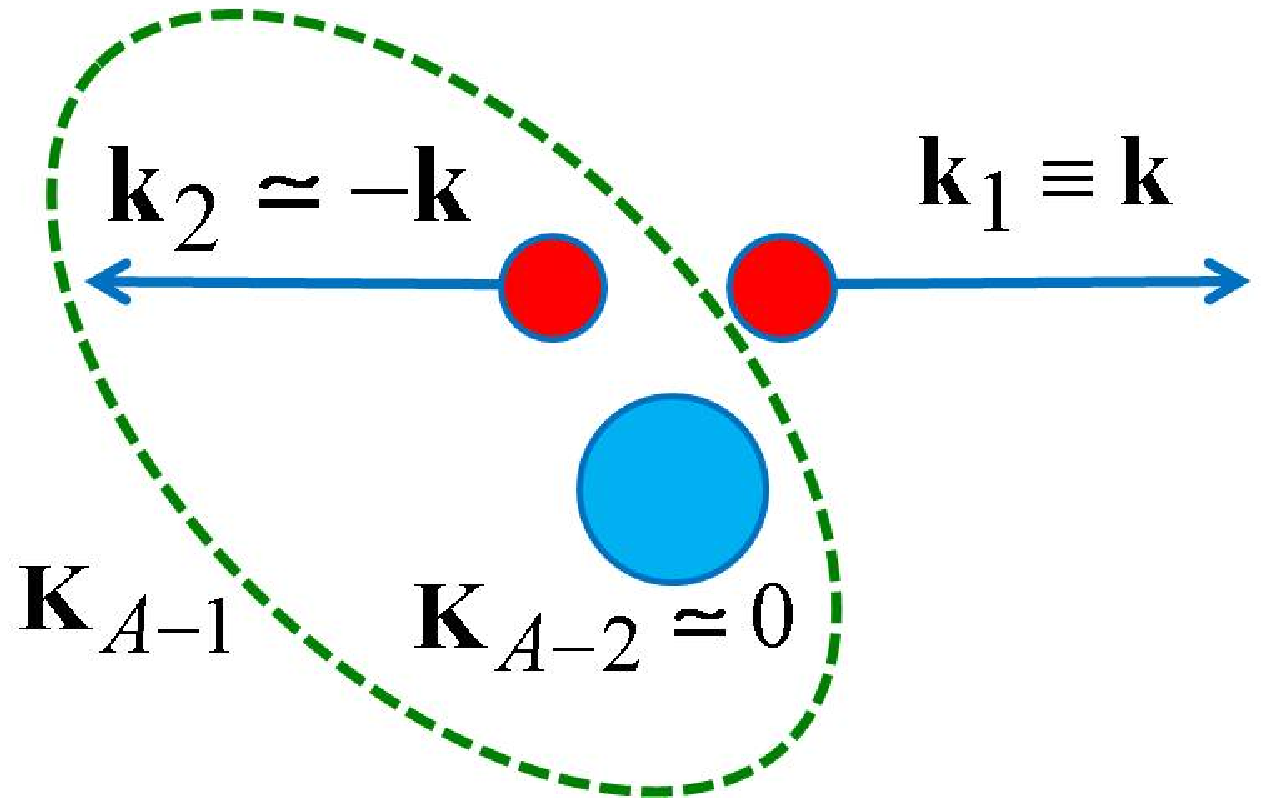}
\caption{The naive two-nucleon correlation model: the high momentum $\textbf{k}_1 \equiv \textbf{k}$ of nucleon '$1$' is entirely balanced by the momentum $\textbf{k}_2 \simeq -\textbf{k}$ of nucleon '$2$', with the system $(A-2)$ being at rest, i.e. with momentum $\textbf{K}_{A-2}\simeq 0$.} \label{Fig_naive}
\end{center}
\end{figure}
In Ref. \cite{mark1,FS}, a first microscopic model leading to Eq. (\ref{Epeak}) has been proposed. It is based upon the assumption that, referring to Fig. \ref{Fig_naive}, the high momentum $\textbf{k}_1 \equiv \textbf{k}$ of nucleon '$1$'
is entirely balanced by the momentum $\textbf{k}_2 \simeq -\textbf{k}$ of nucleon '$2$', whereas the residual spectator system has total momentum $\textbf{K}_{A-2}\simeq 0$. Energy conservation for such a mechanism requires that
\beq \label{E2NC}
    E_{A-1}^*+E_{A-1}^R\simeq \frac{k^2}{2m_N}
\eeq
where
\beq
    E_{A-1}^R\simeq \frac{k^2}{2(A-1)m_N}
\eeq
is the recoil energy of the residual $(A-1)$-nucleon system, whose intrinsic excitation energy could therefore be written as
\beq
    E_{A-1}^*\simeq\frac{A-2}{A-1}\:\frac{k^2}{2m_N}
\eeq
Within such a picture, the nucleon spectral function simply reads as follows
\beq
    P_{2NC}^A(k,E)=\frac{n_1(k)}{4\pi}\:\delta\left(E-E_{thr}^{(2)}-\frac{A-2}{A-1}\:\frac{k^2}{2m_N}\right)
\eeq
where
\beq
    E_{thr}^{(2)}\equiv |E_A|-|E_{A-2}|
\eeq
is the two-nucleon breakup threshold, being $|E_A|$ and $|E_{A-2}|$ the (positive) ground state energies of nucleus $A$ and $(A-2)$.
\subsection{The convolution formula}
\begin{figure}[!h]
\begin{center}
\includegraphics[scale=0.7]{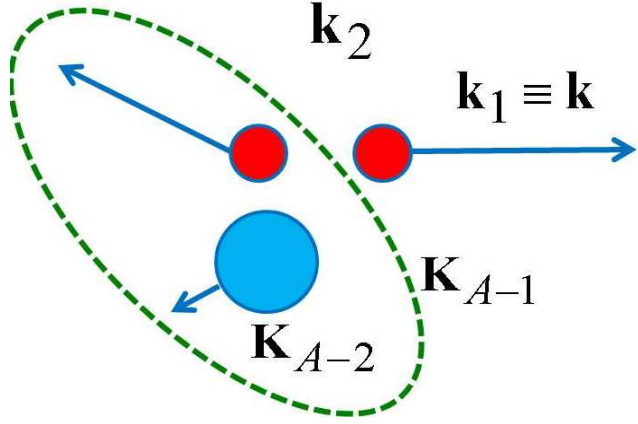}
\caption{The few nucleon correlation model: the high momentum $\textbf{k}_1 \equiv \textbf{k}$ of nucleon '$1$' is entirely balanced by the momentum $\textbf{k}_2$ of nucleon '$2$', and by the momentum $\textbf{K}_{A-2}$ of the residual system $(A-2)$, with $\textbf{K}_{A-2} \ll |\textbf{k}_2|$.} \label{Fig_FNC}
\end{center}
\end{figure}

The naive 2NC model has been implemented in \cite{ciosim,CSFS}, by assuming that the momentum $\textbf{k}_1\equiv \textbf{k}$ of nucleon '$1$' is not fully balanced by the momentum $\textbf{k}_2$ of nucleon '$2$', but also by the residual $(A-2)$ system, as depicted in Fig. \ref{Fig_FNC}; in other words, it is assumed that the spectator system $(A-2)$ moves with small momentum
$\textbf{K}_{A-2}\neq 0$. We will call such a configuration \emph{few nucleon correlation} (FNC) configuration.
\\Introducing the previously defined center of mass, $\textbf{k}_{CM}$, and relative, $\textbf{k}_{rel}$, momenta of the correlated pair as follows
\bey
    \textbf{k}_{CM}&\equiv&\textbf{k}_1+\textbf{k}_2=\textbf{k}+\textbf{k}_2 \\
    \textbf{k}_{rel}&\equiv&\frac{\textbf{k}_1-\textbf{k}_2}{2}=\frac{\textbf{k}-\textbf{k}_2}{2}
\eey
momentum conservation yields
\bey
    \textbf{k}_1+\textbf{k}_2+\textbf{K}_{A-2}=0 \\
    \textbf{K}_{A-2}=-\textbf{k}_{CM}
\eey
and energy conservation reads as follows
\beq
    E-E_{A-1}^f+E_A=E+|E_{A-2}|-\frac{|\textbf{t}_{2,(A-2)}|^2}{2\mu_{2,(A-2)}}-|E_A|=E-\left(E_{thr}^{(2)}+\frac{|\textbf{t}_{2,(A-2)}|^2}{2\mu_{2,(A-2)}}\right)
\eeq
where
\beq \label{Ex1}
    \frac{|\textbf{t}_{2,(A-2)}|^2}{2\mu_{2,(A-2)}}=\frac{(A-2)}{2m_N(A-1)}\:\left[\frac{(A-2)\textbf{k}_2-\textbf{K}_{A-2}}{(A-1)}\right]^2
\eeq
is the energy of the relative motion of particles '$2$' and '$(A-2)$', i.e. the excitation energy $E_{A-1}^*$ of the residual $(A-1)$ system. In Eq. (\ref{Ex1}), eventually,
\beq
    \mu_{2,(A-2)}=\frac{A-2}{A-1}\:m_N
\eeq
is the reduced mass of particles '$2$' and '$(A-2)$'.
\\Given these assumptions, the spectral function, at high $k$ and high $E$, has been obtained in Ref. \cite{ciosim} by assuming that the ground state wave function of nucleus $A$ factorizes as follows
\beq \label{ansazt}
    \Psi_A^0(\{\textbf{r}_i\}_A)\simeq \hat{\mathcal{A}}\left\{\sum_{n,m,f_{A-2}} a_{n,m,f_{A-2}}
    \left[\Phi_n(\textbf{x})\otimes \chi_m(\textbf{y})\right]\otimes \Psi_{A-2}^{f_{A-2}}(\{\textbf{r}_i\}_{A-2})\right\}
\eeq
where
\beq
    \textbf{x}=\textbf{r}_1-\textbf{r}_2 \qquad \qquad \textbf{y}=\textbf{r}_3-\frac{\textbf{r}_1+\textbf{r}_2}{2}
\eeq
are the relative and CM coordinates, respectively, with $\textbf{r}_3\equiv \textbf{R}_{A-2}$, $\hat{\mathcal{A}}$ is a proper antisymmetrization operator, and $\otimes$ is a short-hand notation for the standard Clebsh-Gordan coupling of orbital and spin angular momenta; $\{\Phi_n(\textbf{x}) \}$ and  $\{\chi_m(\textbf{y}) \}$ represent a complete set of states describing the relative and CM motion of the pair, and, eventually, $\Psi_{A-2}^{f_{A-2}}(\{\textbf{r}_i\}_{A-2})$ is the complete set of states describing the $(A-2)$-nucleon system.
\\As already explained, SRC correspond to the high momentum and high removal energy components of the spectral function $P^A(k,E)$. In momentum space, SRC are described by a correlated pair with a very high relative momentum $k_{rel}> k_F$, and a low CM momentum $k_{CM} \lesssim k_F$; the latter condition justifies the choice of the CM motion in $s$-state, so that
\beq \label{ansazt2}
    \Psi_A^0(\{\textbf{r}_i\}_A)\simeq \hat{\mathcal{A}}\left\{\chi_0(\textbf{y}) \sum_{n,f_{A-2}} a_{n,0,f_{A-2}} \left[\Phi(\textbf{x})\otimes \Psi_{A-2}^0(\{\textbf{r}_i\}_{A-2})\right]\right\}
\eeq
where  $\chi_0(\textbf{y})$  describes the low momentum CM wave function in $s$-state.
Since the $s$-wave motion also implies that the system $(A-2)$ is in the ground state or in a low energy excited state,
 one can write, eventually, \cite{ciosim}
\beq \label{ansazt3}
    \Psi_A^0(\{\textbf{r}_i\}_A)\simeq \hat{\mathcal{A}}\left\{\chi_0(\textbf{y})
    \left[\Phi(\textbf{x})\otimes \Psi_{A-2}^{\bar{0}}(\{\textbf{r}_i\}_{A-2})\right]\right\}
\eeq
where
\beq
    \Phi(\textbf{x})=\sum_n\:a_{n00}\Phi_n(\textbf{x})
\eeq
describes the relative motion of the correlated pair in the nuclear medium. In Eq. (\ref{ansazt3}),  $f_{A-2}=\bar{0}$ denotes the excitation spectrum of the system $(A-2)$ which, since, as already pointed out, the CM of the pair involves only low momentum components, have been mainly limited to the ground state and to the (low-lying) excited states corresponding to configurations generated by the removal of two particles from different shell model states of the target.
\\The above formalism leads to the following convolution formula, involving only the relative and CM momentum distributions of the correlated pair
\bey \label{PkE2NC}
    P_{FNC}^A(k,E) &=&\mathcal{N}_2\: \int d\textbf{k}_2 \: d\textbf{K}_{A-2}\:  n_{rel}\left(\frac{|\textbf{k}-\textbf{k}_2|}{2} \right) \: n_{CM}^{soft}(\textbf{k}+\textbf{k}_2)\nonumber \\
    &\times& \:\delta \left( E-E_{thr}^{(2)}- \frac{(A-2)}{2M(A-1)} \:
    \left ( \frac{(A-2)\textbf{k}_2-\textbf{K}_{A-2}}{(A-1)}\right)^2 \right)
    \nonumber\\
        &\times& \: \delta\left( \textbf{k}+\textbf{k}_2+\textbf{K}_{A-2}\right)
\eey
where $n_{rel}$ and $n_{CM}^{soft}$ are the momentum distributions of the relative and center of mass motion of the two nucleons in a correlated pair, respectively, and the other notations, following Fig. \ref{Fig_nucleus}, are self explaining; eventually, the factor $\mathcal{N}_2$ satisfies  the normalization condition given by Eq. (\ref{norm}). We stress once again that in such an approach, the residual $(A-2)$-nucleon system is assumed to be in its ground state, which implies that the CM of the correlated pair moves with low momentum, so that only ''soft'' components of $n_{CM}$ contribute to the spectral function $P_{FNC}^A$, since we have seen in Chapter $2$ that high momentum components of a nucleon are linked to high excitation energies of ($A-1$) and this has to be true also for the system ($1$-$2$)-($A-2$).
\begin{figure}[!h]
\centerline{\includegraphics[scale=0.48]{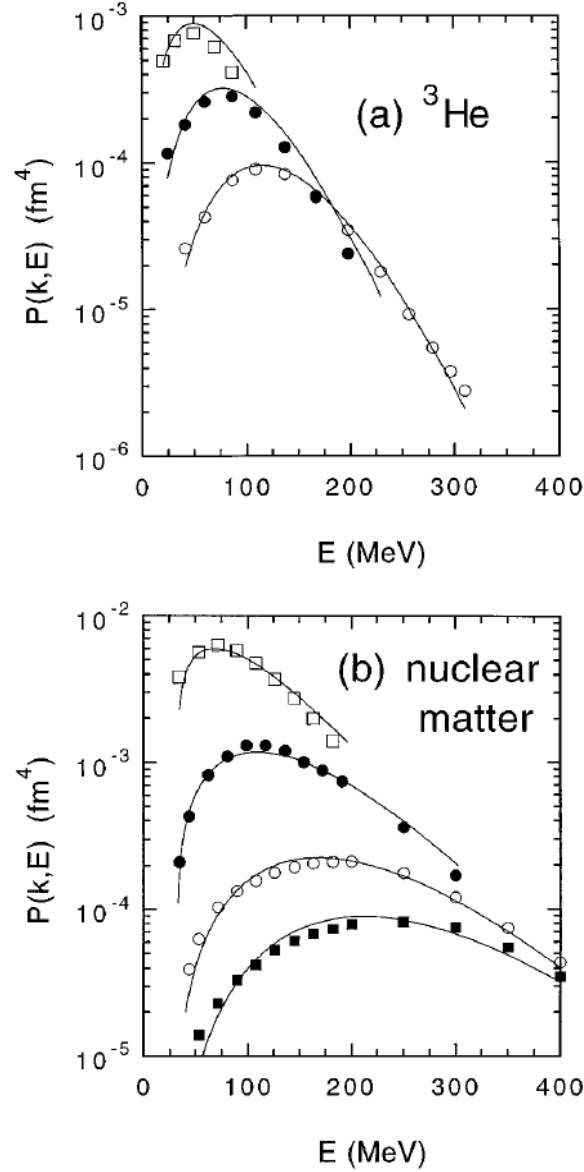}}
\caption{The nucleon spectral function of $^3He$ \cite{CPS} and nuclear matter \cite{BFF,BFF2} versus the removal energy $E$ for various values of the momentum $k$. For $^3He$ (a) the squares, full dots, and open dots correspond to $k=2.2,2.8,3.5\:fm^{-1}$, respectively. For nuclear matter (b) the open squares, full dots, open dots, and full squares correspond to $k=1.5,2.2,3.0,3.5\:fm^{-1}$, respectively. The solid line correspond to the FNC theoretical calculation, Eq. (\ref{PkE2NCf}). After Ref. \cite{ciosim}.}\label{Fig322}
\end{figure}
The naive 2NC model is recovered by placing
\beq
    n_{CM}(\textbf{k}_{CM})=\delta(\textbf{k}_{CM})
\eeq
i.e. by assuming that the spectator nucleus is at rest.
\\Integrating Eq. (\ref{PkE2NC}) over $\textbf{k}_2$ and the angular variables of $\textbf{K}_{A-2}$, one gets
\beq \label{PkE2NCf}
    P_{FNC}^A(k,E)=\mathcal{N}_2\:\frac{2\pi m_N}{k}\int_{K_{A-2}^-}^{K_{A-2}^+}dK_{A-2}\:K_{A-2}\:n_{rel}(k_x^*)\:n_{CM}(K_{A-2})
\eeq
where
\bey
    K_{A-2}^\pm&=&\frac{A-2}{A-1}\:\left|k\pm k_0  \right|\\
    k_0&=&\sqrt{2m_N\:\frac{A-1}{A-2}[E-E_{thr}^{(2)}]}\\
    k_x^*&=&\sqrt{  \frac{Ak^2+(A-2)k_0^2}{2(A-1)}-\frac{AK_{A-2}^2}{4(A-2)}}
\eey
The convolution formula (\ref{PkE2NCf}) has been calculated in Ref. \cite{ciosim} by using the following effective relative and CM nucleon momentum distributions
\beq \label{neff1}
    n_{rel}^{eff}(k_{rel})=C^A\: n_D(k_{rel})
\eeq
and
\beq \label{neff2}
    n_{CM}^{eff}(k_{CM})=\left(\frac{\alpha_{CM}}{\pi}\right)^{3/2}\:e^{-\alpha_{CM}\:k^2_{CM}}
\eeq
respectively,
with the parameter $\alpha_{CM}$ determined as explained in Ref. \cite{ciosim}.
\begin{figure}[!h]
\centerline{\includegraphics[scale=0.77]{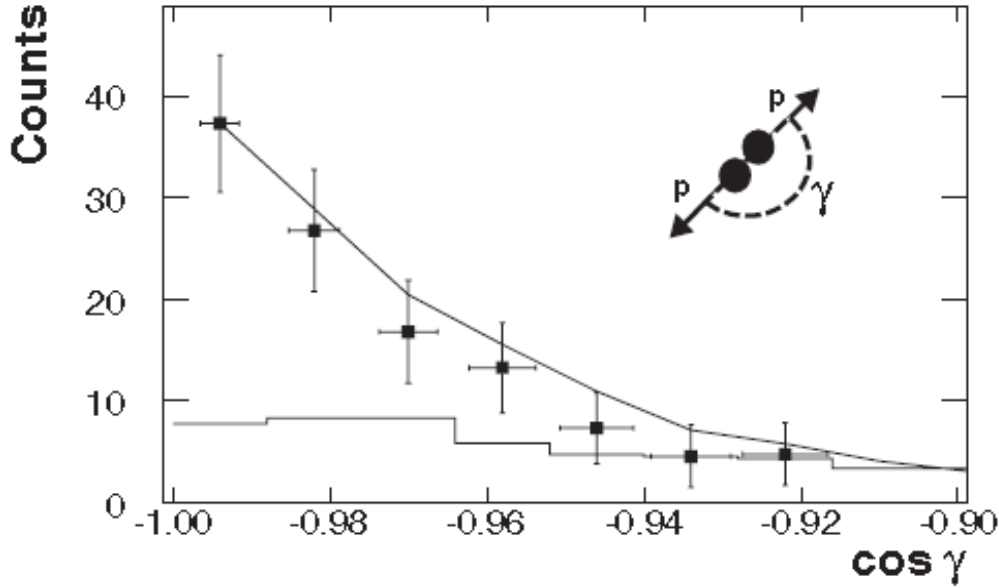}}
\caption{The distribution of the cosine of the opening angle between  the missing momentum $\textbf{p}_{m}$ and $\textbf{p}_{rec}=p_{CM}-\textbf{p}_m$, for the $p_m=0.55\: GeV/c$ kinematics. The histogram shows the distribution of random events. The curve is a simulation of the scattering off a moving pair with a width of $0.136\:GeV/c$ for the pair CM momentum. After Ref. \cite{Shneor}.}\label{FigBB}
\end{figure}
\\Recently, $n_{rel}$ and $n_{CM}$ have been obtained from many-body approaches \cite{Schiavilla,ACM2}, separately for $p$-$n$ and $p$-$n$ pairs, and they quantitatively agree with the forms given by Eqs. (\ref{neff1}) and (\ref{neff2}).
In such a way, the convolution formula (\ref{PkE2NCf}) is completely defined in terms of many-body quantities.
\\Eventually, it should be pointed out that, in the recent BNL experiment on $^{12}C$ target \cite{Shneor}, the CM momentum distribution has been  measured,  finding that it fully agrees with the prediction of Ref. \cite{ciosim}, as shown  by the full line in Fig \ref{FigBB}.
\section{Brueckner-Bethe-Goldstone theory and the validation of the convolution formula}
\begin{figure}[!h]
\centerline{\includegraphics[scale=0.5]{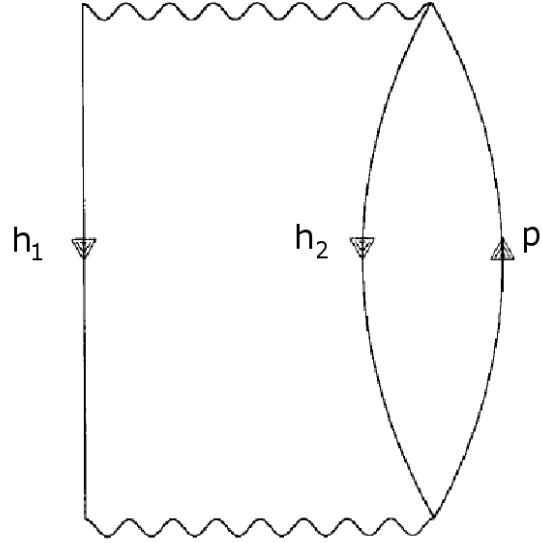}}
\caption{The diagram of the BBG expansion considered in Ref. \cite{BBC} to obtain the high $k$ and $E$ behavior of the spectral function.}\label{Fig_BBG}
\end{figure}

The convolution model can be microscopically derived from the Bruckner-Bethe-Goldstone (BBG) theory \cite{BBC}. Let us discuss this aspect in more detail. \\In nuclear matter the spectral function corresponding to the nucleon self-energy
\beq
    M(k,E)=V(k,E)+\imath\:W(k,E)
\eeq
is given by the well known result \cite{MB}
\beq \label{ppp}
    P^{NM}(k,E)=-\frac{1}{\pi}\:\Im\:\mathcal{G}(k,E)=\frac{1}{\pi}\:\frac{W(k,E)}
    {\left[ -E-k^2/2m_N-V(k,E)\right]^2+W^2(k,E)}
\eeq
where $\mathcal{G}(k,E)$ is the single-particle Green function
\beq
    \mathcal{G}(k,E)=\frac{1}{-E-k^2/2m_N-V(k,E)-\imath W(k,E)} \: .
\eeq
It has be noticed that the real, $V(k,E)$, and imaginary parts, $W(k,E)$, of the self-energy are highly off-shell in the considered energy and momentum ranges. We are interested in the region where $E$ is much greater than the Fermi energy $E_F$; for high $k$ and $E$, one has
\beq
    E+\frac{k^2}{2m_N} \gg |V(k,E)|, |W(k,E)|
\eeq
and Eq. (\ref{ppp}) becomes
\beq \label{nnn}
    P^{NM}(k,E)=\frac{1}{2}\:\sum_{hh'p}\frac{|<kp|G(e(h)+e(h'))|hh'>_a|^2}{\left( E+k^2/2m_N\right)^2}\: \delta \left(E-e(p)+e(h)+e(h')\right)
\eeq
where $p$ denotes a ''particle state'' (outside the Fermi sea), and $h(h')$ a ''hole state'' (inside the Fermi sea), with energies $e(p)$, $e(h)$ and $e(h')$ respectively, and $G$ is the BBG scattering matrix
\bey
    G_{12}(\omega)&=&v+v\frac{Q}{\omega-H_0+\imath\eta}\:G(\omega)=v+v\frac{Q}{\omega-H_0+\imath\eta}\:v\no \\
    &+&  v\frac{Q}{\omega-H_0+\imath\eta}\:v\frac{Q}{\omega-H_0+\imath\eta}\:v+\ldots
\eey
where $Q$ is the Pauli projection operator, which restricts the two nucleons in intermediate states to lie outside the Fermi sea. It should be stressed that Eq. (\ref{nnn}) corresponds to the $1p$-$2h$ diagram of Fig. \ref{Fig_BBG} which, through the G matrix, sums up all ''ladder diagrams'' where the BBG scattering matrix is replaced by the bare NN interaction.
\\Eq. (\ref{nnn}) can be expressed in terms of the \emph{defect wave function}, defined as
\beq
    |\xi_{12}>=|\Psi_{12}>-|\phi_{12}>
\eeq
where $|\Psi_{12}>$ is the correlated two-body wave function, and $|\phi_{12}>$ the uncorrelated one.
After a length algebra, one finds \cite{BBC}
\beq
    P^{NM}(k,E)=\frac{m_N\:\rho^2}{32k}\:\int_{|k-k_0|}^{|k+k_0|}dk_{CM}\:k_{CM}\:n_{CM}^{FG}(k_{CM})\:n_{rel}
    \left( \sqrt{\frac{1}{2}k^2-\frac{1}{4}k_{CM}^2+\frac{1}{2}k_0^2}\right)
\eeq
where
\beq
    \rho=\frac{2}{3\pi^2}k_F^3
\eeq
is the nuclear matter density and
\beq
    k_0=\sqrt{2m_N\:\frac{A-1}{A-2}\:\left(E-E_{thr}^{(2)}\right)}\: .
\eeq
\begin{figure}[!h]
\centerline{\includegraphics[scale=0.5]{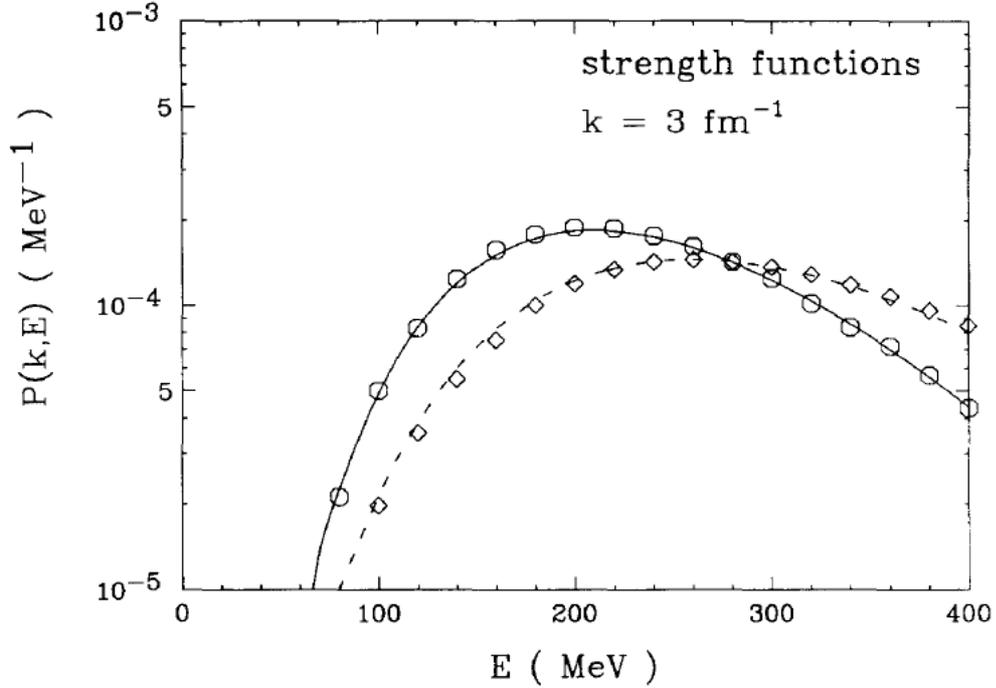}}
\caption{The spectral function obtained from BBG theory plotted at $k=3\:fm^{-1}$ (diamonds). For comparison also shown are the spectral function calculated in BBG theory with the free single particle spectrum (octagons) and the one calculated by the convolution formula (\ref{PkE2NCf}) with the bare nucleon mass (full line), and the the pertinent effective mass (dashed line). After Ref. \cite{BBC}.}
\label{Fig_BBG2}
\end{figure}
It can be seen that the convolution formula (\ref{PkE2NCf}) is recovered. This represents a robust validation of the convolution formula, as shown in Fig. \ref{Fig_BBG2}.
%
%
\section{Many-body validation of the factorization of the nuclear wave function at high momenta}
In this section, we will show that the convolution formula results from a rather general property of the nuclear wave function. Let us recall that the convolution formula has been obtained by assuming that the nuclear wave function $\Psi_A$ factorizes into the wave function of a correlated pair and the wave function of a core of the $(A-2)$-nucleon system. In Ref. \cite{CSunp}, the s-wave three-body Faddeev wave function obtained with $MT/V'$ \cite{MTpot} potential has been used to calculate the following quantity
\beq \label{ratiowf}
    r=\frac{\Psi_A(x,y,\theta)}{\Psi_A(x,y',\theta)}
\eeq
\begin{figure}[!h]
\vskip-1.5cm
\centerline{\includegraphics[scale=1.8]{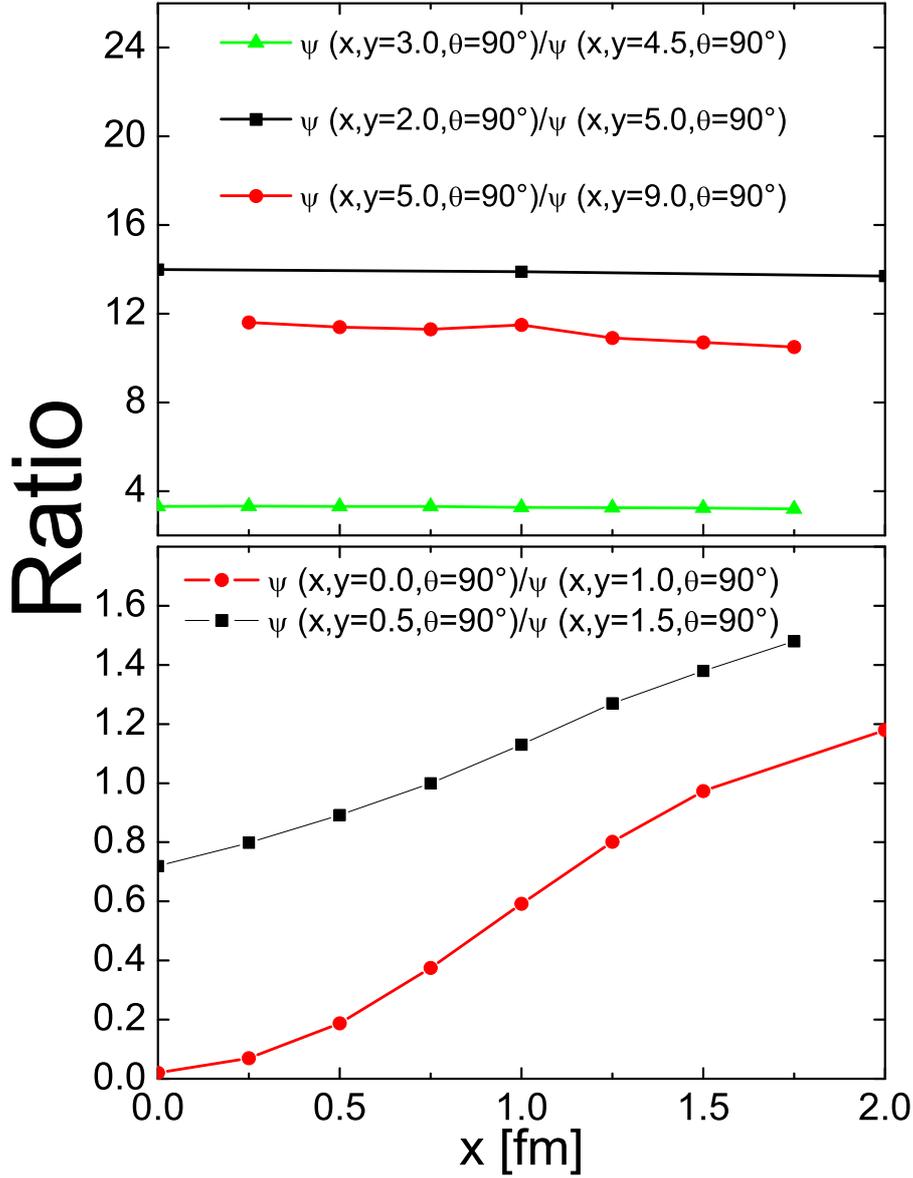}}
\vskip-1.0cm
\caption{Eq. (\ref{ratiowf}) plotted versus the relative coordinate (\ref{xrel}), for $\theta=90°$ and fixed values of $y$ and $y'$. (\emph{Upper panel}) Green triangles: $y=3.0\:fm$, $y'=4.5\:fm$; black squares: $y=2.0\:fm$, $y'=5.0\:fm$; red circles: $y=5.0\:fm$, $y'=9.0\:fm$. (\emph{Lower panel}) Red circles: $y=0.0\:fm$, $y'=1.0\:fm$; black squares: $y=0.5\:fm$, $y'=1.5\:fm$ \cite{CSunp}.}
\label{Fig_wf1}
\end{figure}
where
\beq \label{xrel}
    |\textbf{x}|=|\textbf{r}_1-\textbf{r}_2|
\eeq
is the two-nucleon relative coordinate,
\beq
    |\textbf{y}|=\left|\textbf{r}_3-\frac{\textbf{r}_1+\textbf{r}_2}{2}\right|
\eeq
the CM coordinate, and $\theta$ the angle  between them. If the ratio (\ref{ratiowf}) is plotted versus $|\textbf{x}|$ for fixed values of $|\textbf{y}|$, $|\textbf{y}'|$ and $\theta$, exhibits a constant behavior; this is evidence of the factorization of $\Psi_A(x,y,\theta)$ in the variable $x$ and $y$. As a matter of fact, it can be seen that writing
\beq
    \Psi_A(x,y,\theta)=f(x,\theta)h(y)
\eeq
one gets
\beq
    r=\frac{\Psi_A(x,y,\theta)}{\Psi_A(x,y',\theta)}=\frac{f(x,\theta)h(y)}{f(x,\theta)h(y')}=\frac{h(y)}{h(y')}=const \: .
\eeq
The results presented in the upper panel of Fig. \ref{Fig_wf1} clearly show that, when $|\textbf{y}|$ is large and $|\textbf{x}|$ small, $\Psi_A$ indeed factorizes. On the contrary, as shown in the lower panel of Fig. \ref{Fig_wf1}, for small values of $|\textbf{y}|$ and small values of $|\textbf{x}|$, the nuclear wave function does not factorize. Thus the factorization of the nuclear wave function at high momenta is validated by many-body wave functions.
%
%
%
\section{Three-nucleon correlations} \label{sec:3N}
\subsection{Naive three nucleon correlation models}
We have seen that the high momentum and high removal energy components of the nuclear spectral function are due to \emph{few nucleon correlations} (FNC), in particular in the region around
\beq
    E^*_{A-1}\simeq \frac{A-2}{A-1}\: \frac{k^2}{2m_N} \: .
\eeq
\begin{figure}[!h]
\begin{center}
\includegraphics[scale=0.9]{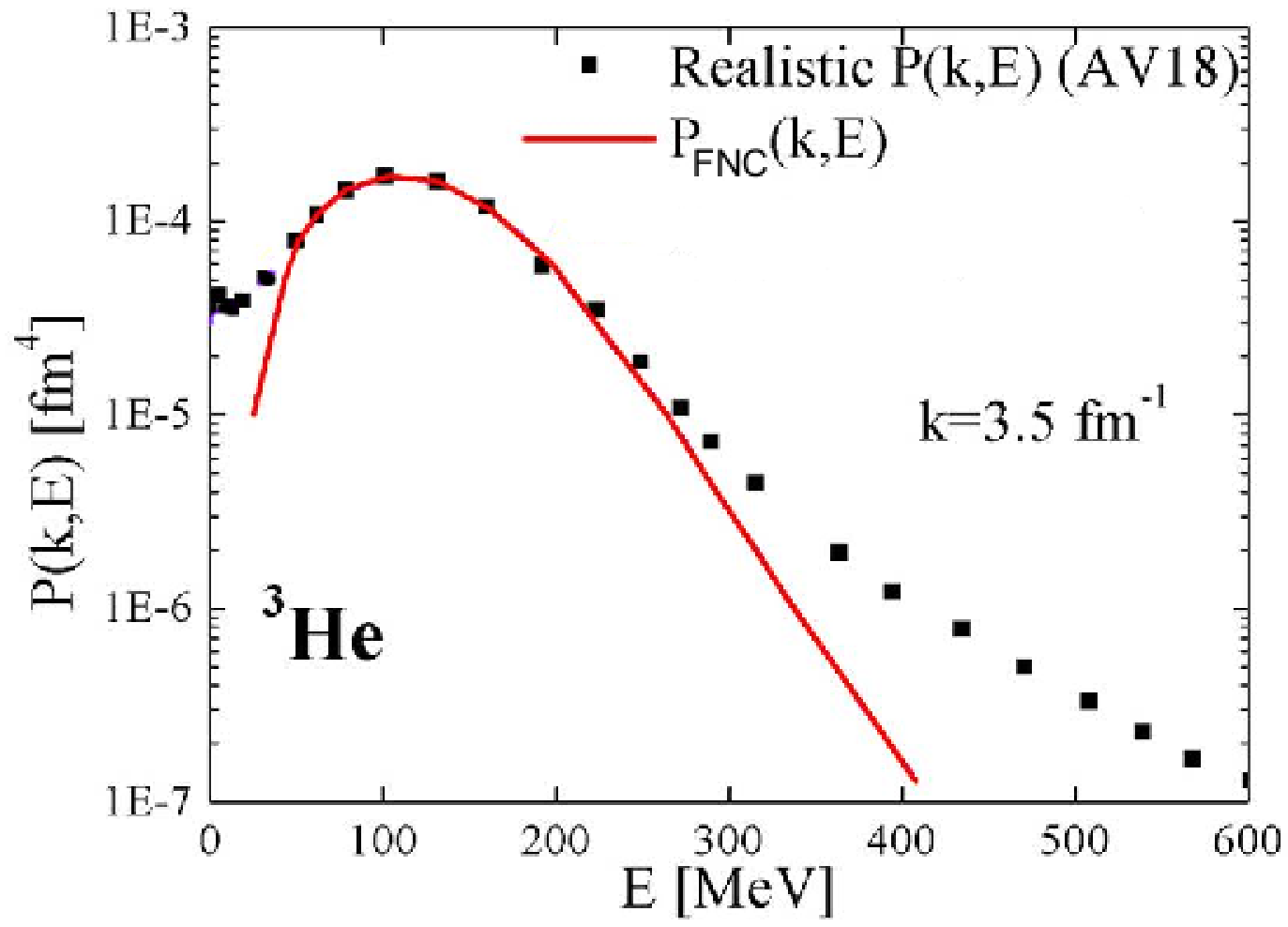}
\caption{The realistic (black squares) spectral function of $^3He$ \cite{CK} compared with the theoretical calculation performed within the convolution formula (\ref{PkE2NCf}), at $k=3.5\:fm^{-1}$. After Ref. \cite{CC}.}\label{Fig333}
\end{center}
\end{figure}
\noindent
FNC cannot however explain the regions at high values of $k$ and large values of the excitation energy
\beq \label{Elarge}
    E^*_{A-1}\gg \frac{A-2}{A-1}\: \frac{k^2}{2m_N}
\eeq
and the region at high values of $k$ and low values of
\beq \label{Esmall}
    E^*_{A-1}\ll \frac{A-2}{A-1}\: \frac{k^2}{2m_N} \: .
\eeq
This can be clearly seen in Fig. \ref{Fig333}. In these regions, one has to consider the effects of three-nucleon correlations (3NC), i.e. configurations characterized by three nucleons which have comparable momenta, and share almost the full momentum of the nucleus
\beq
    \textbf{k}_1+\textbf{k}_2+\textbf{k}_3+\textbf{K}_{A-3}=0
\eeq
i.e.
\beq
    \textbf{k}_1+\textbf{k}_2+\textbf{k}_3\simeq 0 \: .
\eeq
\begin{figure}[!h]
\begin{center}
\includegraphics[scale=0.65]{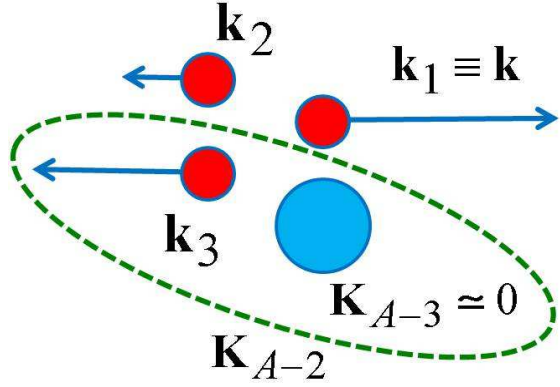}
\caption{The naive 3NC model: the high momentum $\textbf{k}_1 \equiv \textbf{k}$ of nucleon '$1$' is entirely balanced by the momentum $\textbf{k}_2$  of nucleon '$2$' and $\textbf{k}_3$  of nucleon '$3$', with the system $(A-3)$ being at rest, i.e. with momentum $\textbf{K}_{A-3}\simeq 0$. Thus, the momentum $\textbf{K}_{A-2}$ is given by the momentum $\textbf{k}_3$ nucleon '3'.} \label{Fig_3NCo}
\end{center}
\end{figure}
\noindent
The naive 3NC model will be defined as follows: referring to Fig. \ref{Fig_3NCo}, we will consider that the high momentum $\textbf{k}_1\equiv \textbf{k}$ of nucleon '$1$' is completely balanced by the high momenta $\textbf{k}_2$ and $\textbf{k}_3$ of particle '$2$' and '$3$', respectively, with the residual $(A-3)$-nucleon system at rest.
\\Within such a naive model, the residual system $(A-2)$ has momentum
\beq
    \textbf{K}_{A-2}=\textbf{k}_3=-(\textbf{k}_1+\textbf{k}_2)=-\textbf{k}_{CM}
\eeq
and excitation energy
\beq \label{Ex2}
    E_{A-2}^*=\frac{|\textbf{t}_{3,(A-3)}|^2}{2\mu_{3,(A-3)}}=\frac{A-3}{A-2}\:\frac{k_3}{2m_N}
\eeq
given by the relative motion of particle '$3$' and the residual $(A-3)$-nucleon system, with reduced mass
\beq
    \mu_{3,(A-3)}=\frac{A-3}{A-2}\:m_N \: .
\eeq
The excitation energy $E_{A-1}^*$ of the residual system $(A-1)$ can be written as follows
\beq \label{Ex}
    E_{A-1}^*=\frac{|\textbf{t}_{2,(A-2)}|^2}{2\mu_{2,(A-2)}}+\frac{|\textbf{t}_{3,(A-3)}|^2}{2\mu_{3,(A-3)}}
\eeq
i.e. by considering both the contribution arising from 2NC,  given by Eq. (\ref{Ex1}), and the one due to 3NC, given by Eq. (\ref{Ex2}). It can be clearly seen from Fig. \ref{Fig_3NCo}, and it will be explained in more detail in $\S$\ref{sec:y3}, that the assumptions of the naive model for three correlated nucleons refer only to that values of $E^*_{A-1}$  given by Eq. (\ref{Esmall}).
\\The 3NC contribution to the spectral function $P_1^A(k,E)$, given by Eq. (\ref{PkE1}), turns out to be \cite{ciocbm2}
\bey \label{PkE3NC}
    &&P_{3NC}^A(k,E) =\mathcal{N}_3\: \int d\textbf{k}_2 \: d\textbf{K}_{A-2}\:  n_{rel}\left(\frac{|\textbf{k}-\textbf{k}_2|}{2} \right) \: n_{CM}^{hard}(\textbf{k}+\textbf{k}_2)
    \nonumber\\
    &&\times \: \delta \left( E-E_{thr}^{(2)}- \frac{(A-2)}{2m_N(A-1)}
    \left ( \frac{(A-2)\textbf{k}_2-\textbf{K}_{A-2}}{(A-1)}\right)^2 - \frac{A-3}{2m_N(A-2)}\textbf{K}^2_{A-2}\right)\no \\
     &&\times  \: \delta\left( \textbf{k}+\textbf{k}_2+\textbf{K}_{A-2}\right)
\eey
where $\mathcal{N}_3$ is necessary to satisfy the normalization condition (\ref{norm}). It should be pointed out that Eq. (\ref{PkE3NC}) is nothing but Eq. (\ref{PkE2NC}) with  $n_{CM}^{soft}$ replaced by  $n_{CM}^{hard}$.
\begin{figure}[!h]
\begin{center}
\includegraphics[scale=1.0]{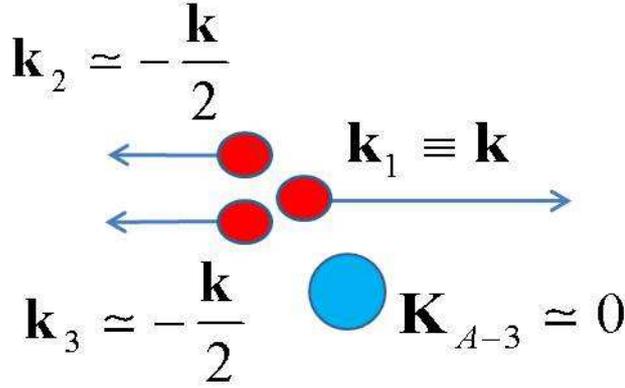}
\caption{First approximation to 3NC: the high momentum $\textbf{k}_1 \equiv \textbf{k}$ of nucleon '$1$' is entirely balanced by the momentum $\textbf{k}_2=\textbf{k}_3=-\textbf{k}/2$  of nucleon '$2$' and '$3$', with the system $(A-3)$ being at rest, i.e. with momentum $\textbf{K}_{A-3}\simeq 0$.} \label{Fig_3NCorr}
\end{center}
\end{figure}
\noindent
\\A first approximation to Eq. (\ref{PkE3NC}) could be obtained by considering
\beq
    \textbf{k}_2=\textbf{k}_3=-\frac{\textbf{k}}{2}
\eeq
with
\beq
    \textbf{K}_{A-2}=-\frac{\textbf{k}}{2}=-\textbf{k}_{CM}
\eeq
and
\beq
    \textbf{K}_{A-3}=0
\eeq
as shown in Fig. \ref{Fig_3NCorr}.
\\In this case, the ''hard'' CM nucleon momentum distribution will read as follows
\beq
    n_{CM}^{hard}(k_{CM})=\delta\left(\textbf{k}_{CM}-\frac{\textbf{k}}{2}\right)
\eeq
and the removal energy will reduce to
\beq
    E=E_{thr}^{(3)}+\frac{A-3}{A-1}\:\frac{k^2}{4m_N^2}\: .
\eeq
Integrating Eq. (\ref{PkE3NC}) over the momentum $\textbf{k}_2=-\textbf{k}-\textbf{K}_{A-2}$, yields
\beq
    P_{3NC}^A(k,E)=\mathcal{N}_3\:n_{rel}\left( \frac{3}{4}\:\textbf{k}\right)\:
    \delta\left(E-E_{thr}^{(3)}-\frac{A-3}{A-1}\:\frac{k^2}{4m_N}\right)\:.
\eeq

\cleardoublepage
\chapter{Inclusive electron scattering off nuclei at high momentum transfer and final state interaction effects: results of calculations} \label{ch:xs}
\section*{Introduction}
Inclusive electron scattering off nuclei at high momentum transfer can provide non trivial information on nuclear wave function. In particular, the kinematic region corresponding to $\xb > 1$ is
\begin{figure}[!h]
\centerline{\includegraphics[scale=0.6]{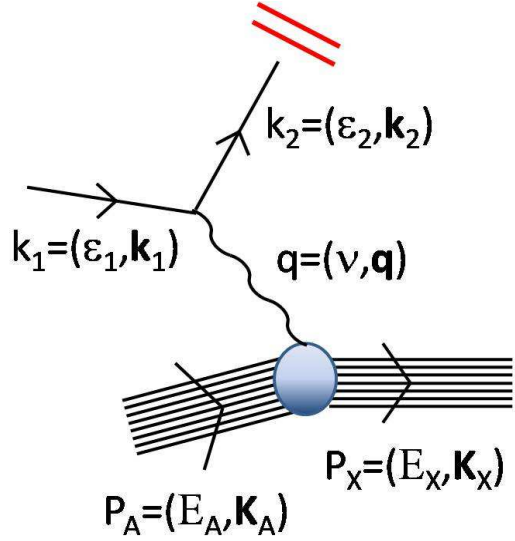}}
\caption{Inclusive electron scattering off a nucleus $A$, in the one photon exchange approximation. In inclusive processes, denoted as $A(e,e')X$, only the scattered electron $e'$ is detected in the final state; $k$, $k'$,  $P_A$ and $P_X$ are the four-momenta of the incoming electron $e$, the scattered electron $e'$, the target nucleus $A$, and the undetected particles, respectively; $q$ is the four-momentum transfer such that $-q^2=Q^2=\textbf{q}^2-\nu^2$, where $\textbf{q}=\textbf{k}_1-\textbf{k}_2$ and $\nu=\epsilon_1-\epsilon_2$ are the three-momentum and the energy transfers, respectively.
}\label{Fig_Fey_incl}
\end{figure}
strongly affected by high momentum and high removal energy components of the nuclear wave function arising from NN SRC.
\begin{figure}[!h]
\begin{center}
\subfigure[\label{Fey_a}]{\includegraphics[scale=0.6]{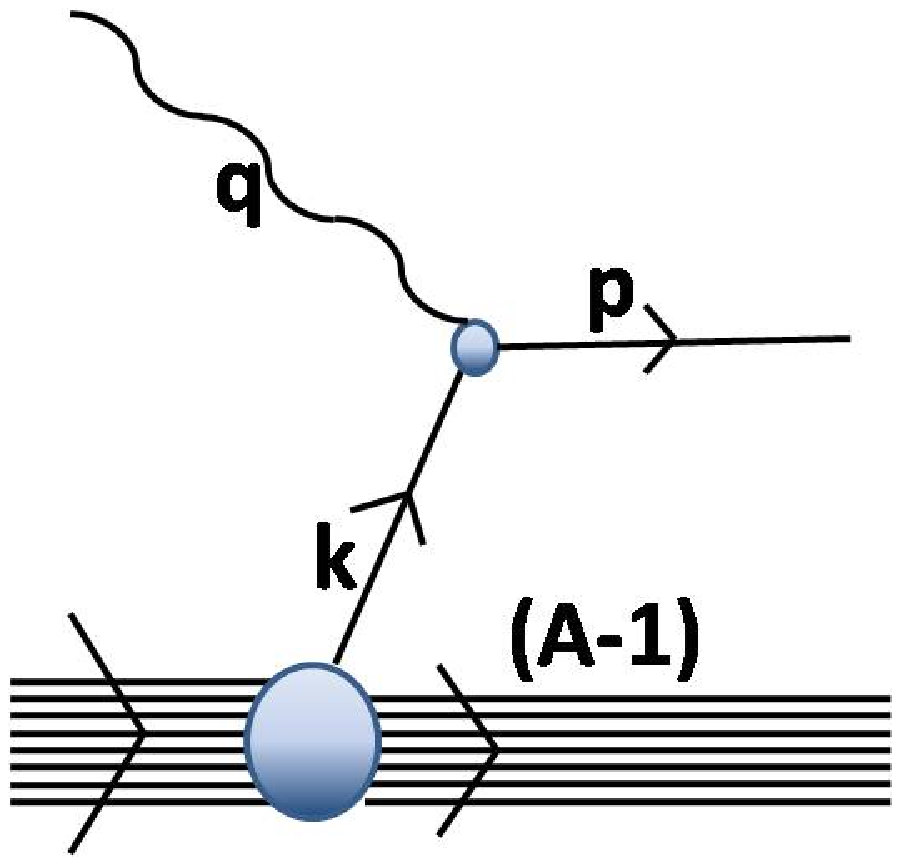}}\qquad
\subfigure[\label{Fey_b}]{\includegraphics[scale=0.6]{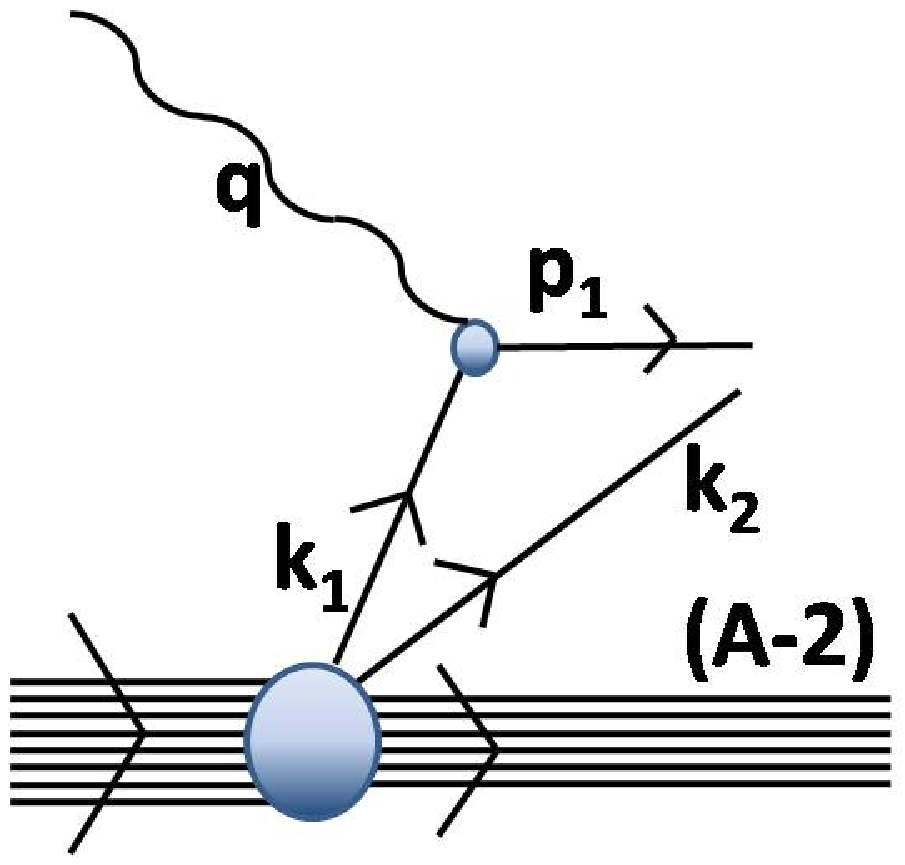}}\\
\subfigure[\label{Fey_d}]{\includegraphics[scale=0.6]{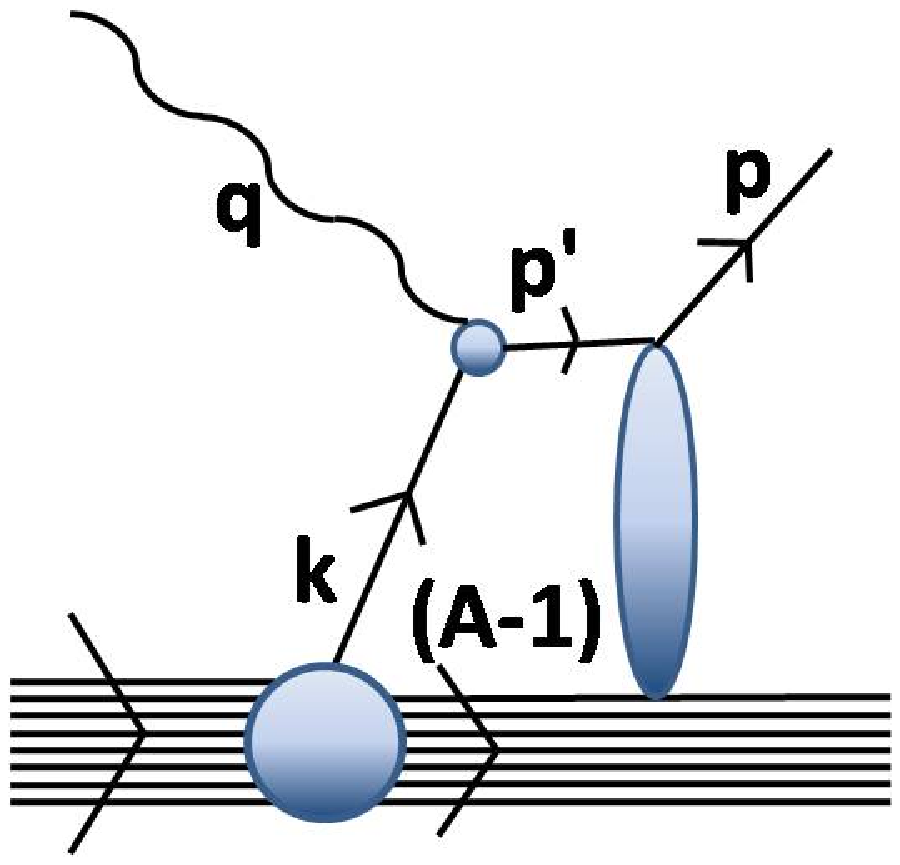}}\qquad
\subfigure[\label{Fey_c}]{\includegraphics[scale=0.6]{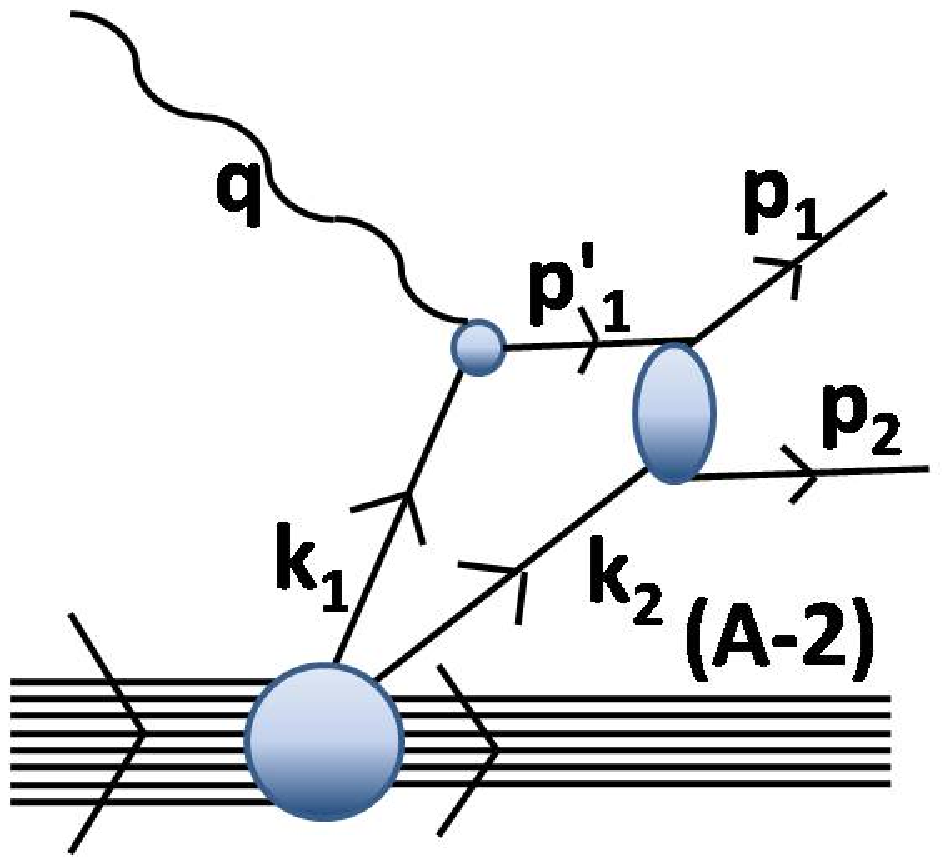}}
\caption{Processes contributing to the $A(e,e')X$ cross section: (a) one-nucleon emission within the IA; (b) virtual photon absorption by a correlated NN pair within the IA; (c) single nucleon rescattering of a nucleon knocked out from shell model states; (d) elastic two-nucleon rescattering between the emitted nucleons of a correlated pair \cite{CSlett}.}\label{Fig_Fey}
\end{center}
\end{figure}
\\The Feynman diagram depicted in Fig.  \ref{Fig_Fey_incl} describes the inclusive $A(e,e')X$ cross section which, as is well known, in one photon exchange has the following form
\be
    \sez=\sigma_{M}\: \left[ W_2^A(Q^2,\nu)+2\tan^2\frac{\theta}{2}
     W_1^A(Q^2,\nu) \right]
\ee
where
\be
\sigma_{M}=\frac{\alpha^2\cos^2\frac{\theta}{2}}{4\epsilon_1^2\:\sin^2\frac{\theta}{2}}
\ee
is the Mott cross section, which describes the scattering from a pointlike nucleon,
\beq
    Q^2=-q^2=\textbf{q}^2-\nu^2=4\epsilon_1\:\epsilon_2\:\sin^2(\theta/2)>0
\eeq
is the squared four-momentum transfer, $\theta$ is the scattering angle and $W_{1,2}$ are the nuclear structure functions.
\\In plane wave impulse approximation (PWIA), shown in Figs.\ref{Fey_a} and \ref{Fey_b}, the latter are given by
    %
\beq
    W^A_i(Q^2,\nu)=\sum_{N=1}^A\int\: d\textbf{k}\:\int\:dE\:P_N(k,E)\:\left[C_iW_1^N(Q^2,\nu')+D_i W_2^N(Q^2,\nu')\right]
\eeq
where $i=\{1,2\}$, $W_{1,2}^N$ are the nucleon structure functions, $C_i$ and $D_i$ are kinematical factors whose explicit expression is given e.g. in \cite{CL}, and, finally,
\beq
    \nu'=\frac{p\cdot Q}{m_N}
\eeq
where $p$ is the four-momentum of the struck off-shell nucleon.
\\According to the value of the invariant mass $W$ produced by the interaction of the virtual photon with a nucleon in the nucleus, the inclusive process $A(e,e')X$ is governed by the following two mechanisms: i) the quasi elastic (qe) process, for which $W=m_N$, and ii) the deep inelastic scattering (DIS), corresponding to $W>m_N$. The basic nuclear  quantity that governs qe and DIS processes at $\xb >1$ is the nucleon spectral function.
\\The PWIA is only the first approximation to the calculation of the cross section, since the final state interaction (FSI) of the struck nucleon with the residual  system ($A-1$) may play an important role. As $Q^2$ increases, the effects due to the FSI in the qe process are expected to decrease, whereas the DIS contribution increases. \\In this Thesis, we will focus only on the qe process. Two types of FSI in the qe process, to be considered in what follows, are shown in Fig. \ref{Fey_d} and \ref{Fey_c}.

%
%
\section{The quasi elastic cross section}\label{sec:QE_XS}
In PWIA, the qe inclusive cross section describing the knock-out by the incoming electron of a nucleon $N$ from the nucleus $A$, can be written as follows \cite{CPS}
\bey
 &&\sigma_2^A (q,\nu)\equiv \frac{d^{2}\sigma^A_{qe}(q,\nu)}{d{\Omega}_2\,d{\nu}} = \no \\
 &&\sum_{N=1}^{A} \int dE \int d{\bf k} \spettN
 \sigma_{eN}(q,\nu,{\bf k},E)\,\delta(\nu+M_{A} -E_N-E_{A-1})\: .\no \\
 \label{X-section}
\eey
The argument of the energy conserving $\delta$-function is
 \beq
    \nu+M_{A} =\sqrt{m_N^2+({\bf k}+{\bf q})^2}+\sqrt{{M_{A-1}^{*^{2}}} +{\bf k}^{2}}
    \label{energycons}
\eeq
where
\beq
    M_{A-1}^*=M_{A-1}+E_{A-1}^*
\eeq
is the mass of the excited residual system, and momentum conservation reads as follows
\beq
    {\bf q}={\bf p}+{\bf P}_{A-1}={\bf p}-{\bf k}\: .
\eeq
Here $\nu = \epsilon_1 -\epsilon_2$ and ${\bf q}= {\bf
k}_1 -{\bf k}_2$ are the energy  and three-momentum transfers, $\sigma_{eN}$ is the elastic electron cross section off a moving off-shell nucleon with  momentum $k\equiv |{\bf k}|$ and removal energy $E$, ${\bf p}$ and ${\bf P}_{A-1}$ are the momenta of the  undetected struck nucleon and the final $(A-1)$ system; eventually, $P_N^A(k,E)$ is the nucleon spectral function discussed in the prevous chapter. From now on, for ease of presentation, only isoscalar nuclei, i.e. with  $P_p^A(k,E)=P_n^A(k,E)\equiv P^A(k,E)$, will be considered.
\\After integrating over $\cos\alpha=(\textbf{k}\cdot \textbf{q})/(kq)$, Eq. (\ref{X-section}) becomes
\bey \label{X-section2}
     \sigma_2^A (q,\nu)&=&2\pi \sum_{N=1}^{A} \int_{E_{min}}^{E_{max}(q,\nu)} dE\, \int_{k_{min}(q,\nu,E)}^{k_{max}(q,\nu,E)} \: kdk \spettN \no \\
 &\times& \: \overline{\sigma}_{eN}(q,\nu,{\bf k},E)\,\left|\frac{\p \nu}{k \p \cos\alpha} \right|^{-1}
\eey
where the limits of integration \cite{CPS4}
\bey
    E_{min}&=&|E_A|-|E_{A-1}|=M_{A-1}+m_N-M_A \\
    E_{max}&=&M_{A}^*-M_A \\
    \label{kmin}
    k_{min}&=&\frac{(\nu+M_A)\:\left|k_{CM}-\left[q-(\nu+M_A)\right]\: \sqrt{\left(
    M_{A-1}+E_{A-1}^*\right)^2+k_{CM}^2} \right|}{M_A^*} \no\\ \\
    k_{max}&=&\frac{(\nu+M_A)\:\left\{k_{CM}+\left[q-(\nu+M_A)\right]\: \sqrt{\left(
    M_{A-1}+E_{A-1}^*\right)^2+k_{CM}^2}\right\}}{M_A^*} \no \\
\eey
are imposed by energy conservation, where
\beq
    M_A^*=\sqrt{\left(\nu+M_A \right)^2-q^2}
\eeq
is the invariant mass,
\beq
    k_{CM}=\frac{\sqrt{\left[ M_A^{*^2}- \left( M_{A-1}+E_{A-1}^{f^*}\right)^2-m_N^2\right]^2-4\left( M_{A-1}+E_{A-1}^*\right)^2m_N^2}}{2M_A^*}
\eeq
and the phase space factor
\beq
\frac{\p \nu}{k \p \cos\alpha}=\frac{q}{\sqrt{m_N^2+q^2+k^2+2kq\cos\alpha}}
\eeq
results from the dependence upon $\cos\alpha$ of the nucleon energy in the final state in the $\delta$-function. Eventually,
\bey
    \bar{\sigma}_{eN}=\frac{\sigma_M}{E_1\:E_2}\: \Bigg\{ \left(\frac{q_\mu^2}{q^2}\right)^2\:
    \left[\frac{(E_1+E_2)^2}{4}\:(F_{1N}^2+\bar{\tau}F^2_{2N})-\frac{q^2}{4}(F_{1N}+F_{2N})^2\right] \no \\
            +\left[\tan^2\frac{\theta}{2}+\frac{q^2_\mu}{2q^2}\right]\:\left[ k^2\sin^2\alpha
    \left(F_{1N}^2+\bar{\tau F_{2N}^2}\right)+\frac{\bar{q}_\mu^2}{2}\:\left(
    F_{1N}+F_{2N}\right)^2 \right]
    \Bigg\} \quad
\eey
is the electron-nucleon cross section for a relativistically moving nucleon, averaged over the polar angle, where $E_1=\sqrt{m_N^2+k^2}$, $q^2_\mu=q^2-\nu^2$, $\bar{q}_\mu^2=q^2-(E_1-E_2)^2$, and $\bar{\tau}=\bar{q}_\mu^2/(4\:m_N^2)$.
\\At high momentum transfer, the quantity
\be
    \left[Z\bar{\sigma}_{ep}+N\bar{\sigma}_{en}\right]\:\left| \frac{\p\nu}{k\p\cos\alpha}\right|^{-1}
\ee
depends very weakly upon $k$, so that it can be taken out of the integral and evaluated, e.g., at $k=k_{min}$. Therefore Eq. (\ref{X-section2}) can be written in the factorized form
\beq \label{xsPWIA}
    \sigma_2^A (q,\nu)=\left[Z\:\bar{\sigma}_{ep}+N\:\bar{\sigma}_{en}\:\left|\overline{\frac{\p \nu}{k \p \cos\alpha} } \right|^{-1}\right]_{(k_{min},E_{min})}\:F^A(q,\nu)
\eeq
where
\beq \label{Funzscala}
    F^A(q,\nu)=2\pi \int_{E_{min}}^{E_{max}(q,\nu)} dE \int_{k_{min}(q,\nu,E)}^{k_{max}(q,\nu,E)} k\:dk\: P^A(k,E)
\eeq
is the nuclear structure function.
%
%
\section{The scaling function}
At high values of the momentum transfer, the rapid falloff of $P^A(k,E)$ with $k$ and $E$ allows the replacement $E_{max}=k_{max}=+\infty$, and Eq. (\ref{Funzscala}) can be written as
\be \label{Funzscala22}
    F^A(q,\nu)=2\pi \int_{E_{min}}^{+\infty} dE \int_{k_{min}(q,\nu,E)}^{+\infty} k\:dk\: P^A(k,E)
\ee
Therefore, the structure function depends upon $q$ and $\nu$ only through $k_{min}$, which is determined from the energy conservation (\ref{energycons}).
\\Let us replace the energy transfer $\nu$ with a generic scaling variable
\be
    Y=Y(q,\nu)
\ee
which is only required to be a function of $q$ and $\nu$ (and any arbitrary constant)  so that, no matter with the specific form of $Y$, the cross section and  the structure function can be expressed  not in terms of the two canonical independent variables $q$ and $\nu$, but, without loss of generality,  in terms of $q$ and $Y=Y(q,\nu)$. Correspondingly, a  scaling function  $F^A(q,Y)$ is introduced, which is nothing but Eq. (\ref{Funzscala22}) with  $\nu$ replaced  everywhere by $Y$; if, under certain conditions, $F^A(q,Y)$ reproduces the asymptotic scaling function
\be \label{FY}
    F^A(Y)=2\pi \int_{E_{min}}^{+\infty} dE \int_{k^\infty_{min}(Y,E)}^{+\infty} k\:dk\: P^A(k,E)
\ee
\emph{Y-scaling} is said to occur and, depending on the physical meaning of $Y$ and $F^A(Y)$, various information on nucleons in nuclei could be obtained.
\\Using Eq. (\ref{decomposition}), the scaling function becomes
\bey \label{scalingfun}
    F^A(q,Y)=2\pi \sum_{\alpha<\alpha_F}\int_{k_{min}(q,Y,E)}^{+\infty} n_{\alpha}(k) +
    2\pi \int_{E_{min}}^{+\infty} dE \int_{k_{min}(q,Y,E)}^{+\infty} k\:dk\: P_1^A(k,E) \no \\
\eey
and it can be trivially cast in the form
\be \label{FqY}
    F^A(q,Y)=f^A(Y)-B^A(q,Y)
\ee
where the longitudinal momentum distribution
\beq \label{longitudinal}
f^A(Y)=2\pi \int_{|Y|}^{+\infty} k\: dk \:n^A(k)
\eeq
is integrated over all excited states of $(A-1)$, whereas  the \emph{binding correction}
\beq \label{binding}
    B^A(q,Y)=2\pi \: \int_{E_{min}}^{+\infty} dE \: \int_{|Y|}^{k_{min}(q,Y,E)} k\:dk\: P^A_1(k,E)
\eeq
on the contrary, is governed through $ k_{min}(q,Y,E)$ by the continuum energy spectrum of the final $(A-1)$ system. The contribution arising from the latter strongly depends by the difference between $Y$ and $k_{min}$ and, therefore, upon the definition of the former. In the Deuteron case, in fact
\beq
    E=E_{min}=2.22\: MeV
\eeq
\beq
    k_{min}(q,Y,E_{min})=|Y|
\eeq
and thus
\beq
    B^D(q,Y)=0
\eeq
\beq
    F^D(q,Y)=f^D(Y) \: .
\eeq
Therefore, when the \emph{binding correction} can be neglected, as in the Deuteron, the quantities  $f^A(Y)$ and  $n^A(k)$ are linked by the relation
\beq
    n^A(k)=-\frac{1}{2\pi Y}\frac{df^A(Y)}{dY}
\eeq
so that, if $f^A(Y)$ could be extracted from the experimental data, $n^A(k)$ could be
determined.
\\Unfortunately, in general, such an extraction is hindered by  the presence of
\beq
    B^A(q,Y) \neq 0
\eeq
which leads to
\beq
    F^A(q,Y) \neq f^A(Y)
\eeq
and
\beq \label{nkFY}
    n^A(k)=-\frac{1}{2\pi Y}\: \left[ \frac{df^A(Y)}{dY}+ \frac{dB^A(Y)}{dY} \right]\: .
\eeq
The factorization given in  Eq. (\ref{FqY}), in the asymptotic limit reads as
\be
F^A(Y)=f^A(Y) -B^A(Y)
\ee
but again, unfortunately, owing to the presence of $B^A(Y)$, $F^A(Y)$ is not related to a momentum distribution, so that,  in principle, the experimental longitudinal momentum distribution $f_{ex}^A(y)$ and, consequently, $n_{ex}^A(k)$, cannot be extracted from the data.
\\In both Eqs. (\ref{FY}) and (\ref{FqY}), the contribution arising from the \emph{binding correction} depends upon the difference between $Y$ and $k_{min}(q,Y,E)$, which could be minimized only by a proper choice of the scaling variable $Y$, such that $k_{min}(q,Y,E)\simeq |Y|$. The resulting  cross section (\ref{X-section}) would depend only upon the nucleon momentum distributions, obtaining, by this way, a direct access to high momentum components generated by SRC \cite{CCrapid}. It is clear that the outlined picture can in principle be modified by the effects of the FSI of the knocked nucleon with the residual $(A-1)$-nucleon system.
%
%
%
%
\section{The final state interaction}
Owing to the decomposition rule (\ref{decomposition}) of the spectral function $P^A(k,E)$, the inclusive cross section can be written in the following form
\beq
    \sigma^A_2=\sigma_0^A+\sigma_1^A
\eeq
where $\sigma_0^A$ describes the transition to the ground and one-hole states of the $(A-1)$-nucleon system, and $\sigma_1^A$ the transition to more complex highly excited configurations.
The diagrams in Figs. \ref{Fey_a} and \ref{Fey_b} refer, respectively, to the contribution of $\sigma_0^A$ and $\sigma_1^A$ in PWIA, which is based upon the assumption that the reaction is well described by the exchange of a single virtual photon with a single nucleon, which does not interact with the remaining $(A-1)$-nucleons; in particular, the diagram \ref{Fey_a} represents the contribution arising from shell model configurations, whereas the diagram \ref{Fey_b} mainly describes $\gamma^*$ absorption by $2p$-$2h$ configurations generated in the target ground state by NN correlations.  It is well known that the PWIA sizably underestimates the cross section, both in light and heavy nuclei, at low values of $\nu$, i.e. at $\xb > 1$, owing to FSI processes which arise from the interaction between the knocked out nucleon and the residual system.
\\In this Thesis, following the approach proposed in Ref. \cite{CSlett}, we take into account two different FSI processes, i.e.: i) the two-nucleon rescattering in the final state when the struck nucleon is a partner of a correlated pair, shown in Fig. \ref{Fey_c}, and ii) the single-nucleon rescattering of the struck nucleon with the mean optical potential of the residual $(A-1)$ system, depicted in Fig. \ref{Fey_d}. Let us discuss in more detail these two mechanisms.
%
%
\subsection{Two-nucleon rescattering}
The basic assumption underlying the convolution formula (\ref{PkE2NCf}), is that two nucleons are locally strongly correlated at short separations, with their CM being apart from the spectator $(A-2)$-nucleon system. 2NC in a nucleus are reminiscent of correlations in the Deuteron; indeed, as discussed in $\S$\ref{sec:nmd}, nucleon momentum distributions of light and complex nuclei, at $k > 1.5 \: fm^{-1}$, turn out to be the properly rescaled version of the Deuteron momentum distribution.
\begin{figure}
\begin{center}
\subfigure[\label{H2_CS}]
{\includegraphics[scale=0.5]{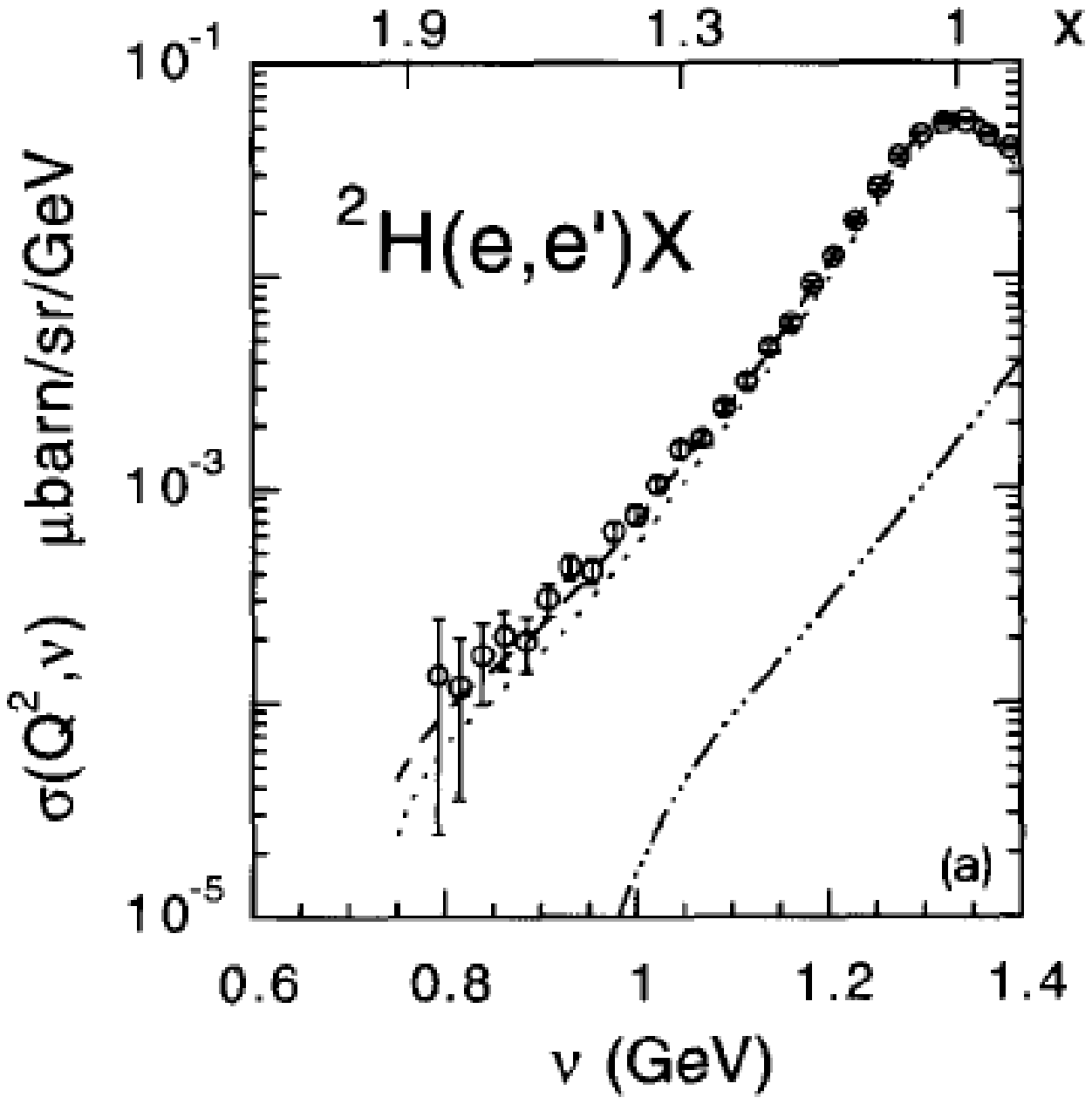}}\:
\subfigure[\label{He4_CS}]{
\includegraphics[scale=0.5]{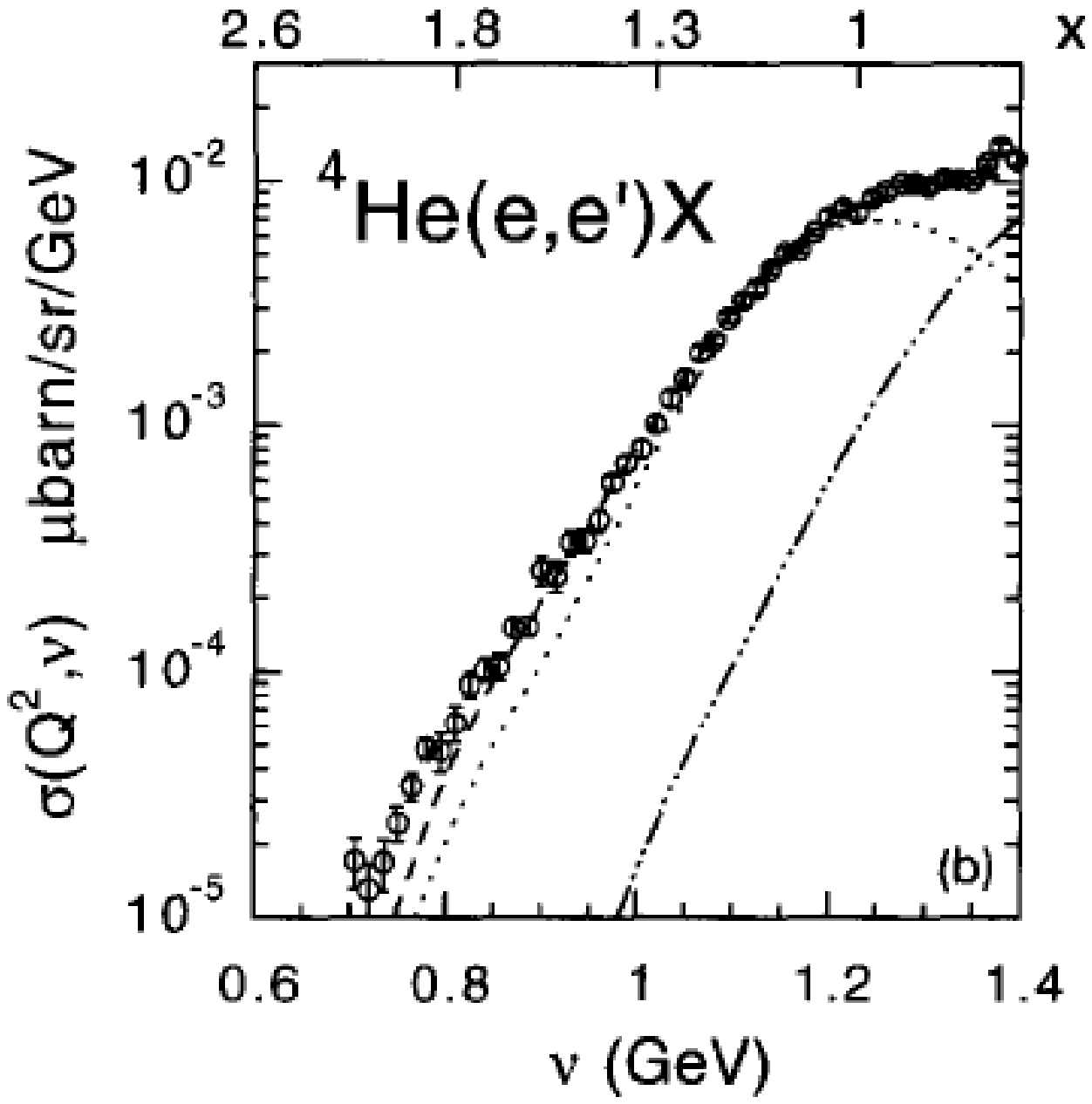}}\\
\subfigure[\label{Fe56_CS}]{
\includegraphics[scale=0.5]{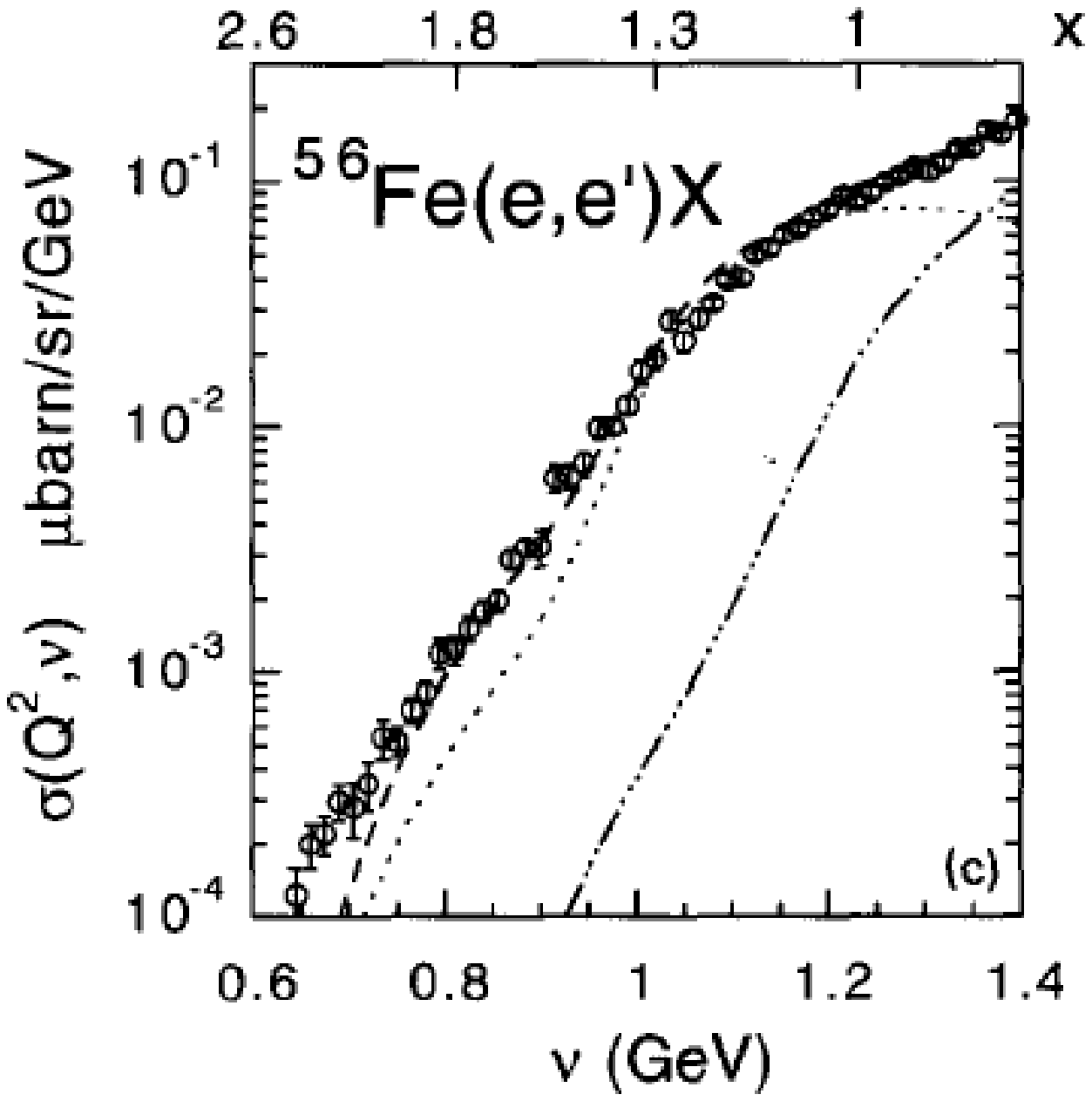}}
\caption{Inclusive cross sections at $Q^2\sim2\:(GeV/c)^2$ versus the energy transfer $\nu$. Calculations have been performed using the free nucleon form factors of Ref. \cite{Galster}, the cc1 presciption of Ref. \cite{DeForest} for $\sigma_{eN}$ and the RSC potential \cite{Reid} for the NN interaction. Dotted line: IA (Figs.\ref{Fey_a})+\ref{Fey_b}; dashed line: IA+ two-nucleon rescattering (Figs. \ref{Fey_a}+\ref{Fey_c});dot-dashed line:contribution from inelastic channels estimated as in Ref.\cite{CDL}. After Ref. \cite{CSlett}.}
\label{CS}
\end{center}
\end{figure}
Therefore, in the 2NC region, the absorption of $\gamma^*$ by a correlated pair  is expected to resemble the one in the Deuteron; if so, such a Deuteron-like picture of the initial state should be extended also to the final state by allowing the two nucleons to elastically rescatter, as depicted in Fig. \ref{Fey_c}.
An important difference from the Deuteron case is that a correlated pair in a nucleus is bound and moves in the field created by the other nucleons.
\\Within the above picture, the contribution of diagrams \ref{Fey_a}-\ref{Fey_c} reads as follows \cite{CSlett}
\bey \label{sigma0PW}
    \sigma_0^A&=&\sum_{\alpha<F} \int d\textbf{k}\:n_\alpha(k)\:\left[Z_\alpha\sigma_{ep}+N_\alpha\sigma_{en} \right]\:\delta\left(\nu+k_\alpha^0-E_{k+q}\right) \\
    \label{sigma12NC}
    \sigma_1^A&=&A\:\sigma_{Mott}\sum_{N_1N_2=n,p}\int d\textbf{k}_{CM}\:n_{CM}^{N_1N_2}(\textbf{k}_{CM})L^{\mu\nu}W_{\mu\nu}^{N_1N_2}
\eey
where
\bey
    E_p&=&\sqrt{m_N^2+\textbf{p}^2}\\
    k^0_\alpha&=&M_A-\sqrt{(M_A+|\epsilon_\alpha|-m_N)^2+k^2}
\eey
$L^{\mu\nu}$ represents the (reduced) leptonic tensor, and $W^{\mu\nu}$ is the hadronic tensor of a correlated pair, which can be written as follows
\bey \label{Wmunu}
    W_{\mu\nu}^{N_1N_2}&=&\sum_{f_{12}}\:\sum_{\beta_{12}}\left[<\beta_{12}|j_{\mu}^{N_1}+j_{\mu}^{N_2}|f_{12}>\right]^*
    \sum_{\beta'_{12}}\left[<\beta'_{12}|j_\nu^{N_1}+j_\nu^{N_2}|f_{12}>\right]\:\no \\
    &\times&\delta\left( \nu+k^0_{CM}-\sqrt{(M_2^{f_{12}})^2+(\textbf{k}_{CM}+\textbf{q})^2}  \right)
\eey
where $j_\mu^N$ is the nucleon current, $k^0_{CM}=M_A-\sqrt{M_{A-2}^2+\textbf{k}^2_{CM}}$, $|\beta_{12}>$ is the relative wave function of the initial state of a correlated pair, and $|f_{12}>$ its continuum final state. Eq. (\ref{sigma12NC}) is based upon the following assumptions on final and initial $A$-nucleon state:
\bey
    |\Psi_A^f>&\sim&\hat{\mathcal A}\left\{|f_{12}>|\textbf{P}_{CM}>|\Psi_{A-2}^f>\right\} \\
    |\Psi_A^0>&\sim&\hat{\mathcal A}\left\{|\beta_{12}>|\mathbf{\chi}^{CM}_{12}>|\Psi_{A-2}^0>\right\}
\eey
where $\hat{\mathcal A}$ is a proper antisymmetrization operator, $|\mathbf{\chi}^{CM}_{12}>$ is the CM wave function of the initial state of a correlated pair, and $|\textbf{P}_{CM}>$ its plane wave final state. It can be seen from Eq. (\ref{Wmunu}) that medium effects on the hadronic tensor of the pair are generated by the energy conserving $\delta$-function, in which the intrinsic energy available to the pair is fixed by its CM four-momentum, and, therefore, by the momentum distribution $n_{CM}^{N_1N_2}$ appearing in Eq. (\ref{sigma12NC}). Even if the CM motion is neglected by placing
\beq
    n_{CM}^{N_1N_2}=\delta(\textbf{k}_{CM})
\eeq
medium effects still would remain through the quantity
\beq
(k_{CM}^0)_{max}=M_A-M_{A-2}
\eeq
which is related to the two-nucleon break up threshold, i.e. the binding of the pair.
\\The inclusive cross section have been calculated in \cite{CSlett} for the Deuteron using the RSC NN potential \cite{Reid}, taking into account the rescattering in S, P and D partial waves; then, using the same two-nucleon amplitudes  $<\beta_{12}|j_\mu^{N_1}+j_\mu^{N_2}|f_{12}>$, the cross section $\sigma_1^A$  have been computed for complex nuclei. The results are shown by the dashed lines in Fig. \ref{CS}: it can be seen that, at $1.3 <\xb < 2$, the process of two-nucleon rescattering brings theoretical predictions in good agreement with experimental data. The most striking aspect of these results is that the same mechanism which explains the Deuteron data, does the same in a complex nucleus, provided the A dependence due to $n_{CM}^{N_1\:N_2}$ and $k^0_{CM}$ (clearly exhibited in Fig. \ref{CS}) is properly considered. It should be pointed out that these results hold for the whole set of kinematics considered in Refs. \cite{Day2,Schutz,Rock,Day}.
%
%
\subsection{Single nucleon rescattering}
\begin{figure}[!h]
\centerline{\includegraphics[scale=0.8]{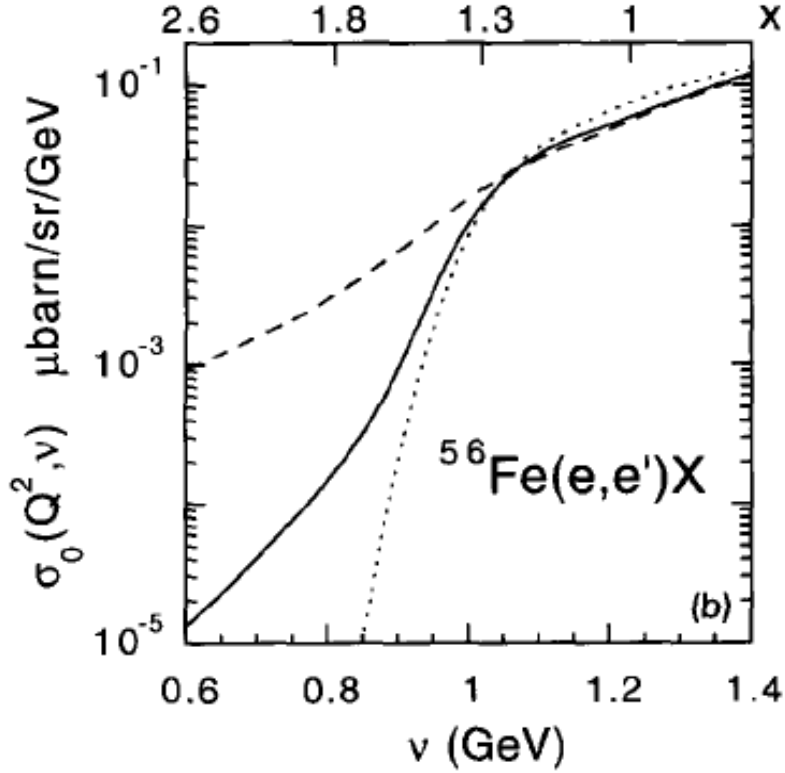}}
\caption{The cross section $\sigma_0$ versus the energy transfer $\nu$. Dotted line: Eq. (\ref{sigma0PW}) (diagram \ref{Fey_a}); dashed line: Eq. (\ref{sigma0op}) (diagrams \ref{Fey_a}+\ref{Fey_d}) calculated using the on-shell potential (\ref{Vopt}); solid line:Eq. (\ref{sigma0op}) (diagrams \ref{Fey_a}+\ref{Fey_d}) calculated using the off-shell potential (\ref{suppr}). After Ref. \cite{CSlett}.}
\label{Fig_suppr}
\end{figure}

The two-nucleon rescattering is not able to describe the experimental data at $x_{Bj} \gtrsim 2$. This fact is not surprising because, at $\xb \gtrsim 2$, more than two nucleons should be involved in the scattering process. This process can be simulated by considering the motion of the nucleon, knocked out from shell model states, in the optical potential generated by the ground state  of the $(A-1)$-nucleon system. Within such an approach, corresponding to diagrams \ref{Fey_a}-\ref{Fey_d}, Eq. (\ref{sigma0PW}) becomes \cite{CSlett}
\beq \label{sigma0op}
    \sigma_0^A=-\sum_{\alpha<F}\int d\textbf{k}\:n_\alpha(k)\:\left[Z_\alpha\sigma_{ep}+N_\alpha\sigma_{en} \right]\:
    \frac{\Im\:V_{opt}}{\left[ \nu + k_\alpha^0 - E_p-\Re\:V_{opt} \right]^2 + \left[\Im\:V_{opt}\right]^2}
\eeq
resulting from the eikonal approximation for the nucleon propagator \cite{Gurvitz}. The optical potential $V_{opt}$ can be cast in the following on-shell form
\beq \label{Vopt}
    V_{opt}=-\rho\:v_N\:\sigma_{NN}\:\frac{\left(\imath + \alpha_{NN}\right)}{2}
\eeq
where $\rho$ is the nuclear density, $v_N$ is the nucleon velocity, $\sigma_{NN}$ is the total NN cross section and $\alpha_{NN}$ is the ratio of the real to the imaginary part of the forward NN scattering amplitude. The imaginary component of Eq. (\ref{Vopt}) describes inelastic processes leading to excitations of the residual system, which, in the high energy regime, mainly correspond to secondary nucleon emissions.
\\It has be pointed out \cite{CSlett}, that the two-nucleon rescattering is not included in the process described by diagram \ref{Fey_d}. As a matter of fact, indeed, the two-nucleon rescattering is not a multiple scattering process, i.e. it does not contribute  to an optical potential, and, particularly, is independent of nuclear density and does not produce any absorption of the outgoing flux.
\\However, treating FSI at $\xb >1$ in terms of on-shell potentials is not justified \cite{mark1,FS,FSDS,Uchiyama}. Indeed, the struck nucleon, with momentum
\beq
    p'^{\,2}\simeq (\nu+m_N-E)^2-(\textbf{k}+\textbf{q})^2
\eeq
can be either on-mass-shell, i.e. with $p'^{\,2}=m_N^2$, or off-mass-shell, namely with $p'^{\,2}\neq m_N^2$, depending on the values of $k$ and $E$. Initial configurations with $k<k_{min}$ give rise to an intermediate off-mass-shell virtual nucleon, so that the use of an on-shell optical potential is unjustified, and can hardly be reconciled with the fact that rescattering amplitudes are expected to decrease with virtuality, because an off-shell nucleon has to rescatter within short times. Therefore, in order to take into account off-shell effects, we have to include in $V_{opt}$ a \emph{suppression factor} of the type
\beq \label{suppr}
    V_{opt}=-\frac{1}{2}\rho\:v_N\:\sigma_{NN}\:\left( \imath+\alpha_{NN}\right)\:e^{-\delta|M^2-p'^2|} \: .
\eeq
The differences between the use of an on-shell or off-shell potential are shown in Fig. \ref{Fig_suppr}.  The parameter $\delta$ appearing in Eq. (\ref{suppr}) is the same for all kinematics considered.
\section{Inclusive cross sections: results of calculations}
\begin{figure}[!h]
\centerline{\includegraphics[scale=0.6]{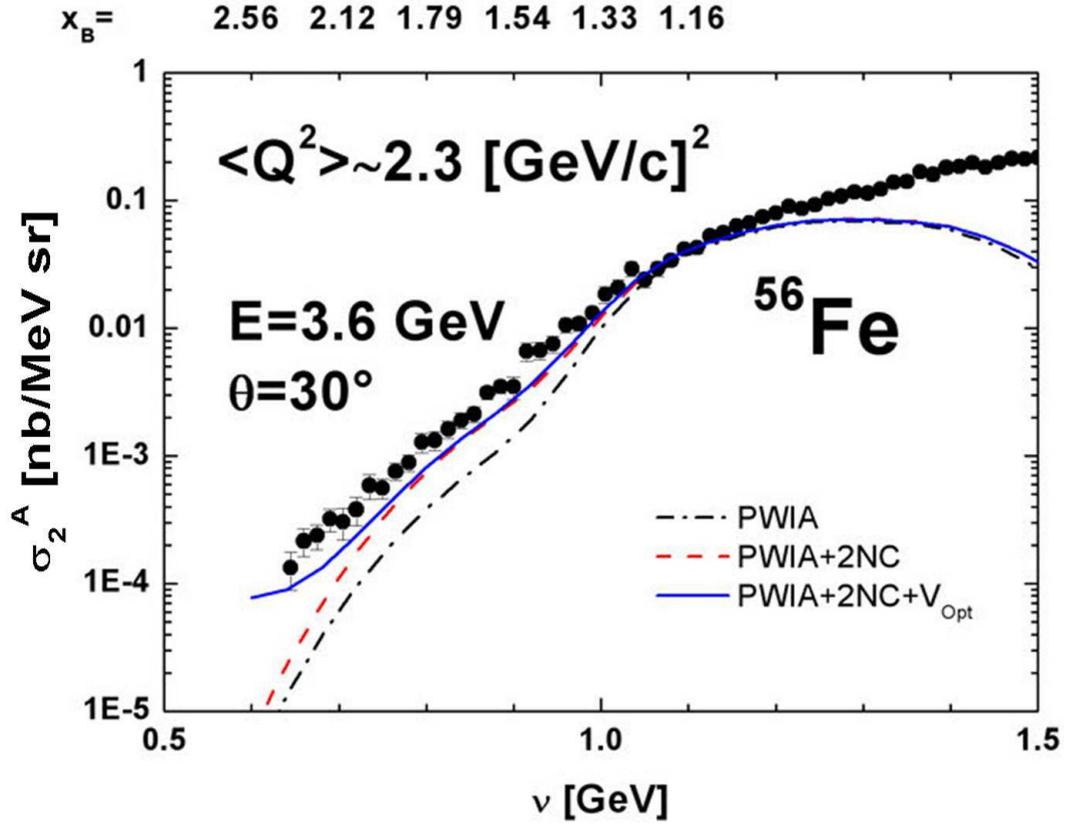}}
\caption{The experimental inclusive cross section $^{56}Fe(e,e')X$ \cite{Day} \emph{vs.} the energy transfer $\nu$, compared with theoretical calculations which include SRC and FSI. Dot-dashed line: PWIA, Eq. (\ref{xsPWIA}); dashed line: PWIA + FSI of the correlated struck nucleon with the correlated partner, Eqs. (\ref{sigma0PW}) and (\ref{sigma12NC}); solid line: the same as dashed red line plus the FSI of the shell model struck nucleon with the mean field of the residual ($A-1$)-nucleon system, Eqs. (\ref{sigma0op}) and (\ref{sigma12NC}). After Ref. \cite{Trieste}.}
\label{Fig_30}
\end{figure}

\begin{figure}[!h]
\centerline{\includegraphics[scale=0.6]{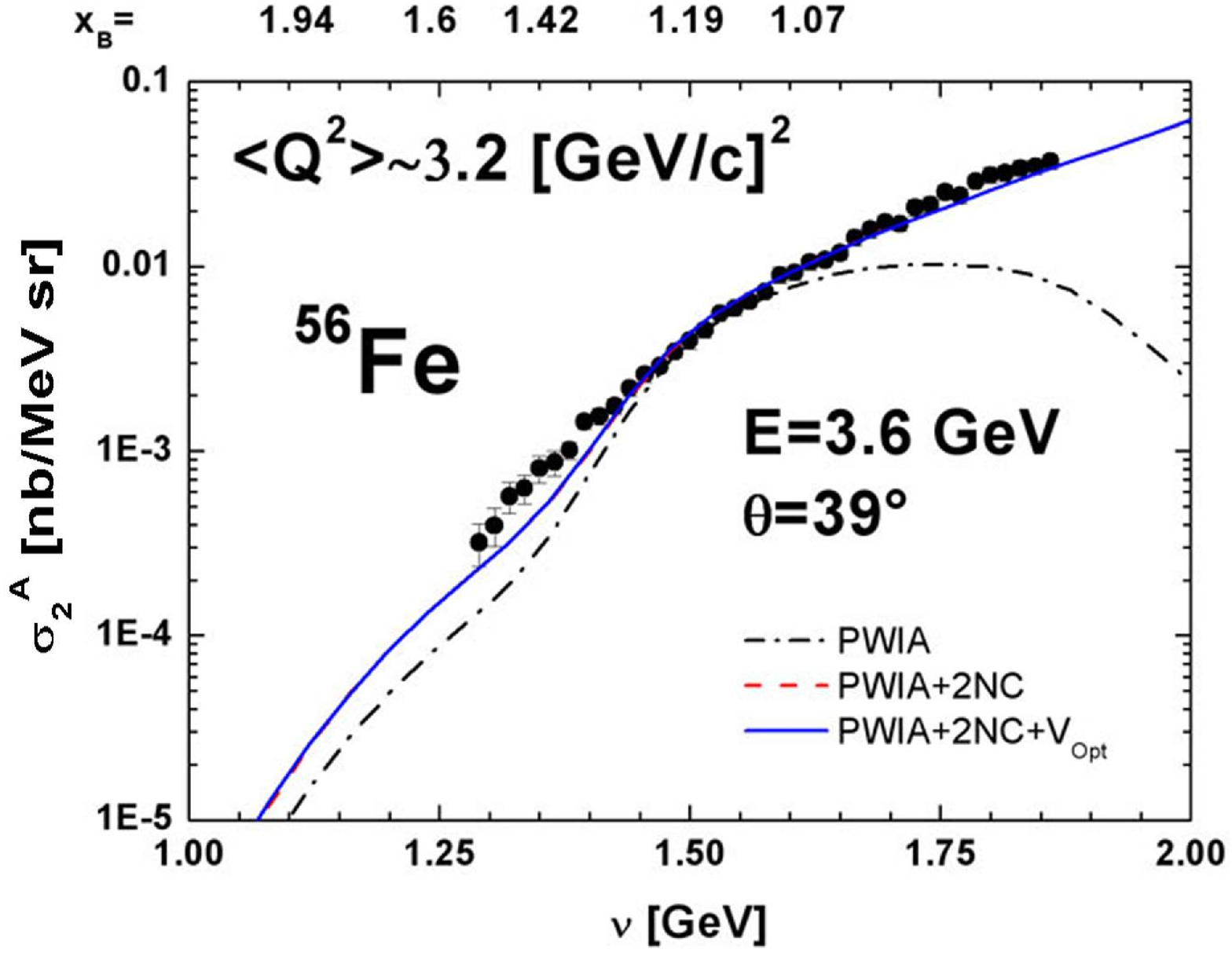}}
\caption{The same as in Fig. \ref{Fig_30}, for different kinematic conditions. After Ref. \cite{Trieste}.}
\label{Fig_39}
\end{figure}

\begin{figure}[!h]
\centerline{\includegraphics[scale=1.6]{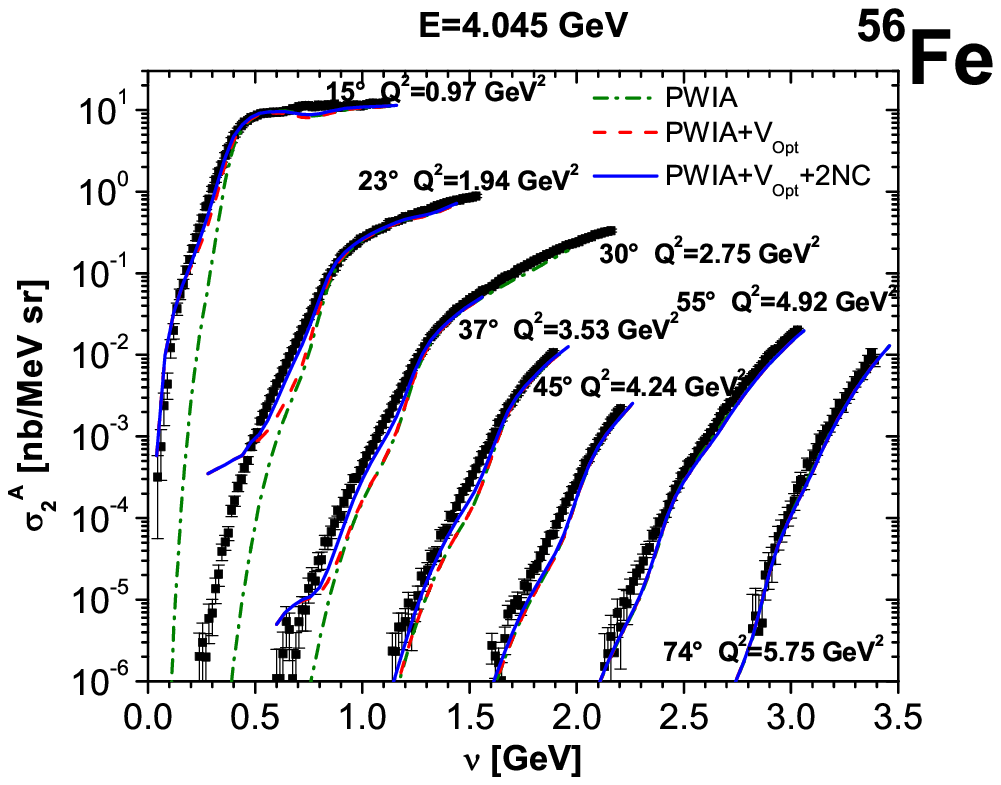}}
\caption{The same as in Fig. \ref{Fig_30}. Experimental data from Ref. \cite{arrington}.}
\label{Fig_CEBAF_F}
\end{figure}

\begin{figure}[!h]
\centerline{\includegraphics[scale=1.6]{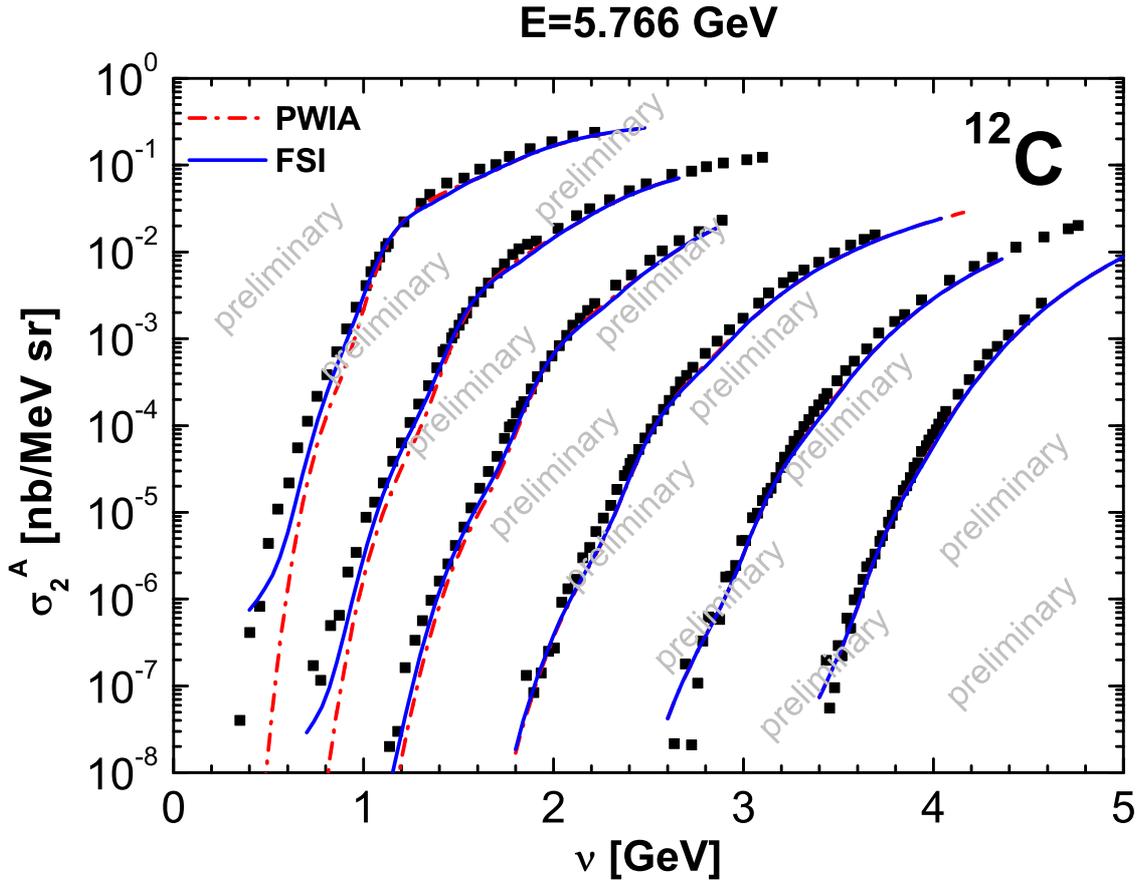}}
\caption{The experimental inclusive cross section $^{56}Fe(e,e')X$ \cite{fomin} \emph{vs.} the energy transfer $\nu$, compared with our preliminary calculations which include SRC and FSI. Dot-dashed line: PWIA, Eq. (\ref{xsPWIA}); solid line:  PWIA plus the FSI of the shell model struck nucleon with the mean field of the residual ($A-1$)-nucleon system and the FSI of the correlated struck nucleon with the correlated partner, Eqs. (\ref{sigma0op}) and (\ref{sigma12NC}).}
\label{Fig_DAY_C}
\end{figure}

The results of our calculations, including the contributions from all diagrams depicted in Fig. \ref{Fig_Fey}, are shown in Figs. \ref{Fig_30} and \ref{Fig_39}.
It can be seen that the PWIA overestimates the experimental data at $\xb > 1.5$, whereas the inclusion of the FSI produces a good agreement between theoretical calculations and experimental data. In the region $1.5\leq\xb\leq2$, the FSI is mainly due to the two-nucleon rescattering whereas, at $\xb>2$, the contribution from the optical potential, which mocks up three-nucleon correlation effects, becomes important. Other results are shown in Figs. \ref{Fig_CEBAF_F} and \ref{Fig_DAY_C}, leading to the same conclusions. Results of the same quality were previously obtained in Ref. \cite{CFW1,CFW2,CSlett}.

\cleardoublepage
\chapter{A novel approach to scaling phenomena in inclusive scattering: mean field, correlations and proper scaling variables.}
\section*{Introduction}
The inclusive cross section can be analyzed in terms of scaling function and scaling variable.
\\The scaling variable $Y$ arises from the energy conservation law which, for an inclusive process, reads as follows
\beq \label{energy}
\nu+M_A=\sqrt{(M_{A-1}+E_{A-1}^*)^2+\textbf{k}^2}+\sqrt{m_N^2+(\textbf{k}+\textbf{q})^2}
\eeq
being
\bey
M_A &=& \mbox{mass of the target nucleus} \\
M_{A-1}&=&\mbox{mass of the residual nucleus $(A-1)$} \\
E_{A-1}^* &=& \mbox{intrinsic excitation energy of the residual system} \\
m_N &=& \mbox{mass of the knocked out nucleon}
\eey
As mentioned in Chapter \ref{ch:PkE_nmd}, a proper choice of the scaling variable $Y$ could minimize the contribution arising from the \emph{binding correction}, allowing a direct link between the scaling function and the nucleon momentum distribution.
Within such an approach, the new inclusive cross section for electron scattering off nuclei can be written only in terms of nucleon momentum distributions, so that inclusive scattering becomes a powerful tool to investigate NN SRC in nuclei.
In what follows, it will be illustrated in detail how the dependence of $k_{min}$ upon $E_{A-1}^*$ gives rise to the \emph{binding correction}, and how to a different definition of the scaling variable $Y$.
%
%
%
\section{The mean field scaling variable}
The final state interaction of the struck nucleon invalidates the PWIA but, in spite of that,  an approach was developed in the past to reduce the effects from both the \emph{binding corrections} and FSI \cite{CPS3,CPS4}; the traditional approach to \emph{Y-scaling} is based upon the traditional scaling variable  $Y\equiv y$, obtained by placing
\beq
    k=|y|
\eeq
\beq
    \cos\alpha=\frac{\textbf{k}\cdot \textbf{q}}{kq}= 1
\eeq
\beq
    E_{A-1}^*=0
\eeq
in the energy conservation law given by Eq. (\ref{energy}), obtaining
\beq \label{energy_y}
\nu+M_A=\sqrt{M_{A-1}^2+y^2}+\sqrt{m_N^2+(y+q)^2}
\eeq
whose solution is \cite{CPS}
\beq
    y=\frac{ -q\Delta \pm \sqrt{q^2\Delta^2-M_{A}^{*^2}
    \left[4 \left( \nu+M_A\right)^2-\Delta^2 \right]}} {2M_A^{*^2}}
\eeq
with
\beq
    M_A^{*^2}=\left( \nu + M_A \right)^2 - q^2
\eeq
and
\beq
    \Delta=M_A^{*^2}+M_{A-1}^2-m_N^2 \: .
\eeq
Within such an approach, $y$ represents the minimum longitudinal momentum of a nucleon having the minimum value of the removal energy
\beq
    E=E_{min}+E_{A-1}^*=E_{min}=m_N+M_{A-1}-M_A\: .
\eeq
At high values of $q$, one has
\beq
    \lim_{q \rightarrow \infty}k_{min}(q,y,E) \equiv k_{min}^\infty(y,E)=|y-(E-E_{min})|
\eeq
so that, when $E=E_{min}$,
\bey
    k_{min}^\infty(y,E)&=&|y| \\
    B^A(q,y)&=&2\pi \: \int_{E_{min}}^{+\infty} dE \: \int_{|y|}^{k_{min}(q,y,E)} k\:dk\: P^A_1(k,E)=0
\eey
and thus the scaling function reduces to
\beq
    F^A(y)=f^A(y)=\int_{|y|}^{+\infty} k\: dk \:n^A(k)
\eeq
explicitly showing scaling in $y$.
\begin{figure}[!h]
\centerline{\includegraphics[scale=1.3]{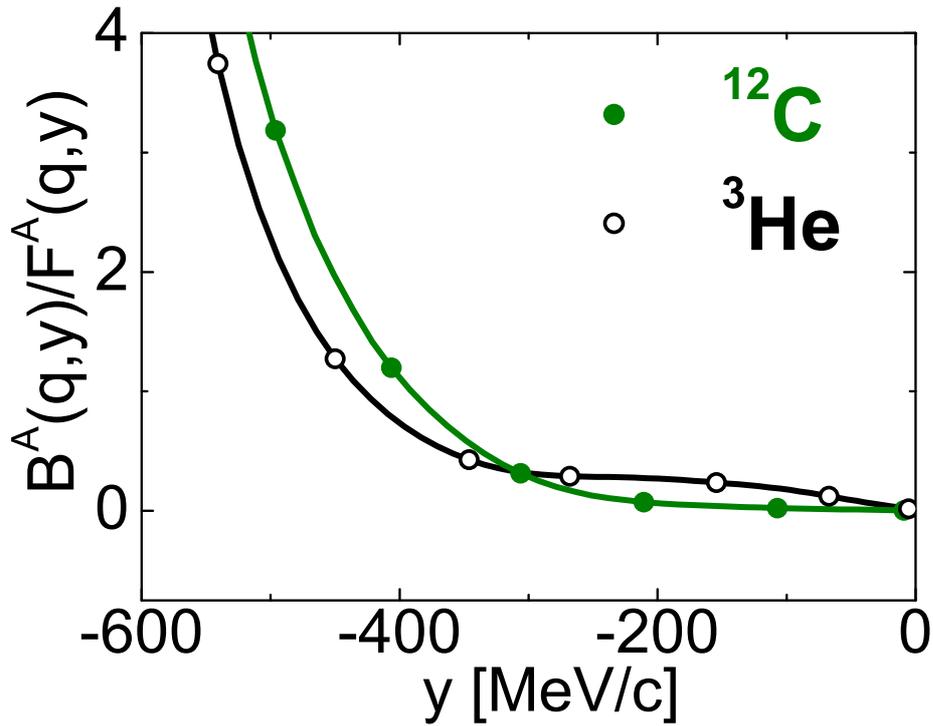}}
\caption{The ratio of the \emph{binding correction} $B^A(q,y)$ (Eq. (\ref{binding})) to the scaling function $F^A(q,y)$ (Eq. (\ref{Funzscala})) for $^3He$ (open dots) and $^{12}C$ (full dots), calculated using the scaling variable $y$. After Ref. \cite{CCrapid}.}\label{Fig3.2}
\end{figure}
\noindent
\\Unfortunately, this occurs only in the Deuteron, whereas in the general case, for $A>2$,  the excitation energy  $E_{A-1}^*$ of the residual system is different from zero and
\beq
    E=E_{min}+E_{A-1}^*>E_{min}
\eeq
leading to
\beq
    B^A(q,y)>0
\eeq
and thus to the relation  $F^A(y) \ne f^A(y)$, given by Eq. (\ref{FqY}).
\\To illustrate the relevant role played by the \emph{binding correction} in the traditional approach to \emph{Y-scaling}, the ratio
\beq \label{ratio}
    \frac{B^A(q,y)}{F^A(q,y)}=\frac{B^A(q,y)}{f^A(y)-B^A(q,y)}
\eeq
which represents the deviation of the scaling function $F^A(q,y)$ from the longitudinal momentum distribution $f^A(y)$, is shown in Fig. \ref{Fig3.2}, plotted versus the scaling variable $y$. It can be seen that, at high values of $|y|$, the effects from binding are very large whereas, at low values of $|y|$, binding effects can be neglected.
\begin{figure}[!h]
\centerline{\includegraphics[scale=0.9]{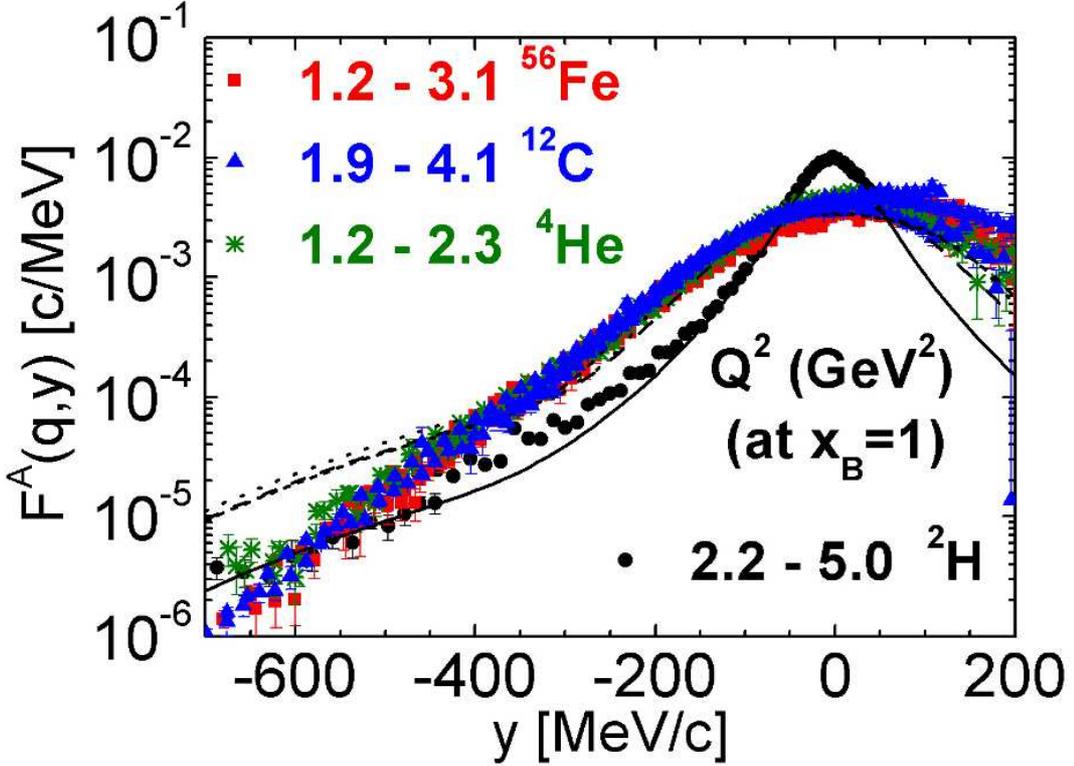}}
\caption{The experimental scaling function $F^A_{exp}(q,y)$
of $^4He$, $^{12}C$, and $^{56}Fe$ obtained from the experimental data of Refs. \cite{arrington,Schutz2}. The longitudinal momentum distributions (Eq. (\ref{longitudinal})) of $^{2}H$ (full line), $^4He$ (long-dashed),  $^{12}C$ (dashed) and  $^{56}Fe$ (dotted) are also shown. After Ref. \cite{CCrapid}.}\label{Fig3.3}
\end{figure}
\\Moreover, the experimental scaling function
\beq
    F_{ex}^A(q,y)=\frac{\sigma_{ex}^A(q,y)}{\left[(Z s_{ep}+N s_{en})\:\frac{E_p}{q}\right]_{k_{min},E_{min}}}
\eeq
plotted versus the scaling variable $y$, as shown in Fig.  \ref{Fig3.3}, confirms that the scaling function strongly differs from the longitudinal momentum distribution, and therefore does not exhibit any proportionality to the Deuteron scaling function $f^D(y)$. Therefore it should be pointed out that, when expressed in terms of $y$, a comparison between experimental and theoretical scaling functions requires the knowledge of the nucleon spectral function, generated by the main role played by $B^A(q,y)$.
\\Moreover, the experimental scaling function  exhibits a strong $q$ dependence owing to the FSI and binding effects, and differs from the asymptotic scaling function $F_{ex}^A(y)$.  The latter, however, has been  obtained in Ref. \cite{CPS3,CPS4}  by extrapolating to $q \rightarrow \infty$ the available values of $F^A_{ex}(q,y)$, on the basis that FSI can be represented as a  power series in $1/q$,  and dies  out at large $q^2$,  a conclusion that has been reached by   various authors (see e.g. \cite{rinawest}).
\\It is therefore the dependence of $k_{min}$ upon $E_{A-1}^{*}$ that gives rise to the binding effect.  This is an unavoidable defect of the usual approach to \emph{Y-scaling}, based on
the scaling variable $y$; except for the trivial case of the Deuteron, in fact, in a complex nucleus the final $(A-1)$-nucleon system can be left in all possible excited states, including the continuum but, by definition, the traditional scaling variable $y$ can only be identified with the longitudinal momentum of weakly bound, shell nucleons (${E_{A-1}^*} \sim 0-20\, MeV$). The longitudinal momentum for such nucleons is very different from the  strongly bound, correlated nucleons (${E_{A-1}^*} \sim 50-200 \,MeV$), and this explains why, at large values of  $|y|$, the scaling function is not related to the longitudinal momentum of strongly bound  correlated nucleons,  whose contributions almost entirely exhaust the behavior of the scaling function.
\\As stressed in Refs. \cite{CW,CFW1,CFW2}, to establish a global link between experimental data  and longitudinal momentum components, one has to conceive a scaling variable {\it that could equally well represent longitudinal momenta of both weakly bound and strongly bound nucleons, so that the binding correction could be minimized}.
\\The experimental longitudinal  momentum distribution $f_{ex}^A(y)$ has thereby been obtained by adding to  $F_{ex}^A(y)$ the \emph{binding correction} $B^A(y)$ evaluated theoretically, as shown in Fig. \ref{Fig_fy}, and $n_{ex}^A(k)$ has been obtained by Eq. (\ref{nkFY}). Such a procedure affects the final results in terms of large errors on  the extracted
momentum distributions, particularly at large values of $k$; in spite of  these errors, the extracted momentum distributions, at $k\gtrsim 1.5-2\,\,fm^{-1}$, turned out to be larger  by orders of magnitude from the prediction of mean field approaches, and in qualitative agreement with realistic many-body calculations that include SRC.
\\In order to make  the extraction of $f_{ex}^A(y)$ as independent as possible from theoretical \emph{binding corrections}, in Ref. \cite{CW} is thus necessary to introduce another scaling variable, incorporating relevant physical dynamical effects left out in the definition of $y$.
\begin{figure}[!h]
\centerline{\includegraphics[scale=0.9]{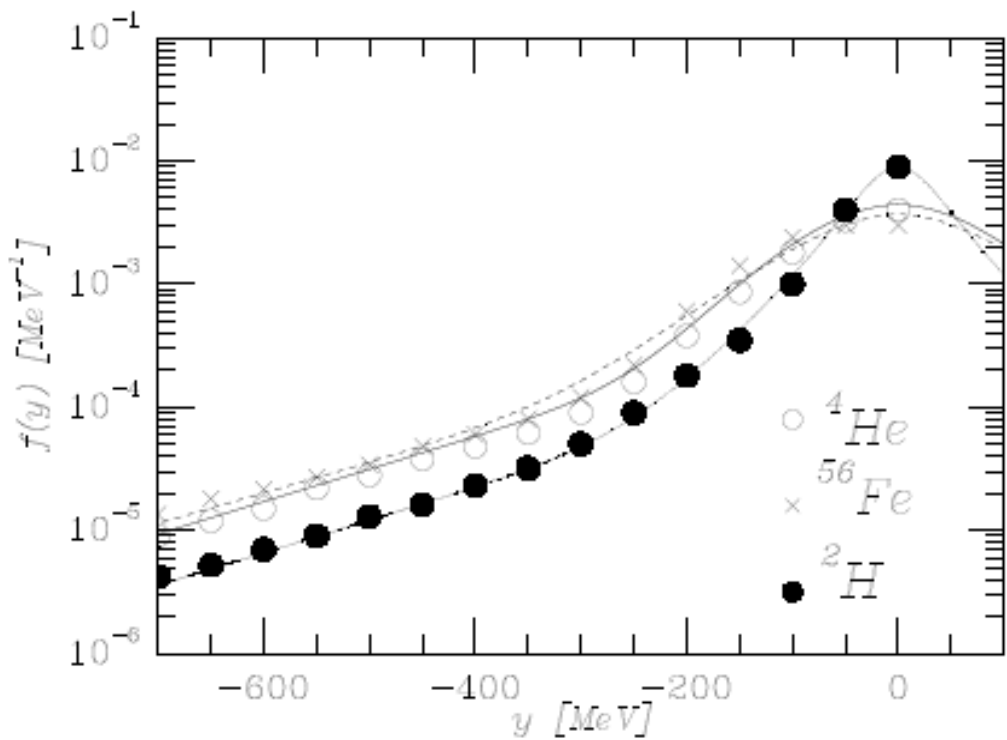}}
\caption{The longitudinal momentum distribution for $^2H$ (dotted line), $^4He$ (full line) and $^{56}Fe$ (dashed line) corresponding to a parametrization obtained in Ref. \cite{CFW1}.}
\label{Fig_fy}
\end{figure}
%
%
\section{Two-nucleon correlation scaling variable}
\begin{figure}[!h]
\centerline{\includegraphics[scale=0.6]{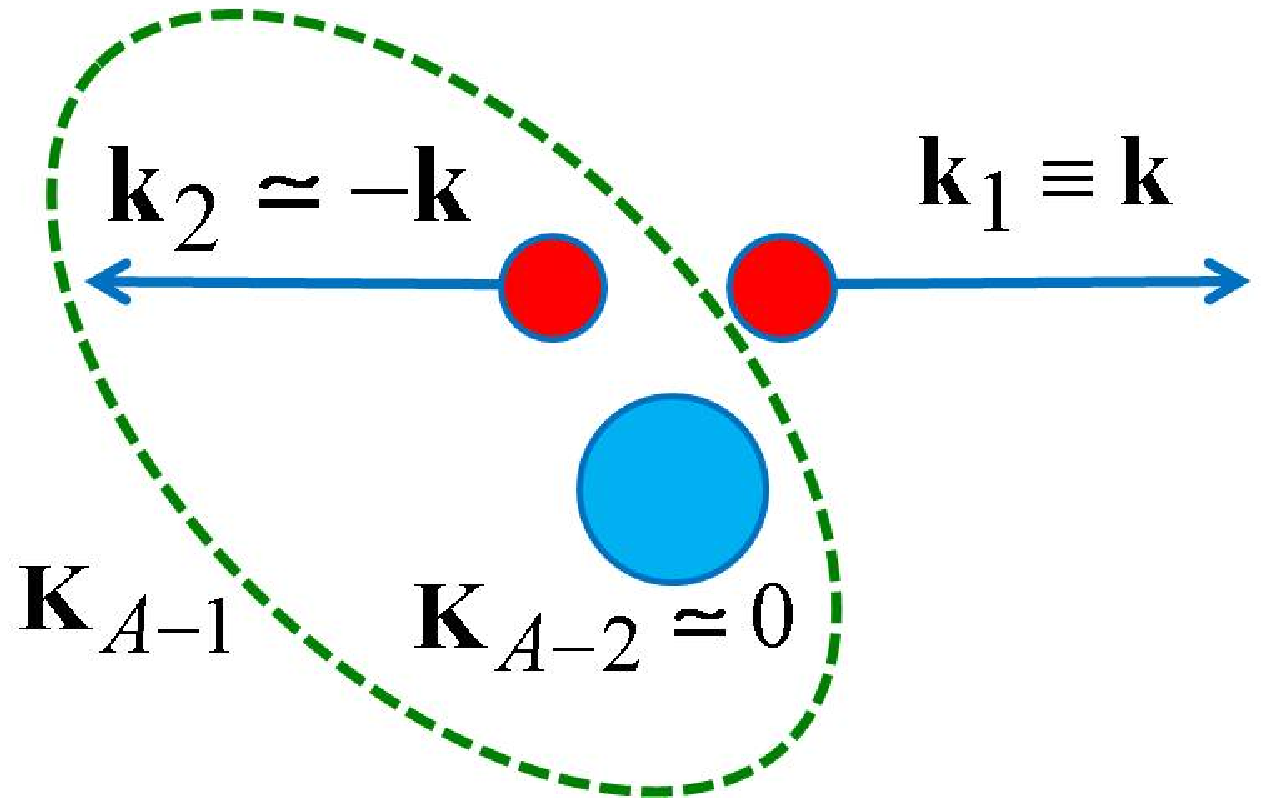}}
\caption{2NC correlations in a nucleus A: the high
momentum $\textbf{k}_1 \equiv \textbf{k}$ of  nucleon "1" is almost completely balanced
by the momentum $\textbf{k}_2\simeq -\textbf{k}$ of the partner nucleon "2",
 whereas the residual system moves with low momentum $\textbf{K}_{A-2}$. Momentum conservation is $\sum_1^A\:\textbf{k}_i=\textbf{k}_1+\textbf{k}_2+\textbf{K}_{A-2}=0$.}\label{Fig3.1}
\end{figure}

\begin{figure}[!hbp]
\centerline{\includegraphics[scale=1.3]{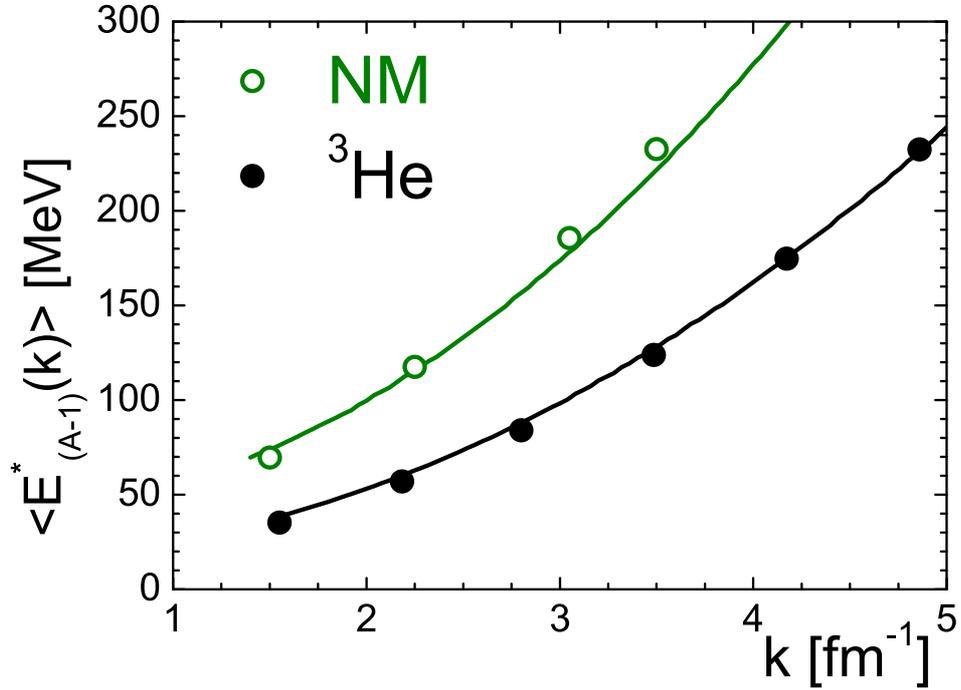}}
\caption{The average value of $E_{A-1}^*(k)$  [Eq. (\ref{average})] calculated for nuclear matter with the spectral function of Ref. \cite{BFF,BFF2} (open dots), and for $^3He$ with the spectral function from the Pisa wave functions \cite{pisa} (full dots). The full lines are obtained with the spectral function of the few-nucleon correlation model of Ref. \cite{ciosim}.}\label{Fig_energy}
\end{figure}

2NC are defined, as previously explained in $\S$\ref{sec:2Nuc}, as those nucleon configurations shown in Fig.  \ref{Fig3.1} \cite{FSS2}: momentum conservation in the ground state of the target nucleus
\beq
    \sum_{i=1}^A \textbf{k}_i=0
\eeq
is almost entirely exhausted by two correlated nucleons with high momenta, the $(A-2)$-nucleon system acting mainly as a spectator, moving with very low momentum.
The intrinsic excitation energy of the $(A-1)$-nucleon system is in this case
\beq
E_{A-1}^*=\frac{(A-2)}{(A-1)}\:\frac{(\textbf{k}_2-\textbf{K}_{A-2})^2}{2m_N}
\eeq
which becomes
\beq
E_{A-1}^*=\frac{(A-2)}{(A-1)}\:\frac{k^2}{2m_N}
\eeq
in the naive 2NC model, i.e. the model based upon the assumption $\textbf{K}_{A-2}=0$.
Since high excitation states of the final $(A-1)$-nucleon system are generated by SRC in the ground state of the target nucleus, the traditional (mean field) scaling variable $y$ does not incorporate, by definition, SRC effects, for it is
obtained by placing $E_{A-1}^*=0$ in the energy conservation law (\ref{energy}). Motivated by this observation, in Ref. \cite{CW}, a new scaling variable, $Y\equiv y_{CW} \equiv y_2$ has been introduced, by setting
\beq
    k=|y_2|
\eeq
\beq
    \cos\alpha=\frac{\textbf{k}\cdot \textbf{q}}{kq}= 1
\eeq
\beq
    E_{A-1}^*=<E_{A-1}^*(k)>_{2NC}
\eeq
in Eq. (\ref{energy}), which in this case reads as follows
\beq \label{energy_y2}    
\nu+M_A=\sqrt{(M_{A-2}+m_N+<E_{A-1}^*(k)>_{2NC})^2+y_2^2}+\sqrt{m_N^2+(y_2+q)^2} \: .
\eeq
The inclusion of the term
\beq
    <E_{A-1}^*(k)>_{2NC}= \frac{1}{n^A(k)}\int P_{2NC}^A(k,E_{A-1}^*) E_{A-1}^*d\,E_{A-1}^*
\label{average}
 \eeq
makes the scaling variable $y_2$ to properly depend upon  the momentum dependence of the average excitation energy  of $(A-1)$, generated by two-nucleon correlations. Here
\be
    E_{A-1}^*=E-E_{thr}^{(2)}
\ee
where
\be
    E_{thr}^{(2)}=M_{A-2}+2m_N-M_A
\ee
is the threshold energy for two-particle emission.
\\The quantity in Eq. (\ref{average}) has been calculated using  a realistic  spectral function for nuclear matter and
$^3He$.  The results are  presented in Fig.  \ref{Fig_energy}, where they are  compared with the prediction of  the spectral function of the \emph{few nucleon correlation} (FNC) model of Ref. \cite{ciosim}, according to which
 \begin{equation}
E_{A-1}^{*}({\bf k}, {\bf K}_{CM})= \frac{A-2}{A-1}{\frac{1}{2m_N}}\left[{\bf {k}} -\frac
{A-1}{A-2}{\bf {K}}_{CM}\right]^2
\label{twelve}
 \end{equation}
where ${\bf {K}}_{CM}$ is the CM momentum of a correlated pair. In view of the very good  agreement between the FNC model and the exact many-body results for  nuclear matter and $^3He$, the former has been used  to calculate $<E_{A-1}^*(k)>_{2NC}$ for nuclei with $3< A <\infty$.
\\The values shown  in Fig.  \ref{Fig_energy}  can be interpolated  by
 \be
    <E_{A-1}^{*}(k)>_{2NC} = \frac{A-2}{A-1}\:T_N +b_A-{c_A}|{\bf k}|
    \label{Estar}
 \ee
where
\be
    T_N=\sqrt{m_N^2+k^2}-m_N
\ee
 and
\beq \label{ba}
    b_A=\frac{\int d\textbf{k}_{CM}\frac{A-1}{A-2} \:\frac{\textbf{k}^2_{CM}}{2M}\: P_1(k,E)}
    {\int d\textbf{k}_{CM} P_1(k,E)}
\eeq
\beq
\label{ca}
    c_A= \frac{\frac{1}{M} \int d\textbf{k}_{CM} \:\textbf{k} \cdot \textbf{k}_{CM}\: P_1(k,E)}
    {\int d\textbf{k}_{CM} P_1(k,E)}
\eeq
result from the CM motion of the pair.
\\The values of $b_A$ and $c_A$ used in our calculations are listed in Table \ref{tab:ba_ca}.
\begin{table} [!h]
\large
\begin{center}
\begin{tabular}{||p{2.5cm}||*{2}{c|}|}
\hline
\hline
Nucleus & $b_A$ (MeV)& $c_A$ \\
\hline
$^{3}He$ & $-2.94$ & $-0.03$\\
\hline
$NM$ & $37.3$ & $0.04$\\
\hline
\hline
\end{tabular}
\caption{Values of the parameters appearing in Eqs. (\ref{ba}) and (\ref{ca}) for complex nuclei and nuclear matter.}
\label{tab:ba_ca}
\end{center}
\end{table}
\\For a complex nucleus and not too large values of $y_2$, a solution for Eq. (\ref{energy_y2}) can be written as
\be
    y_2= -\frac{\widetilde q}{2} + \frac{\nu_A}{2W_A} \sqrt{{W_A^2}-{ 4m_N^2}}
\label{y2}
 \ee
where
\beq
    \nu_A = \nu +  \widetilde{M}_D
\eeq
\beq
    \widetilde{M}_D = 2m_N -E_{th}^{(2)} - b_A + <E_{gr}>
\eeq
\beq
    \widetilde{q} = q+c_A{\nu}_A
\eeq
\beq
    \rm{W}_{A}^{2}= {\nu_A}^2 - {\bf q}^2 =\widetilde M_D^{2} + 2 \nu \widetilde M_D - Q^{2}
\eeq
In order to counterbalance the effects of $<E_{A-1}^*(k)>_{2NC}$ at low  $|y_2|$, in the definition of
\be
    M_{A-1}^*=M_A+<E_{A-1}^*(k)>_{2NC}-<E_{gr}>
\ee
has been added the value $<E_{gr}>$, fixed by  the Koltun sum rule (see \cite{CW,CFW1,CFW2}).
\\In the Deuteron case
\bey
    y_2 = y = -\frac{q}{2} + \frac{\nu_D}{2W_D}\: \sqrt {W_D^2 - {4 m_N^2}}
\eey
 with
\be
    {\nu_D} = \nu + M_{D}
\ee
and  where
\be
    {\rm W}_{D}^{2} =  {\nu_D}^2 - {\bf q}^2 = {M_D}^2 + 2{\nu}M_D- Q^2
\ee
is the Deuteron invariant mass.
\\For small values of $|y_2|$, such that
\be
    {\frac{A-2}{A-1}\left(\sqrt{y_{2}^2+m_N^2}-m_N\right) +b_A-{c_A}{|y_{2}|}} \ll <E_{gr}>
\ee
the  variable $y$, representing the longitudinal momentum of a weakly bound nucleon,
is recovered.
\\Therefore $y_2$ effectively takes into account the $k$ dependence of  $E_{A-1}^*$, both at low and high values of $y_2$, and  interpolates between the correlation and the single-particle regions; it   can be interpreted as the {\it minimum longitudinal  momentum of a nucleon that, at high values of $y_{2}$,  has removal energy $<E_{A-1}^*>_{2NC}$ and is partner of a  correlated two-nucleon pair with effective mass $\widetilde M_D$.}
\\It should be pointed out that, in our calculations, the fourth-order equation resulting from Eq. (\ref{energy_y2}) has been solved exactly; this, together with the relativistic extension of the definition of the mean excitation energy, is necessary to extend $y_2$ to high values.
\begin{figure}[!h]
\centerline{\includegraphics[scale=1.3]{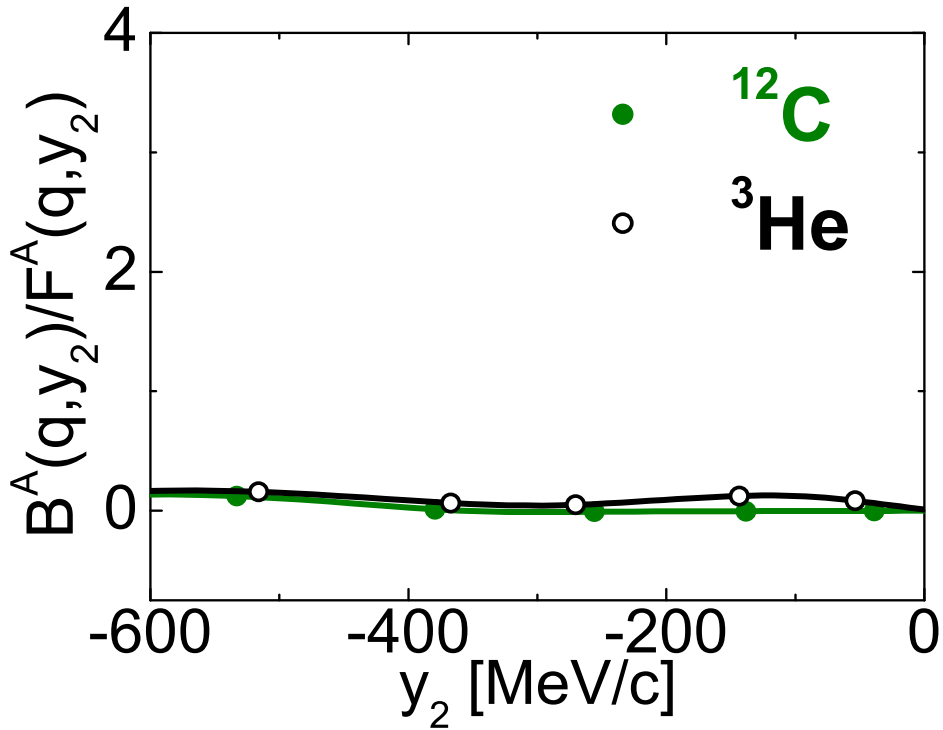}}
\caption{The ratio of the \emph{binding correction} $B^A(q,y_2)$ (Eq. (\ref{binding})) to the scaling function $F^A(q,y_2)$ (Eq. (\ref{Funzscala})) for $^3He$ (open dots) and $^{12}C$ (full dots), calculated using the scaling variable $y_2$. After Ref. \cite{CCrapid}.}\label{Fig3.4}
\end{figure}
\\Within such an approach to \emph{Y-scaling}, the effects due to the binding are strongly suppressed, as clearly illustrated in Fig.  \ref{Fig3.4}, where the ratio given by Eq. (\ref{ratio}) vanishes in the whole region of $y_2$ considered; the main feature of $y_2$, in fact, is that
\be
    k_{min}(q,y_2,E) \simeq \vert y_{2} \vert
\ee
which leads to
\be
B^A(q,y_{2})\simeq 0
\ee
with  the following two relevant consequences:
\begin{enumerate}
\item the relation
\be
    F^A(q,y_{2}) \simeq f^A(y_{2})=\int_{|y_2|}^{+\infty} k\:dk\: n^A(k)
\ee
holds, thus, plotting the data in terms of $y_2$ can provide a direct access to the nucleon momentum distributions, and so to SRC in nuclei;
\item as mentioned in $\S \: 2.3$, many-body calculations show that at high momenta, $k \gtrsim 2\:fm^{-1}$, all nucleon momentum distributions are simply rescaled version of the Deuteron one, i.e.
\be
    n^A(k) \simeq C^{A} n^D(k)
\ee
where $C^A$ is a constant; as a consequence, one would expect that, at high  values of $|y_{2}|$, $F^A(q,y_{2})$  will behave in the same way in the Deuteron and in complex nuclei,  so that, accordingly,
\be
 f^A(y_{2}) \simeq C^A f^D(y_{2})\: .
\ee
At the same time, on the contrary, at low values of $|y_{2}|$, $F^A(q,y_{2})$ should exhibit an $A$ dependence generated by the different asymptotic behavior of the nuclear wave functions in configuration space.
\end{enumerate}
\begin{figure}[!h]
\centerline{\includegraphics[scale=1.25]{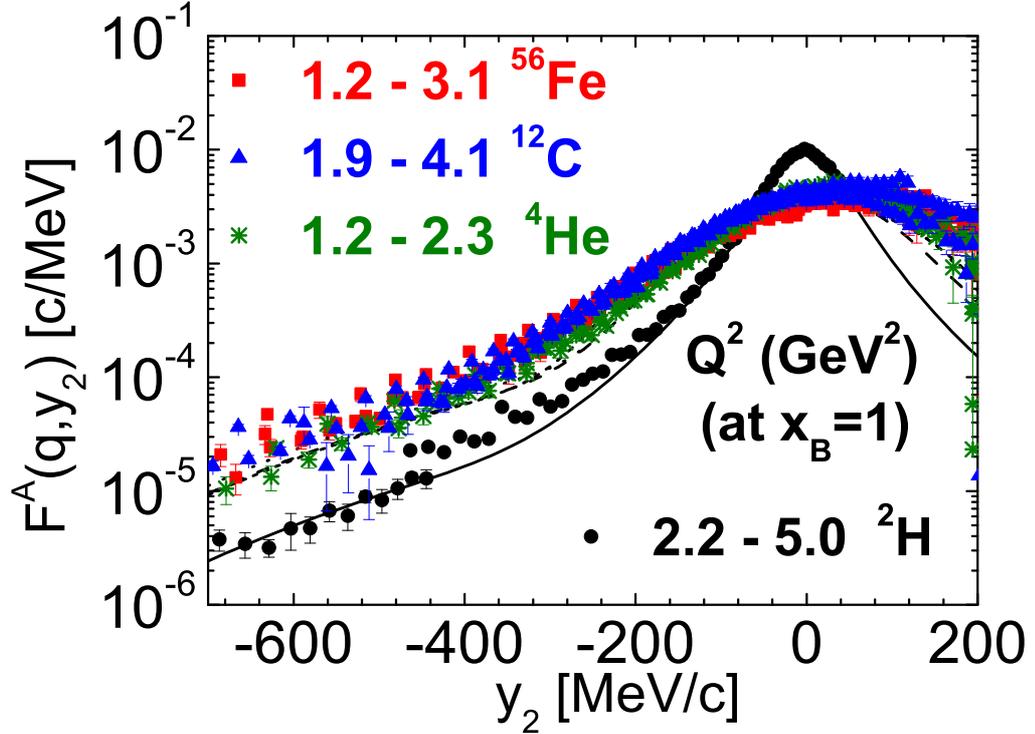}}
\caption{The experimental scaling function $F^A_{exp}(q,y)$
of $^4He$, $^{12}C$, and $^{56}Fe$ obtained from the experimental data of Refs. \cite{arrington,Schutz2}. The longitudinal momentum distributions (Eq. (\ref{longitudinal})) of $^{2}H$ (full line), $^4He$ (long-dashed),  $^{12}C$ (dashed) and  $^{56}Fe$ (dotted) are also shown. After Ref. \cite{CCrapid}.}\label{Fig3.5}
\end{figure}
\begin{figure}[!htp]
\centering{\includegraphics[scale=1.2]{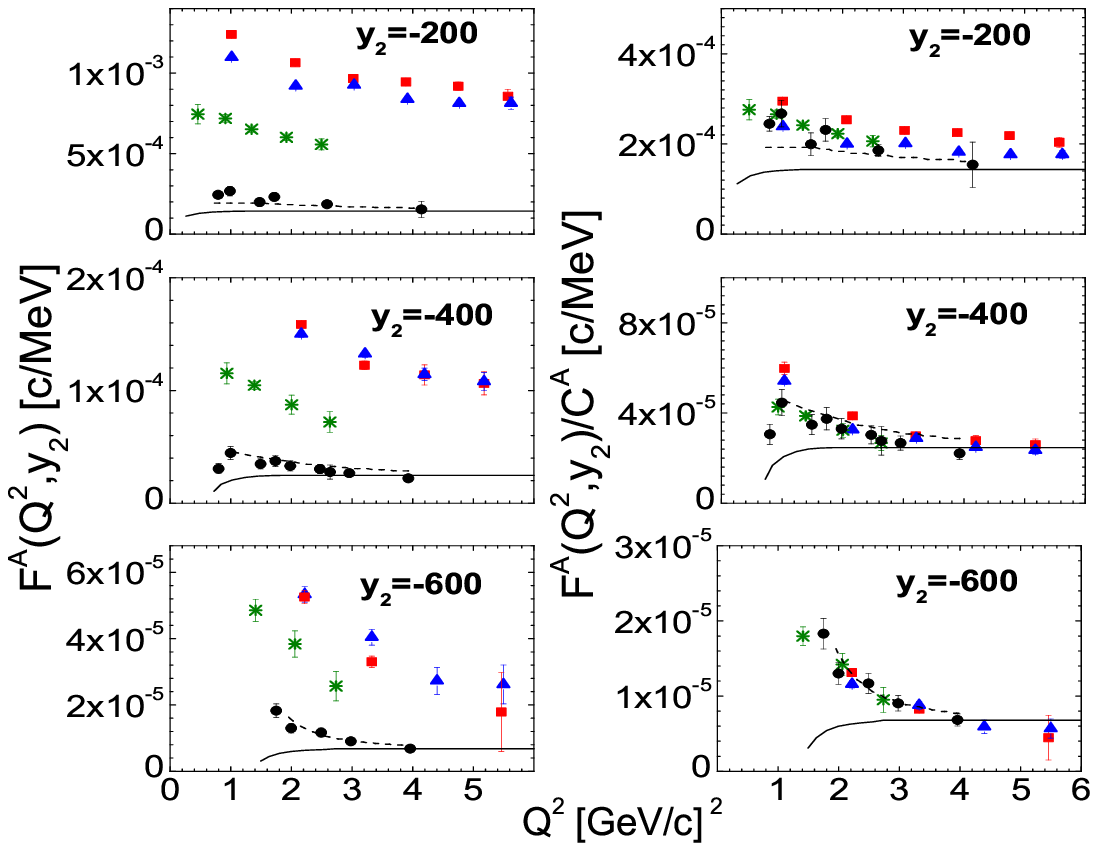}}
\caption{The scaling function $F^A(Q^2,y_{2})$ from   the lower panel  of Fig.  \ref{Fig3.5} plotted  {\it vs.} $Q^2$ at fixed values of $y_{2}$ ($^4He$-{\it asterisks }, $^{12}C$-{\it triangles}, $^{56}Fe$-{\it squares}).
In the right panel the data for $^4He$, $^{12}C$ and $^{56}Fe$ have been divided by the constants $C^4=2.7$,   $C^{12}=4.0$ and $C^{56}=4.6$, respectively. The theoretical curves refer to $^2H$ and represent the PWIA results ({\it full}) and the results that include the FSI ({\it dashed}), both obtained with the AV18 interaction \cite{WSS}.
Scaling variables are in MeV/c. After Refs. \cite{CC,CCrapid}}
  \label{Fig3.6}
\end{figure}
This is fully confirmed by Fig.  \ref{Fig3.5}, where the scaling function $F^A(q,y_2)$ obtained from available experimental data  on $^4He$, $^{12}C$ and $^{56}Fe$, is plotted  versus the scaling variable $y_2$; it can be seen that, at high values of $|y_2|$, $F^A(q,y_{2})$ scales exactly to $f^A(y_{2})$. This is even better demonstrated in Fig.  \ref{Fig3.6}, where  the scaling function $F^A(q,y_2)$ is plotted versus $Q^2$ at fixed values of $y_2$. By this way, the scaling behavior of $F^A(q,y_2)$ is better illustrated. In the same Figure, in order to analyze more quantitatively the scaling behavior of $F^A(q,y_2)$, the latter has been plotted together with the theoretical scaling function for $A=2$, calculated in PWIA (solid line), and taking FSI into account (dashed line).
The left panel clearly illustrates that, due to FSI effects, scaling is violated and approached  from the top,
and not from the bottom, as predicted by the PWIA. However, scaling violation seems to exhibit a $Q^2$ dependence  which is very similar in Deuteron and in complex nuclei. This is illustrated in more details in the right panel of the Figure, which shows the quantity
\be
    F^A(Q^2,y_{2})/C^A \simeq F^D(Q^2,y_{2})
\ee
where the $A$ dependent constants $C^A$ have been chosen so as to make the experimental scaling function for a nucleus $A$ to coincide as much as possible with the Deuteron scaling function $F^D(Q^2,y_{2})$. It clearly appears that the scaling functions of heavy and light nuclei  scale to the Deuteron scaling function; moreover the values obtained for $C^A$ turn out to be in agreement, within the statistical errors, with the theoretical predictions of Ref. \cite{mark1}, as well as  with the experimental results on the ratio \cite{FSS2}
\beq \label{ratioDeut}
    R(\xb,Q^2) = \frac{2}{4}\:\frac{\sigma_2^A (\xb,Q^2)}{ \sigma_2^D (\xb,Q^2)}
\eeq
\begin{figure}[!h]
\centerline{\includegraphics[scale=0.65]{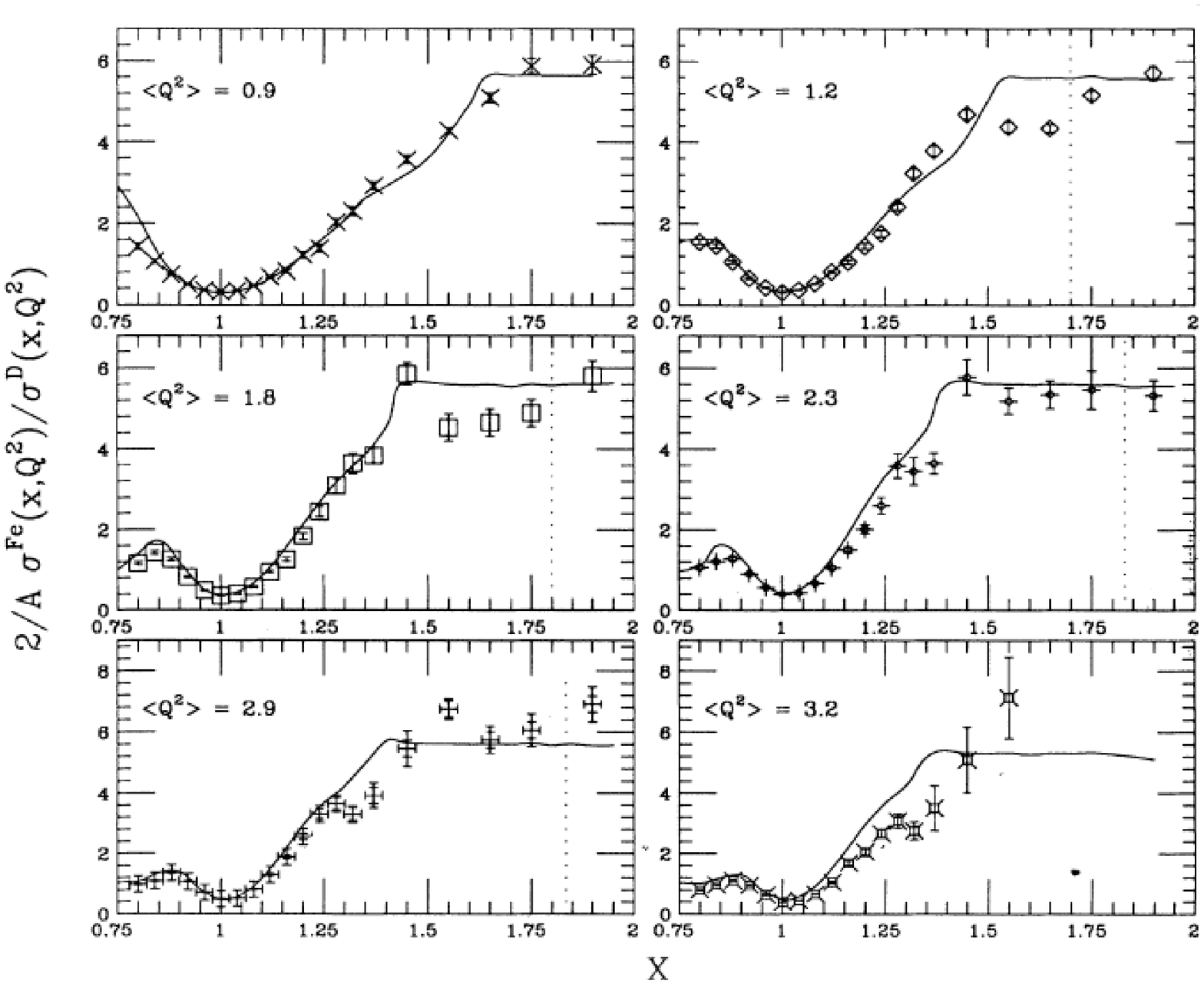}}
\caption{Eq. (\ref{ratioDeut}) \emph{vs.} the scaling variable $X\equiv \xb$ for $A=56$ at six different values of $Q^2$. The average $Q^2$ is given for each frame. To the right of the vertical dashed line those data correspond to a final state less $50\:MeV$ greater than the Deuteron rest mass. The solid line is a calculation based on the nuclear spectral function of Ref. \cite{ZS}. After Ref. \cite{FSS2}.}
\label{Fig_rafioFFS}
\end{figure}
\noindent shown in Fig. \ref{Fig_rafioFFS}; it is also important to stress that, although FSI are very relevant, they appear to be similar in Deuteron and in a nucleus A, which is evidence that, in the SRC region, FSI are mainly restricted to the correlated pair.
%
%
\section{Three-nucleon-correlation scaling variable}\label{sec:y3}
\begin{figure}[!h]
\centering%
\subfigure[\label{Fig3.7}]%
{\includegraphics[scale=0.8]{./immagini/3NC.eps}}\qquad
\subfigure[\label{Fig3.8}]%
{\includegraphics[scale=0.8]{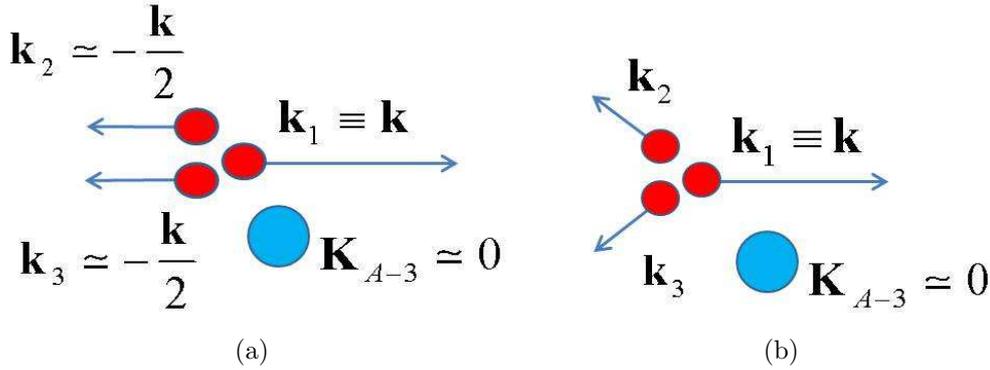}}
\caption{Two types of  3NC  configurations which are present in the spectral function
of a nucleus A; they  correspond to: (a) high momentum $k$ and low removal energy $E$, and (b) high momentum $k$ and high removal energy $E$.}
\end{figure}

Three-nucleon correlations also contribute, in principle, to the inclusive cross section for $A \geq 3$. 3NC correspond, as previously explained in $\S$\ref{sec:3N}, to those three-nucleon configurations in which the high momentum $\textbf{k}_1 \equiv \textbf{k}$ of nucleon "1" is almost entirely balanced by the momenta $\textbf{k}_2$ and $\textbf{k}_3$ of nucleons "2" and "3", respectively.
\\The excitation energy of the $(A-1)$-nucleon system is given in this case by (Cf. Eq. (\ref{Ex})) \cite{CC}
\beq \label{Energy3NC}
E_{A-1}^*=\frac{\left(\textbf{k}_2-\textbf{k}_3\right)^2}{m_N} + \frac{A-3}{A-1}\:\frac{\left[\left(\textbf{k}_2+\textbf{k}_3\right)-2\textbf{K}_{A-3}\right]^2}{4m_N}
\eeq
which shows that, whereas 2NC are directly linked to high values of excitation energies
\beq
    E_{A-1}^* \simeq \frac{(A-2)}{(A-1)}\: \frac{k^2}{2m_N}
\eeq
high momentum components due to 3NC may lead both to low and to high values of $E_{A-1}^*$, as shown in the examples of Figs. \ref{Fig3.7} and \ref{Fig3.8}, respectively.
\\In the configuration of Fig.  \ref{Fig3.7}, the momentum $\textbf{k}_1 \equiv \textbf{k}$ of nucleon $"1"$ is almost entirely balanced by nucleons $"2"$ and $"3"$, with momenta $\textbf{k}_2\simeq\textbf{k}_3\simeq-\textbf{k}/2$, and one has
\beq \label{E3NC}
E_{A-1}^*=\frac{A-3}{A-1}\:\frac{k^2}{4m_N}\: .
\eeq
In the configuration of Fig.  \ref{Fig3.8},
\beq
    k_2=k_3=-\frac{|\textbf{k}|}{2}\cos\left(\frac{\theta}{2}\right)
\eeq
with
\beq
    \cos\theta=-\frac{(\textbf{k}_2\cdot \textbf{k}_3)}{(k_2 k_3)}
\eeq
and
$E_{A-1}^*$ could be very large when $k_2$ and $k_3$ are large.
\begin{figure}[!h]
\centerline{\includegraphics[scale=0.9]{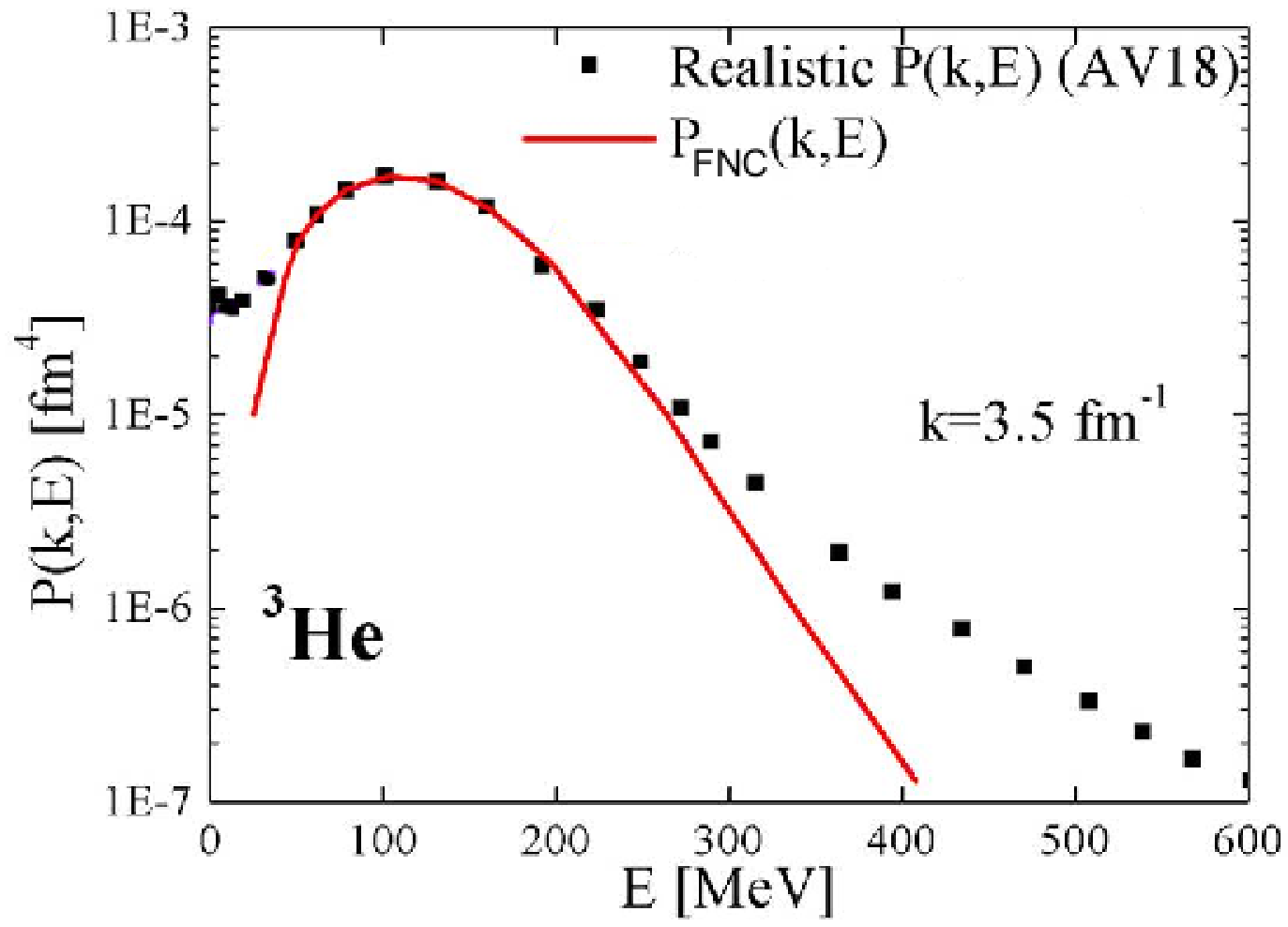}}
\vskip -0.5cm \caption{The  spectral function of $^3He$ \emph{vs.} the removal energy $E$, at $k=3.5\: fm^{-1}$ \cite{CK}. The realistic spectral function corresponding to the Pisa wave function (squares) is compared with the FNC model (full line) given by Eq. (\ref{PkE2NCf}), at $k=3.5\:fm^{-1}$. After Ref. \cite{CC}.}
\label{FigHe3}
\vskip 0.5cm
\end{figure}
\\Let us investigate the presence and relevance of 3NC configurations in the spectral function of the 3N system, for which the Schr$\ddot{o}$dinger equation has been  solved exactly. When $A=3$, 3NC of the type shown in Fig.   \ref{Fig3.7} lead to  $E_2^*=0$ (cf. Eq. (\ref{E3NC})).
\\In Fig.  \ref{FigHe3}, as already illustrated in $\S$\ref{sec:3N}, the realistic spectral function of $^3He$ obtained \cite{CK} using the Pisa wave function \cite{pisa} corresponding to the AV18 interaction \cite{WSS} (full squares), is compared with the predictions of the FNC model (solid line) given by the convolution formula (\ref{PkE2NCf}) \cite{CC}. It can be seen that the FNC spectral function reproduces the exact one in a wide range of removal energies ($50 \lesssim E \lesssim 250\: MeV$), but fails at very low and very high values of E, where the effects from 3NC are expected to provide an appreciable contribution.
\\It is clear  from Fig.  \ref{FigHe3} that 3NC of the type shown in Fig.  \ref{Fig3.8} can  hardly be present at $k<3.5 fm^{-1}$ and  $E \leq 300\: MeV$,  so that it is legitimate  to ask ourselves whether these 3NC can show up in available  experimental data. To answer this question, let us now consider the maximum value of the removal energy achieved in the experiments, which is also the upper limit of integration in Eq. (\ref{Funzscala}), i.e.
\beq \label{emax}
E_{max}(q,\nu)=\sqrt{(\nu+M_A)^2-q^2}\: .
 \eeq
\begin{figure}[!h]
\centerline{\includegraphics[scale=1.2]{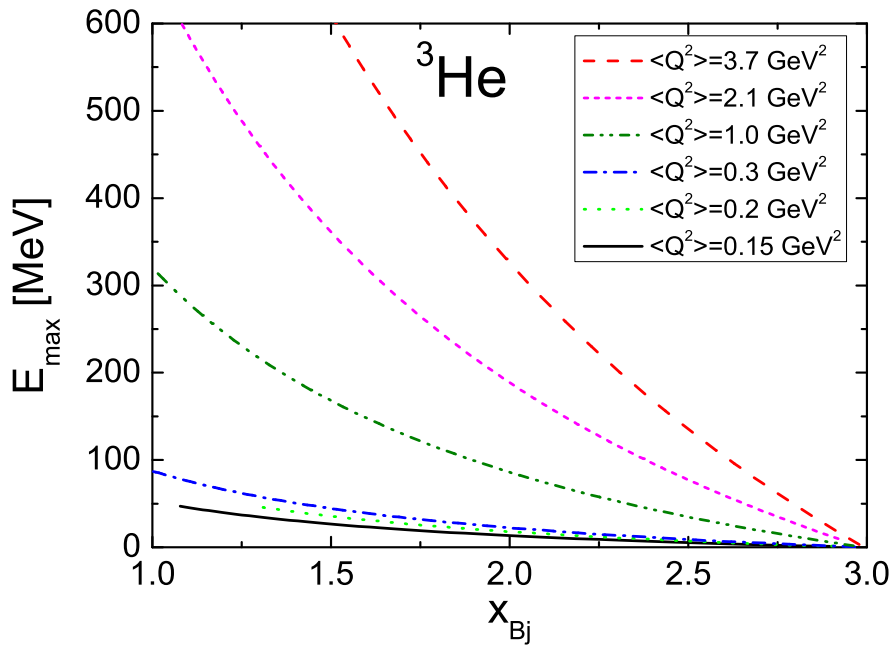}}
\caption{The maximum value of the removal energy $E_{max}$ (Eq. (\ref{emax}))
available in inclusive q.e. scattering off $^3He$ plotted\emph{ vs.} $x_{Bj}$, at increasing
values of $Q^2$ shown in the inset. After Ref. \cite{CC}.}\label{Fig9}
\vskip 0.5cm
\end{figure} \noindent
In Fig.  \ref{Fig9},  we show  the value of $E_{max}(q,\nu)$ plotted versus the Bjorken scaling variable in the region $1 \leq x_{Bj} \leq 3$, in correspondence of a set of values
of $\nu$ and $q$ typical of available experimental data on $^3He$.
It can be seen from Figs. \ref{Fig3.6} and \ref{FigHe3} that, in the region $ 2\leq x_{Bj} \leq 3$, only  3NC configurations of the type shown in Fig.  \ref{Fig3.7} can contribute to present $A(e,e')X$ kinematics; for this reason we will consider, for the time being, only this type of 3NC.
\\Therefore the 3NC scaling variable $Y\equiv y_3$ is obtained by placing
\beq
    k=|y_3|
\eeq
\beq
    \cos\alpha=\frac{\textbf{k}\cdot \textbf{q}}{kq}= 1
\eeq
\beq
    E_{A-1}^*=<E_{A-1}^*(k)>_{3NC}
\eeq
in the energy conservation (\ref{energy}), which becomes
\beq \label{energy_y3}    
\nu+M_A=\sqrt{(M_{A-3}+2m_N+<E_{A-1}^*(k)>_{3NC})^2+y_3^2}+\sqrt{m_N^2+(y_3+q)^2}
\eeq
where the excitation energy of the residual system is (cf. (Eq.\ref{E3NC}))
\beq
    <E_{A-1}^*(k)>_{3NC}=\frac{A-3}{A-1}\:\frac{k^2}{4m_N}
\label{average3NC}
 \eeq
and corresponds to the 3NC configuration of type \ref{Fig3.7}.
\\Even in this case, Eq. (\ref{energy_y3}) has been solved exactly in our calculations.
%
%
\section{Domain of existence of the three scaling variables}\label{sec:domain}
In the previous sections, we have obtained three different scaling variables, $y$, $y_2$ and $y_3$, by placing different values of $E_{A-1}^*$ in Eq.  (\ref{energy}), namely  $E_{A-1}^*=0$, $E_{A-1}^*=<E_{A-1}^*(k)>_{2NC}$, and $E_{A-1}^*=<E_{A-1}^*(k)>_{3NC}$, respectively.
\begin{figure}[!h]
\centerline{\includegraphics[scale=1.2]{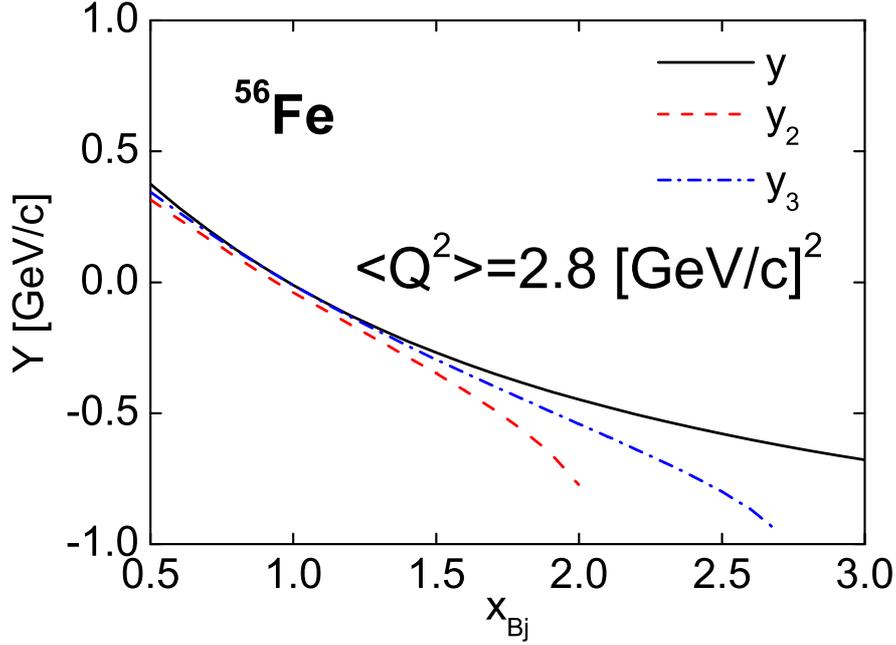}}
\caption{The scaling variables $y$, $y_2$ and $y_3$ {\it vs.} $x_{Bj}$ for
the nucleus $^{56}Fe$. After Ref. \cite{CC}.}\label{Fig10}
\end{figure}
\\In Fig.  \ref{Fig10}, the values of  $y$, $y_2$ and $y_3$ are plotted versus $x_{Bj}$, in the case of $^{56}Fe$, for a fixed value of $Q^2$. It should be pointed out that the magnitude of
\beq \label{xb}
    \xb=\frac{Q^2}{2m_N\nu}=A\:\frac{Q^2}{2M_A\nu}
\eeq
cannot be larger than the number of nucleons in a given nucleus, as clearly results from the definition of the invariant mass
\beq
    W^2=-Q^2+M_A^2+2M_A\nu=Q^2 \: \left( -1 + \frac{M_A}{xm_N} \right) + M_A^2 \geq M_A^2
\eeq
which leads to
\beq \label{xblim}
    x_{Bj} \leq \frac{M_A}{m_N}\simeq A\: .
\eeq
This result is really important in order to explain the different behaviors of the three scaling variables which, as already discussed in the previous sections, differ only in the definition of $E_{A-1}^*$ used in Eq. (\ref{energy}). Let us analyze this point in more detail.
\\The 2NC scaling variable (\ref{y2}), requires that
\beq
    W_A^2-4m_N^2 \geq 0 \: .
\eeq
By approximating $\widetilde{M}_D\sim 2m_N$, and owing to Eq. (\ref{xb}),  one gets
\beq
    -Q^2+2\:\frac{Q^2}{x} \geq 1
\eeq
and thus
\beq
    -Q^2\:\left(1-\frac{2}{x}\right) >1
\eeq
which means
%
%
\beq
    \xb \lesssim 2 \: .
\eeq
The scaling variable $y_2$ is therefore defined only for $x_{Bj}\leq 2$. Indeed $y_2$ represents 2NC in heavy nuclei resembling the ones acting in Deuteron and, by placing $A=2$ in Eq. (\ref{xblim}), the same domain of existence is recovered.
%
%
%
%
%
In the same way, the domain of the other two scaling variables $y$ and $y_3$, turns out to exist in the range of $\xb\leq A$ and $\xb\leq 3$, respectively. Indeed, the scaling variable $y$ describes the mean field configuration, whereas the scaling variable $y_3$ describes 3NC as in $^3He$; thus, by placing $A$ and $A=3$ in Eq. (\ref{xblim}), the same domains are recovered.
\\The different values of $E_{A-1}^*$ used in Eq. (\ref{energy}), impose therefore different limits in the domain of existence of the three scaling variables used in our calculations, due to the different configurations of nucleons in nuclei they refer to.
%
%
\section{A new approach to the treatment of the inclusive cross section}
\begin{figure}
\begin{center}
\includegraphics[scale=2.0]{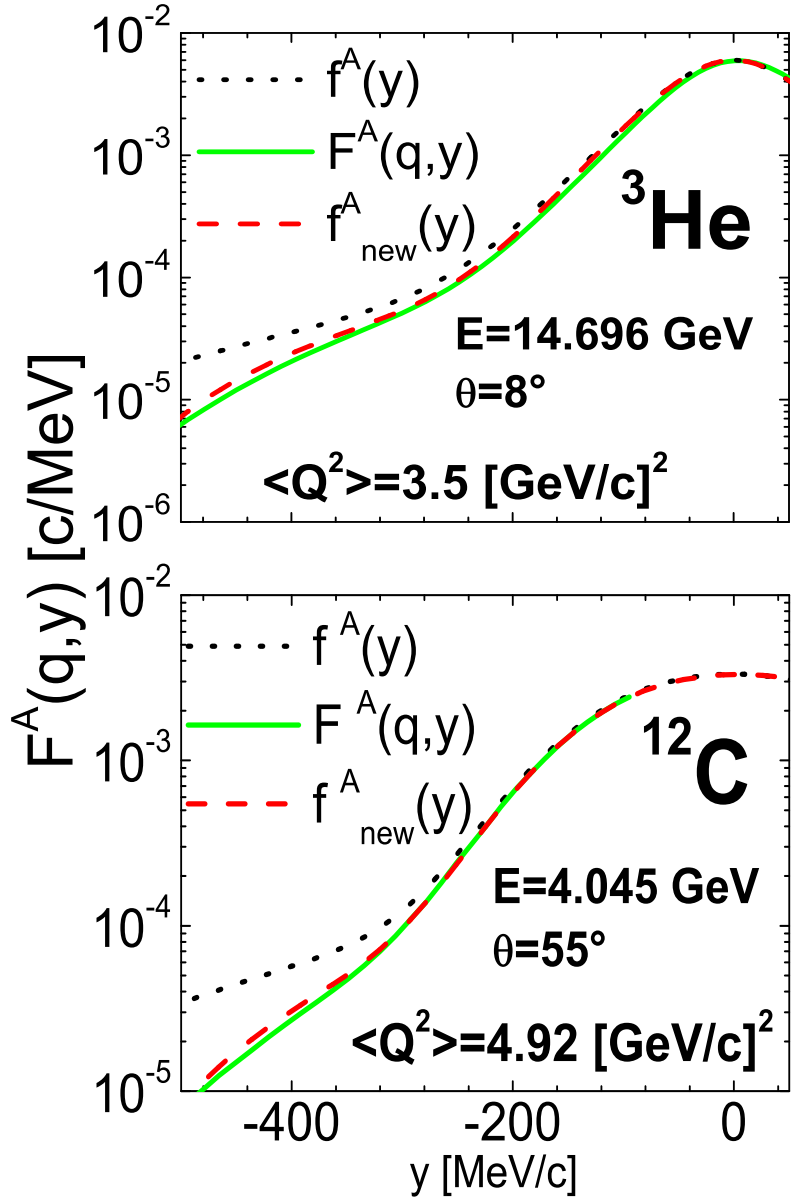}
\end{center}
\vskip-1.2cm
\caption{Different definitions of the scaling function \emph{vs.} the scaling variable $y$. Dotted line: the longitudinal momentum distribution (\ref{longitudinal}); solid line: the scaling function (\ref{Fyqold}), calculated with the spectral function; dashed line: the new scaling function (\ref{Fyqnew}), calculated with the nucleon momentum distributions.}
\label{Fig_fyq_fy2}
\end{figure}

Let us recall, as already discussed in $\S$\ref{sec:QE_XS}, that, at high $Q^2$, the calculation of the quasi elastic inclusive cross section depends upon the scaling function
\beq \label{Funzscala2}
        F^A(q,y)=2\pi \int_{E_{min}}^{+\infty} dE \int_{k_{min}(q,y,E)}^{+\infty} k\:dk\: P^A(k,E)
\eeq
whose calculation requires, due to the presence of the nucleon spectral function $P(k,E)$ (Eq. (\ref{PkE})), the knowledge of the entire energy spectrum of the $(A-1)$-nucleon system. Owing to the decomposition rule (\ref{decomposition}), the contributions from different final nuclear states could be explicitly separated out, writing
\bey \label{Fyqold}
  F^A(q,y)&=& f_0(y)+F_2(y,q)+F_3(y,q)
\eey
where
\beq \label{f0}
    f^A_0(q,y)=2\pi\int_{|y|}^{+\infty} n_0^A(k)\: k\:dk
\eeq
describes the shell model contribution,
\beq \label{F2yq}
    F^A_2(q,y)=2\pi\int_{E_{min}}^{+\infty}dE\int_{k_{min}(y,q,E)}^{+\infty}k\:dk\:P_{2NC}(k,E)
\eeq
the 2NC pair contribution and, eventually,
\beq    \label{F3yq}
    F^A_3(q,y)=2\pi\int_{E_{min}}^{+\infty}dE\int_{k_{min}(y,q,E)}^{+\infty}k\:dk\:P_{3NC}(k,E)
\eeq
the 3NC contribution.
\\As already pointed out in the previous sections, the scaling variables $y$, $y_2$ and $y_3$ effectively take into account the energy $E_{A-1}^*$ of the residual system; by this way,  the effects due to the \emph{binding correction} (\ref{binding}) are strongly suppressed, and a direct link between the scaling function and the nucleon momentum distributions can be established.
\\The original idea of our novel approach \cite{CC} to $A(e,e')X$ processes, is based upon the replacement of Eq. (\ref{Fyqold}) by the following one
\bey \label{Fyqnew}
  F^A_{new}(y,q)\equiv f^A_{new}(y) &=& f_0(y)+f_2(y_2)+f_3(y_3)
\eey
with $f_0(y)$ given by Eq. (\ref{f0}), and Eqs. (\ref{F2yq}) and  (\ref{F3yq}) replaced by
\beq \label{f2}
    f^A_2(y_2)=2\pi\int_{|y_2|}^{+\infty} n_2^A(k)\: k dk
\eeq
and
\beq
    \label{f3}
    f^A_3(y_3)=2\pi\int_{|y_3|}^{+\infty} n_3^A(k)\: k dk
\eeq
respectively. It should be pointed out that $y_2=y_2(q,y)$ and $y_3=y_3(q,y)$, with $y=y(q,\nu)$.
\\In Eq. (\ref{f2}), we use the nucleon momentum distribution
\beq \label{soft}
    n_2^A(k)=\int d\textbf{k}_{CM}\: n_{rel}(\textbf{k}+\textbf{k}_{CM})\:n_{CM}^{soft}(\textbf{k}_{CM})
\eeq
which describes the virtual photon absorption by a 2N correlated pair and, in Eq. (\ref{f3}), the nucleon momentum distribution
\beq \label{hard}
n_3^A(k)=\int d\textbf{k}_{CM}\: n_{rel}(\textbf{k}+\textbf{k}_{CM})\:n_{CM}^{hard}(\textbf{k}_{CM})
\eeq
which corresponds to the contribution arising from 3NC; for shell model nucleons the usual nucleon momentum distribution (see Eq. (\ref{PkEsm2}))
\beq
n_0^A(k) = \frac{1}{4\pi A}\: \sum_{\alpha<\alpha_F} A_\alpha\:\widetilde{n}_{\alpha}(k)
\eeq
defined in $\S$\ref{sec:P0}, is adopted.
\\These three quantities should satisfy the normalization  condition (\ref{nmdnorm}), i.e.
\beq
    \int_0^\infty k^2\:dk\:n_0^A(k)+    \int_0^\infty k^2\:dk\:n_2^A(k)+    \int_0^\infty k^2\:dk\:n_3^A(k)=1 \:.
\eeq
Due to the difficulties in the calculation of the nucleon momentum distribution (\ref{hard}), until now we have considered only the independent particle shell model and 2NC components in Eq. (\ref{Fyqnew}). We have thus compared the longitudinal momentum distribution (\ref{longitudinal}), where $E=E_{min}$, with Eq. (\ref{Fyqnew}), where $E=E_{min}<E^*_{A-1}(k)>_{2NC}$, and with the exact calculation of Eq. (\ref{Fyqold}) in terms of the spectral function, finding, as shown in Fig. \ref{Fig_fyq_fy2},  a good agreement between the two  approaches.
\\It appears, therefore, that inclusive cross sections can be calculated only by using momentum distributions, provided the excitation energy of the ($A-1$) system is effectively taken into account in the lower limit of integration in $k$, i.e. by using the scaling variable $y_2$.
\\The same conclusions can be reached also by writing the scaling function in a more general form, \emph{viz}.
\beq
    F^A(q,Y)=\int_{E_{min}}^{+\infty} dE \int_{|Y|}^{+\infty} d^3k\: P^A(k,E)\:.
\eeq
The scaling variables $y$, $y_2$ and $y_3$ allow us to write
\beq
    P_0^A(k,E)=n_0^A(k)\:\delta(E-E_{min})
\eeq
\beq
    P_2^A(k,E)=n_2^A(k)\:\delta(E-<E(k)>_{2NC})
\eeq
\beq
    P_3^A(k,E)=n_3^A(k)\:\delta(E-<E(k)>_{3NC})\:.
\eeq

\cleardoublepage
%
%
\chapter{Results of calculations of the inclusive cross section ratios} \label{chap:results}
\section*{Introduction}
In what follows, the results of our calculations of the inclusive cross section ratios shown in $\S$\ref{sec:inclusive} will be presented.
%
%
%
%
\section{Inclusive cross section ratios in PWIA}
Let us recall that the experimental inclusive cross section ratio (\ref{ratio_inclusive}) of nucleus $A$ to the nucleus $^3He$ was defined in $\S$\ref{sec:inclusive} as follows
\beq \label{ratio_inclusive2}
    r(A,^3He)=\frac{2\sigma_{ep}+\sigma_{en}}{Z\sigma_{ep}+N\sigma_{en}}\:\frac{\sigma(A)}{\sigma(^3He)}
\eeq
which, within our novel approach, reduces to the following ratio
\beq \label{ratio_incl_new}
    r(A,^3He)=\frac{2\sigma_{ep}+\sigma_{en}}{Z\sigma_{ep}+N\sigma_{en}}\:\frac{F^A_{new}(y^A,q)}{F^{^3He}_{new}(y^{^3He},q)}\: .
\eeq
\begin{figure}[!h]
\centerline{\includegraphics[scale=1.6]{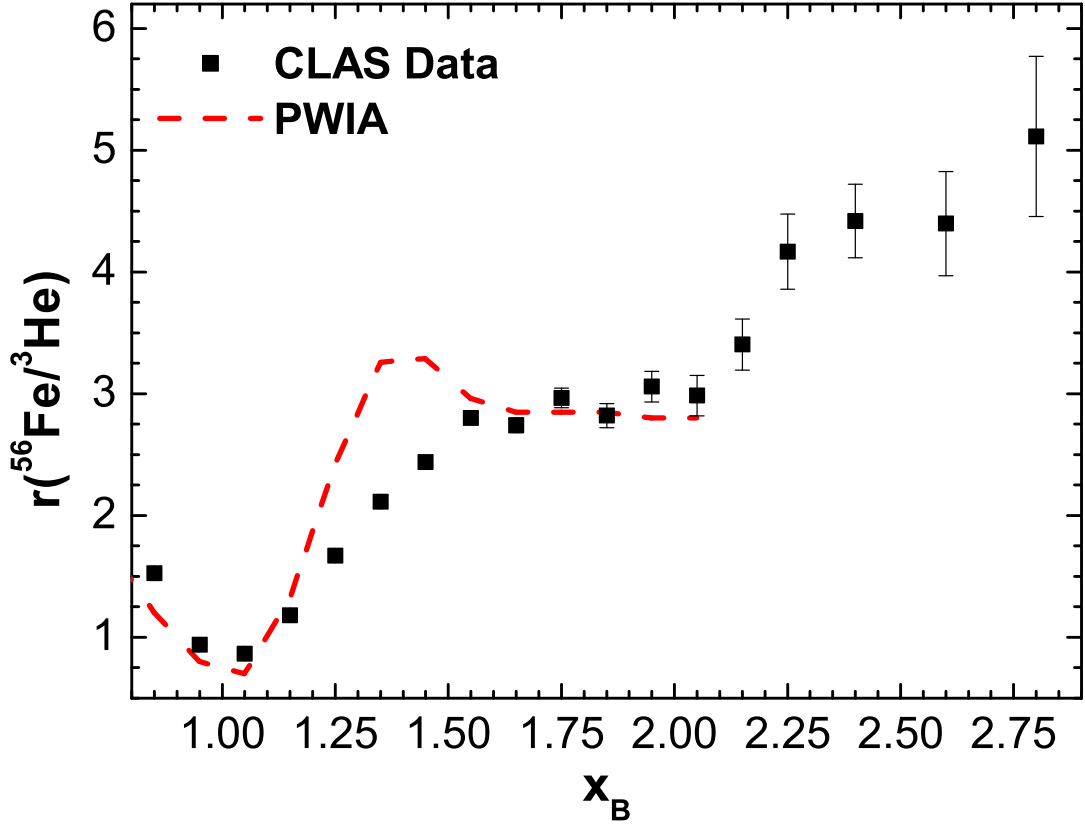}}
\caption{The experimental cross section ratio shown in Fig. \ref{Fig_inclusive} compared with our preliminary PWIA theoretical results, for $A=56$.}
\label{Fig_ratioPW}
\end{figure} \noindent
Our preliminary results of the PWIA ratio, for $A=56$ are shown in Fig. \ref{Fig_ratioPW}; it exhibits a good agreement with CLAS data only for $1.5 \lesssim x_{Bj}\lesssim 2$, i.e. in the region of 2NC; on the contrary, at $x_{Bj}\lesssim 1.5$ the PWIA does not lead to satisfactory results.
\\This fact agrees with the results already shown in Fig. \ref{Fig3.6}: in the region of 2NC the data of heavy nuclei scale to the Deuteron ones, and thus FSI effects vanish in the ratio $r(A/^3He)$, leading to the first plateaux; indeed in the 2NC region, Eq. (\ref{ratio_incl_new}) reads as follows
\beq \label{ratio_2NC}
    r(A,^3He)\simeq \frac{2\sigma_{ep}+\sigma_{en}}{Z\sigma_{ep}+N\sigma_{en}}\:
    \frac{\int_{|y_2^A|}^\infty k\:dk\: n^A_2(k)}{\int_{|y_2^{^3He}|}^\infty k\:dk\: n^{^3He}_2(k)}\simeq \frac{C^A}{C^{3^He}}=const \: .
\eeq
In the kinematical region at $x_{Bj}\lesssim 1.5$, on the contrary, the ratio exhibits a strong sensitivity upon the $A$ dependent FSI of the knocked nucleon with the residual system, and this is the reason of the disagreement between the experimental ratio and our calculations performed in PWIA.
%
%
\section{FSI and distorted nucleon momentum distribution}
\begin{figure}[h]
\begin{center}
\subfigure[\label{Fey_H2d}]{\includegraphics[scale=0.45]{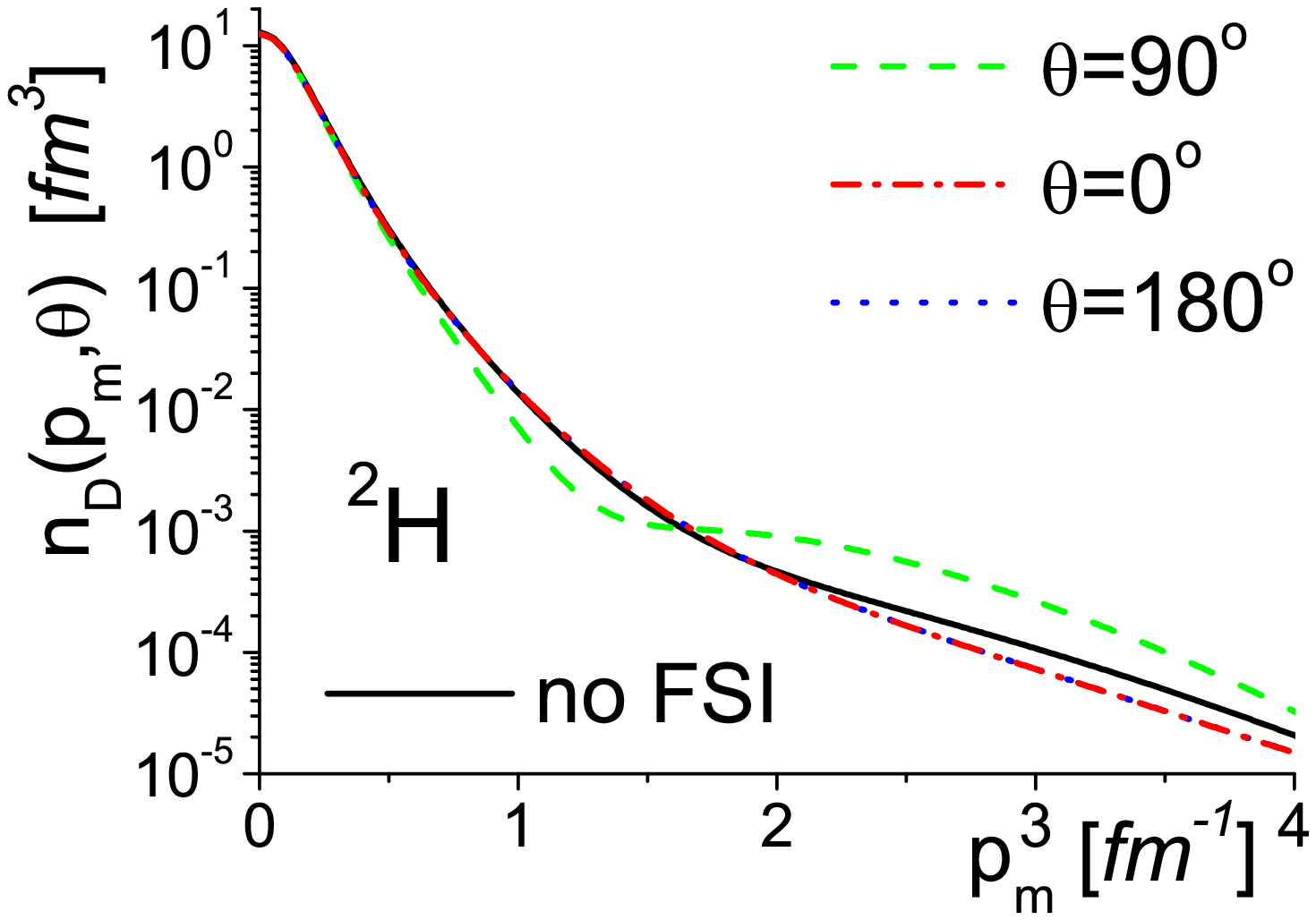}}\qquad
\subfigure[\label{Fey_He4d}]{\includegraphics[scale=0.45]{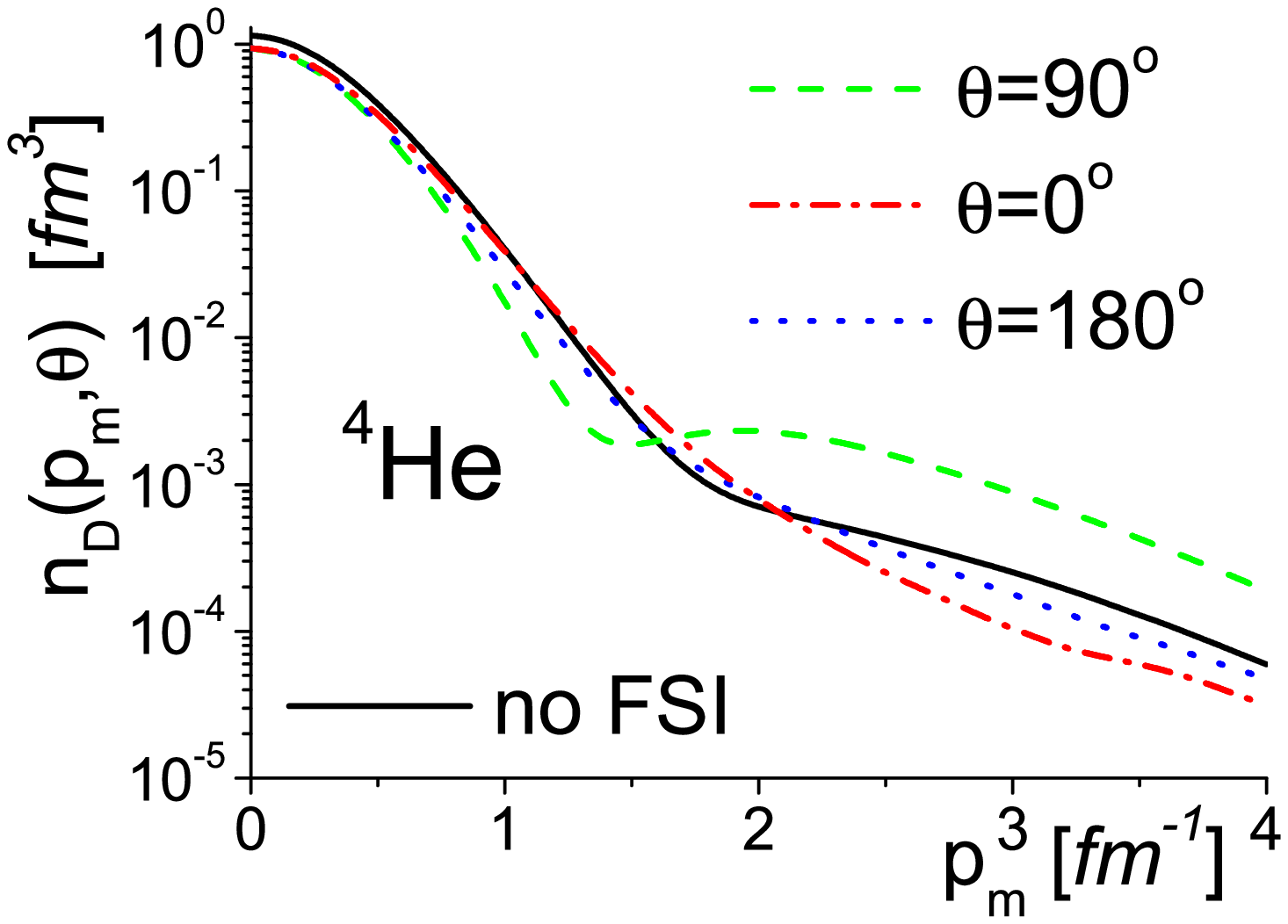}}\\
\subfigure[\label{Fey_O16d}]{\includegraphics[scale=0.45]{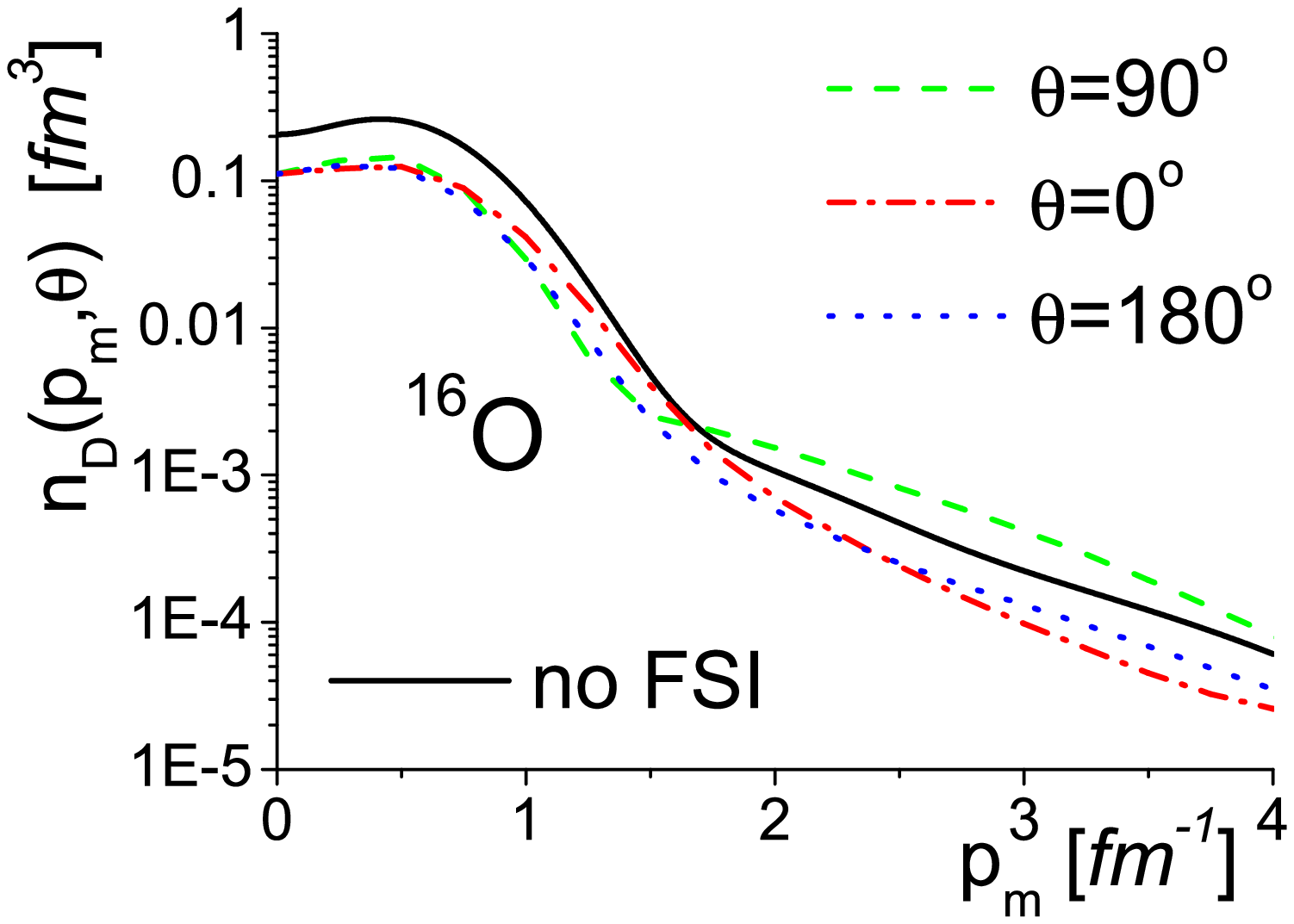}}
\caption{The distorted nucleon momentum distributions of Eq. (\ref{disssss}) versus the momentum $|\textbf{p}_m|$, for (a) $^2H$, (b) $^4He$ and (c) $^{16}O$. Dashed line: $\theta=90°$; dot-dashed red line: $\theta=0°$; dotted line: $\theta=180°$; solid line: no FSI taken into account. After Ref. \cite{tutti}.}\label{Fig_nmddis}
\end{center}
\end{figure}
In order to include FSI effects within our new model for inclusive scattering, we will now introduce the distorted nucleon momentum distributions, defined as follows \cite{CKT}
\begin{figure}[!h]
\begin{center}
\subfigure[\label{Fey_He90}]{\includegraphics[scale=0.5]{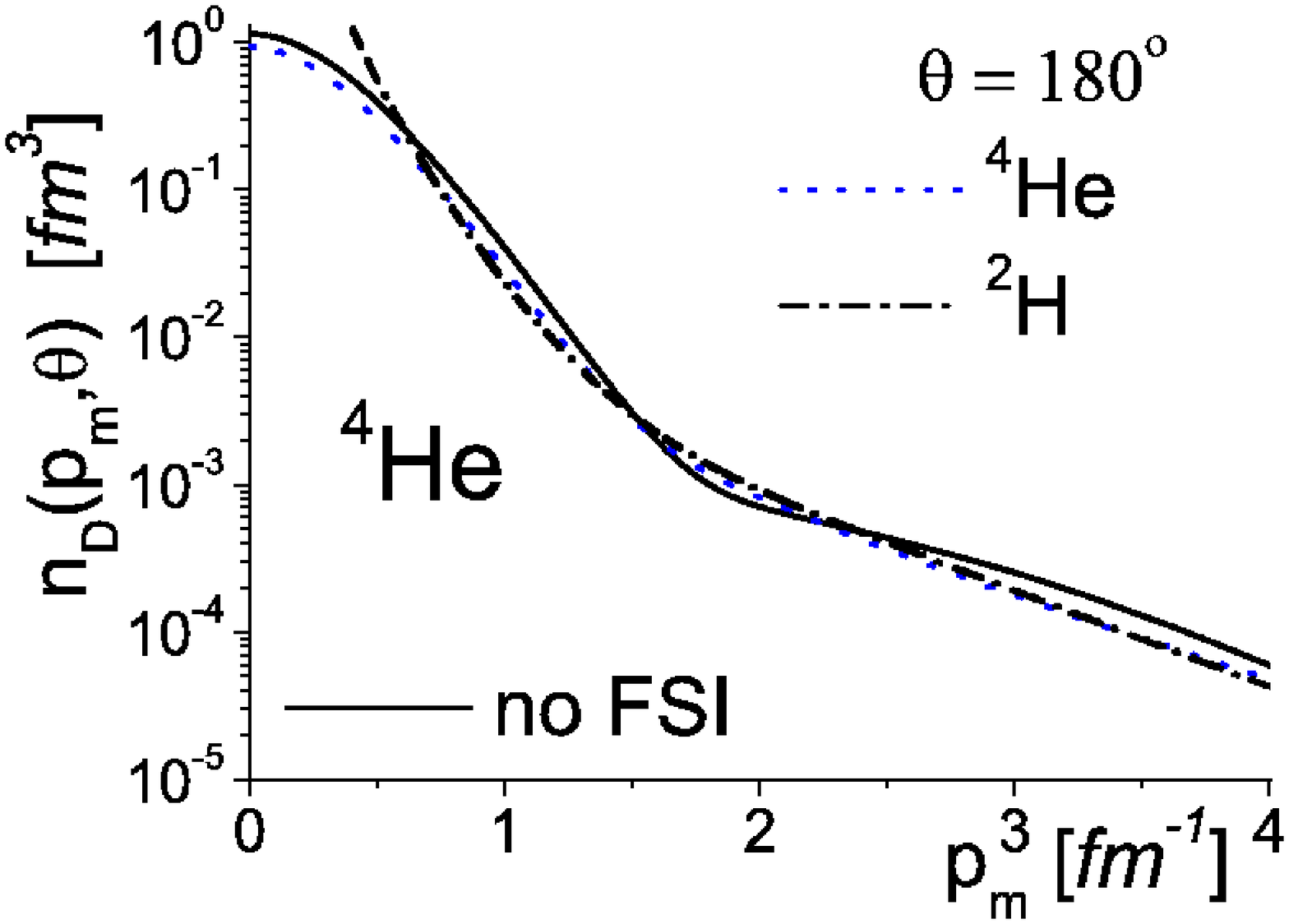}}\\
\subfigure[\label{Fey_He180}]{\includegraphics[scale=0.5]{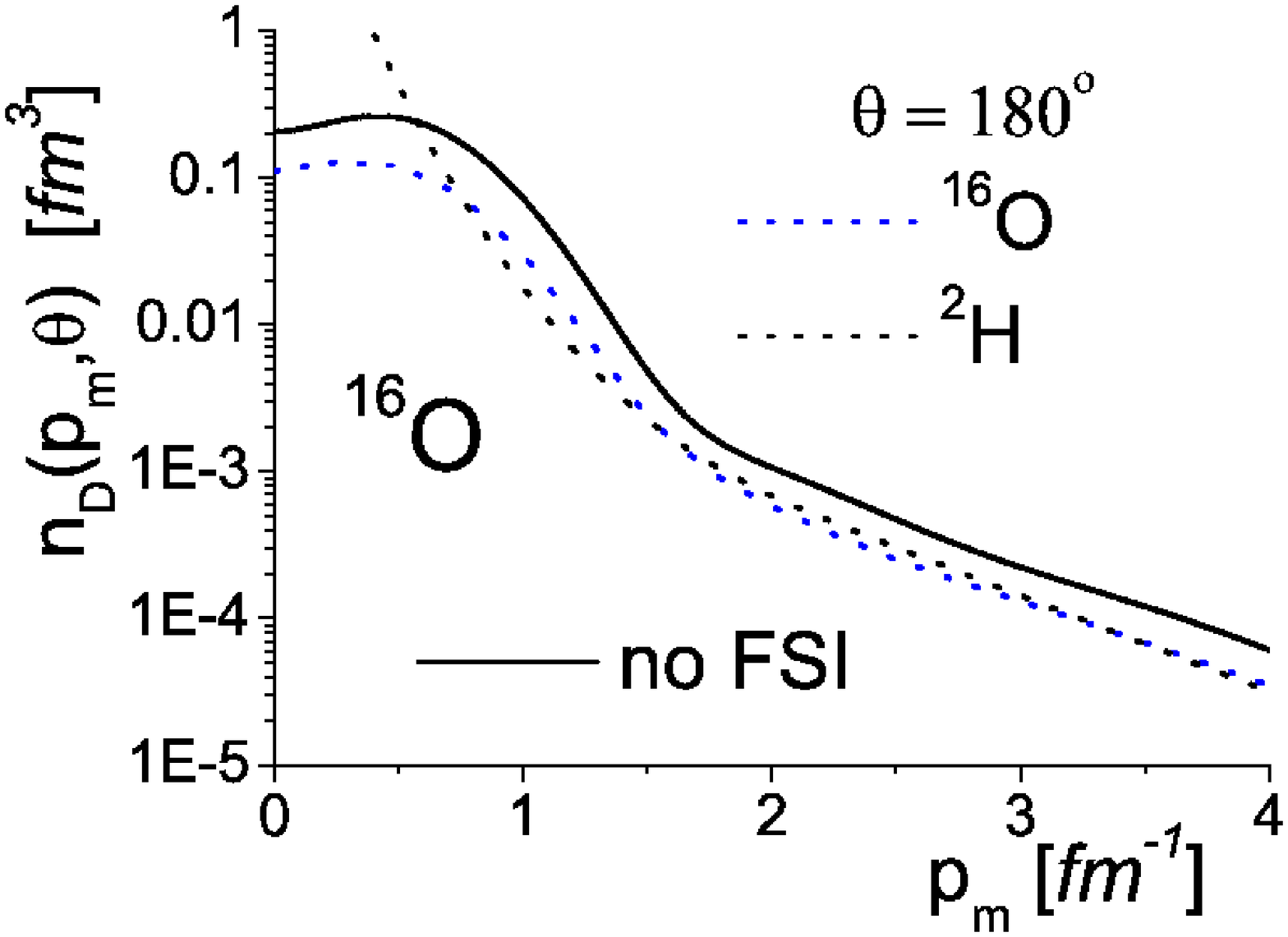}}
\caption{Dashed line: the distorted nucleon momentum distribution of $^4He$ and $^{16}O$ at $\theta=180°$; dotted line: the rescaled Deuteron momentum distribution at the same angle; full line: the undistorted momentum distribution. The same behavior appears at $\theta=0°$ and $\theta=90°$. After Ref. \cite{tutti}.}\label{Fig_He4dis}
\end{center}
\end{figure}
\beq \label{disssss}
    n_D^A(p_m,\theta)=\int d\textbf{r}\:d\textbf{r}'\:e^{\imath\:\textbf{p}_m\cdot (\textbf{r}-\textbf{r}')}\:\rho_D(\textbf{r},\textbf{r}')\: .
\eeq
The quantity $\textbf{p}_m$ appearing in Eq. (\ref{disssss}), is the \emph{missing momentum}
\beq
    \textbf{p}_m=\textbf{q}-\textbf{k}
\eeq
defined in terms of the three-momentum transfer $\textbf{q}$ and the momentum of the knocked out nucleon \textbf{k} (see $\S$\ref{sec:Hig}), with $\theta$ being the angle between them;  $\rho_D(\textbf{r},\textbf{r}')$ is the distorted one-body mixed density matrix, i.e. the quantity \cite{MCT}
\beq \label{rhoD}
    \rho_D(\textbf{r},\textbf{r}')=\frac{<\phi_0|\hat{F}^\dag\: S^\dag\: \hat{\rho}(\textbf{r},\textbf{r}')\:S\:\hat{F}|\phi_0>}{<\phi_0|\phi_0>}\: .
\eeq
In Eq. (\ref{rhoD}), $\phi_0$ is the mean field wave function,  $\Psi_A^v=\hat{F}\:\phi_0$ is the realistic correlated wave function,  $\hat{S}$ is the operator which takes into account FSI, and, eventually, $\hat{\rho}(\textbf{r},\textbf{r}')$ is the one-body density matrix operator.
When $\hat{S}=1$, PWIA is recovered, and the \emph{missing momentum} equals the nucleon momentum before interaction.
\\The distorted momentum distributions have been calculated in Ref. \cite{MCT} adopting the eikonal Glauber representation for the quantity $\hat{S}$, namely
\beq
    \hat{S}=\prod_{j=2}^A \: G(\textbf{b}_1-\textbf{b}_j,z_1-z_j)
\eeq
where
\beq
    G(\textbf{b}_1-\textbf{b}_j,z_1-z_j))1-\theta(z_j-z_1)\:\Gamma(\textbf{b}_1-\textbf{b}_j)
\eeq
with $\Gamma(\textbf{b}_1-\textbf{b}_j)$ being the usual Glauber profile function, i.e.
\beq
    \Gamma(\textbf{b}_1-\textbf{b}_j)=\sigma_{tot}^{NN}\:\frac{1}{4\pi b_0^2}\:e^{-b^2/b_0^2}\: .
\eeq
Here, $\textbf{r}=\{\textbf{b}_1,z_1\}$ is the coordinate of the struck nucleon with transverse and longitudinal coordinates  $\textbf{b}_1$ and $z_1$ (the axis $z$ is along the direction of the struck nucleon), $\sigma_{tot}^{NN}$ is the total nucleon-nucleon cross section, $b_0$ is the slope parameter of the total NN elastic cross section, and $\theta(z_j-z_1)$ ensures that the struck nucleon interacts only in the forward direction. \\Calculations similar of the ones of Ref. \cite{MCT} for $^4He$ have been performed for closed shell nuclei $^{16}O$ and $^{40}Ca$ \cite{tutti}. The results for the three nuclei are presented in Figs. \ref{Fig_nmddis} and \ref{Fig_He4dis}, where they are compared with the distorted momentum distributions of the Deuteron. It is amazing to see that even in the case of the distorted momentum distribution a sort of \emph{Deuteron scaling} is observed at high values of the \emph{missing momentum}, where
\beq
    n_D^A(p_m,\theta) \simeq \widetilde{C}^A\: n_D^{^2H}(p_m,\theta)\: .
\eeq
\section{Inclusive cross section ratios with FSI}
Let us replace in Eq. (\ref{Fyqnew}) the undistorted nucleon momentum distributions with the distorted ones; we obtain
\bey \label{Fyqdist}
  f^A_{D}(y)&=& f_{0D}(y)+f_{2D}(y_2)+f_{3D}(y_3)
\eey
where
\begin{figure}[!h]
\centerline{\includegraphics[scale=1.6]{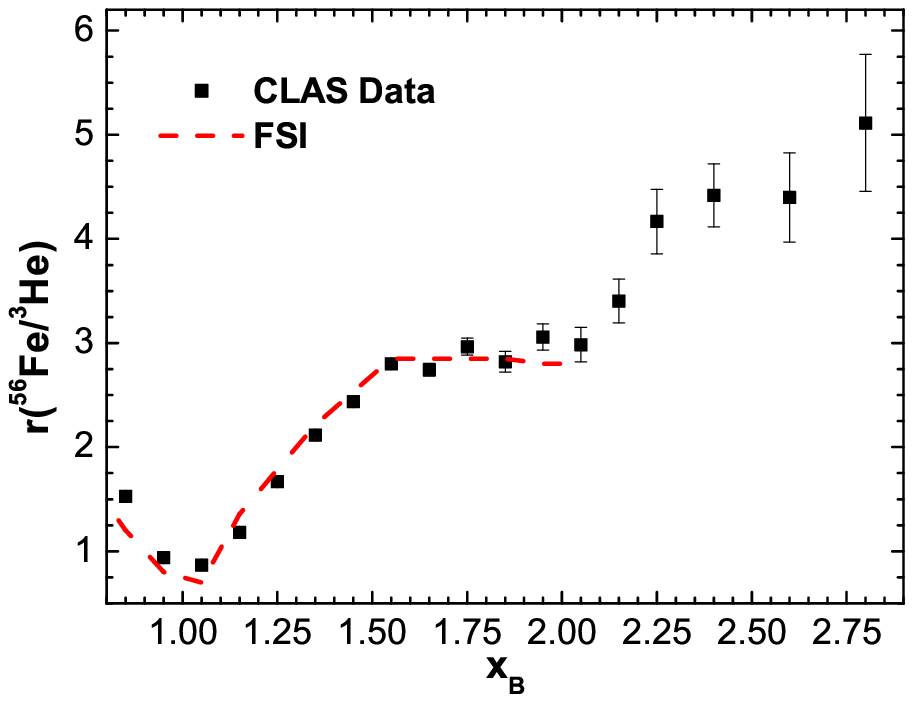}}
\caption{The same as in Fig. \ref{Fig_ratioPW}, with theoretical calculations performed with Eq. (\ref{Fyqdist}). After Ref. \cite{CC}.}
\label{Fig_ratioFSI}
\end{figure}
\bey
    \label{f0d}
    f^A_{0D}(y)= 2\pi\int_{|y|}^{+\infty} |\textbf{p}_m| \:d|\textbf{p}_m| \: \int  n_{0D}^A(|\textbf{p}_m|,\theta)\:d\theta  \\
    \label{f2d}
    f^A_{2D}(y_2)=2\pi\int_{|y_2|}^{+\infty}  |\textbf{p}_m| \:d|\textbf{p}_m| \: \int  n_{2D}^A(|\textbf{p}_m|,\theta)\:d\theta\\
    \label{f3d}
    f^A_{3D}(y_3)=\pi\int_{|y_3|}^{+\infty} |\textbf{p}_m| \:d|\textbf{p}_m| \: \int  n_{3D}^A(|\textbf{p}_m|,\theta)\:d\theta \: .
\eey
By including FSI effects, the preliminary results shown in Fig. \ref{Fig_ratioFSI} are obtained, showing a good agreement with CLAS data, both at $x_{Bj}\lesssim 1.5$ and in the 2NC region ($1.5 \lesssim x_{Bj}\lesssim 2$), where Eq. (\ref{ratio_2NC}) reads now as
\beq \label{ratioD_2NC}
    r(A,^3He)\simeq \frac{\widetilde{C}^A}{\widetilde{C}^{^3He}}=const \: .
\eeq
\section{3NC nucleon momentum distributions}
\begin{figure}[!h]
\centerline{\includegraphics[scale=1.2]{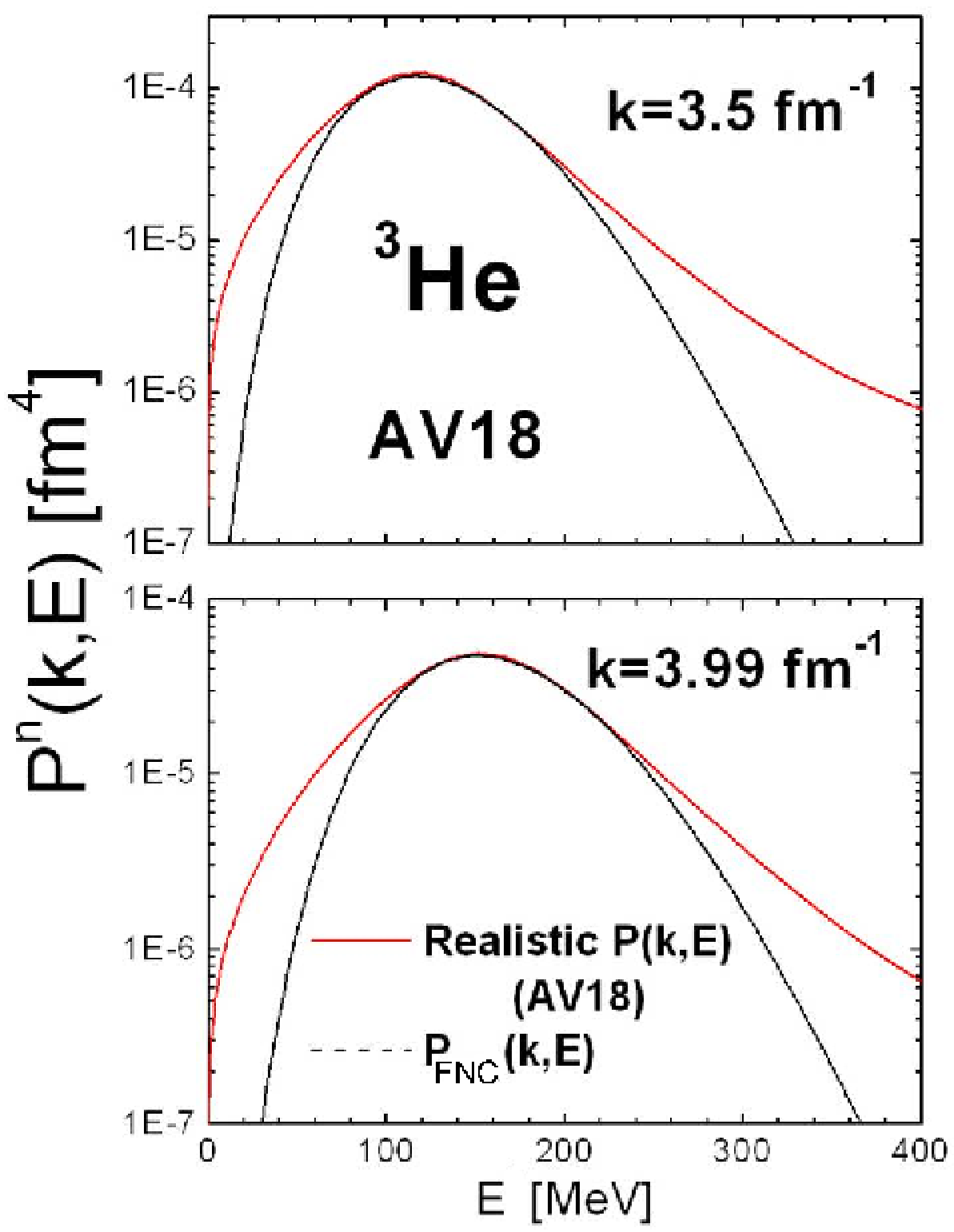}}
\caption{The neutron spectral function for $^3He$ \emph{vs.} the removal energy $E$, for two different fixed values of the momentum $k$.}
\label{Fig_neutron}
\end{figure}

\begin{figure}[!h]
\centerline{\includegraphics[scale=1.2]{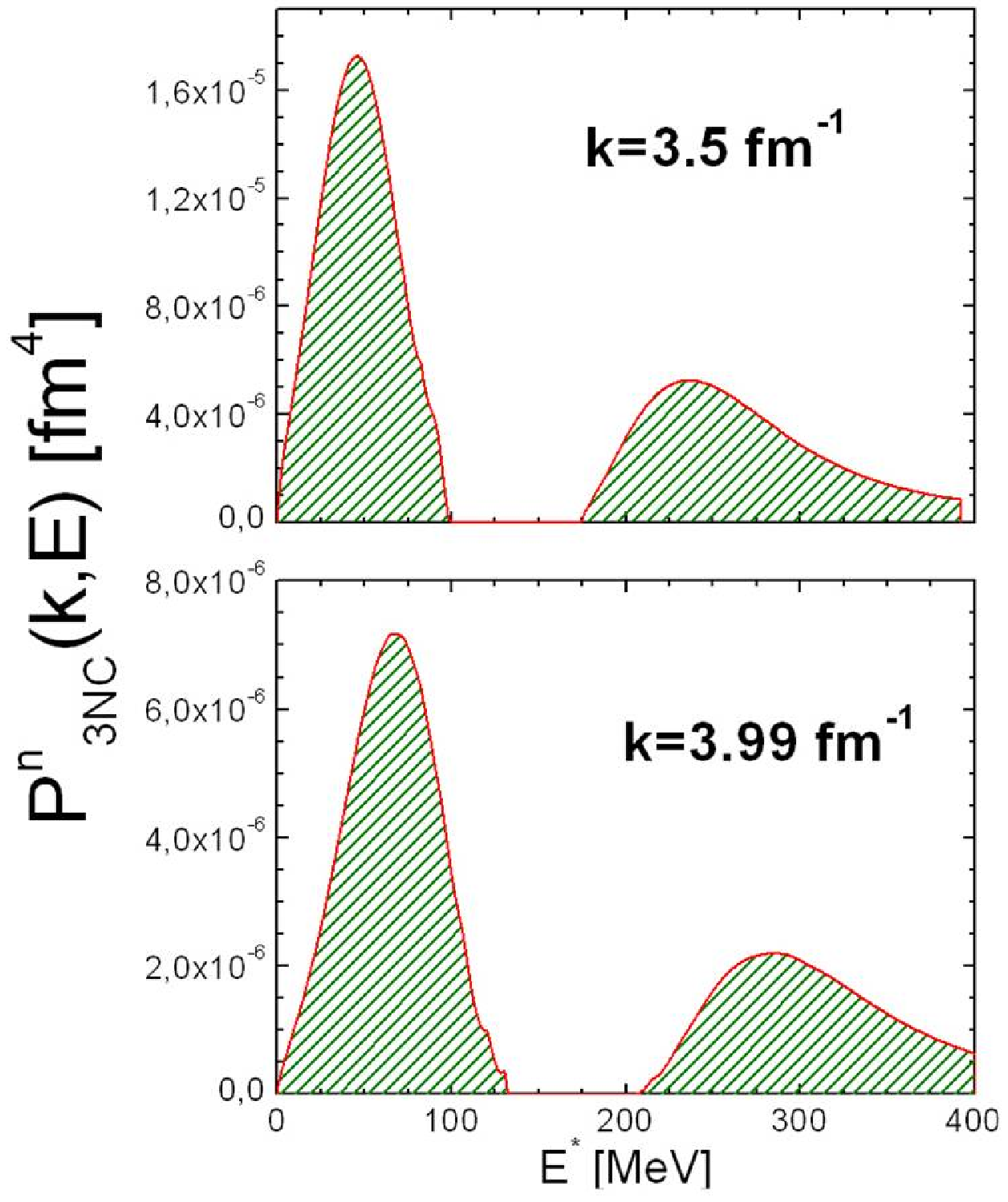}}
\caption{The 3NC contribution to the spectral function, extracted by subtracting from the exact spectral function the FNC one, both shown in Fig. \ref{Fig_neutron}, plotted versus the energy $E^*=E-E_{thr}^{(2)}$.}
\label{Fig_n3}
\end{figure}

In order to include 3NC effects, and to extend our calculations in the region of $2 \lesssim x_{Bj} \lesssim 3$, we need the 3NC nucleon momentum distributions $n_3^A(k)$, which are, to date,  completely unknown. So the problem arises of how to determine them. Herebelow the following suggestion is presented.
\\In Fig. \ref{Fig_neutron}, the exact neutron spectral function of $^3He$ \cite{CK} and the one calculated within the FNC model (\ref{PkE2NCf}) are compared. The original idea of our approach is to subtract from the exact spectral function the FNC one, obtaining, by this way, the component of the spectral function due to 3NC, i.e. $P_{3NC}(k,E)$. \\The corresponding nucleon momentum distributions are nothing but the integral of the spectral function
\beq
    n^A_3(k)=4\pi\:\int_{E_{min}}^{+\infty} P^A_{3NC}(k,E)\,dE
\eeq
so that the green area shown in Fig. \ref{Fig_n3} yields $n^A_3(k)$.
\\Calculations of 3NC effects in the region $2 \lesssim x_{Bj} \lesssim 3$ are in still in progress, and will be reported elsewhere \cite{ciocbm2}, but the results of our preliminary calculations show that
\beq
    n_2^A(k) \gg n_3^A(k)
\eeq
which is in agreements with the finding summarized in Fig.\ref{Fig_pizza}, where the independent low momentum shell model nucleons and the 2NC high momentum nucleons almost exhaust the description of nucleons in a nucleus $A$.
\\Moreover, in order to explain the second plateaux appearing in the inclusive cross section ratios (\ref{Fig_inclusive}), we should expect that $n_3^A(k)$ shows a \emph{$^3He$ scaling behavior}, i.e.
\beq
    n_3^A(k)\simeq C_3^A\: n_3^{^3He}(k)
\eeq
so that
\beq
    \frac{2\sigma_{ep}+\sigma_{en}}{Z\sigma_{ep}+N\sigma_{en}}\:\frac{F^A_{new}(y_A,q)}{F^{^3He}_{new}(y_3,q)} \simeq C_3^A = const
\eeq
in the 3NC region, at $2 \lesssim x_{Bj} \lesssim 3$. 

\cleardoublepage
\chapter*{Summary and conclusions}
\addcontentsline{toc}{chapter}{Summary and conclusions}
In the first part of this Thesis, we have illustrated the theoretical techniques used to solve the many-body problem of nuclei and, in particular, the theoretical problems encountered in finding a solution at high values of the nucleon momentum. From a theoretical point of view, we have justified an approximation based only upon  two- and three-body potentials, with higher order potentials being neglected. We have then pointed out the necessity of taking into account short range nucleon nucleon correlations (SRC) for a complete description of nuclei, and the necessity of a deep knowledge of them for answering different questions in various fields of the modern physics.
\\In the second and original part of this Thesis, we have focused on the way SRC manifest themselves in the high momentum components of one- and two-nucleon momentum distributions; moreover, we have stressed out that, due to SRC, and in particular to the dominance of $n$-$p$  pairs in a nucleus $A$, the nucleon momentum distribution for a nucleus $A$ is nothing but the Deuteron momentum distribution rescaled by a constant $C^A$, a behavior known as  \emph{Deuteron Scaling}.
\\We have then illustrated our novel approach to the study of SRC effects in inclusive lepton scattering off nuclei, based upon the introduction of the proper scaling variables  $y$, $y_2$ and $y_3$; these take effectively into account the excitation energy $E_{A-1}^*$ of the residual system in different ways, allowing us to describe the $A(e,e')X$ cross section in terms of the corresponding momentum distributions generated by 2NC and 3NC. 
\\We have shown that, in the region of 2NC, our new approach, in terms of nucleon momentum distributions, results to be a good approximation of the one based upon the use of the spectral function. Moreover, we have illustrated, that the experimental scaling function, in the 2NC region, not only scales to the Deuteron scaling function, but also exhibits $A$ independent final state interaction (FSI) effects, mostly due to the FSI in the correlated pair.
\\Calculations in the region of 3NC requires the knowledge of the nucleon momentum distributions $n_3^A(k)$, and we have proposed an approach to obtain the three-nucleon momentum distributions from the knowledge of the full and the 2NC spectral function.
\\To sum up, we have demonstrated that the usual approach to scaling, based on the variable $y$, is not very useful
as far as the investigation of the short range structure of nuclei is concerned. On the contrary, the direct, global, and $A$-independent link between the  scaling function $F^A(Q^2,y_{2})$ and the longitudinal momentum distributions $f^A(y)$ allows one to obtain information on the general behavior of the high momentum components in nuclei, which are governed by SRC.
\\Eventually, we have demonstrated that, within our novel approach, the inclusive cross section ratio $r(A/A')$ reduces to the scaling function ratio of nuclei $A$ and $A'$.  Our preliminary calculations of the scaling function ratio, performed for $A=56$ and $A'=3$, show in PWIA a good agreement with CLAS data only for $1.5 \lesssim x_{Bj}\lesssim 2$, i.e. in the region of 2NC; on the contrary, at $x_{Bj}\lesssim 1.5$, the PWIA does not lead to satisfactory results. This is not a surprising result, indeed, in the region of 2NC the data of heavy nuclei scale to the Deuteron ones, and thus FSI effects vanish in the ratio $r(A/A')$, leading to the first plateaux; in the kinematical region at $x_{Bj}\lesssim 1.5$, on the opposite, the ratio exhibits a strong sensitivity upon the $A$-dependent FSI of the nucleon  knocked out from mean field states with the residual system. Including explicitly these FSI effects in the mean field contribution, we obtained a good reproduction of the experimental plateaux attributed to 2NC.
\\Calculations including the 3NC configurations, which are necessary in order to extend  our comparison with the CLAS experimental data to the region $2 \lesssim x_{Bj} \lesssim 3$,  are in progress, and will be presented elsewhere \cite{ciocbm2}.

\cleardoublepage
%
\bibliographystyle{h-physrev3}   
\bibliography{main}         
\addcontentsline{toc}{chapter}{Bibliography}
\cleardoublepage
\pagestyle{plain}
\chapter*{Acknowledgements}
\addcontentsline{toc}{chapter}{Acknowledgements}
I started talking when I was three years old. Before then I was silent, but really curious. I just wanted to observe people and the world around me. Then I started to talk. Now, sometimes, I talk too much. But that is another story. When I started talking I continued to observe everything around me. My grandparents always reminded me: ''Your forehead was never plain!'' I was always exploring and observing, and so tables, chairs, and everything else over $80\:cm$ tall, were my worst enemies.
\\This is one of the reasons I decided to study Physics. One of the reasons I embarked on such an incredible journey.  ''\emph{To stand on the shoulders of giants}''.
\\During the first years I discovered shapes and positions, motion and objects I had never seen before. All incredible, but all in black $\&$ white. In the last three years,  I finally discovered the wonderful colors of the world around me. And that was amazing.
\\But nothing of this could ever have been possible without some special people I would like to now thank.
\\First of all, my supervisor Prof. Claudio Ciofi degli Atti.  All my gratitude for his continuous support during these years of study and research. He was always there, listening to my doubts, helping me when I got confused, talking about my ideas and proofreading this Thesis. In the last years, he also showed me some wonderful pieces of the world. But, most of all, he believed in me.
\\A special thanks to the team of Perugia, spread all over the world: Prof. Sergio Scopetta, Dr. Massimiliano Alvioli and Veronica Palli, Prof. Leonid Kaptari and Prof. Hiko Morita. They all were always ready to help me and answer my questions.
\\I am forever indebted to my family and Silvio Pietro for their understanding, endless patience, and encouragement when it was most required. I know sometimes (just sometimes!) it is not an easy task to stand beside me, especially when I'm sure I can dance on a thread by taking the world in my hands and people on my shoulders. You never left me alone and I never lacked your never-ending love.
\\A big thank you to all the people I met during this incredible journey, and a special thank you to my friend Sara. She walked among the stars while I was flying through these wonderful colors; and she shared with me these last days of craziness in writing our Ph.D. Thesis.
\\I cannot mention all the people who come to mind. But this journey could not have been so wonderful without every single person who shared with me even just one laugh or one smile. So I can only thank you all for your smiles, your support and your love. And last, but not least...for these wonderful colors: thank you all!
\\I do not know what there will be after these shapes and colors. I am not so silent, now. But I'm always really curious.

\cleardoublepage


\end{document}